\pdfoutput=1
\documentclass[12pt]{article}
\usepackage{amsmath}
\usepackage{amssymb,bm}
\usepackage{graphicx}
\usepackage{enumerate}
\usepackage{natbib}
\usepackage{url} 
\usepackage{multirow}
\usepackage{hyperref}
\usepackage{caption}
\usepackage{subcaption}
\usepackage{float}

\newcommand{\blind}{0}
\newcommand{\indep}{\rotatebox[origin=c]{90}{$\models$}}

\def\bftheta{\bm{\theta}}

\def \boldeta {{\boldsymbol{\eta}}}
\def \boldt {{\boldsymbol{t}}}

\def \bftheta {{\boldsymbol{\theta}}}
\def \bfGamma {{\boldsymbol{\Gamma}}}
\def \bfgamma {{\boldsymbol{\gamma}}}
\def \bfalpha {{\boldsymbol{\alpha}}}

\def \bfG {{\mathbf{G}}}

\begin{document}

\def\spacingset#1{\renewcommand{\baselinestretch}%
{#1}\small\normalsize} \spacingset{1}


\if0\blind
{
  \title{\bf Bayesian Weighted Mendelian Randomization for Causal Inference based on Summary Statistics}
  \author{Jia Zhao\\
    School of Mathematical Sciences, Beijing Normal University\\
    Department of Mathematics,\\
    The Hong Kong University of Science and Technology\\
    Jingsi Ming\\
    Department of Mathematics,\\
    The Hong Kong University of Science and Technology\\
    Xianghong Hu\\
    Department of Mathematics, Hong Kong Baptist University\\
    Department of Mathematics,\\ 
    Southern University of Science and Technology\\
    Gang Chen\\
    The WeGene Company\\
    Jin Liu\\
    Centre for Quantitative Medicine, Duke-NUS Medical School\\
    Can Yang\\
    Department of Mathematics,\\
    The Hong Kong University of Science and Technology\\
    }
  \maketitle
} \fi

\if1\blind
{
  \bigskip
  \bigskip
  \bigskip
  \begin{center}
    {\LARGE\bf Title}
\end{center}
  \medskip
} \fi

\begin{abstract}
The results from Genome-Wide Association Studies (GWAS) on thousands of phenotypes provide an unprecedented opportunity to infer the causal effect of one phenotype (exposure) on another (outcome). Mendelian randomization (MR), an instrumental variable (IV) method, has been introduced for causal inference using GWAS data. Due to the polygenic architecture of complex traits/diseases and the ubiquity of pleiotropy, however, MR has many unique challenges compared to conventional IV methods. We propose a Bayesian weighted Mendelian randomization (BWMR) for causal inference to address these challenges. In our BWMR model, the uncertainty of weak effects owing to polygenicity has been taken into account and the violation of IV assumption due to pleiotropy has been addressed through outlier detection by Bayesian weighting. To make the causal inference based on BWMR computationally stable and efficient, we developed a variational expectation-maximization (VEM) algorithm. Moreover, we have also derived an exact closed-form formula to correct the posterior covariance which is often underestimated in variational inference. Through comprehensive simulation studies, we evaluated the performance of BWMR, demonstrating the advantage of BWMR over its competitors. Then we applied BWMR to make causal inference between 130 metabolites and 93 complex human traits, uncovering novel causal relationship between exposure and outcome traits. The BWMR software is available at \url{https://github.com/jiazhao97/BWMR}
\end{abstract}

\noindent%
\vfill

\newpage
\spacingset{1.45} 
\section{Introduction}
\label{sec:intro}
Determination of the causal effect of a risk factor (exposure) on a complex trait or disease (outcome) is critical for health management and medical intervention. Random controlled trial (RCT) is often considered as the golden standard for causal inference. When the evidence from RCT is lacking, Mendelian Randomization (MR) \citep{Katan1986,Evans2015} was proposed to mimic RCT using natural genetic variations for causal inference. The idea is that the genotypes are randomly assigned from one generation to next generation according to Mendelian Laws of Inheritance. Therefore, genotypes which should be unrelated to confounding factors can serve as instrumental variables (IVs) \citep{Lawlor2008,Baiocchi2014}, helping to eliminate the possibility of reverse causality.

In recent years, MR becomes more and more popular because Genome-Wide Association Studies (GWAS) have been performed on thousands of phenotypes. In particular, summary statistics from GWAS are available through public gateways (e.g., {\url{https://www.ebi.ac.uk/gwas/downloads/summary-statistics}}). These data sets contain very rich information, such as reference allele frequency of single nucleotide polymorphisms (SNPs), the effect size of a SNP on the phenotype and its standard error, such that MR can be performed without accessing individual-level GWAS data. 

From the statistical point of view, MR can be viewed as an instrumental variable method. However, due to the complexity of human genetics (e.g., polygenicity \citep{Visscher2017}, pleiotropy \citep{Solovieff2013,Yang2015} and linkage disequilibrium (LD) in human genome \citep{Consortium2012} and complicated data processing and sharing (e.g., sample overlap in multiple GWAS), MR has several unique challenges compared with conventional IV methods. First, in the presence of polygenic architecture, there exist many weak SNP-exposure effects rather than strong effects only. The uncertainty of estimated weak effects needs to be taken into account. Second, based on current GWAS results, it is often observed that a SNP can affect both exposure and outcome traits. This phenomenon is referred as ``pleiotropy'' or ``horizontal pleiotropy''. The ubiquity of horizontal pleiotropy easily makes the assumption in classical IV methods invalid, resulting in many false positives if horizontal pleiotropy is not taken into account. Third, a number of potential risks may be involved in summary statistics-based MR, such as, different LD patterns in exposure and outcome traits, selection bias of SNP-exposure effects  and other bias due to overlapped samples.

To address these challenges, much efforts have been devoted in MR recently. To name a few, PRESSO has been proposed to handle horizontal pleiotropy \citep{Verbanck2018}. A sampling strategy is used in PRESSO to detect outliers due to horizontal pleiotropy and then inverse variance weighted (IVW) regression \citep{Burgess2013} is applied to estimate causal effects. However, estimation uncertainty of SNP-exposure effects is ignored in the PRESSO model. It is also not computationally efficient owing to the sampling strategy. By assuming that horizontal pleiotropic effects are unknown constant, Egger \citep{Bowden2015} extends the IVW method by introducing an additional intercept term. Despite this improvement over IVW, Egger suffers from high estimation error in presence of many weak horizontal pleiotropic effects as its assumption fails in this situation. Moreover, Egger would be biased when there exist a few large pleiotropic effects (i.e., outliers). GSMR \citep{Zhu2018} improves existing methods by introducing outlier detection to remove the influence of large pleiotropic effects and take the linkage disequilibrium between IVs into account. However, GSMR adopts an ad-hoc outlier detection procedure and it does not take weak pleiotropic effects into consideration. Therefore, its type I error rate can be inflated in real data analysis. RAPS is a newly developed method \citep{Zhao2018}, aiming to improve statistical power for causal inference by including weak effects in GWAS and removing outliers due to horizontal pleiotropy. Although the theoretical property of RAPS has been established, its accompanying algorithm is numerically unstable, resulting in unreliable estimated causal effects (see our experimental results in supplementary Figs. 43-45). 

In this paper, we propose a method named `Bayesian Weighted Mendelian Randomization (BMWR)' for causal inference using summary statistics from GWAS. In BWMR, uncertainty of estimated weak effects in GWAS and influence of horizontal pleiotropy have been addressed in a unified statistical framework. To make BWMR computationally efficient and stable, we developed a variational expectation-maximization algorithm to infer the posterior mean of the causal effects. Importantly, we further derived an exact closed-form formula to correct the posterior covariance which was often underestimated in variational inference. Through comprehensive simulation studies, we showed that BMWR was computationally stable and statistically efficient, compared to existing related methods. Then we applied BWMR to real data analysis, further demonstrating the statistical properties of BWMR and 
revealing novel causal relationships between exposure and outcome traits.

\section{Methods}
\label{sec:meth}
\subsection{Model}
We primarily focus on estimating the causal effect of an exposure $X$ on an outcome $Y$ with unknown confounding factors $U$ in the framework of MR. To perform the standard MR, an instrumental variable method for causal inference, the following three criteria ensuring the validity of IVs should be satisfied:
\begin{itemize}
  \item Relevance criterion: IVs are associated with the exposure $X$.
  \item Effective random assignment criterion: IVs are independent of confounders $U$.
  \item Exclusion restriction criterion: IVs only affect the outcome $Y$ through the exposure $X$.
\end{itemize}
Assuming the validity of IVs, we begin with the following linear model for MR:
\begin{equation}
X = \sum\limits_{j=1}^N \gamma_jG_j+\eta_X U+\epsilon_X, Y = \beta X+\eta_Y U+\epsilon_Y,
\label{eq: linear-model-1}
\end{equation}
where $G_j\,(j=1,2,...,N)$ are independent SNPs serving as IVs, $\gamma_j\,(j=1,2,...,N)$ are SNP-exposure effects, $\eta_X$ and $\eta_Y$ are effects of confounders $U$ on exposure $X$ and outcome $Y$, $\epsilon_X$ and $\epsilon_Y$ are independent error terms, i.e., $(\epsilon_X, \epsilon_Y)\indep(G_1,G_2,...,G_N,U)$, $\epsilon_X\indep\epsilon_Y$, and $\beta$ denotes the causal effect of interest. Eq. (\ref{eq: linear-model-1}) implies
\begin{equation}\label{eq: linear-model-2}
Y=\sum^N_{j=1}\beta \gamma_j G_j + (\beta \eta_x + \eta_y)U + \beta\epsilon_x + \epsilon_y.
\end{equation}
Let $\Gamma_j$ be the effect of $G_j$ on outcome $Y$. From Eq. (\ref{eq: linear-model-2}), we have
\begin{equation}
\Gamma_j=\beta\gamma_j,\,j=1,2,...,N.
\label{eq: original-model}
\end{equation}
This equation implies that we can estimate $\beta$ based on $\Gamma_j$ and $\gamma_j$. 

In practice, we do not know $\Gamma_j$ and $\gamma_j$, but have their estimates from summary statistics of GWAS. Let $\{\hat{\gamma}_j, {\sigma}_{X_j}, p_{X_j}\}$ and $\{\hat{\Gamma}_j, {\sigma}_{Y_j}, p_{Y_j}\}$ be estimated effect sizes of $G_j$, standard errors and $p$-values for both exposure $X$ and outcome $Y$, respectively. Because of the large sample size of GWAS, we ignore the uncertainty in estimating ${\sigma}_{X_j},{\sigma}_{Y_j}$. To ensure the relevance criterion, we first select SNPs which are significantly associated with exposure $X$ (e.g., ${p}_{X_j} \leq 5\times 10^{-8}$). We further apply LD clumping to those selected SNPs to ensure the independence of IVs \citep{Purcell2007}. The remaining summary statistics are denoted as $D=\{\boldsymbol{\hat{\gamma}},\boldsymbol{\hat{\Gamma}},\boldsymbol{\sigma}_X,\boldsymbol{\sigma}_Y\}$ serving as the input data of our model, where $\boldsymbol{\hat{\gamma}}=(\hat{\gamma}_1,...,\hat{\gamma}_N),\hat{\boldsymbol{\Gamma}}=(\hat{\Gamma}_1,...,\hat{\Gamma}_N),\boldsymbol{\sigma}_X=(\sigma_{X_1},...,\sigma_{X_N}),\boldsymbol{\sigma}_Y=(\sigma_{Y_1},...,\sigma_{Y_N})$. Note that the relationship between estimated effect sizes and the corresponding true effect sizes is given in the following probabilistic model:
\begin{equation}
\hat{\gamma}_j|\gamma_j \sim \mathcal{N}(\gamma_j, \sigma_{X_j}^2),\,
\hat{\Gamma}_j|\Gamma_j \sim \mathcal{N}(\Gamma_j, \sigma_{Y_j}^2),\,
j=1,2,...,N.
\label{eq: GWAS-stat}
\end{equation}
According to Mendel's Law, it is also reasonable to assume that effective random assignment criterion is satisfied in real-data application. However, the exclusion restriction criterion does not often hold because of the ubiquity of horizontal pleitropy (i.e., IVs may have direct effects on outcome $Y$).

To address this challenge, we notice that Eq. (\ref{eq: original-model}) can be relaxed as 
\begin{equation}
\Gamma_j = \beta\gamma_j+\alpha_j,\,j=1,2,...,N,
\label{eq: MR core}
\end{equation}
by assuming that 
\begin{equation}
\alpha_j\overset{\text{i.i.d.}}{\sim}\mathcal{N}(0,\tau^2),\,
(\alpha_1,\alpha_2,..,\alpha_N)\indep(\gamma_1,\gamma_2,...,\gamma_N),
\label{eq: alpha-gamma}
\end{equation}
where $\tau^2$ is an unknown parameter to be estimated from data.
In such a way, estimating $\beta$ from Eq. (\ref{eq: MR core})
can be viewed as a noisy version of estimation from Eq. (\ref{eq: original-model}), where $\alpha_1,\dots,\alpha_N$ are viewed with independent random noise. In fact, the linear model corresponding to Eq. (\ref{eq: MR core}) is 
\begin{align}\label{eq: MR-noise}
X = \sum\limits_{j=1}^N \gamma_jG_j+\eta_X U+\epsilon_X, Y = \beta X+\sum_{j=1}^N \alpha_jG_j+\eta_Y U+\epsilon_Y.
\end{align}
The presence of term $\alpha_j G_j$ indicates that the exclusion restriction criterion can be relaxed as long as the direct effect $\alpha_j$ satisfies (\ref{eq: alpha-gamma}). The nonzero variance $\tau^2$ naturally accounts for the influence of weak pleiotropic effects $\alpha_j$, $j=1,\dots,N$.

Combining Eqs. (\ref{eq: GWAS-stat},\ref{eq: MR core},\ref{eq: alpha-gamma}), we have 
\begin{equation}
\hat{\gamma}_j|\gamma_j \sim \mathcal{N}(\gamma_j, \sigma_{X_j}^2),
\hat{\Gamma}_j|\beta,\gamma_j \sim \mathcal{N}(\beta\gamma_j, \sigma_{Y_j}^2+\tau^2).
\label{eq: bwmr-1}
\end{equation}
If we treat $\gamma_j\, (j=1,\dots,N)$ as model parameters, then the number of model parameters is going to increase as the number of samples $\{\hat{\gamma_j},\hat{\Gamma}_j\}$ increases. Therefore, we assign a prior distribution on $\gamma_j$,
\begin{equation}
\gamma_j\overset{\text{i.i.d.}}{\sim}\mathcal{N}(0,\sigma^2),
\label{eq: bwmr-2}
\end{equation}
with an unknown parameter $\sigma^2$ to be estimated from data. To facilitate statistical inference in Bayesian framework, we also assign a non-informative prior on $\beta$
\begin{equation}
\beta\sim\mathcal{N}(0,\sigma_0^2).
\label{eq: bwmr-3}
\end{equation}
When $\sigma_0\rightarrow\infty$, the result from Bayesian inference on $\beta$ will naturally converge to the inference result based on maximum likelihood estimation. Let $\boldsymbol{\gamma}$ be the collection of $\{\gamma_1,\dots,\gamma_N\}$. Combining Eq. (\ref{eq: bwmr-1}) with Eq. (\ref{eq: bwmr-2}) and Eq. (\ref{eq: bwmr-3}), our probabilistic model becomes
\begin{equation}
\begin{aligned}\label{bwmr-temp}
& p(\hat{\boldsymbol\gamma}, \hat{\boldsymbol\Gamma},\beta, {\boldsymbol{\gamma}} | {\boldsymbol{\sigma}}_X^2, {\boldsymbol{\sigma}}_Y^2,  \tau^2,\sigma^2, \sigma_0^2) \\
= &  
\mathcal{N}(\beta|0,\sigma_0^2)
\prod\limits_{j=1}^N \mathcal{N}(\gamma_j|0,\sigma^2) \\
& \prod\limits_{j=1}^N \mathcal{N}(\hat{\gamma}_j|\gamma_j,\sigma_{X_j}^2)
\prod\limits_{j=1}^N \mathcal{N}(\hat{\Gamma}_j|\beta\gamma_j,\sigma_{Y_j}^2+\tau^2).
\end{aligned}
\end{equation}

In the presence of strong horizontal pleiotropy, the direct effect $\alpha_j$ in Eq. (\ref{eq: MR-noise}) may become extremely large and thus it does not satisfy Eq. (\ref{eq: alpha-gamma}). To ensure that our model works well in this case, we cast those $\alpha_j$ as outliers. As inspired by reweighed probabilistic models \citep{Wang2016}, here we propose a strategy to guarantee the robustness of the above model. Let $\boldsymbol{w}=[w_1,\dots,w_N]$, where $w_j\in\{0,1\}$ is the weight corresponding to the $j$-th observation. We would like to assign $w_j=0$ if $\alpha_j$  deviates from model (\ref{bwmr-temp}), and $w_j = 1$ otherwise. With this consideration, we assume 
\begin{align*}
w_j|\pi_1 \overset{\text{i.i.d.}}{\sim} \mbox{Bernoulli}(\pi_1), \pi_1\sim\mbox{Beta}(a_0,1),
\end{align*}
where $a_0$ is specified as 100 throughout this paper, preferring there is a small proportion of outliers.
Then we reformulate model (\ref{bwmr-temp}) as 
\begin{equation}
\begin{aligned}
& p(\hat{\boldsymbol\gamma}, \hat{\boldsymbol\Gamma}, \beta, \pi_1, {\boldsymbol{\gamma}}, \boldsymbol{w} | {\boldsymbol{\sigma}}_X^2, {\boldsymbol{\sigma}}_Y^2, \tau^2,\sigma^2, \sigma_0^2, a_0) \\
= &  
\frac{1}{A}\mathcal{N}(\beta|0,\sigma_0^2) \Pr(\pi_1|a_0)
\prod\limits_{j=1}^N \mathcal{N}(\gamma_j|0,\sigma^2)
\prod\limits_{j=1}^N \Pr(w_j|\pi_1)
\\
& \prod\limits_{j=1}^N \mathcal{N}(\hat{\gamma}_j|\gamma_j,\sigma_{X_j}^2)
\prod\limits_{j=1}^N \mathcal{N}(\hat{\Gamma}_j|\beta\gamma_j,\sigma_{Y_j}^2+\tau^2)^{w_j}.
\end{aligned}
\label{eq: BWMR-l}
\end{equation}
where $A$ is the normalization constant to ensure (\ref{eq: BWMR-l}) is a valid probability model.

To sum up, we refer to our model (\ref{eq: BWMR-l}) as Bayesian Weighted Mendelian Randomization (BWMR). BWMR takes summary statistics $D=\{\hat{\boldsymbol{\gamma}},\hat{\boldsymbol{\Gamma}},\boldsymbol{\sigma}_X,\boldsymbol{\sigma}_Y\}$ as its input, aiming to provide the posterior mean and variance on causal effect $\beta$, in which $\mathbf{z}=\{\beta,\pi_1,\boldsymbol{w},\boldsymbol{\gamma}\}$ is the collection of latent variables, $\boldsymbol{\theta}=\{\tau^2,\sigma^2\}$ is the set of the model parameters to be optimized, and $\boldsymbol{h}_0=\{\sigma_0=1\times 10^6,a_0=100\}$ is the collection of fixed hyper-parameters.

\subsection{Algorithm}

To provide accurate statistical inference, we are interested in posterior distribution of the latent variables:
\begin{align*}
p(\mathbf{z}|D,\boldsymbol{\theta},\boldsymbol{h}_0)=\frac{p(\hat{\boldsymbol\gamma}, {\hat{\boldsymbol{\Gamma}}}, \mathbf{z}| {\boldsymbol{\sigma}}_X^2, {\boldsymbol{\sigma}}_Y^2, \boldsymbol{\theta}, \boldsymbol{h}_0)}{p(\hat{\boldsymbol\gamma}, {\hat{\boldsymbol{\Gamma}}} | {\boldsymbol{\sigma}}_X^2, {\boldsymbol{\sigma}}_Y^2, \boldsymbol{\theta}, \boldsymbol{h}_0)},
\end{align*}
where 
\begin{align*}
p(\hat{\boldsymbol\gamma}, {\hat{\boldsymbol{\Gamma}}} | {\boldsymbol{\sigma}}_X^2, {\boldsymbol{\sigma}}_Y^2, \boldsymbol{\theta}, \boldsymbol{h}_0) 
= \int_{\mathbf{z}}p(\hat{\boldsymbol\gamma}, {\hat{\boldsymbol{\Gamma}}}, \mathbf{z}| {\boldsymbol{\sigma}}_X^2, {\boldsymbol{\sigma}}_Y^2, \boldsymbol{\theta}, \boldsymbol{h}_0)\mbox{d}\mathbf{z}.
\end{align*}
However, exact evaluation of the posterior distribution is very challenging because the integration is intractable.

Instead, we propose a variational expectation-maximization (VEM) algorithm to approximate the posterior. Let $q(\mathbf{z})$ be a variational distribution. The logarithm of the marginal likelihood can be written as
  \begin{align*}
  & \log p(\hat{\boldsymbol\gamma}, {\hat{\boldsymbol{\Gamma}}} | {\boldsymbol{\sigma}}_X^2, {\boldsymbol{\sigma}}_Y^2, \boldsymbol{\theta}, \boldsymbol{h}_0)\\
  & = \mathbb{E}_{q(\mathbf{z})}[\log p(\hat{\boldsymbol\gamma}, {\hat{\boldsymbol{\Gamma}}} | {\boldsymbol{\sigma}}_X^2, {\boldsymbol{\sigma}}_Y^2, \boldsymbol{\theta}, \boldsymbol{h}_0)]\\
  & = \mathbb{E}_{q(\mathbf{z})}\left[\log\frac{p(\hat{\boldsymbol\gamma}, {\hat{\boldsymbol{\Gamma}}}, \mathbf{z}| {\boldsymbol{\sigma}}_X^2, {\boldsymbol{\sigma}}_Y^2, \boldsymbol{\theta}, \boldsymbol{h}_0)}{q(\mathbf{z})}
  + \log\frac{q(\mathbf{z})}{p(\mathbf{z}|D,\boldsymbol{\theta},\boldsymbol{h}_0)}\right]\\
  &= \mathcal{L}(q,\boldsymbol{\theta})+KL(q(\mathbf{z})\|p(\mathbf{z}|D,\boldsymbol{\theta},\boldsymbol{h}_0)),\\
  \end{align*}
  where 
  \begin{align*}
  & \mathcal{L}(q;\boldsymbol{\theta})
  :=\mathbb{E}_{q(\mathbf{z})}\left[\log\frac{p(\hat{\boldsymbol\gamma}, {\hat{\boldsymbol{\Gamma}}}, \mathbf{z}| {\boldsymbol{\sigma}}_X^2, {\boldsymbol{\sigma}}_Y^2, \boldsymbol{\theta}, \boldsymbol{h}_0)}{q(\mathbf{z})}\right],\\
  & KL(q(\mathbf{z})\|p(\mathbf{z}|D,\boldsymbol{\theta},\boldsymbol{h}_0))
  = \mathbb{E}_{q(\mathbf{z})}\left[\log\frac{q(\mathbf{z})}{p(\mathbf{z}|D,\boldsymbol{\theta},\boldsymbol{h}_0)}\right].
  \end{align*}
Given that the Kullback-Leibler (KL) divergence $KL(\cdot\|\cdot)$ is non-negative, $\mathcal{L}(q;\boldsymbol{\theta})$ is an evidence lower bound (ELBO) of marginal likelihood. Thus, maximization of ELBO $\mathcal{L}(q;\boldsymbol{\theta})$ w.r.t. variational distribution $q$ and parameters $\boldsymbol{\theta}$ is commonly referred to as E-step and M-step: In the E-step, variational distribution $q$ is updated to approximate the true posterior distribution. In the M-step, the set of model parameters $\bftheta$ is optimized to increase EBLO and thus increase the marginal likelihood.

We adopt mean-field variational Bayes (MFVB) and assume that $q(\mathbf{z};\boldsymbol{\eta})$ can be factorized as
\begin{equation*}
q(\mathbf{z};\boldsymbol{\eta}) = q(\beta;\boldsymbol{\eta})q(\pi_1;\boldsymbol{\eta})\prod\limits_{j=1}^N q(\gamma_j;\boldsymbol{\eta})\prod\limits_{j=1}^N q(w_j;\boldsymbol{\eta}).
\label{eq: MF assumption}
\end{equation*}
where $\boldsymbol{\eta}$ collects all the variational parameters.
Without further assuming any specific forms of the variational distributions, the optimal variational distributions in the E-step can be obtained naturally as
\begin{equation*}
\begin{aligned}
& q(\beta|\mu_\beta, \sigma^2_{\beta})=\mathcal{N}(\beta | \mu_{\beta}, \sigma_{\beta}^2), q(\boldsymbol{\gamma}|\{\mu_{\gamma_j}, \sigma_{\gamma_j}^2\})=\prod\limits_{j=1}^N\mathcal{N}(\gamma_j | \mu_{\gamma_j}, \sigma_{\gamma_j}^2),\\
& q(\boldsymbol{w}|\pi_{w_j})=\mbox{Bernoulli}(w_j|\{\pi_{w_j}\}),q(\pi_1|a,b)=\mbox{Beta}(\pi_1| a, b),
\end{aligned}
\label{eq: MF-variational dist}
\end{equation*}
with the updating equations
$$\begin{aligned}
\sigma_{\beta}^2 = &\left[ \frac{1}{\sigma_0^2} + \sum\limits_{j=1}^N\frac{\pi_{w_j}(\mu_{\gamma_j}^2+\sigma_{\gamma_j}^2)}{\sigma_{Y_j}^2+\tau^2} \right]^{-1}\\
\mu_{\beta} =& \left( \sum\limits_{j=1}^N \frac{\pi_{w_j}\mu_{\gamma_j}\hat{\Gamma}_j}{\sigma_{Y_j}^2+\tau^2} \right)\sigma_{\beta}^2,\\
\sigma_{\gamma_j}^2 =& \left[ \frac{1}{\sigma_{X_j}^2} + \frac{\pi_{w_j}(\mu_{\beta}^2+\sigma_{\beta}^2)}{\sigma_{Y_j}^2+\tau^2} + \frac{1}{\sigma^2} \right]^{-1},\\
\mu_{\gamma_j} =& \left( \frac{\hat{\gamma}_j}{\sigma_{X_j}^2} + \frac{\pi_{w_j}\mu_{\beta}\hat{\Gamma}_j}{\sigma_{Y_j}^2+\tau^2} \right)\sigma_{\gamma_j}^2,\\
a=&a_0+\sum\limits_{j=1}^N \pi_{w_j},
b = N+1-\sum\limits_{j=1}^N \pi_{w_j},
\pi_{w_j} = \frac{q_{j1}}{q_{j0}+q_{j1}},\\
q_{j0} =& \exp[\psi(b) - \psi(a+b)],\\
q_{j1} = & \exp\left[
    -\frac{\log(2\pi)}{2} - \frac{\log(\sigma_{Y_j}^2+\tau^2)}{2} + \psi(a) - \psi(a+b) \right] \\
    & +\exp\left[- \frac{(\mu_{\beta}^2+\sigma_{\beta}^2)(\mu_{\gamma_j}^2+\sigma_{\gamma_j}^2)-2\mu_{\beta}\mu_{\gamma_j}\hat{\Gamma}_j+\hat{\Gamma}_j^2}{2(\sigma_{Y_j}^2+\tau^2)}\right],
\end{aligned}$$
where variational parameters are collected in $$\boldsymbol{\eta}=\{\mu_\beta,\sigma^2_\beta; a, b; \{\mu_{\gamma_j},\sigma^2_{\gamma_j}\}^N_{j=1}; \{\pi_{w_j}\}^N_{j=1}\},$$ and $\psi(\cdot)$ represents the digamma function. In the M-step, by setting the derivative of ELBO w.r.t. $\sigma^2$ to zero, the updating equation for $\sigma^2$ can be easily obtained as
$$\sigma^2 = \sum\limits_{j=1}^N (\mu_{\gamma_j}^2+\sigma_{\gamma_j}^2)/N.$$
Derivation of the updating equation for $\tau^2$ is much more technical. We have made use of a number of tricks in convex optimization and obtain a closed-form updating equation for $\tau^2$ as
{\scriptsize{
\begin{align*}
\tau^2 = \sqrt{\dfrac{\sum\limits_{j=1}^N \dfrac{\pi_{w_j}[(\mu_{\beta}^2+\sigma_{\beta}^2)(\mu_{\gamma_j}^2+\sigma_{\gamma_j}^2)-2\mu_{\beta}\mu_{\gamma_j}\hat{\Gamma}_j+\hat{\Gamma}_j^2](\tau^{(\mathrm{old}))})^4}{[\sigma_{Y_j}^2+(\tau^{(\mathrm{old}))})^2]^2}}
{\sum\limits_{j=1}^N \frac{\pi_{w_j}}{\sigma_{Y_j}^2 + (\tau^{(\mathrm{old})})^2}}},
\end{align*}}}
where $(\tau^{(\mathrm{old}))})^2$ is the estimate of $\tau^2$ at previous iteration.  Importantly, all updating approaches in VEM algorithm have closed-forms, ensuring the efficiency and stability of our proposed algorithm. The details of the above derivation are given in the supplementary document. After convergence of the VEM algorithm, we can obtain the set of optimized variational parameters $\boldsymbol{\eta}^*$ and the set of estimated model parameters $\hat{\boldsymbol{\theta}}$. The posterior mean of $\beta$ has been naturally given by $\mu^*_\beta$, which is very accurate as shown later. However, the posterior variance of $\beta$ given by MFVB, i.e., $(\sigma^2_\beta)^*$, often underestimates the true posterior variance. We address this issue in next section.

\subsection{Inference}

As inspired by the linear response methods \citep{Giordano2015,Giordano2018}, we propose a closed-from formula to correct the underestimated posterior variance, yielding an accurate inference for $\beta$. 

For notation convenience, we denote $p_0(\mathbf{z})$ as the true posterior of interest. We define a perturbation of the true posterior as
\begin{equation*}
p_t(\mathbf{z}) := p_0(\mathbf{z})\exp(t\beta-C(t)),
\label{eq: perturbation}
\end{equation*}
where $t$ is a scalar and $C(t)$ normalizes $p_t(\mathbf{z})$. When $t=0$, $p_t(\mathbf{z})$ becomes the unperturbed posterior $p_0(\mathbf{z})$. Clearly, $p_t(\mathbf{z})$ forms an exponential family parameterized by $t$. By the property of the cumulant generating function of exponential families, we have
\begin{align*}
\mathbb{E}_{p_t}[\beta]=\frac{\mbox{d}C(t)}{\mbox{d}t},\,
\mbox{Var}_{p_t}[\beta]=\frac{\mbox{d}^2C(t)}{\mbox{d}t^2},
\end{align*}
implying that the posterior variance of $\beta$ can be written as
\begin{align}\label{TruePosteriorVar}
\mbox{Var}_{p_0}[\beta]
=\left.\frac{\mbox{d}\mathbb{E}_{p_t}[\beta]}{\mbox{d}t}\right|_{t=0}.
\end{align}
This implies that the sensitivity of posterior mean at $t=0$ can provide information about the true posterior variance.

Now we introduce the key idea to approximate the posterior variance in Eq. (\ref{TruePosteriorVar}) from MFVB. Let $q_t(\mathbf{z})$ be the mean-field approximation to $p_t(\mathbf{z})$, i.e., $q_t(\mathbf{z}):= q(\mathbf{z};\boldsymbol{\eta}^*_t)=\arg \min_{q} KL(q(\mathbf{z};\boldsymbol{\eta})||p_t(\mathbf{z}))$. Since MFVB often provides accurate inference on posterior mean \citep{Blei2017}, we assume the following conditions hold:
\begin{equation}\label{Condition}
\begin{aligned}
&\mbox{Condition 1: }\mathbb{E}_{q_t}[\beta] \approx \mathbb{E}_{p_t}[\beta] \mbox{ for all } t, \mbox{ and }\\
&\mbox{Condition 2: }\frac{d \mathbb{E}_{q_t}[\beta]}{dt}\Big|_{t=0} \approx \frac{\mathbb{E}_{p_t}[\beta]}{dt}\Big|_{t=0}.
\end{aligned}
\end{equation}
Condition 1 says that MFVB can provide good approximations to posterior means of all the perturbations. Condition 2 further requires the accuracy of the first order approximation at $t=0$. As we shall show in the supplementary document, the following equation holds:
\begin{equation}\label{MFVB_approx}
\frac{d \mathbb{E}_{q_t}[\beta]}{dt}\Big|_{t=0} = -\mathbf{g}^T\mathbf{H}^{-1} \mathbf{g},
\end{equation}
where $$\mathbf{g}=\left.\frac{\mbox{d}\mathbb{E}_{q(\mathbf{z};\boldsymbol{\eta})}[\beta]}{\mbox{d}\boldsymbol{\eta}^T}\right|_{\boldsymbol{\eta}=\boldsymbol{\eta}^*_0},\,
\mathbf{H}=\left.\frac{\partial\mathcal{L}(q,\boldsymbol{\theta})}{\partial\boldsymbol{\eta}\partial\boldsymbol{\eta}^T}\right|_{\boldsymbol{\eta}=\boldsymbol{\eta}^*_0} $$
with $\boldsymbol{\eta}^*_0$ being the set of optimal variational parameters obtained at $t=0$. Note that $\boldsymbol{\eta}^*_0$ is exactly the parameter set $\boldsymbol{\eta}^*$ obtained by the MFVB algorithm in Section 2.2.
Consequently, combining Eqs. (\ref{TruePosteriorVar},\ref{Condition},\ref{MFVB_approx}), the posterior variance of $\beta$ can be approximated by
\begin{equation*}
\mbox{Var}_{p_0}[\beta]
=\left.\frac{\mbox{d}\mathbb{E}_{p_t}[\beta]}{\mbox{d}t}\right|_{t=0}\approx \left.\frac{\mbox{d}\mathbb{E}_{q_t}[\beta]}{\mbox{d}t}\right|_{t=0}=-\mathbf{g}^T\mathbf{H}^{-1} \mathbf{g}.
\end{equation*}
To provide statistical inference on $\beta$ as what many MR methods offered, we specify $\sigma_0$ in Eq. (\ref{eq: bwmr-3}) as $1\times 10^6$ throughout this paper. In such a way, BWMR can provide the estimate of $\beta$ using $\mu^*_\beta$, its standard error using $\sqrt{\mbox{Var}_{p_0}[\beta]}$ and corresponding $p$-value. We shall use simulation study to evaluate whether the above proposed method indeed provides an accurate inference.

\section{Results}
\subsection{Simulation study}
To closely mimic real data analysis, we simulated individual-level data, and then obtained $(\hat{\gamma}_j,\sigma_{X_j},p_{X_j})$ and $(\hat{\Gamma}_j,\sigma_{Y_j},p_{Y_j})$ by simple linear regression. We also included the IVs selection procedure in the simulation study to investigate the potential effect of selection bias. Let $\bfG_1\in \mathbb{R}^{n_1\times N_0}$ and $\bfG_2\in \mathbb{R}^{n_2\times N_0}$ be the individual-level genotype data for exposure $X$ and outcome $Y$, where $N_0$ was the number of genotyped SNPs, $n_1$ and $n_2$ were sample sizes of exposure data and outcome data, respectively. The columns of $\bfG_1$ and $\bfG_2$ were generated independently such that these independent SNPs could serve as IVs. The minor allele frequencies of the SNPs were drawn from the uniform distribution $U[0.05,0.5]$. The vectors of effect sizes $\bfgamma$ and $\bfGamma=\beta \bfgamma + \bfalpha$ were simulated based on a four-group model \citep{Chung2014} to investigate the influence of horizontal pleiotropy. Let $\mathcal{G}_{00},\mathcal{G}_{10},\mathcal{G}_{01}$, and $\mathcal{G}_{11}$ denote these four groups of SNPs, and $\pi_{00},\pi_{10},\pi_{01}$ and $\pi_{11}$ denote the proportions of SNPs in these four groups, respectively. The four groups of SNPs were given as follows.
\begin{itemize}
  \item In $\mathcal{G}_{00}$, the SNPs affected neither exposure $X$ nor outcome $Y$, i.e., $\gamma_j=0$, $\alpha_j=0$. The SNPs in this group were irrelevant, serving as noise IVs. 
  \item In $\mathcal{G}_{10}$, the SNPs affected exposure $X$ only, i.e., $\gamma_j\neq 0$ and $\alpha_j=0$. The SNPs included in this group served as valid IVs in the framework of MR.
  \item In $\mathcal{G}_{01}$, the SNPs affected outcome $Y$ only, i.e., $\gamma_j=0$ and $\alpha_j\neq 0$. The SNPs in this group are noise IVs.
  \item In $\mathcal{G}_{11}$, the SNPs directly affected both exposure $X$ and outcome $Y$, i.e., $\gamma_j,\alpha_j\neq 0$. This group was used to mimic the phenomenon of horizontal pleiotropy. The SNPs in this group could be selected as IVs, but they were actually invalid.
\end{itemize}
For nonzero $\gamma_j$ and $\alpha_j$, they were simulated from $\mathcal{N}(0,\sigma^2_\gamma)$ and $\mathcal{N}(0,\sigma^2_\alpha)$. Given the vectors of effect sizes, we generated the vectors of phenotypes as $\mathbf{x} = \bfG_1 \bfgamma + \boldsymbol{\varepsilon}_1$ and $\mathbf{y} = \bfG_2 \bfGamma + \boldsymbol{\varepsilon}_2$, where $\boldsymbol{\varepsilon}_1$ and $\boldsymbol{\varepsilon}_2$ were independent noises from $\mathcal{N}(0,\sigma^2_{\epsilon_1})$ and $\mathcal{N}(0, \sigma^2_{\epsilon_2})$, respectively. To evaluate the type I error and statistical power of different methods, we varied $\beta\in\{0.0,0.1,0.2,0.3,0.4,0.5\}$ and group proportions $\{\pi_{00},\pi_{10},\pi_{01},\pi_{11}\}$ while controlled the two signal-noise-ratios ($SNR_1:=\sqrt{\mathrm{Var}(\bfG_1 \bfgamma)/\mathrm{Var}(\boldsymbol{\varepsilon}_1)}$ and $SNR_2:=\sqrt{\mathrm{Var}(\bfG_2 \bfGamma)/\mathrm{Var}(\boldsymbol{\varepsilon}_2)}$) at 1:1 by specifying the variance parameters $\{\sigma^2_\gamma,\sigma^2_\alpha,\sigma^2_{\epsilon_1}, \sigma^2_{\epsilon_2}\}$. Throughout the simulation study, we set sample sizes for exposure and outcome data as $n_1=n2=5,000$ and the number of total SNPs as $N_0=10,000$.

With individual-level data $\{\bfG_1, \bfG_2,\mathbf{x}, \mathbf{y}\}$, we obtained the summary-level statistics $\hat{\gamma}_j, \hat{\Gamma}_j$, with their standard errors $\sigma_{X_j},\sigma_{Y_j}$ and $p$-values $p_{X_j},p_{Y_j},\,j=1,2,...,N_0$, by regressing $\mathbf{x}$ on each column of $\bfG_1$ and $\mathbf{y}$ on each column of $\bfG_2$, respectively. Then we selected SNPs as IVs if their $p$-values were smaller than a given threshold at $1\times 10^{-5}$. As a result, the input data set for four MR methods (i.e., BWMR, Egger, GSMR, and RAPS) was given as $D=\{\hat{\gamma}_j,\sigma^2_{X_j},\hat{\Gamma}_j,\sigma_{Y_j}|p_{X_j}\leq 1\times 10^{-5}\}$. In our simulation study, we  observed that all the MR methods gave biased estimates of $\beta$ (see results in Fig. 24 of the supplementary document). In fact, this phenomenon should be attributed to winner's curse in the context of GWAS \citep{Zhong2008} or selection bias in statistical literature \citep{Efron2011}. Briefly speaking, $\hat{\gamma}_j$ in the input data set $D$ is selected based on $p_{X_j}\leq 1\times 10^{-5}$, and thus $\mathbb{E}[\hat{\gamma}_j|\gamma_j; p_{X_j}\leq 1\times 10^{-5}]\neq \mathbb{E}[\hat{\gamma}_j|\gamma_j]=\gamma_j$, i.e., the extracted effect size $\hat{\gamma}_j$ would be a biased estimate of $\gamma_j$ due to the selection process ($p_{X_j}< 1\times 10^{-5}$).
To avoid selection bias, we simulated two independent datasets for the exposure (i.e., $\mathbf{x} = \mathbf{G}_1 \bfgamma + \boldsymbol{\varepsilon}_1$ and $\tilde{\mathbf{x}} = \tilde{\mathbf{G}}_1 \bfgamma + \tilde{\boldsymbol{\varepsilon}}_1$ with equal sample sizes $n_1=\tilde{n}_1$). We used one data set (i.e., $\{\mathbf{G}_1,\mathbf{x}\}$) to compute $p$-values for SNP selection and used another data set (i.e., $\{\tilde{\mathbf{G}}_1,\tilde{\mathbf{x}}\}$) to extract the corresponding estimated SNP-exposure effects and its standard error $\{\hat\gamma_j,\sigma_{X_j}\}$. In such a way, the issue of selection bias was avoided in our simulation study. 

\begin{figure*}[!htbp]
  \centering
  \includegraphics[width=0.36\textwidth]{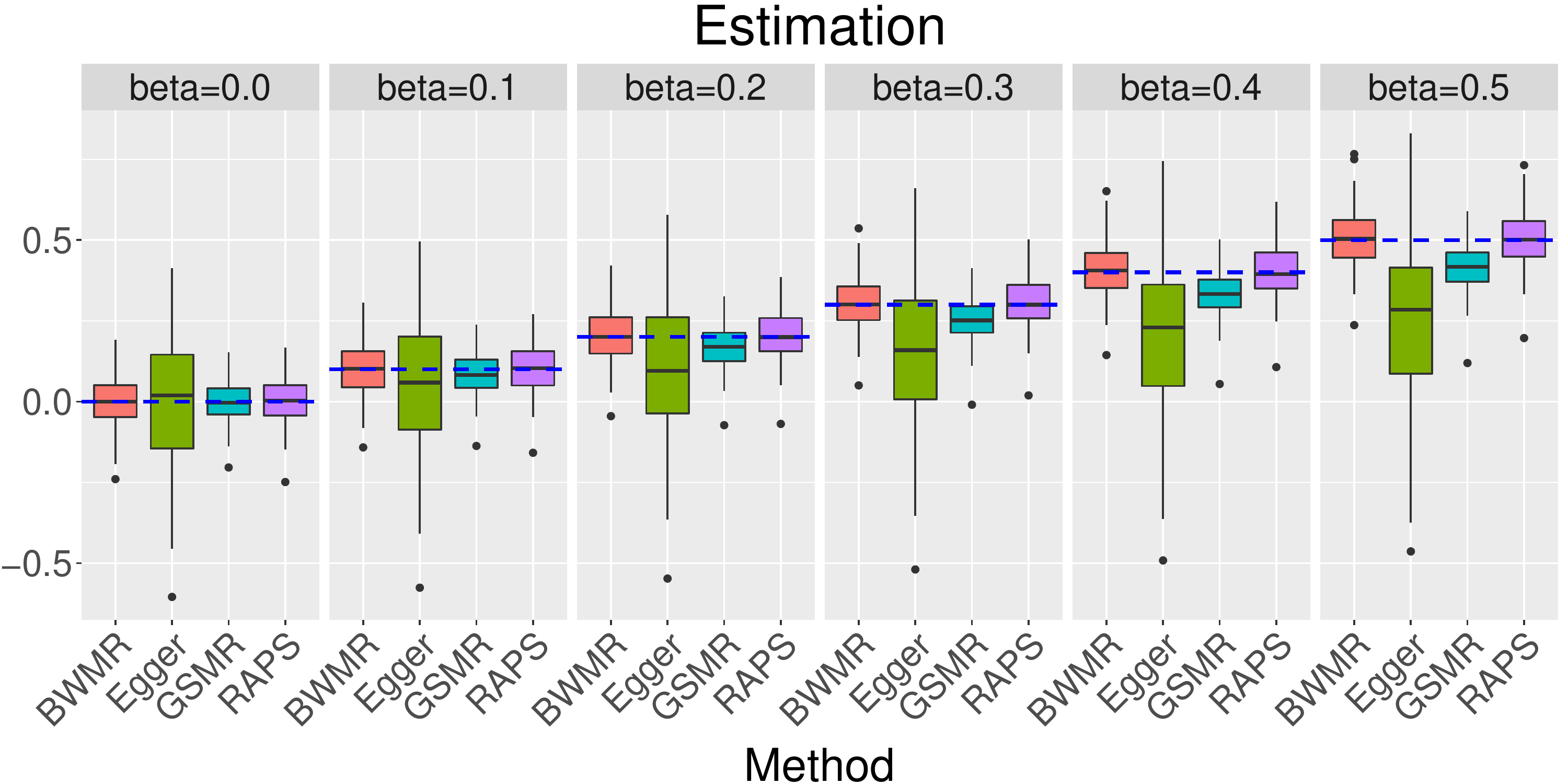}\quad
  \includegraphics[width=0.19\textwidth]{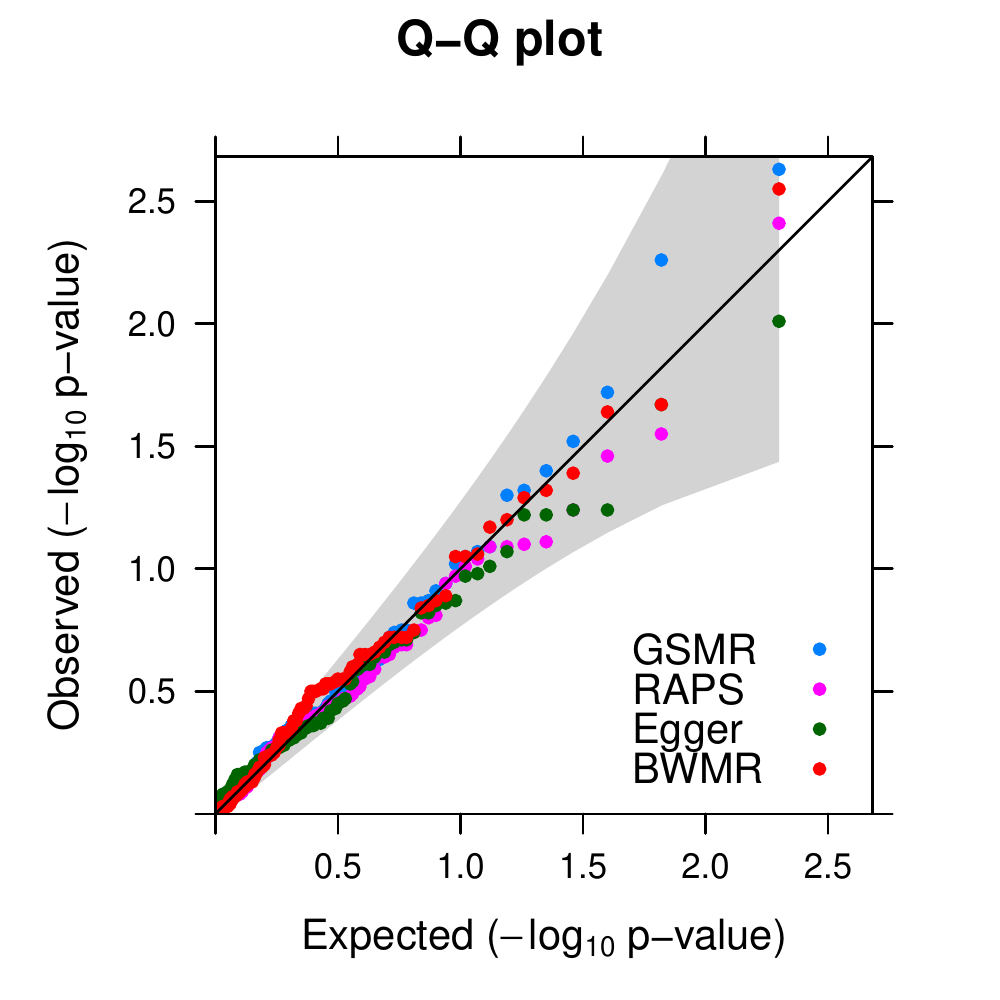}\quad
  \includegraphics[width=0.18\textwidth]{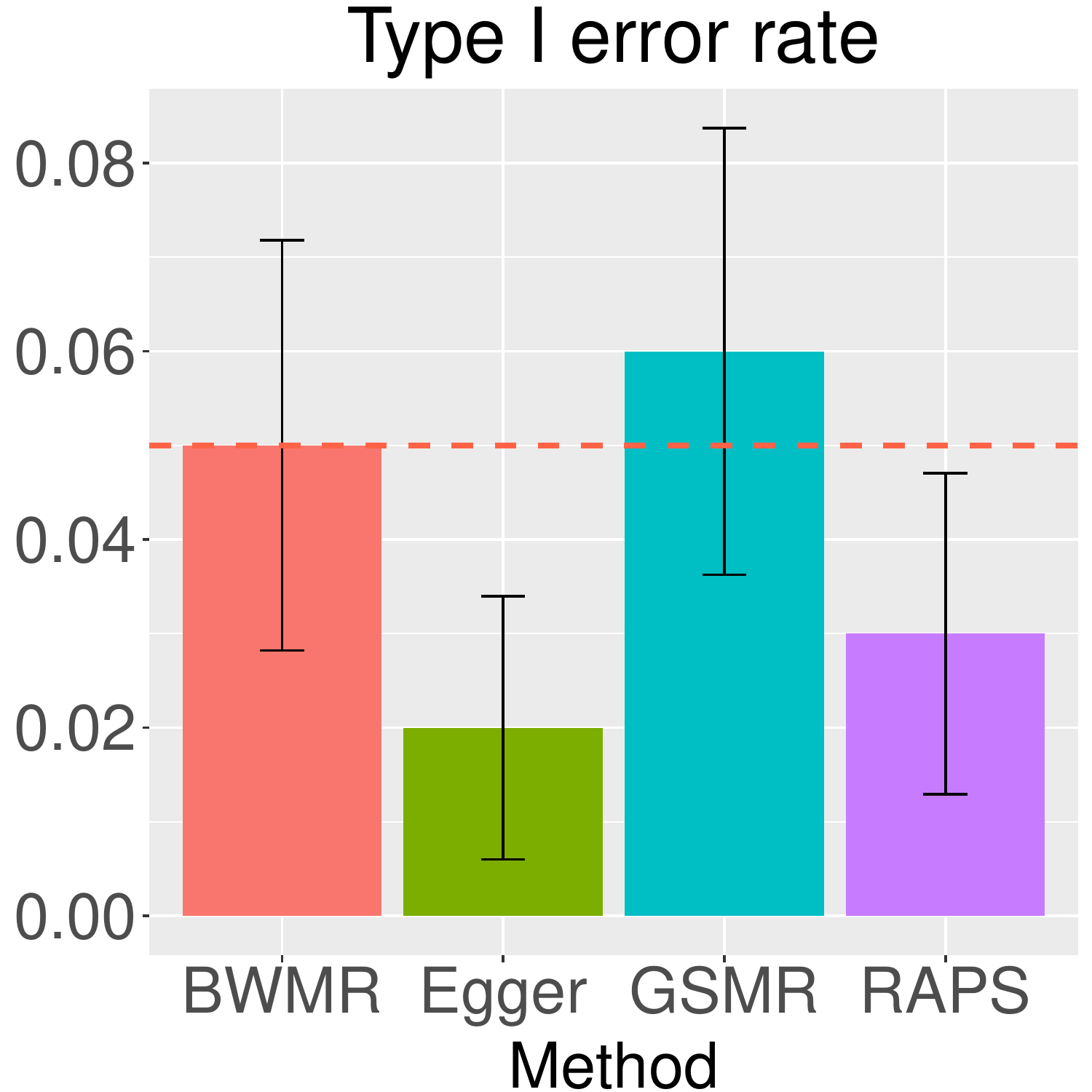}\quad
  \includegraphics[width=0.18\textwidth]{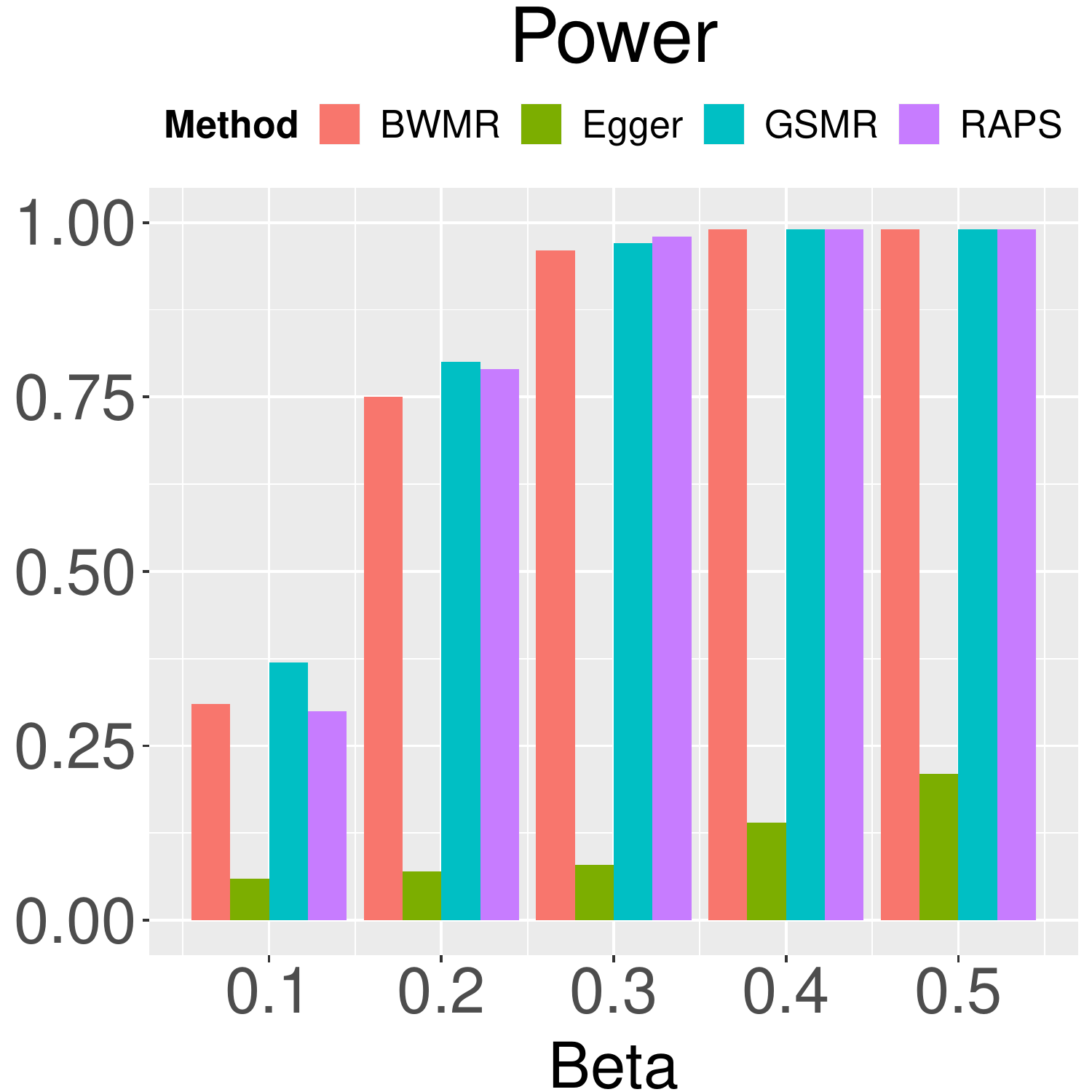}\quad\\

  \includegraphics[width=0.36\textwidth]{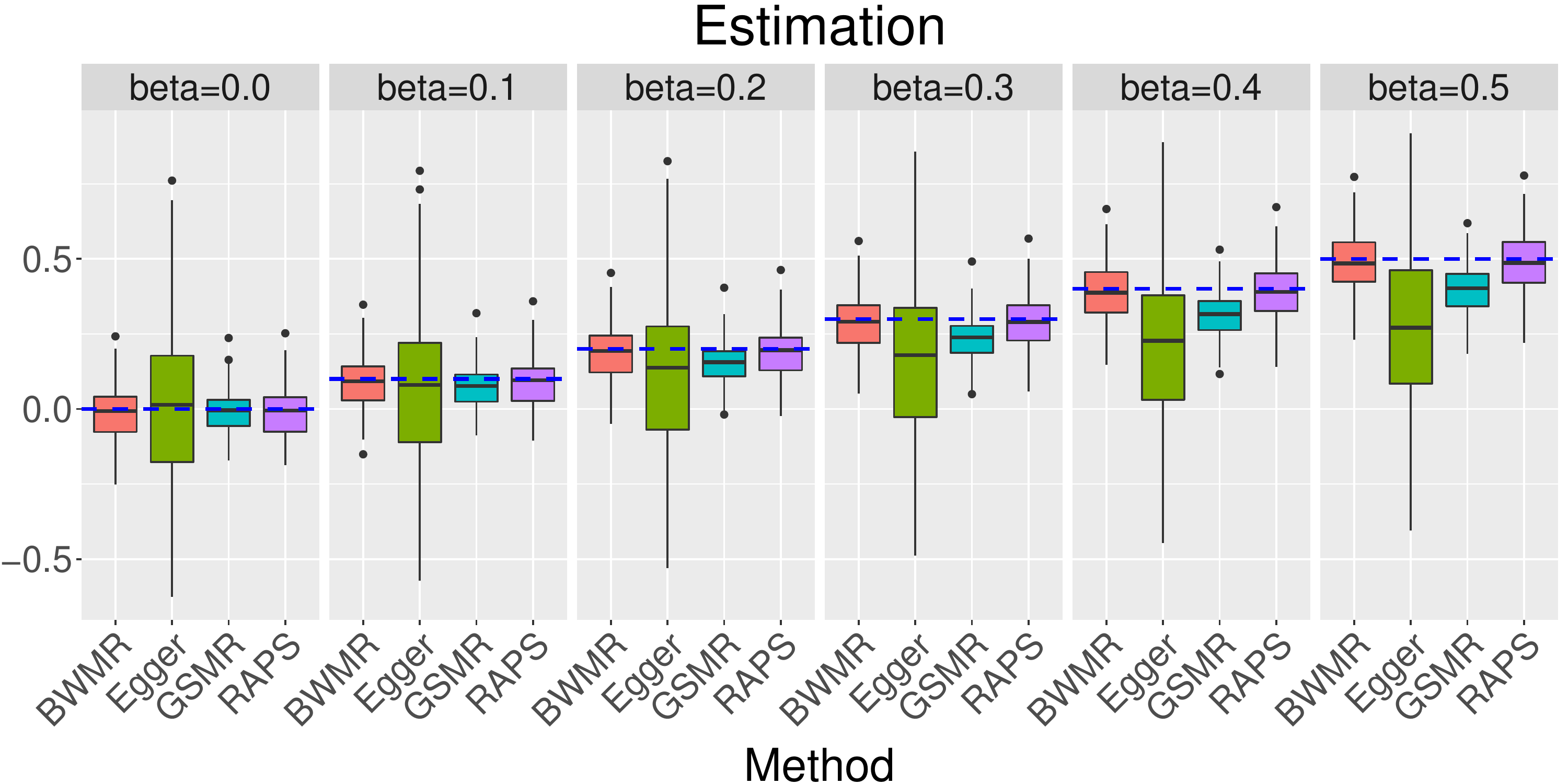}\quad
  \includegraphics[width=0.19\textwidth]{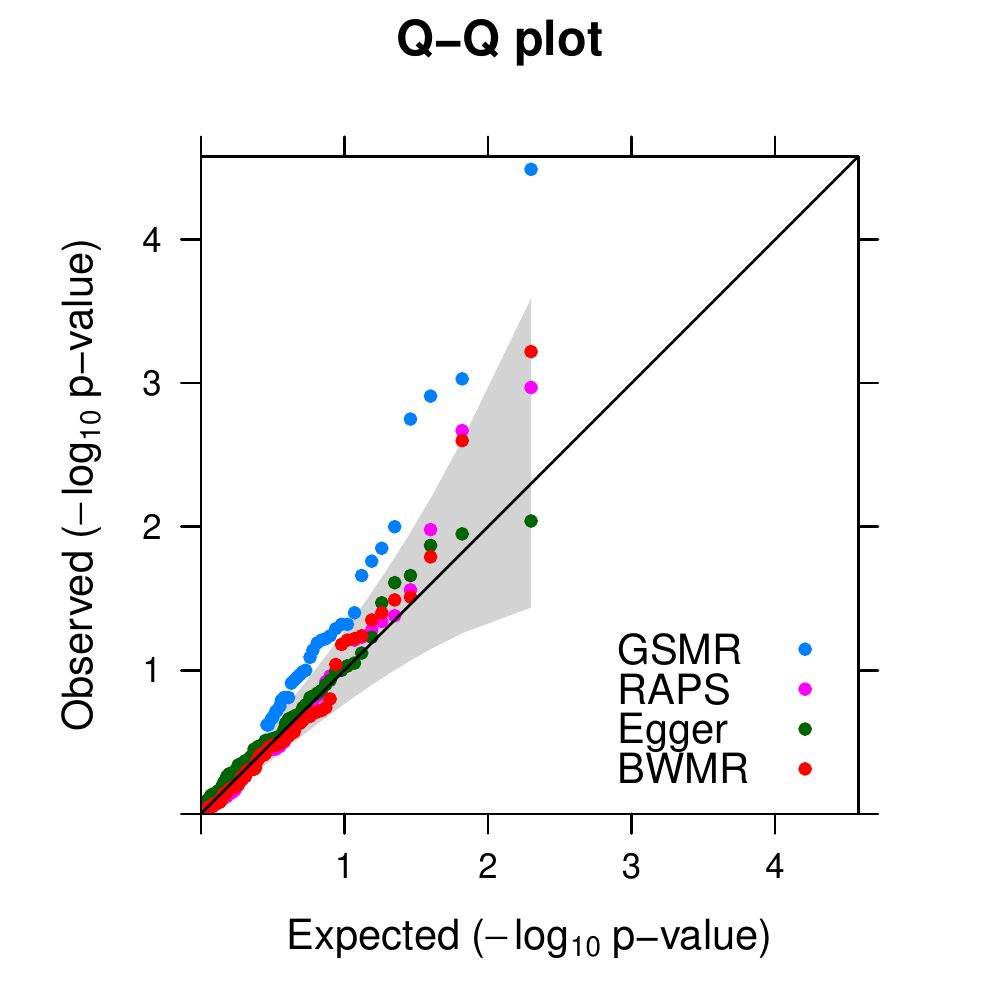}\quad
  \includegraphics[width=0.18\textwidth]{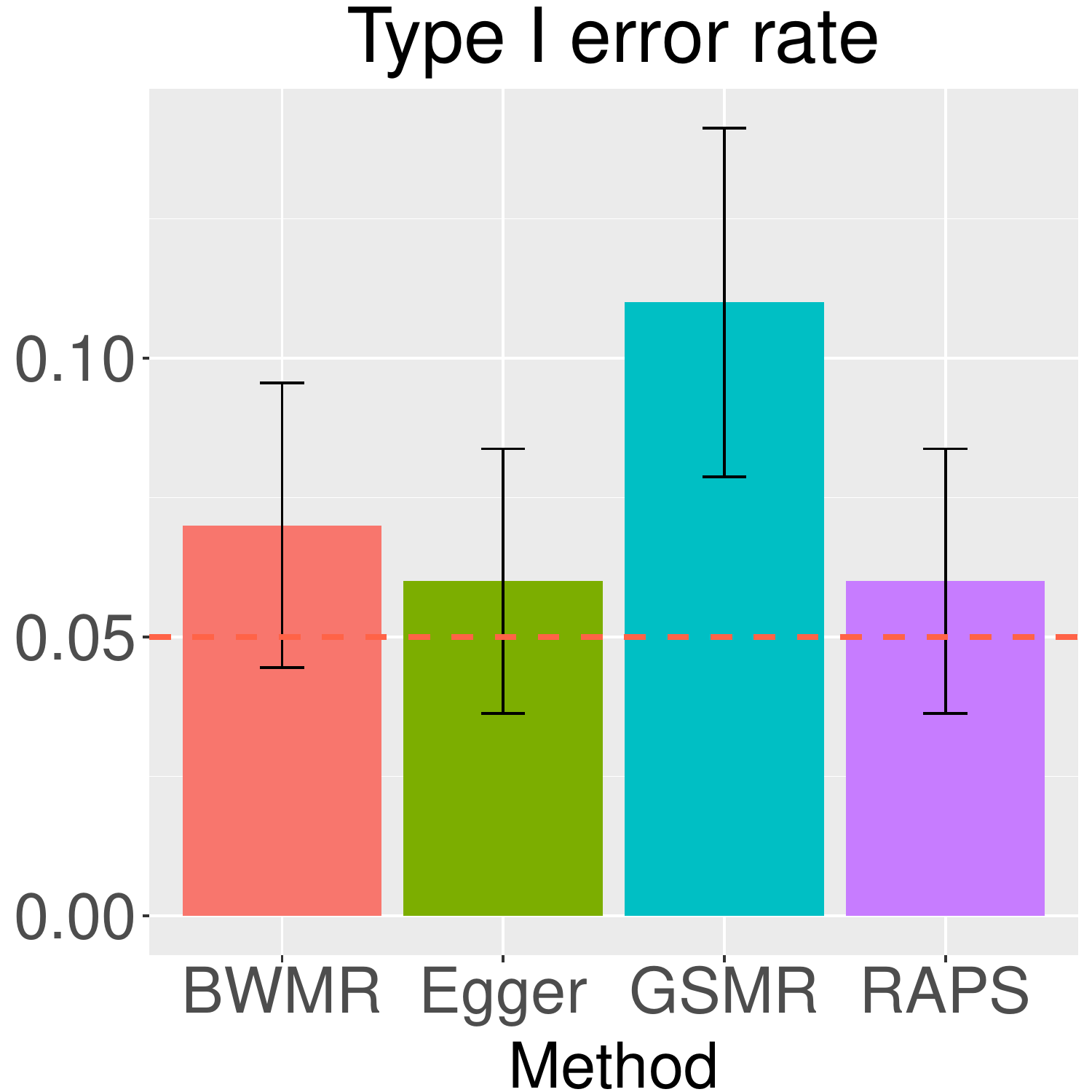}\quad
  \includegraphics[width=0.18\textwidth]{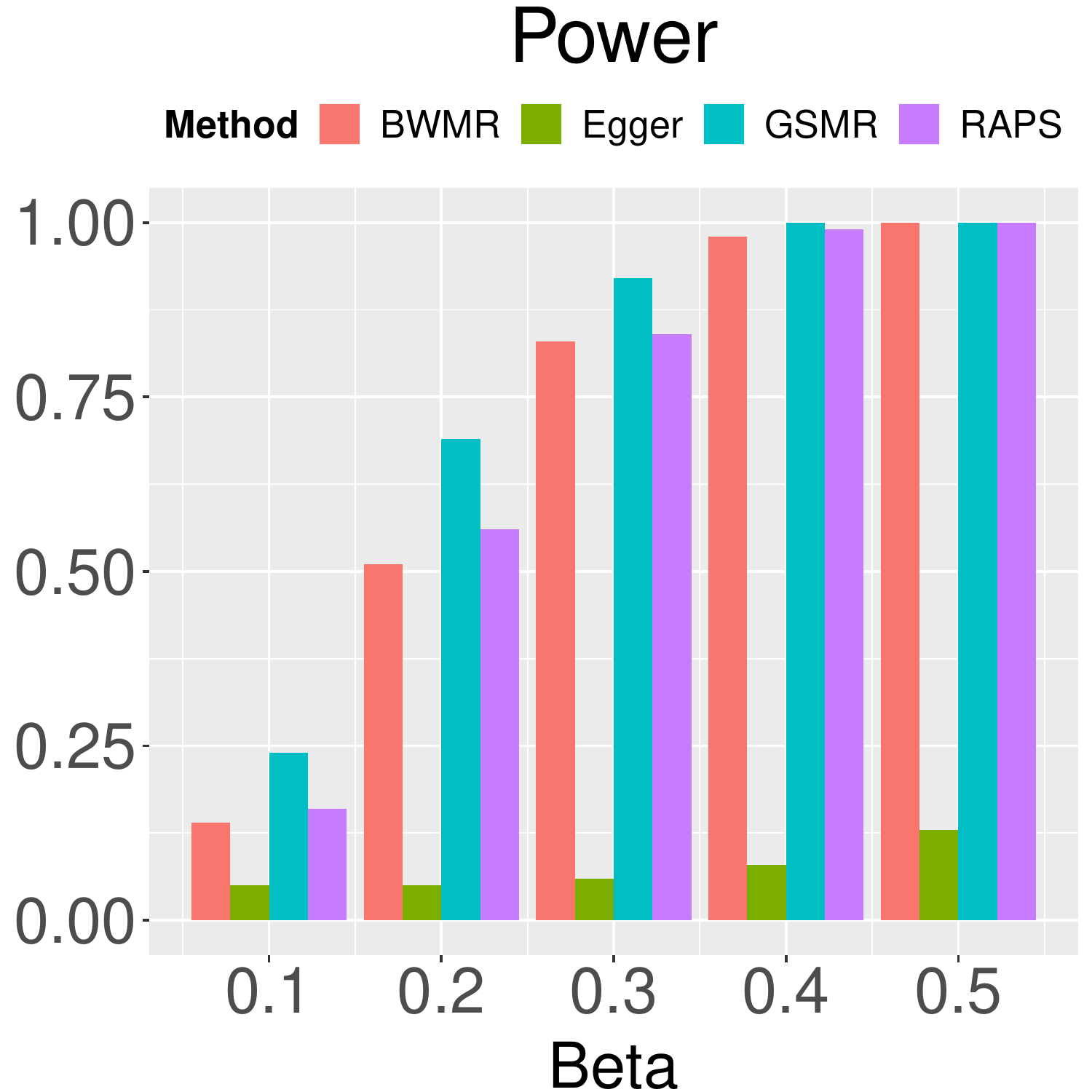}\quad\\

  \includegraphics[width=0.36\textwidth]{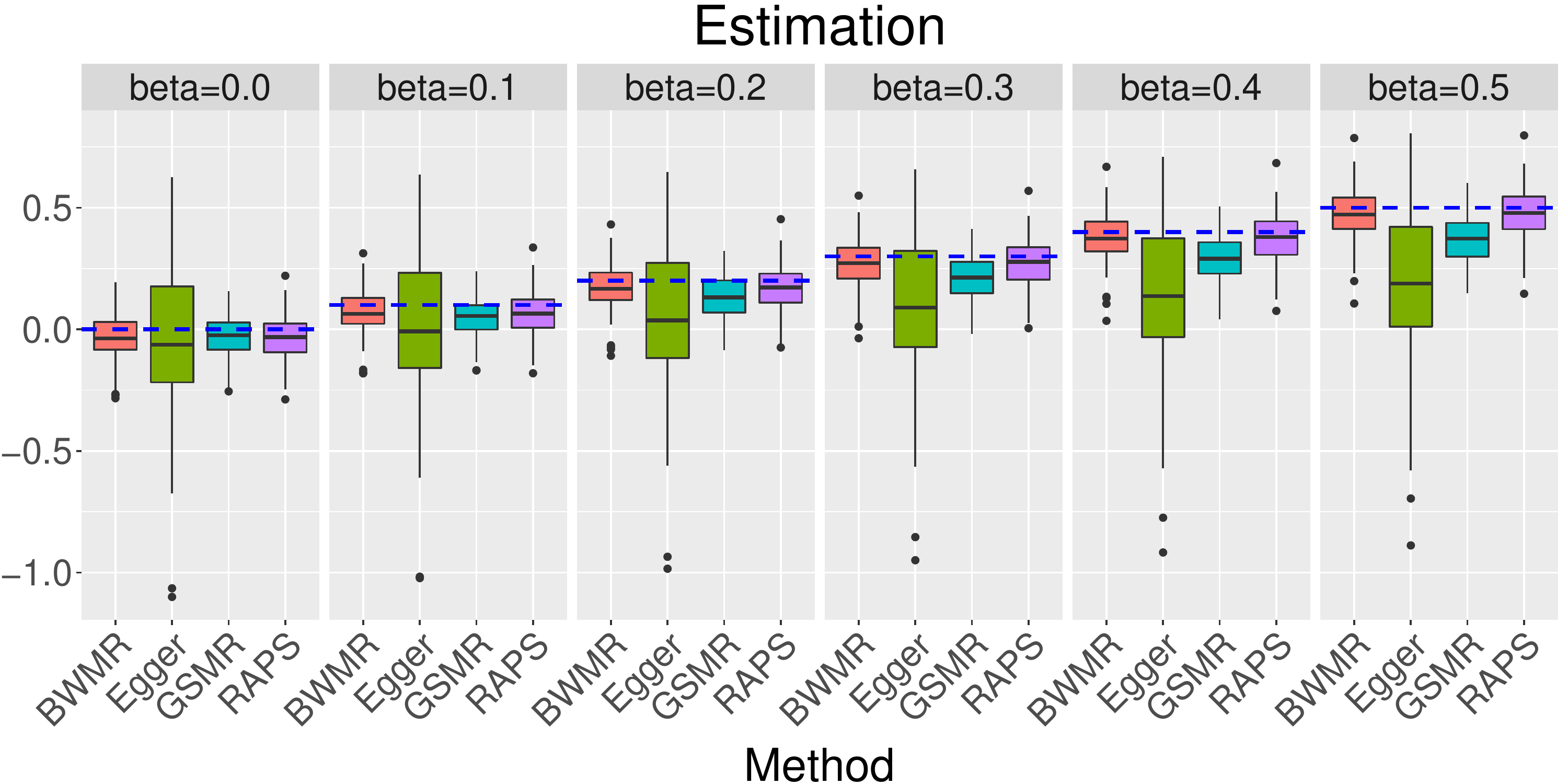}\quad
  \includegraphics[width=0.19\textwidth]{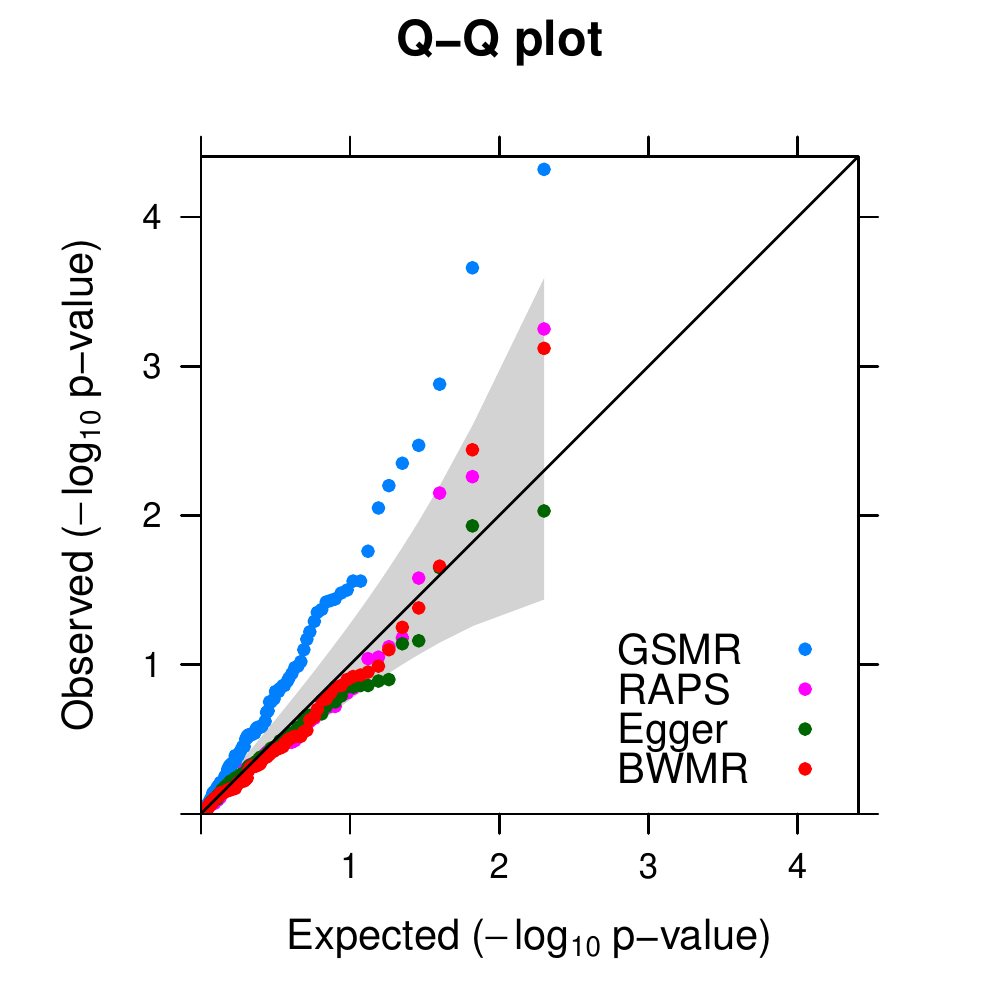}\quad
  \includegraphics[width=0.18\textwidth]{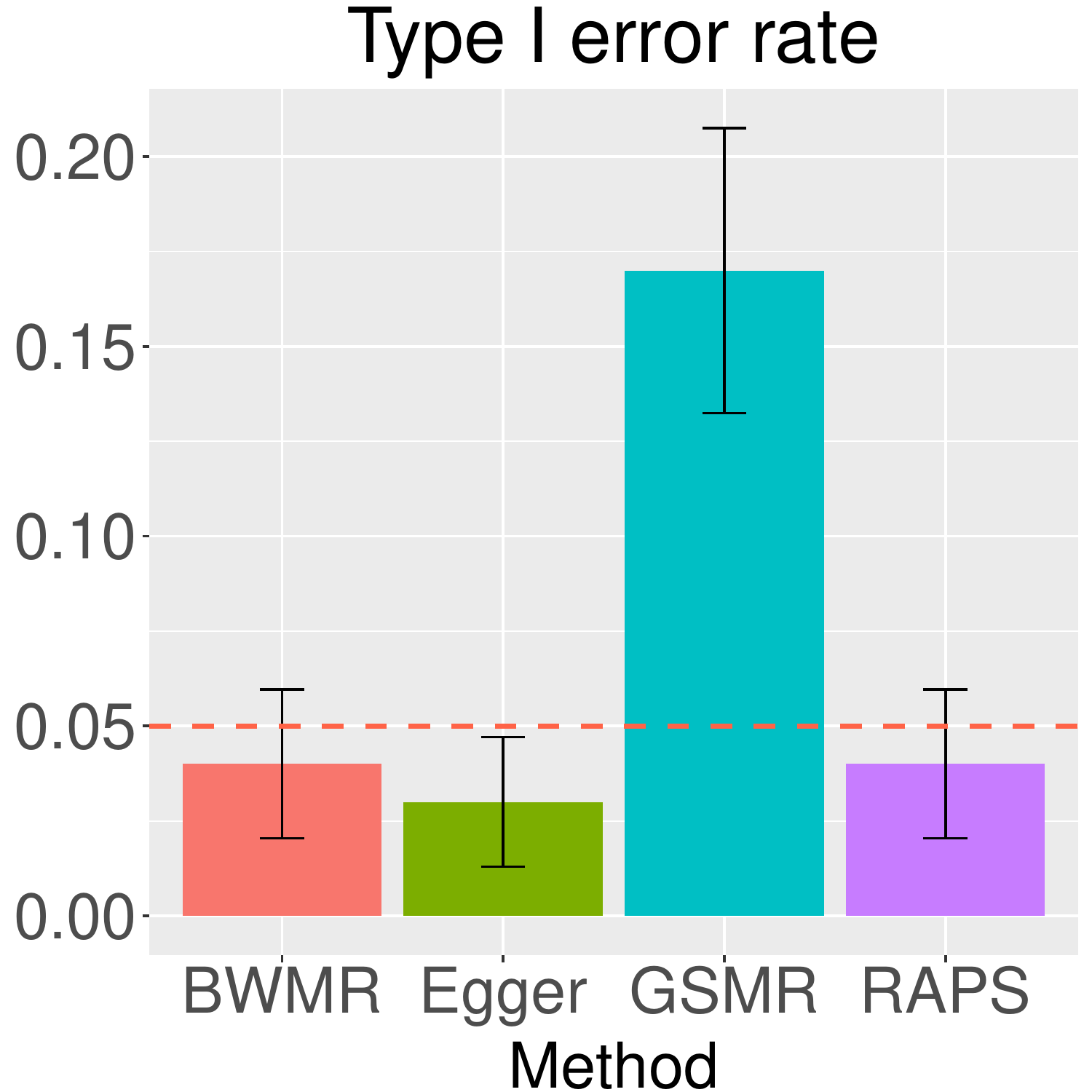}\quad
  \includegraphics[width=0.18\textwidth]{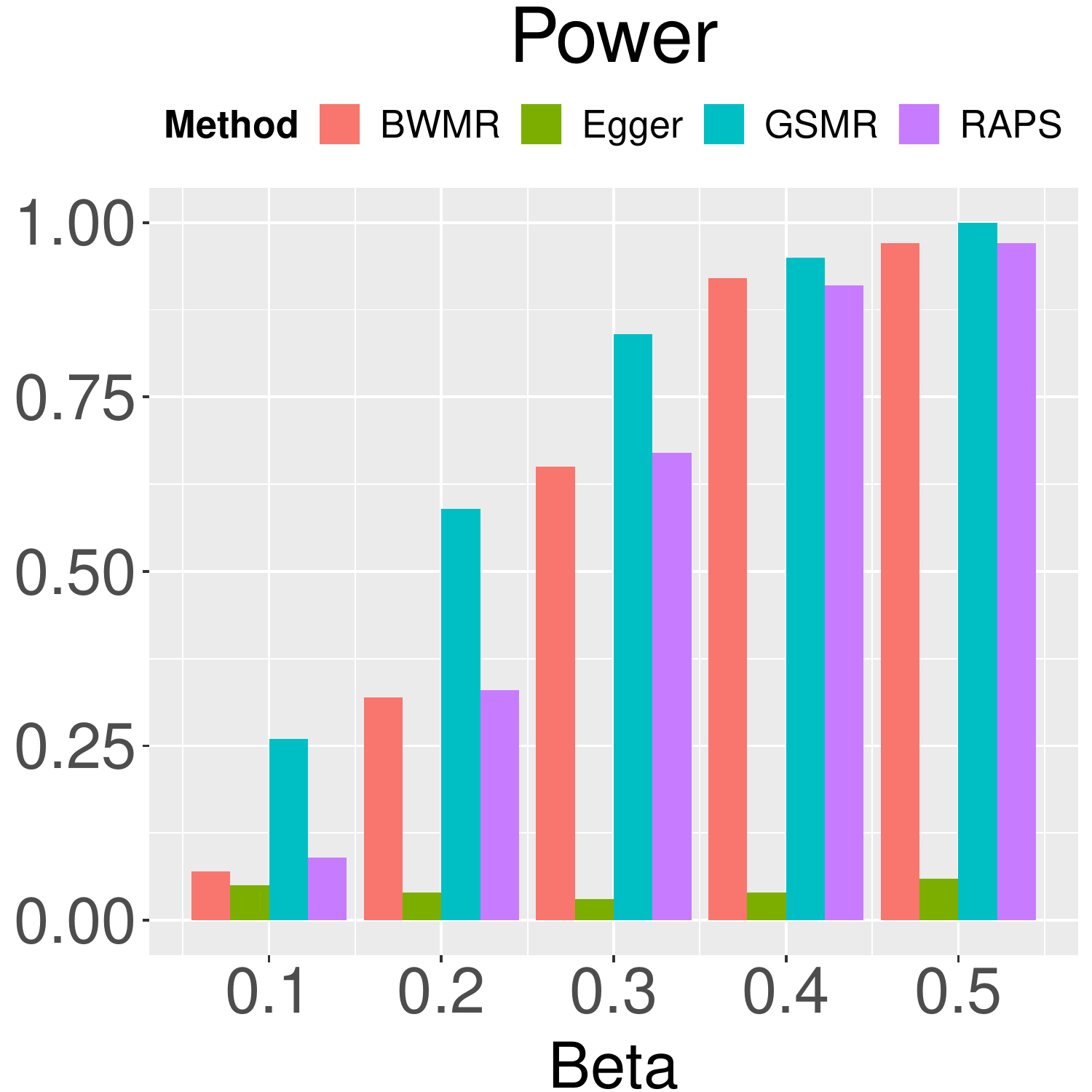}\quad\\
  \caption{Comparison of MR methods affected by horizontal pleiotropy (the proportions of horizontal pleiotropic IVs were specified as $\frac{\pi_{11}}{\pi_{10}+\pi_{11}}=20\%,\,50\%$, and $80\%$, in the top, middle and bottom panel, respectively). The parameters were chosen to be $N_0=10,000$, $n_1=n_2=5,000$, $\beta\in\{0.0,0.1,0.2,0.3,0.4,0.5\}$, $SNR_1=SNR_2=1:1$, and $p\mbox{-value threshold}=1\times 10^{-5}$. We evaluated the empirical type I error rate and power by controlling type I error rates at the nominal level $0.05$. The results were summarized from 100 replications.}
  \label{fig: simu-individual pleiotropy}
\end{figure*}

To investigate the performance of BWMR and three other related MR methods on the influence of horizontal pleiotropy, we varied the propotion $\frac{\pi_{11}}{\pi_{10}+\pi_{11}}\in\{0.2,0.5,0.8\}$, approximately controlling the proportion of IVs affected by horizontal pleiotropy among all IVs at 20\%, 50\%, 80\% respectively. We observed that BWMR showed satisfactory performance in terms of estimation accuracy, type I error control and statistical power, as shown in Fig. \ref{fig: simu-individual pleiotropy}. The satisfactory performance of BWMR can be attributed to its following properties: 1. adaptive use of variance component $\tau^2$ to account for weak pleiotropic effects; 2. robustness to strong horizontal pleiotropy guaranteed by the Bayesian weighting scheme; 3. stable and efficient algorithm for parameter estimation and inference. From the simulation study, we also observed that GSMR seemed to have the highest statistical power among all the MR methods. However, the $p$-value of GSMR was also observed to be inflated in the qq-plot when a higher proportion of IVs  (e.g., 50\%, 80\%) affected by horizontal pleiotropy (see supplementary Fig. S20). Thus, the type I error rate of GSMR was unable to be controlled at the nominal level. This is because GSMR ignored the weak pleiotropic effects and underestimated the standard error of the casual effect. Egger was found to be the most conservative one among the four MR methods. Its estimation was observed to have the largest variance because the intercept term introduced in Egger was unable to model the weak pleiotropic effects which should not be simply treated as a constant. Among the four MR methods, BWMR and RAPS had similar performance in this simulation study, but RAPS was found to be numerically not stable in the next real data analysis part. To have a better understanding of the four MR methods, we provided a detailed discussion about their relationship in the supplementary document (see our supplementary note). To make the simulation results easily reproducible, we have made the code of our simulation study available at 
\url{https://github.com/jiazhao97/sim-BWMR}.

\subsection{Real Data Analysis}
\subsubsection{Materials}
To investigate the influence of selection bias on real data analysis and the reliability of the four MR methods, we collected three different GWASs of four global lipids, i.e., high-density lipoprotein (HDL) cholesterol (HDL-C), low-density lipoprotein (LDL) cholesterol (LDL-C), triglycerides (TG) and total cholesterol (TC). For narrative convenience, we shall refer them as Data-A \citep{Willer2013}, Data-B \citep{Kettunen2016} and Data-C \citep{klarin2018genetics} (the details of the three GWASs are given in the supplementary document). It is worthwhile to note that Data-C only includes SNPs significantly associated with global lipids. 

Besides the four global lipids in Data-A, Data-B, and Data-C, we further collected GWAS summary statistics, covering $126$ metabolites \citep{Kettunen2016} and $93$ human complex traits. These metabolites include lipoprotein (e.g. very-low-density lipoprotein (VLDL) and LDL), fasting glucose, Vitamin D levels and serum urate. The $93$ human complex traits include anthropometric traits (e.g. body mass index (BMI)), cardiovascular measures (e.g. coronary artery disease (CAD)), immune system disorders (e.g. atopic dermatitis), metabolic traits (e.g. dyslipidemia), neurodegenerative diseases (e.g. Alzheimer’s disease (AD)), psychiatric disorders (e.g. bipolar disorder), social traits (e.g. intelligence) and other complex traits (e.g. breast cancer). The details of summary statistics are given in the supplementary document.

\begin{figure}[!htbp]
  \centering
  \includegraphics[width=0.23\textwidth]{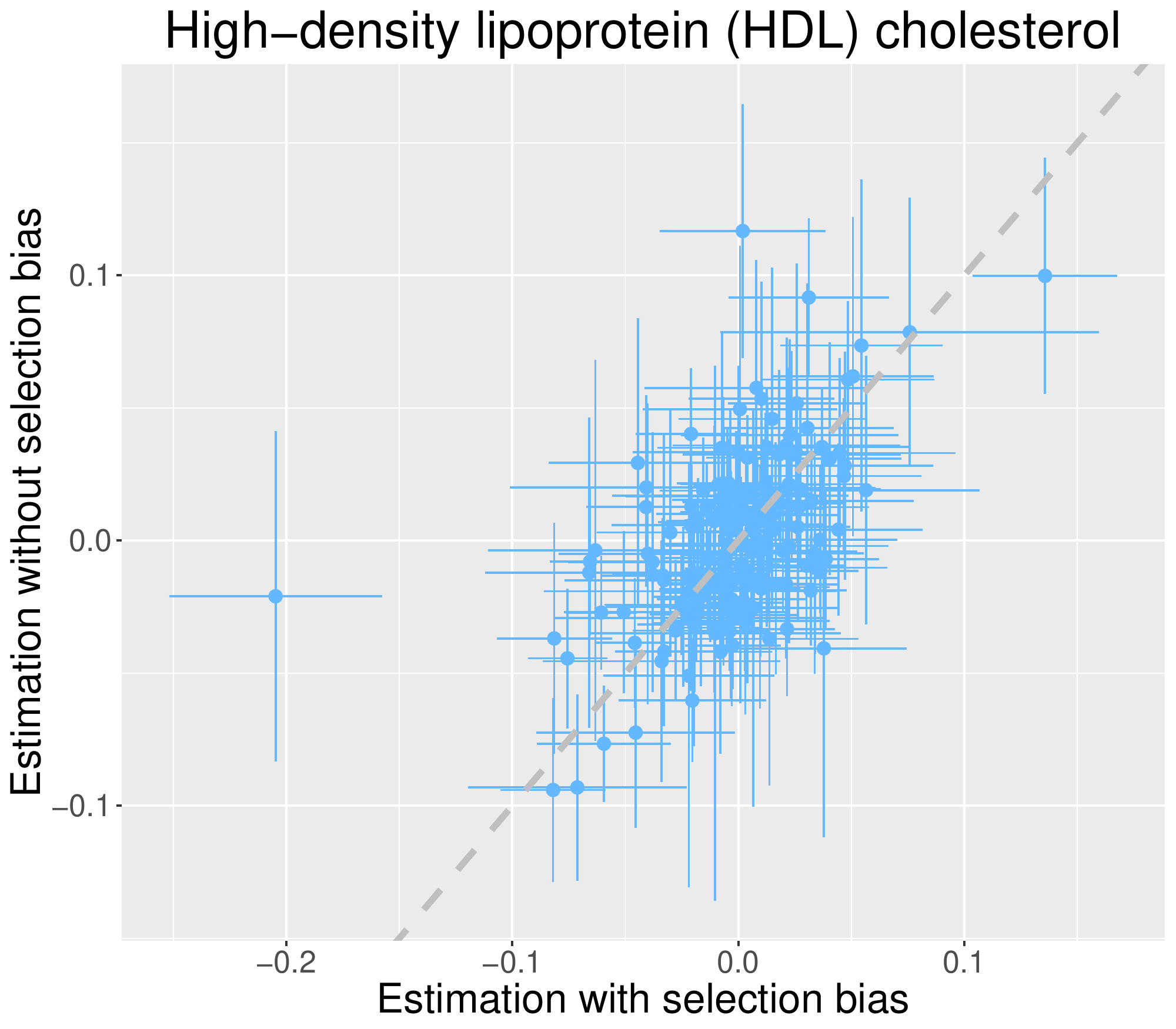}
  \includegraphics[width=0.23\textwidth]{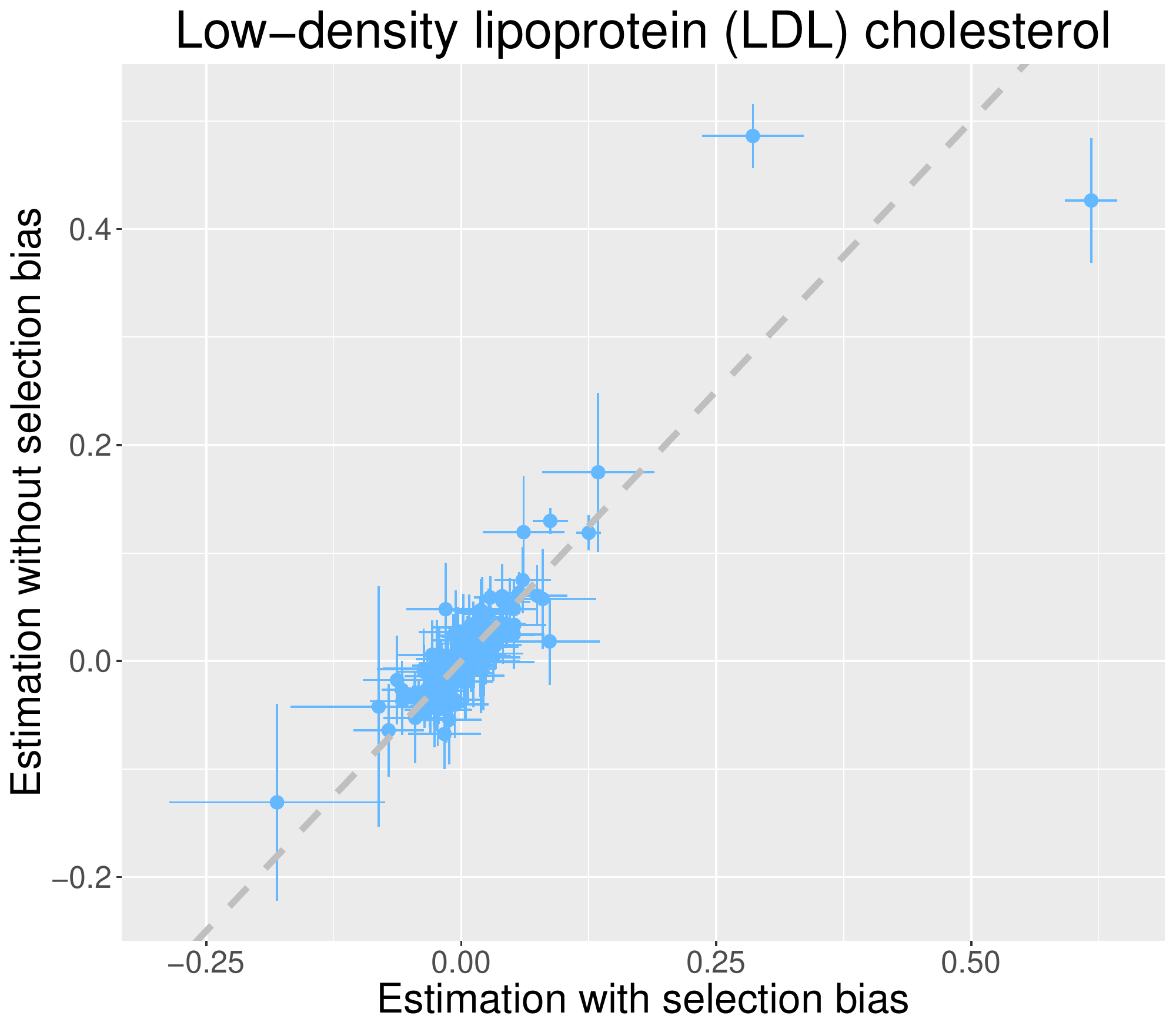}
  \includegraphics[width=0.23\textwidth]{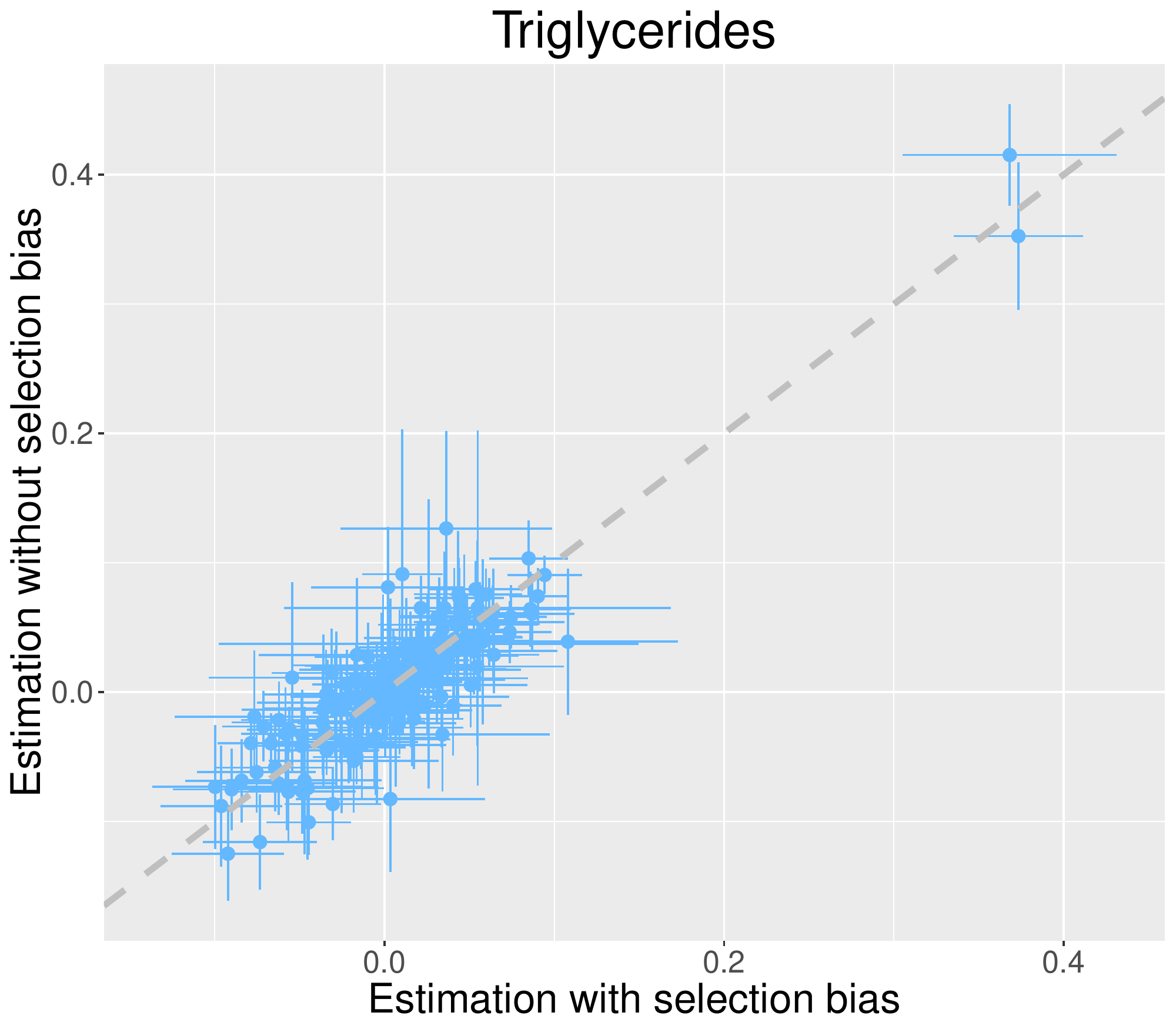}
  \includegraphics[width=0.23\textwidth]{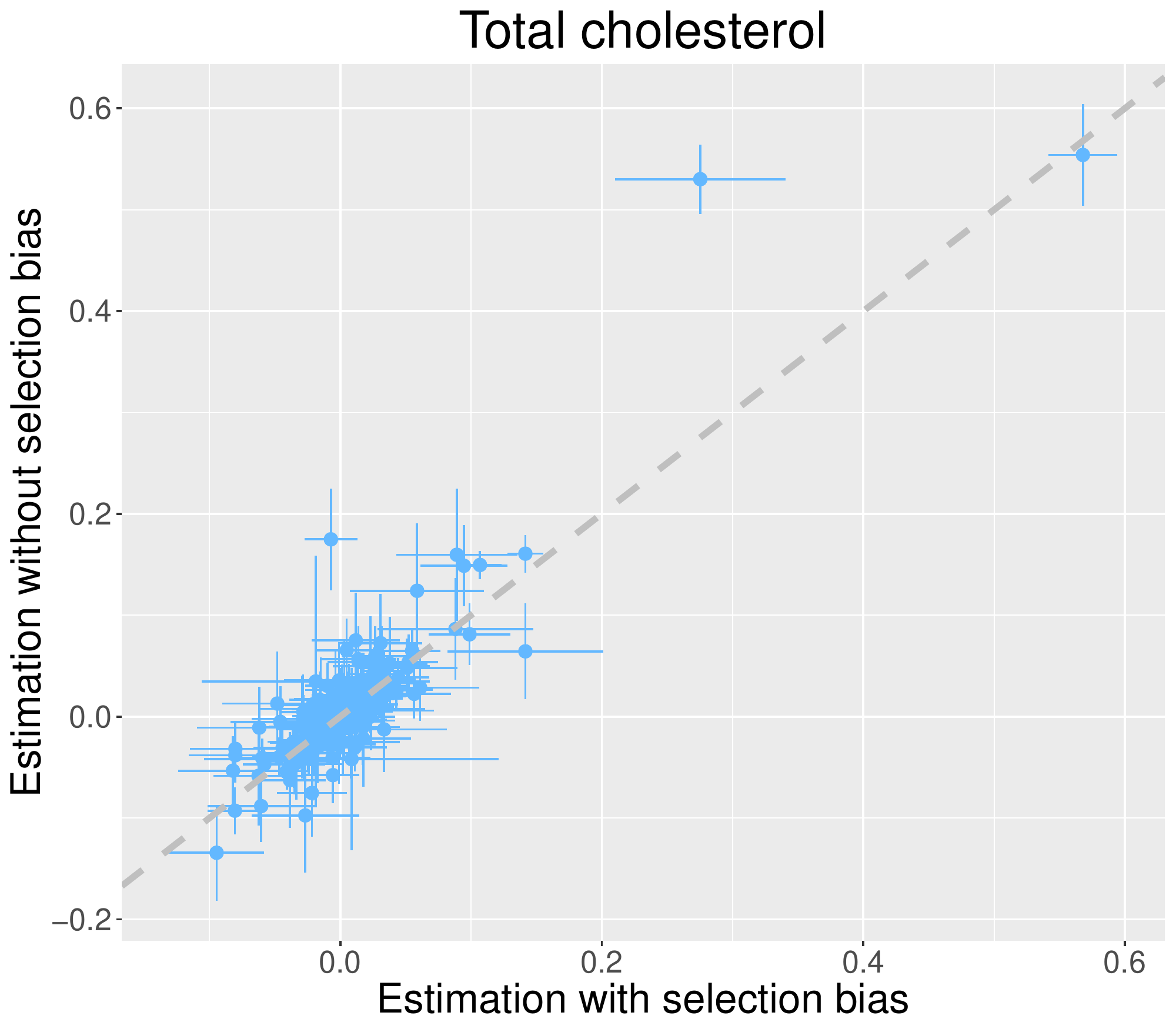}
  \caption{Comparisons of analysis results provided by BWMR with selection bias ($x$-axis) and without selection bias ($y$-axis). The dots represent estimated causal effect sizes $\hat{\beta}$ and the bars represent their standard errors $\mathrm{se}(\hat{\beta})$. The diagonal is indicated by the dashed line.}
  \label{fig: global-lipids-selectionbias}
\end{figure}

\subsubsection{Examination of selection bias}
In this subsection, we consider estimating the causal effect $\beta$ of each of four global lipids (i.e., HDL-C, LDL-C, TG and TC) on 93 complex traits. Given Data-A and Data-B, to avoid potential selection bias, we used one exposure data set (e.g., Data-A) to select IVs and extracted the corresponding estimated effect sizes and their standard errors in another exposure data set (e.g., Data-B), i.e., $\{\hat{\gamma}_j,\sigma_{X_j}\}$. Based on the selected IVs, we also obtained effect sizes and their standard errors from the outcome data sets, i.e., $\{\hat{\Gamma}_j,\sigma_{Y_j}\}$. The analysis based on this type of input data was expected to provide unbiased estimation results. To investigate selection bias, we selected IVs and extracted the estimated effect sizes and standard errors using the same data set (e.g., Data-A). Therefore, this type of input data was expected to suffer from selection bias. We applied BWMR to both unbiased and biased input data sets, and compared the analysis results to evaluate the influence of selection bias. The results from BWMR are shown in Fig. \ref{fig: global-lipids-selectionbias}, and the results from other three methods are shown in the supplementary Figs. 26, 27 and 28. Although the analysis results were expected to be different, no systematic discrepancies between biased results and unbiased results were observed. As we illustrated in simulation study, the influence of selection bias decreases as the sample size increases (see supplementary Fig. 25). Note that the sample sizes of Data-A and Data-B are very large ($n_A=188,577$ and $n_B=24,925$, respectively). The sample sizes are large enough such that the influence of selection bias would be ignorable. As a result, we concluded that selection bias might not introduce much bias when we were analyze the causal effects between those metabolites and complex traits in this paper.

\subsubsection{Consistency of BWMR's analysis results}
Before applying BWMR to infer the causal effects between hundreds of metabolites and complex traits, we first checked the reliability of BWMR's analysis results on the four global lipids. To do so, we explored the third data set, i.e., Data-C, to select IVs, and then used Data-A and Data-B to extract exposure effect sizes and their standard errors, denoted as $\{\hat{\gamma}^{(A)}_j,\sigma^{(A)}_{X_j}\}$ and $\{\hat{\gamma}^{(B)}_j,\sigma^{(B)}_{X_j}\}$, and then extracted the outcome effect sizes and their standard errors $\{\hat{\Gamma}_j,\sigma_{Y_j}\}$. After that, we applied BWMR to $\{\hat{\gamma}^{(A)}_j,\sigma^{(A)}_{X_j}, \hat{\Gamma}_j,\sigma_{Y_j}\}$ and $\{\hat{\gamma}^{(B)}_j,\sigma^{(B)}_{X_j}, \hat{\Gamma}_j,\sigma_{Y_j}\}$. The analysis results are shown in Fig. \ref{fig: global-lipids-diffdata}. The results of other three MR methods are shown in supplementary Figs. 29, 30 and 31. We can see clearly that the results based on Data-A and Data-B agree well with each other, indicating that estimating the causal relationship between metabolites and complex traits by BWMR can be quite reliable.

\begin{figure}[!htbp]
  \centering
  \includegraphics[width=0.23\textwidth]{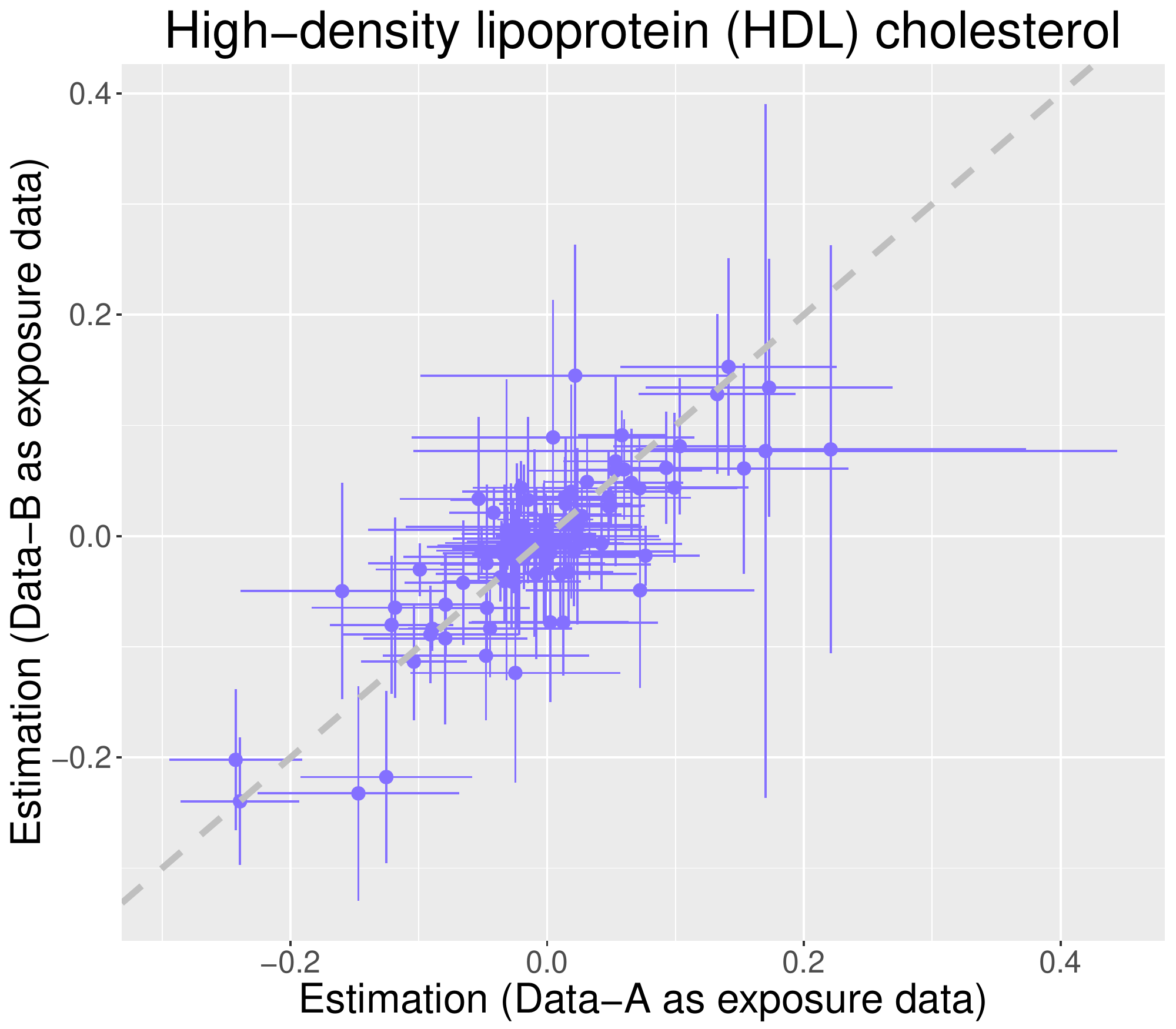}
  \includegraphics[width=0.23\textwidth]{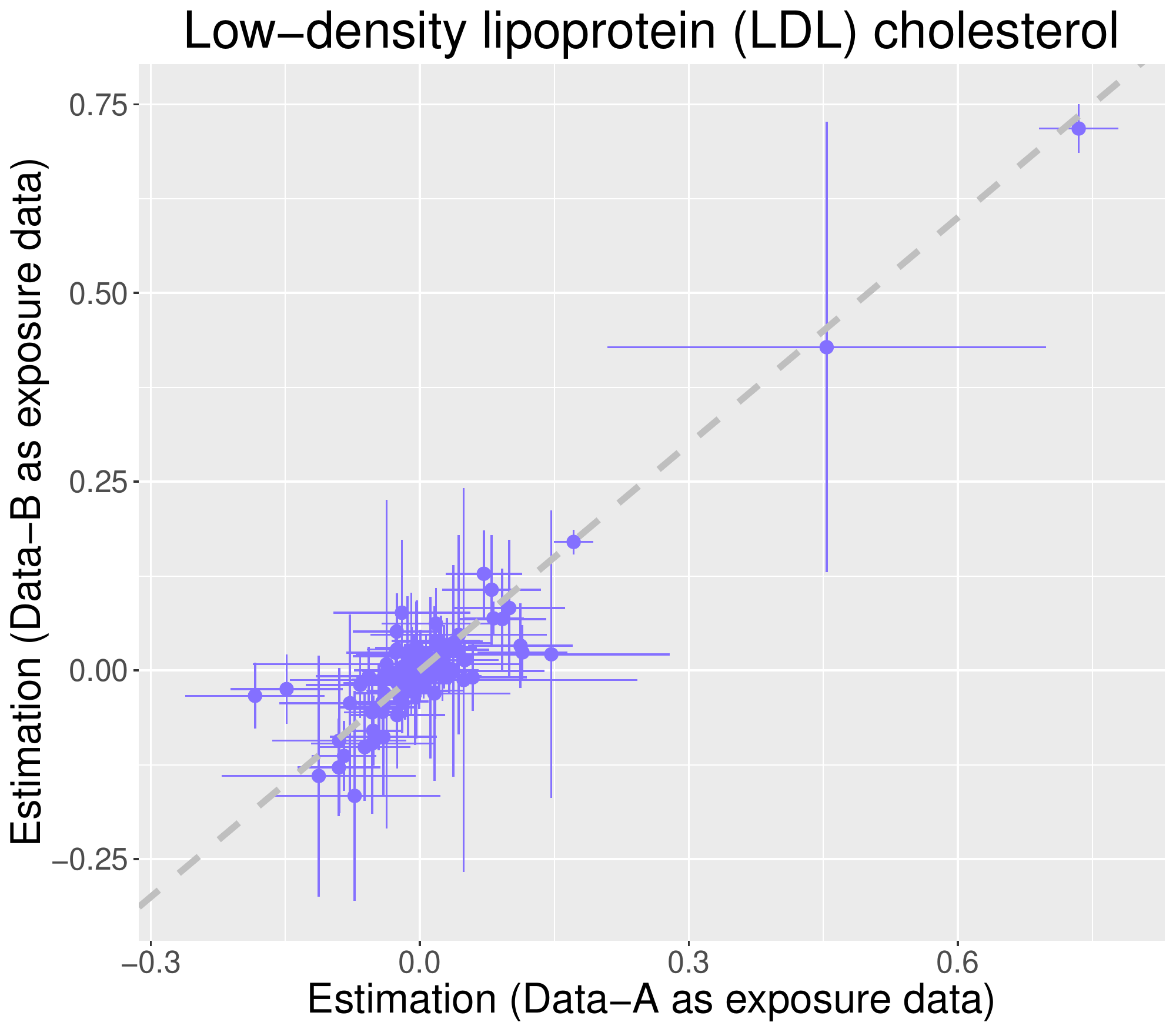}
  \includegraphics[width=0.23\textwidth]{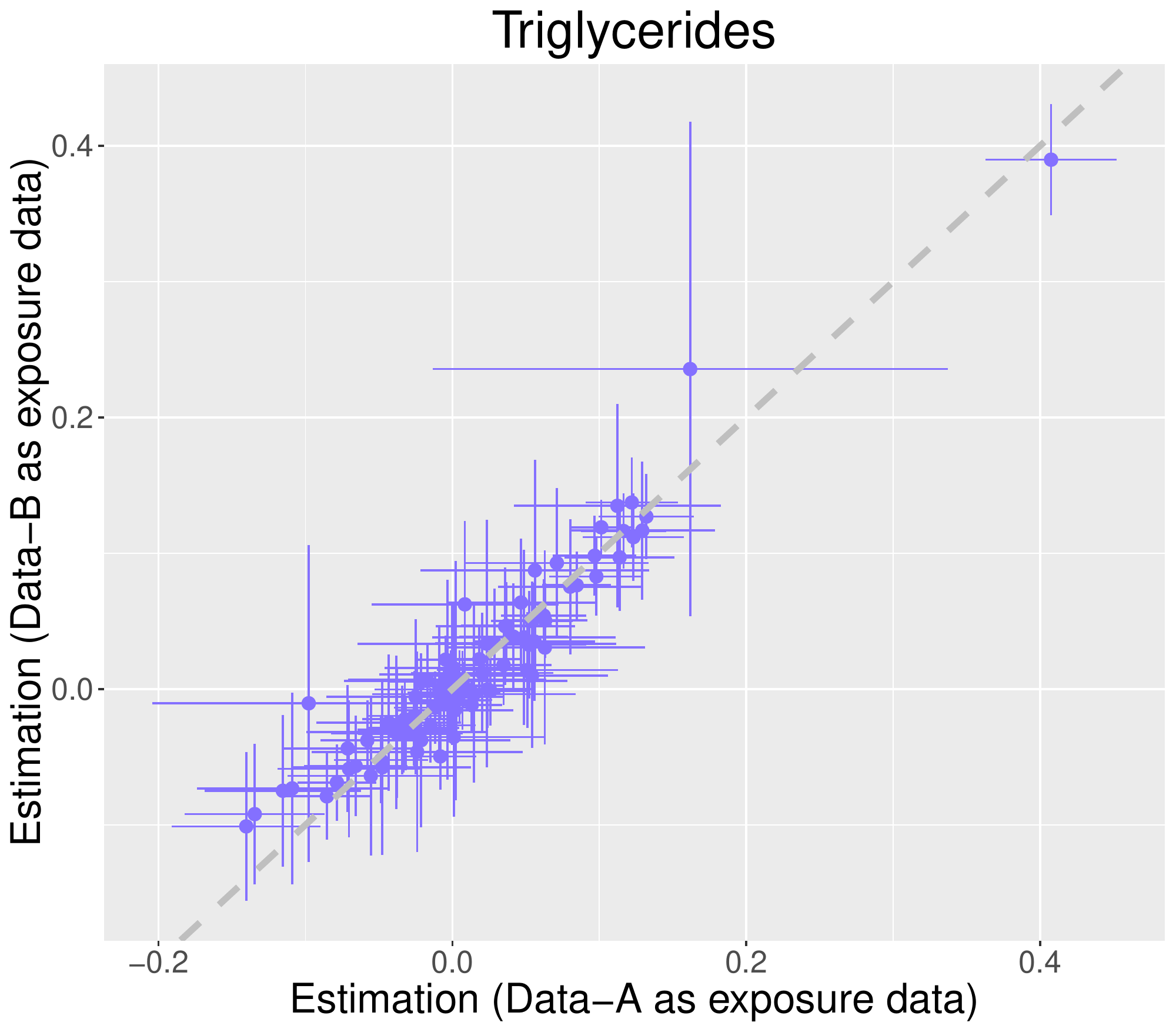}
  \includegraphics[width=0.23\textwidth]{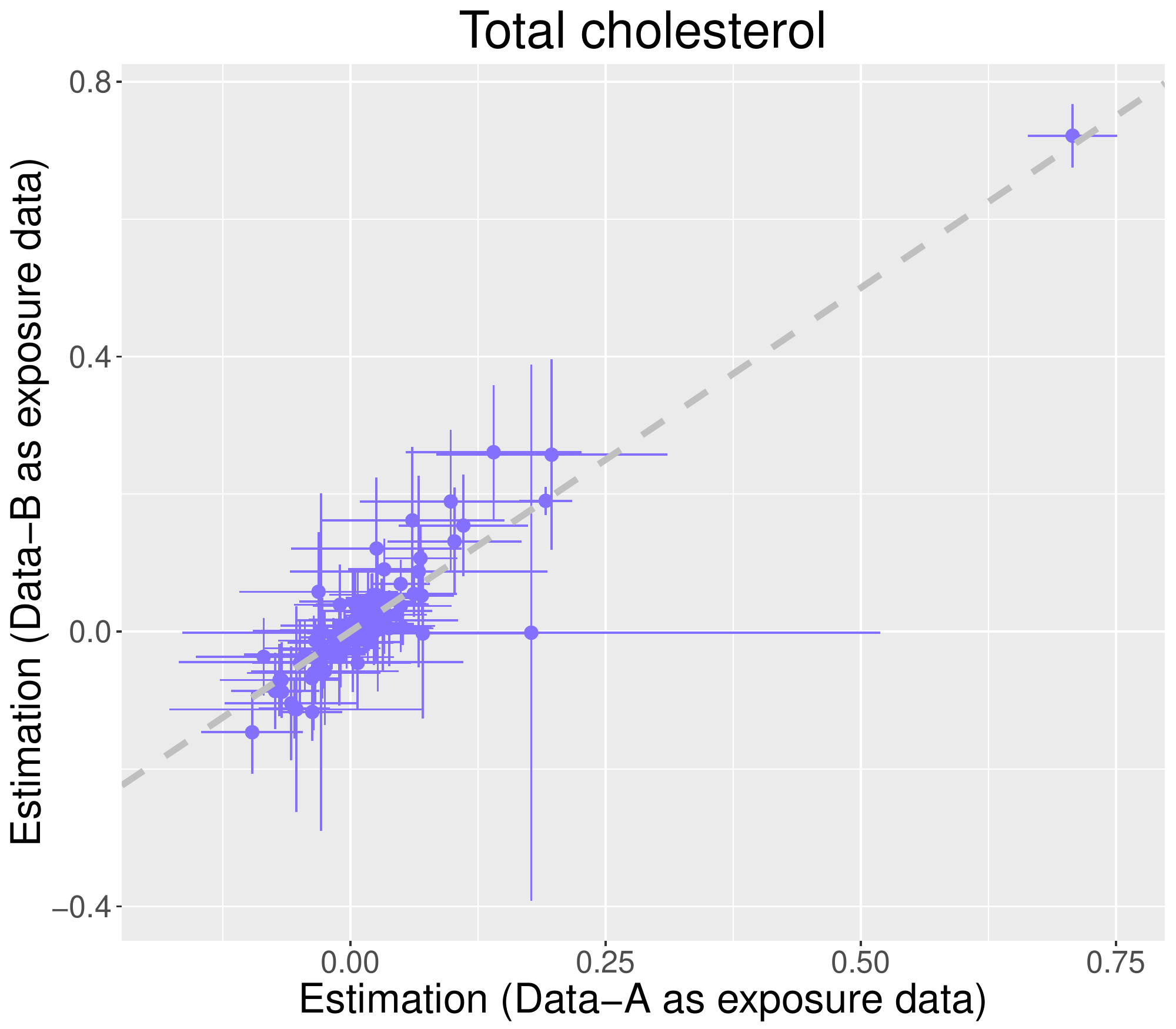}
  \caption{Comparisons of analysis results provided by BWMR using Data-A as exposure data ($x$-axis) and using Data-B as exposure data ($y$-axis). The dots represent estimated causal effect sizes $\hat{\beta}$ and the bars represent their standard errors $\mathrm{se}(\hat{\beta})$. The diagonal is indicated by the dashed line.}
  \label{fig: global-lipids-diffdata}
\end{figure}

\subsubsection{MR-based causal inference between metabolites and complex traits}

\begin{figure}[!htbp]
  \centering
  \includegraphics[width=0.22\textwidth]{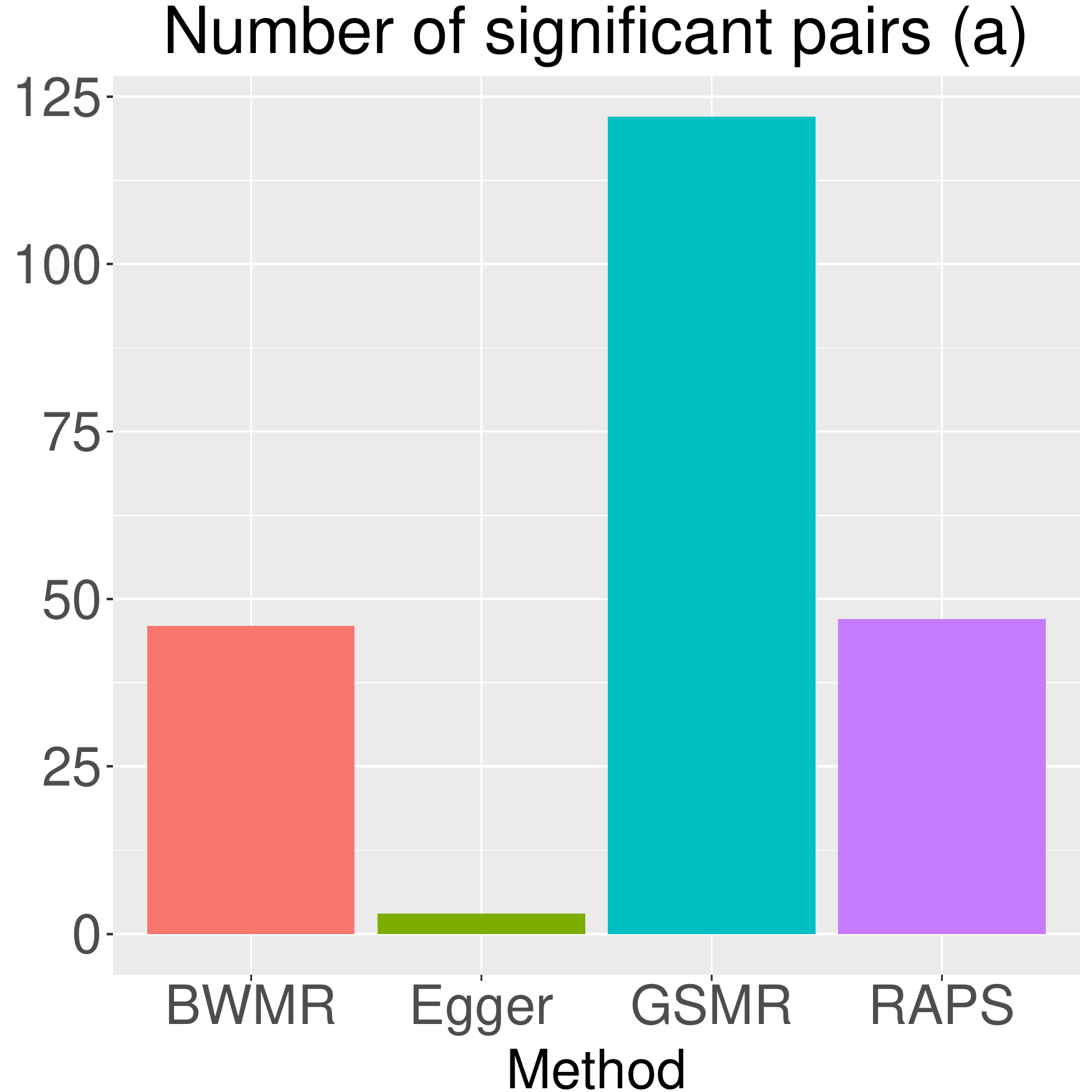}
  \includegraphics[width=0.22\textwidth]{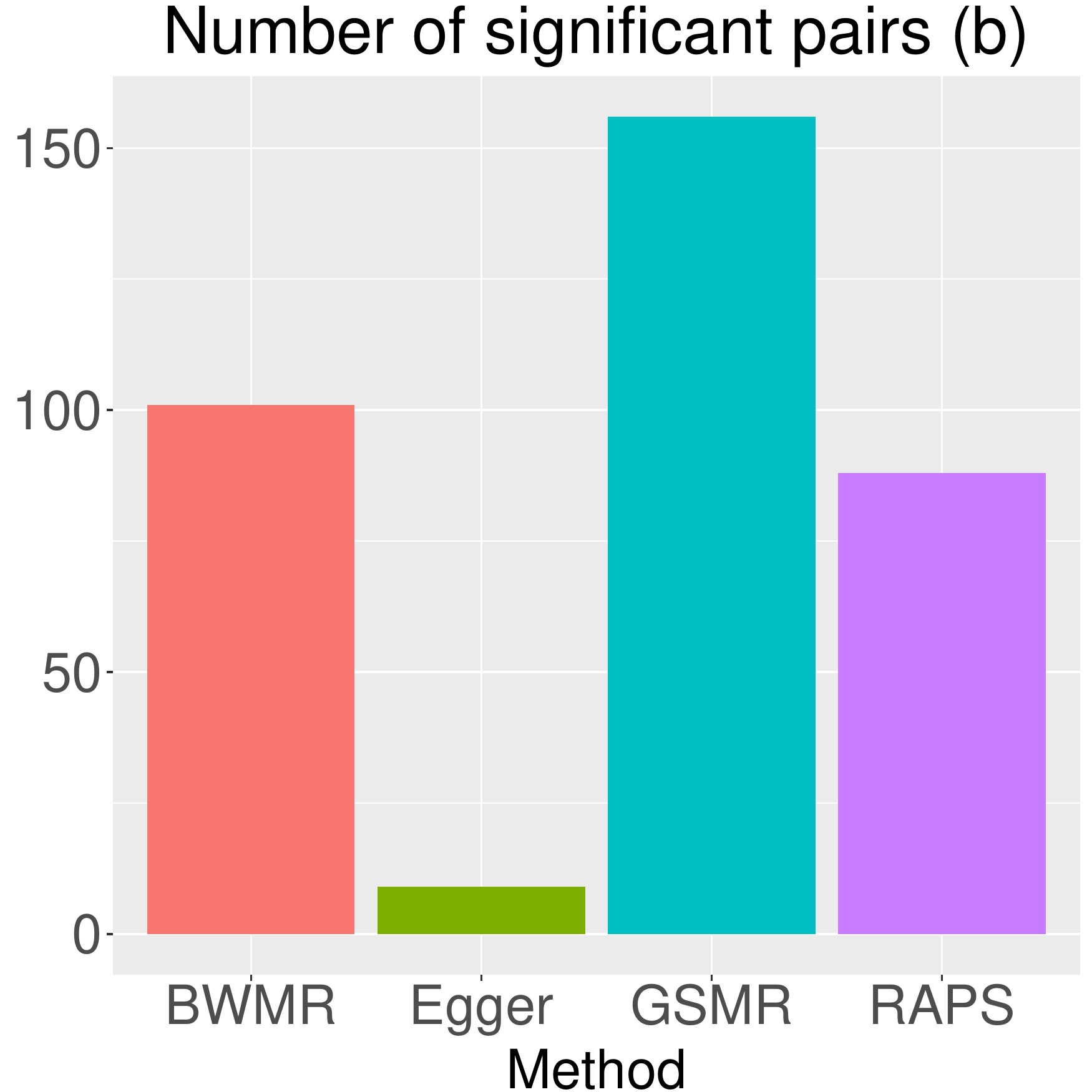}
  \includegraphics[width=0.23\textwidth]{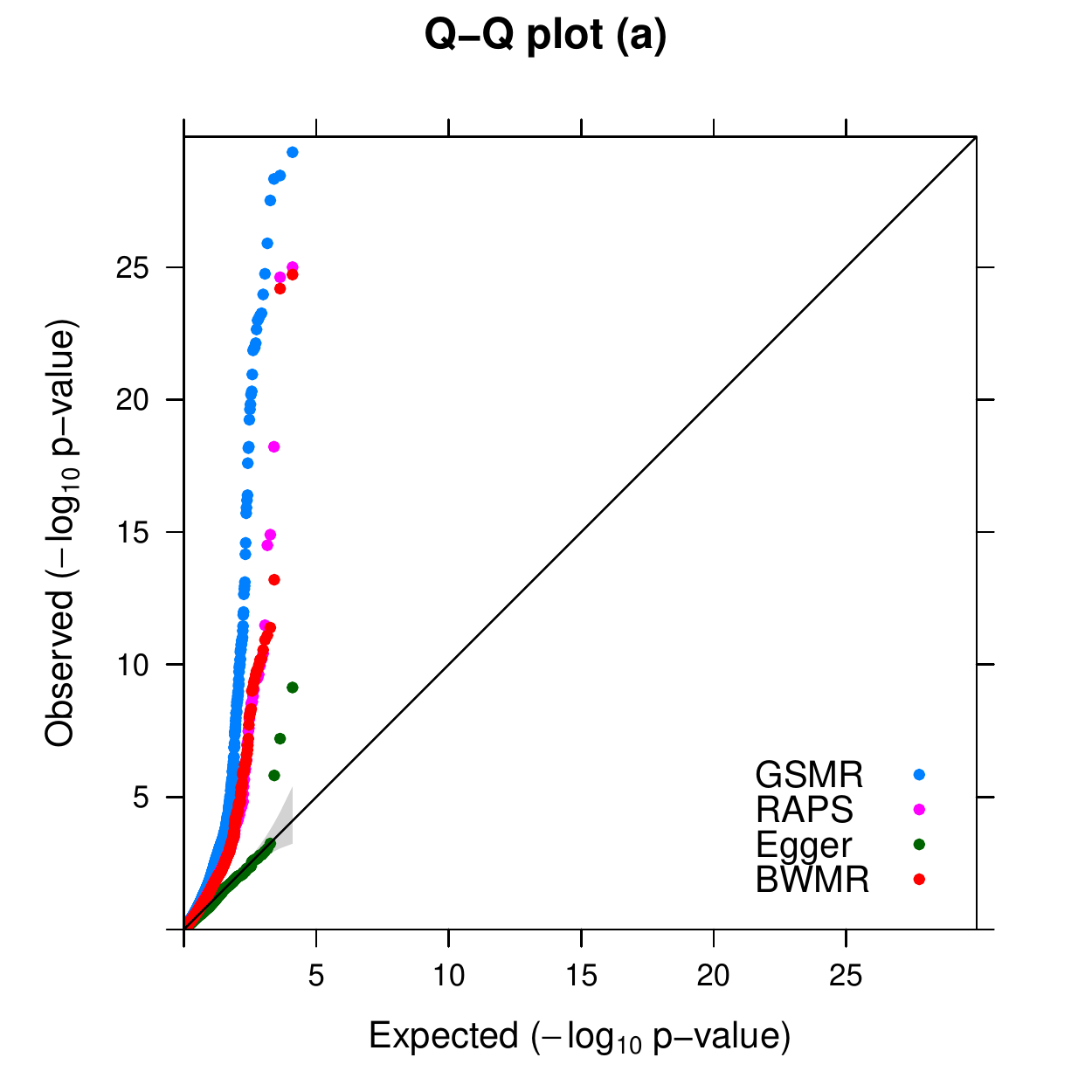}
  \includegraphics[width=0.23\textwidth]{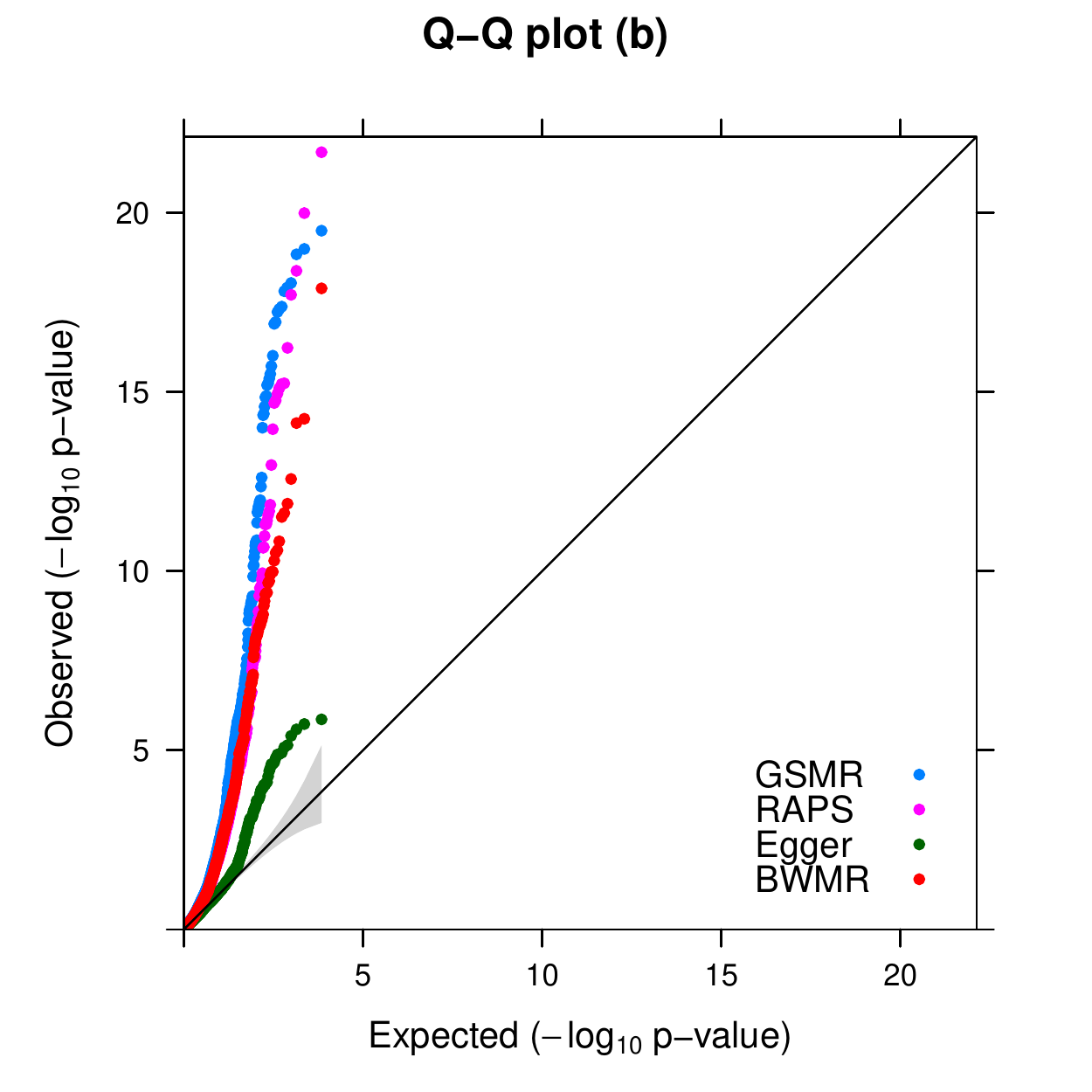}
  \caption{Left: numbers of significant causal effects of metabolites on complex traits; Middle Left: numbers of significant causal effects of complex traits on metabolites; Middle Right: QQ plot of $p$-values based on estimates of causal effects of metabolites on complex traits; Right: QQ plot of $p$-values based on estimates of causal effects of complex traits on metabolites.}
  \label{fig: n_sig-qqplot}
\end{figure}

After carefully examining the selection bias issue and consistency of four MR methods, it is ready for us to apply BWMR and the other three MR methods to estimate: (a) the causal effects of metabolites on complex human traits, and (b) the causal effects of complex traits on metabolites. The analysis results are summarized in supplementary Figs. 37-49.

We observed a protective effect of HDL-C against the trait dyslipidemia ($\hat{\beta}=-0.20$, $p\mbox{-value}=1.36\times 10^{-5}$), and positive effects of LDL-C ($\hat{\beta}=0.62$, $p\mbox{-value}=2.64\times 10^{-128}$) and TG ($\hat{\beta}=0.37$, $p\mbox{-value}=1.26\times 10^{-22}$) on dyslipidemia. Note that the diagnostic criterion for dyslipidemia includes abnormally low level of HDL-C and abnormally high levels of LDL-C and / or TG \citep{Teramoto2007}. Our observations are consistent with the diagnostic criterion, indicating the reliability of our real data analysis results.

Besides those identified significant causal effects between metabolites and dyslipidemia, we identified 46, 101 significant associations based on Bonferroni correction at level 0.05 in (a), (b) respectively. Among these significant findings, some previous research outcomes are successfully replicated in our analysis. For example, we observed LDL-C had a significant risk effect on CAD ($\hat{\beta}=0.13$, $p\mbox{-value}=1.90\times 10^{-25}$) as confirmed by RCTs \citep{Ference2017}. Similarly, positive causal effects of metabolites included in the class of HDL on CAD were also observed. Serum urate was found to have a positive causal effect on gout ($\hat{\beta}=0.27$, $p\mbox{-value}=1.49\times 10^{-36}$), supporting the result of a previous individual participant data analysis \citep{Dalbeth2018}. Importantly, our results also provided several new insights. For example, we found AD positively affected 47 metabolites (e.g., apoB, metabolites included in LDL and VLDL). As expected, the effect of AD on HDL-C was opposite to its corresponding effect on LDL-C, i.e., a negative effect of AD on HDL-C ($\hat{\beta}=-0.13$, $p\mbox{-value}=9.34\times 10^{-6}$) was identified. We also observed significant causal effects of BMI on 37 metabolites, including 30 positive effects (e.g., effects on TG, serum urate, and metabolites included in VLVL) and seven negative effects on metabolites included in HDL. Interestingly, although AD and BMI were found to significantly affect many metabolites, no metabolites were observed to have causal effects on AD or BMI. Additionally, we found that glycated hemoglobin levels had positive effects on TC ($\hat{\beta}=0.22$, $p\mbox{-value}=4.93\times 10^{-6}$) and LDL-C ($\hat{\beta}=0.19$, $p\mbox{-value}=1.01\times 10^{-5}$). On the contrary, height was found to have negative effects on TC ($\hat{\beta}=-0.07$, $p\mbox{-value}=2.16\times 10^{-10}$) and LDL-C ($\hat{\beta}=-0.05$, $p\mbox{-value}=2.29\times 10^{-7}$).

To have a better understanding of the statistical properties of four MR methods, we compared the numbers of causal associations identified by four MR methods after Bonferroni correction, and conducted qq-plots for $p$-value from the four methods, displayed in Fig. \ref{fig: n_sig-qqplot}. Consistent with our simulation study, the power of GSMR seems to be the highest among the four MR methods. As we explained in simulation, the high power of GSMR comes along with an expensive price, that is, its type I error rate can not be controlled at the nominal level. To be more specific, we offered real data illustrative examples in supplementary Figs. 32, 30. As we can see more clearly, the inflated type I error of GSMR can be attributed to the ignored weak pleiotropic effects and the ad-hoc procedure for outlier detection. Egger was observed to have the lowest statistical power among the four methods. Although the intercept term may help to correct for the bias caused by horizontal pleiotropic effects, the presence of weak pleiotropic effects leads to a large variance of Egger's estimation. In addition, we found that Egger was not robust to large pleiotropic effects (i.e., outliers), and would suffer from improper outlier detection as shown in supplementary Figs. 34 and 35. It seems that the results of BWMR and RAPS are quite similar to each other. However, we found RAPS was numerically unstable in real data analysis (see supplementary Figs. 43-45), while BWMR provided stable estimates.

\section{Conclusion}
We have introduced a statistical approach, BWMR, for causal inference based on summary statistics from GWAS. BWMR can not only accounts for the uncertainty of estimated weak effects and weak horizontal pleiotropic effects, but also adaptively detect outliers due to a few large horizontal pleiotropic effects. Through comprehensive simulations and real data analysis, BWMR is shown to be statistically efficient and computationally stable. As more summary data will become publicly available, BWMR is believed to be of widely use in the future.

\newpage{}
\begin{center}
{\large\bf SUPPLEMENTARY MATERIAL}
\end{center}

\begin{description}

\item[Supplementary Document.pdf] The document includes the detailed derivation of the algorithm for BWMR, comparison of MR methods, more simulation results, sources of GWAS summary statistics datasets used in the this paper, and more real data analysis results. (PDF)

\item[R-package for BWMR] R-package 'BWMR'. The package containing the functions used in BWMR and an example of applying BWMR to real data is available at \url{https://github.com/jiazhao97/BWMR}.

\end{description}

\newpage{}
\renewcommand{\thesubsection}{\Alph{subsection}}
\section*{Supplementary Document}

\subsection{The variational EM algorithm}
Let $D=\{\hat{\boldsymbol{\gamma}},\hat{\boldsymbol{\Gamma}},\boldsymbol{\sigma}_X,\boldsymbol{\sigma}_Y\}$ denote the summary statistics serving as the input data, $\mathbf{z}=\{\beta,\pi_1,\boldsymbol{w},\boldsymbol{\gamma}\}$ be the collection of latent variables, $\boldsymbol{\theta}=\{\tau^2,\sigma^2\}$ be the set of the model parameters to be optimized and $\boldsymbol{h}_0=\{\sigma_0=1\times 10^6,a_0=100\}$ be the collection of fixed hyparameters. Then according to BWMR, the logarithm of complete-data
likelihood is given as
$$
  \begin{aligned}
  & \log\Pr(\hat{\boldsymbol\gamma}, {\hat{\boldsymbol{\Gamma}}}, \mathbf{z} | {\boldsymbol{\sigma}}_X^2, {\boldsymbol{\sigma}}_Y^2, \boldsymbol{\theta}, \boldsymbol{h}_0) \\
  = & \sum\limits_{j=1}^N \left[-\frac{\log(\sigma_{X_j}^2)}{2} - \frac{(\hat{\gamma}_j-\gamma_j)^2}{2\sigma_{X_j}^2}\right] \\
  & + \sum\limits_{j=1}^N w_j\left[ -\frac{\log(2\pi)}{2} - \frac{\log(\sigma_{Y_j}^2+\tau^2)}{2} - \frac{(\hat{\Gamma}_j-\beta\gamma_j)^2}{2(\sigma_{Y_j}^2+\tau^2)}\right] \\
  & - \frac{\beta^2}{2\sigma_0^2} + \sum\limits_{j=1}^N \left[ -\frac{\log(\sigma^2)}{2} - \frac{\gamma_j^2}{2\sigma^2} \right] \\
  & + \sum\limits_{j=1}^N \left[ w_j\log(\pi_1) + (1-w_j)\log(1-\pi_1) \right] \\
  & + (a_0-1)\log(\pi_1) + \mbox{constant}.
  \end{aligned}
$$
To provide accurate statistical inference, we are interested in posterior distribution of the latent variables:
\begin{align*}
\Pr(\mathbf{z}|D,\boldsymbol{\theta},\boldsymbol{h}_0)=\frac{\Pr(\hat{\boldsymbol\gamma}, {\hat{\boldsymbol{\Gamma}}}, \mathbf{z}| {\boldsymbol{\sigma}}_X^2, {\boldsymbol{\sigma}}_Y^2, \boldsymbol{\theta}, \boldsymbol{h}_0)}{\Pr(\hat{\boldsymbol\gamma}, {\hat{\boldsymbol{\Gamma}}} | {\boldsymbol{\sigma}}_X^2, {\boldsymbol{\sigma}}_Y^2, \boldsymbol{\theta}, \boldsymbol{h}_0)},
\end{align*}
where 
\begin{align*}
\Pr(\hat{\boldsymbol\gamma}, {\hat{\boldsymbol{\Gamma}}} | {\boldsymbol{\sigma}}_X^2, {\boldsymbol{\sigma}}_Y^2, \boldsymbol{\theta}, \boldsymbol{h}_0) 
= \int_{\mathbf{z}}\Pr(\hat{\boldsymbol\gamma}, {\hat{\boldsymbol{\Gamma}}}, \mathbf{z}| {\boldsymbol{\sigma}}_X^2, {\boldsymbol{\sigma}}_Y^2, \boldsymbol{\theta}, \boldsymbol{h}_0)\mbox{d}\mathbf{z}.
\end{align*}
However, exact evaluation of the posterior distribution is very challenging because the integration is intractable.

Instead, we propose a variational expectation-maximization (VEM) algorithm to approximate the posterior. Let $q(\mathbf{z})$ be a variational distribution. The logarithm of the marginal likelihood can be written as
  \begin{align*}
  & \log\Pr(\hat{\boldsymbol\gamma}, {\hat{\boldsymbol{\Gamma}}} | {\boldsymbol{\sigma}}_X^2, {\boldsymbol{\sigma}}_Y^2, \boldsymbol{\theta}, \boldsymbol{h}_0)\\
  & = \mathbb{E}_{q(\mathbf{z})}[\log\Pr(\hat{\boldsymbol\gamma}, {\hat{\boldsymbol{\Gamma}}} | {\boldsymbol{\sigma}}_X^2, {\boldsymbol{\sigma}}_Y^2, \boldsymbol{\theta}, \boldsymbol{h}_0)]\\
  & = \mathbb{E}_{q(\mathbf{z})}\left[\log\frac{\Pr(\hat{\boldsymbol\gamma}, {\hat{\boldsymbol{\Gamma}}}, \mathbf{z}| {\boldsymbol{\sigma}}_X^2, {\boldsymbol{\sigma}}_Y^2, \boldsymbol{\theta}, \boldsymbol{h}_0)}{q(\mathbf{z})}
  + \log\frac{q(\mathbf{z})}{\Pr(\mathbf{z}|D,\boldsymbol{\theta},\boldsymbol{h}_0)}\right]\\
  &= \mathcal{L}(q,\boldsymbol{\theta})+\mbox{D}_{KL}(q(\mathbf{z})\|\Pr(\mathbf{z}|D,\boldsymbol{\theta},\boldsymbol{h}_0)),\\
  \end{align*}
  where 
  $$\begin{aligned}
  & \mathcal{L}(q,\boldsymbol{\theta})
  :=\mathbb{E}_{q(\mathbf{z})}\left[\log\frac{\Pr(\hat{\boldsymbol\gamma}, {\hat{\boldsymbol{\Gamma}}}, \mathbf{z}| {\boldsymbol{\sigma}}_X^2, {\boldsymbol{\sigma}}_Y^2, \boldsymbol{\theta}, \boldsymbol{h}_0)}{q(\mathbf{z})}\right],\\
  & \mbox{D}_{KL}(q(\mathbf{z})\|\Pr(\mathbf{z}|D,\boldsymbol{\theta},\boldsymbol{h}_0))
  = \mathbb{E}_{q(\mathbf{z})}\left[\log\frac{q(\mathbf{z})}{\Pr(\mathbf{z}|D,\boldsymbol{\theta},\boldsymbol{h}_0)}\right].
  \end{aligned}$$
Given that the Kullback-Leibler (KL) divergence $\mbox{D}_{KL}(\cdot\|\cdot)$ is non-negative, $\mathcal{L}(q;\boldsymbol{\theta})$ is an evidence lower bound (ELBO) of marginal likelihood. Thus, maximization of ELBO $\mathcal{L}(q;\boldsymbol{\theta})$ w.r.t. variational distribution $q$ and parameters $\boldsymbol{\theta}$ is commonly referred to as E-step and M-step: In the E-step, variational distribution $q$ is updated to approximate the true posterior distribution. In the M-step, the set of model parameters $\bftheta$ is optimized to increase EBLO and thus increase the marginal likelihood.

\subsubsection{E-step}
We adopt mean-field variational Bayes (MFVB) and assume that $q(\mathbf{z})$ can be factorized as
$$
q(\mathbf{z}) = q(\beta)q(\pi_1)\prod\limits_{j=1}^N q(\gamma_j)\prod\limits_{j=1}^N q(w_j).
$$
Then to optimize ELBO $\mathcal{L}(q,\boldsymbol{\theta})$ w.r.t. $q(\beta)$, we re-arrange it into the terms with and without $q(\beta)$:
\begin{equation}
\begin{aligned}
& \mathcal{L}(q,\boldsymbol{\theta})\\
=& \mathbb{E}_{q(\mathbf{z})}\left[\log\frac{\Pr(\hat{\boldsymbol\gamma}, {\hat{\boldsymbol{\Gamma}}}, \mathbf{z}| {\boldsymbol{\sigma}}_X^2, {\boldsymbol{\sigma}}_Y^2, \boldsymbol{\theta}, \boldsymbol{h}_0)}{q(\mathbf{z})}\right]\\
=& \int q(\beta)q(\pi_1)\prod\limits_{j=1}^N q(\gamma_j)\prod\limits_{j=1}^N q(w_j)[\log\Pr(\hat{\boldsymbol\gamma}, {\hat{\boldsymbol{\Gamma}}}, \mathbf{z}| {\boldsymbol{\sigma}}_X^2, {\boldsymbol{\sigma}}_Y^2, \boldsymbol{\theta}, \boldsymbol{h}_0)-\log q(\beta)]\mbox{d}\mathbf{z}\\
&- \int q(\beta)q(\pi_1)\prod\limits_{j=1}^N q(\gamma_j)\prod\limits_{j=1}^N q(w_j)[\log(\pi_1)+\sum\limits_{j=1}^N(\log q(\gamma_j)+\log q(w_j))]\mbox{d}\mathbf{z}\\
=& \int q(\beta)\left[\mathbb{E}_{q_{-\beta}}(\log\Pr(\hat{\boldsymbol\gamma}, {\hat{\boldsymbol{\Gamma}}}, \mathbf{z}| {\boldsymbol{\sigma}}_X^2, {\boldsymbol{\sigma}}_Y^2, \boldsymbol{\theta}, \boldsymbol{h}_0))-\log q(\beta)\right]\mbox{d}\beta + \mbox{constant},
\end{aligned}
\label{eq: Estep-elbo-1}
\end{equation}
where $q_{-\beta}$ denotes $q(\pi_1)\prod\limits_{j=1}^N q(\gamma_j)\prod\limits_{j=1}^N q(w_j)$ and the constant does not relate to $q(\beta)$. By defining a new distribution 
$$\tilde{p}(\beta)\propto\mathbb{E}_{q_{-\beta}}(\log\Pr(\hat{\boldsymbol\gamma}, {\hat{\boldsymbol{\Gamma}}}, \mathbf{z}| {\boldsymbol{\sigma}}_X^2, {\boldsymbol{\sigma}}_Y^2, \boldsymbol{\theta}, \boldsymbol{h}_0)),$$
we can further summarize Eq. (\ref{eq: Estep-elbo-1}) as
\begin{equation}
\mathcal{L}(q,\boldsymbol{\theta})
=-\mbox{D}_{KL}(q(\beta)\|\tilde{p}(\beta))+\mbox{constant}.
\label{eq: Estep-elbo-2}
\end{equation}
From Eq. (\ref{eq: Estep-elbo-2}) we can see that the optimal $q(\beta)$ is given as
$$q(\beta)=\tilde{p}(\beta)\propto\mathbb{E}_{q_{-\beta}}[\log\Pr(\hat{\boldsymbol\gamma}, {\hat{\boldsymbol{\Gamma}}}, \mathbf{z}| {\boldsymbol{\sigma}}_X^2, {\boldsymbol{\sigma}}_Y^2, \boldsymbol{\theta}, \boldsymbol{h}_0)].$$
Thus, 
$$
  \begin{aligned}
  \log q(\beta)
  = & \mathbb{E}_{q_{-\beta}}[\log\Pr(\hat{\boldsymbol\gamma}, {\hat{\boldsymbol{\Gamma}}}, \mathbf{z}| {\boldsymbol{\sigma}}_X^2, {\boldsymbol{\sigma}}_Y^2, \boldsymbol{\theta}, \boldsymbol{h}_0)] + \mbox{constant} \\
  = & \mathbb{E}_{q_{-\beta}}\left[ - \frac{\beta^2}{2\sigma_0^2} - \sum\limits_{j=1}^N \frac{w_j}{2(\sigma_{Y_j}^2+\tau^2)}(\hat{\Gamma}_j-\beta\gamma_j)^2 \right] + \mbox{constant} \\
  = & \left[ - \frac{1}{2\sigma_0^2} - \sum\limits_{j=1}^N \frac{\mathbb{E}_q(w_j\gamma_j^2)}{2(\sigma_{Y_j}^2+\tau^2)} \right]\beta^2 + \left( \sum\limits_{j=1}^N \frac{\mathbb{E}_q(w_j\gamma_j)\hat{\Gamma}_j}{\sigma_{Y_j}^2+\tau^2} \right)\beta + \mbox{constant},
  \end{aligned}
$$
Because $\log q(\beta)$ is a quadratic form, we know $q(\beta)$ is the density function of an Gaussian distribution $q(\beta)=\mathcal{N}(\mu_{\beta},\sigma_{\beta}^2)$:
$$
  \sigma_{\beta}^2 = \left[ \frac{1}{\sigma_0^2} + \sum\limits_{j=1}^N\frac{\mathbb{E}(w_j\gamma_j^2)}{\sigma_{Y_j}^2+\tau^2} \right]^{-1},\,
  \mu_{\beta} = \left[  \sum\limits_{j=1}^N \frac{\mathbb{E}(w_j\gamma_j)\hat{\Gamma}_j}{\sigma_{Y_j}^2+\tau^2} \right]\sigma_{\beta}^2.
$$ 
Similarly, the optimal solution of $q(\gamma_j)$ is given by 
$$
  \begin{aligned}
  \log q(\gamma_j)
  = & \mathbb{E}_{q_{-\gamma_j}}[\log\Pr(\hat{\boldsymbol\gamma}, {\hat{\boldsymbol{\Gamma}}}, \mathbf{z}| {\boldsymbol{\sigma}}_X^2, {\boldsymbol{\sigma}}_Y^2, \boldsymbol{\theta}, \boldsymbol{h}_0)] + \mbox{constant} \\
  = & \mathbb{E}_{q_{-\gamma_j}}\left[ -\frac{1}{2\sigma_{X_j}^2}(\hat{\gamma}_j-\gamma_j)^2
  - \frac{w_j}{2(\sigma_{Y_j}^2+\tau^2)}(\hat{\Gamma}_j-\beta\gamma_j)^2
  - \frac{1}{2\sigma^2}\gamma_j^2 \right] + \mbox{constant} \\
  = & -\frac{\gamma_j^2-2\hat{\gamma}_j\gamma_j + \hat{\gamma}_j^2}{2\sigma_{X_j}^2}
  - \frac{\mathbb{E}(w_j\beta^2)\gamma_j^2-2\hat{\Gamma}_j\mathbb{E}(w_j\beta)\gamma_j}{2(\sigma_{Y_j}^2+\tau^2)}
  -\frac{\gamma_j^2}{2\sigma^2} + \mbox{constant} \\
  = & \left[ -\frac{1}{2\sigma_{X_j}^2} - \frac{\mathbb{E}_q(w_j\beta^2)}{2(\sigma_{Y_j}^2+\tau^2)} - \frac{1}{2\sigma^2} \right]\gamma_j^2
  + \left[ \frac{\hat{\gamma}_j}{\sigma_{X_j}^2} + \frac{\mathbb{E}_q(w_j\beta)\hat{\Gamma}_j}{\sigma_{Y_j}^2+\tau^2} \right]\gamma_j + \mbox{constant}.
  \end{aligned}
$$   
The quadratic form of $\log q(\gamma_j)$ indicates that $q(\gamma_j)$ is the density of a Gaussian distribution, i.e., $q(\gamma_j)=\mathcal{N}(\mu_{\gamma_j}, \sigma_{\gamma_j}^2)$ with
$$
  \sigma_{\gamma_j}^2 = \left[ \frac{1}{\sigma_{X_j}^2} + \frac{\mathbb{E}(w_j\beta^2)}{\sigma_{Y_j}^2+\tau^2} + \frac{1}{\sigma^2} \right]^{-1},
  \mu_{\gamma_j}^2 = \left[ \frac{\hat{\gamma}_j}{\sigma_{X_j}^2} + \frac{\mathbb{E}(w_j\beta)\hat{\Gamma}_j}{\sigma_{Y_j}^2+\tau^2} \right]\sigma_{\gamma_j}^2.
$$   

For latent variable $\pi_1$, accordingly, we have
$$
  \begin{aligned}
  q(\pi_1) 
  = & \mathbb{E}_{q_{-\pi_1}}[\log\Pr(\hat{\boldsymbol\gamma}, {\hat{\boldsymbol{\Gamma}}}, \mathbf{z}| {\boldsymbol{\sigma}}_X^2, {\boldsymbol{\sigma}}_Y^2, \boldsymbol{\theta}, \boldsymbol{h}_0)] + \mbox{constant}  \\
  = & \mathbb{E}_{q_{-\pi_1}}\left\{ \sum\limits_{j=1}^N [w_j\log\pi_1 + (1-w_j)\log(1-\pi_1)] + (a_0-1)\log\pi_1 \right\} + \mbox{constant} \\
  = & \left[ a_0-1 + \sum\limits_{j=1}^N \mathbb{E}_q(w_j) \right]\log(\pi_1) + \left[ N-\sum\limits_{j=1}^N \mathbb{E}_q(w_j) \right]\log(1-\pi_1) + \mbox{constant}.
  \end{aligned}
$$  
The form of $\log q(\pi_1)$ suggests that $q(\pi_1)$ is the density of a Beta distribution, i.e., $\mbox{Beta}(\pi_1|a,b)$:
$$a=a_0+\sum\limits_{j=1}^N \mathbb{E}_q(w_j),
b = N+1-\sum\limits_{j=1}^N \mathbb{E}_q(w_j).$$

Eventually, for latent variable $w_j$, according to coordinate ascent MFVB,
$$
  \begin{aligned}
  q(w_j)
  = & \mathbb{E}_{q_{-w_j}}[\log\Pr(\hat{\boldsymbol\gamma}, {\hat{\boldsymbol{\Gamma}}}, \mathbf{z}| {\boldsymbol{\sigma}}_X^2, {\boldsymbol{\sigma}}_Y^2, \boldsymbol{\theta}, \boldsymbol{h}_0)] + \mbox{constant} \\
  = & \mathbb{E}_{q_{-w_j}} \left\{
  w_j\left[ -\frac12\log(2\pi) - \frac12\log(\sigma_{Y_j}^2+\tau^2) - \frac{1}{2(\sigma_{Y_j}^2+\tau^2)}(\hat{\Gamma}_j-\beta\gamma_j)^2 \right] \right\} \\
  & + \mathbb{E}_{q_{-w_j}}\left[ w_j\log\pi_1 + (1-w_j)\log(1-\pi_1) \right] + \mbox{constant}\\
  = & w_j\left[ -\frac12\log(2\pi) - \frac12\log(\sigma_{Y_j}^2+\tau^2) - \frac{\mathbb{E}_q(\hat{\Gamma}_j-\beta\gamma_j)^2 }{2(\sigma_{Y_j}^2+\tau^2)} \right] \\
  & + w_j\mathbb{E}_q[\log\pi_1] + (1-w_j)\mathbb{E}[\log(1-\pi_1)] + \mbox{constant}.
  \end{aligned}
$$
Notice that $w_j$ is binary and the constant has no connection with $w_j$, here we derive
$$
    \dfrac{q(w_j=0)}{q(w_j=1)}
    =\dfrac{\exp\{\mathbb{E}[\log(1-\pi_1)]\}}{\exp\left\{-\frac12\log(2\pi) - \frac12\log(\sigma_{Y_j}^2+\tau^2) - \frac{\mathbb{E}_q(\hat{\Gamma}_j-\beta\gamma_j)^2}{2(\sigma_{Y_j}^2+\tau^2)} + \mathbb{E}_q[\log\pi_1]\right\}}.
$$
Thus, $q(w_j)$ is the density of the following Bernoulli distribution: $q(w_j)=\mbox{Bernoulli}(\pi_{w_j})$ with
$$\pi_{w_j} = q(w_j=1) = \mathbb{E}_q(w_j) = \frac{q(w_j=1)}{q(w_j=0)+q(w_j=1)}.$$

Considering the form of the optimal variational distribution, the expectations of latent variables are written as
$$\begin{aligned}
& \mathbb{E}_q(\beta)=\mu_{\beta},\,\mathbb{E}_q(\beta^2)=\mu_{\beta}^2+\sigma_{\beta}^2,\\
& \mathbb{E}_q(\gamma_j)=\mu_{\gamma_j},\,\mathbb{E}_q(\gamma_j^2)=\mu_{\gamma_j}^2+\sigma_{\gamma_j}^2, \\
& \mathbb{E}_q[\log(\pi_1)]=\psi(a)-\psi(a+b),\,\mathbb{E}_q[\log(1-\pi_1)]=\psi(b)-\psi(a+b),\\
& \mathbb{E}_q(w_j)=\pi_{w_j},
\end{aligned}$$
where $\psi(\cdot)$ represents the digamma function. Noticing the independence bewteen latent variables under the variational posterior distribution, finally we derive the updating equations for the variational E-step:
$$\begin{aligned}
& \sigma_{\beta}^2 = \left[ \frac{1}{\sigma_0^2} + \sum\limits_{j=1}^N\frac{\pi_{w_j}(\mu_{\gamma_j}^2+\sigma_{\gamma_j}^2)}{\sigma_{Y_j}^2+\tau^2} \right]^{-1},\,
\mu_{\beta} = \left( \sum\limits_{j=1}^N \frac{\pi_{w_j}\mu_{\gamma_j}\hat{\Gamma}_j}{\sigma_{Y_j}^2+\tau^2} \right)\sigma_{\beta}^2,\\
& \sigma_{\gamma_j}^2 = \left[ \frac{1}{\sigma_{X_j}^2} + \frac{\pi_{w_j}(\mu_{\beta}^2+\sigma_{\beta}^2)}{\sigma_{Y_j}^2+\tau^2} + \frac{1}{\sigma^2} \right]^{-1},\,
\mu_{\gamma_j} = \left( \frac{\hat{\gamma}_j}{\sigma_{X_j}^2} + \frac{\pi_{w_j}\mu_{\beta}\hat{\Gamma}_j}{\sigma_{Y_j}^2+\tau^2} \right)\sigma_{\gamma_j}^2,\\
& a=\alpha+\sum\limits_{j=1}^N \pi_{w_j},\,
b = N+1-\sum\limits_{j=1}^N \pi_{w_j},\\
& \pi_{w_j} = \frac{q_{j1}}{q_{j0}+q_{j1}},
\end{aligned}$$
where
$$\begin{aligned}
    q_{j0} = & \exp[\psi(b) - \psi(a+b)], \\
    q_{j1} = & \exp\left[
    -\frac12\log(2\pi) - \frac12\log(\sigma_{Y_j}^2+\tau^2) - \frac{(\mu_{\beta}^2+\sigma_{\beta}^2)(\mu_{\gamma_j}^2+\sigma_{\gamma_j}^2)-2\mu_{\beta}\mu_{\gamma_j}\hat{\Gamma}_j+\hat{\Gamma}_j^2}{2(\sigma_{Y_j}^2+\tau^2)}
    + \psi(a) - \psi(a+b)
    \right].
\end{aligned}$$

\subsubsection{M-step}
We first evaluate the ELBO $\mathcal{L}=\mathcal{L}(q,\boldsymbol{\theta})$.
$$
  \begin{aligned}
  \mathcal{L} = & \mathbb{E}_q[\log\Pr(\hat{{\boldsymbol{\gamma}}}, \hat{{\boldsymbol{\Gamma}}}, \mathbf{z} | {\boldsymbol{\sigma}}_X^2, {\boldsymbol{\sigma}}_Y^2, \boldsymbol{\theta}, \boldsymbol{h}_0)] - \mathbb{E}_q[\log q(\mathbf{z})] \\
  = & \mathbb{E}_q[\log\Pr(\hat{{\boldsymbol{\gamma}}} |{\boldsymbol{\sigma}}_X^2, {\boldsymbol{\gamma}})] + \mathbb{E}_q[\log\Pr(\hat{{\boldsymbol{\Gamma}}}  |{\boldsymbol{\sigma}}_{Y}^2, \beta, {\boldsymbol{\gamma}}, {\boldsymbol{w}}, \tau^2)] \\
  & + \mathbb{E}_q[\log\Pr(\beta|\sigma_0^2)] + \mathbb{E}_q[\log\Pr({\boldsymbol{\gamma}} | \sigma^2)]
  + \mathbb{E}_q[\log\Pr({\boldsymbol{w}} | \pi_1)] + \mathbb{E}_q[\log\Pr(\pi_1|a_0)] \\
    & - \mathbb{E}_q[\log q(\beta)] - \mathbb{E}_q[\log q({\boldsymbol{\gamma}})] - \mathbb{E}_q[\log q({\boldsymbol{w}})] - \mathbb{E}_q[\log q(\pi_1)],
  \end{aligned}
$$
where 
$$
  \begin{aligned}
    & \mathbb{E}_q[\log\Pr(\hat{{\boldsymbol{\gamma}}}  | {\boldsymbol{\sigma}}_X^2, {\boldsymbol{\gamma}})] \\
    = & \mathbb{E}_q\left[ \sum\limits_{j=1}^N \log\mathcal{N}(\hat{\gamma}_j | \gamma_j, \sigma_{X_j}^2) \right] \\
    = & \mathbb{E}_q\left\{ \sum\limits_{j=1}^N [-\frac12\log(\sigma_{X_j}^2) - \frac{1}{2\sigma_{X_j}^2}(\hat{\gamma}_j-\gamma_j)^2] \right\} + \mbox{constant} \\
    = & \sum\limits_{j=1}^N \left\{-\frac12\log(\sigma_{X_j}^2) - \frac{1}{2\sigma_{X_j}^2}[(\hat{\gamma}_j - \mu_{\gamma_j})^2 + \sigma_{\gamma_j}^2]\right\} + \mbox{constant}, \\
    ~\\
    & \mathbb{E}_q[\log\Pr(\hat{{\boldsymbol{\Gamma}}}  |{\boldsymbol{\sigma}}_{Y}^2, \beta, {\boldsymbol{\gamma}}, {\boldsymbol{w}}, \tau^2)] \\
    = & \mathbb{E}_q\left[ \log\prod\limits_{j=1}^N \mathcal{N}(\hat{\Gamma}_j | \beta\gamma_j, \sigma_{Y_j}^2+\tau^2)^{w_j} \right] \\
    = & \mathbb{E}_q\left\{\sum\limits_{j=1}^N w_j\left[ -\frac12\log(2\pi) - \frac12\log(\sigma_{Y_j}^2+\tau^2) - \frac{(\hat{\Gamma}_j-\beta\gamma_j)^2}{2(\sigma_{Y_j}^2+\tau^2)} \right] \right\} + \mbox{constant} \\
    = & \sum\limits_{j=1}^N \pi_{w_j}\left[ -\frac12\log(2\pi) - \frac12\log(\sigma_{Y_j}^2+\tau^2) \right]
    + \sum\limits_{j=1}^N \pi_{w_j}\left[ - \frac{(\mu_{\beta}^2+\sigma_{\beta}^2)(\mu_{\gamma_j}^2+\sigma_{\gamma_j}^2)-2\mu_{\beta}\mu_{\gamma_j}\hat{\Gamma}_j+\hat{\Gamma}_j^2}{2(\sigma_{Y_j}^2+\tau^2)} \right]\\
    & + \mbox{constant}, \\
    ~\\
    & \mathbb{E}_q[\log\Pr(\beta | \sigma_0^2)] \\
    = & \mathbb{E}_q\left[ \log\mathcal{N}(\beta | 0, \sigma_0^2) \right] \\
    = & \mathbb{E}_q\left[ -\frac12\log(\sigma_0^2) - \frac{1}{2\sigma_0^2}\beta^2 \right] + \mbox{constant} \\
    = & - \frac{1}{2\sigma_0^2}(\mu_{\beta}^2+\sigma_{\beta}^2) + \mbox{constant}, \\
  \end{aligned}
$$
$$
  \begin{aligned} 
    & \mathbb{E}_q[\log\Pr({\boldsymbol{\gamma}} | \sigma^2)] \\
    = & \mathbb{E}_q\left[ \sum\limits_{j=1}^N \log\mathcal{N}(\gamma_j | 0, \sigma^2) \right] \\
    = & \mathbb{E}_q\left\{ \sum\limits_{j=1}^N \left[ -\frac12\log(\sigma^2) - \frac{1}{2\sigma^2}\gamma_j^2 \right] \right\} + \mbox{constant} \\
    = & -\frac{N}{2}\log(\sigma^2) - \frac{1}{2\sigma^2}\sum\limits_{j=1}^N(\mu_{\gamma_j}^2+\sigma_{\gamma_j}^2) + \mbox{constant}, \\
    ~\\
    \mathbb{E}_q[\log\Pr({\boldsymbol{w}} | \pi_1)]
    = & \mathbb{E}_q\left[ \sum\limits_{j=1}^N \log\mbox{Bernoulli}(w_j | \pi_1) \right] \\
    = & \mathbb{E}_q\left\{ \sum\limits_{j=1}^N \left[ w_j\log(\pi_1) + (1-w_j)\log(1-\pi_1) \right] \right\} \\
    = & \sum\limits_{j=1}^N \{\pi_{w_j}[\psi(a)-\psi(a+b)] + (1-\pi_{w_j})[\psi(b)-\psi(a+b)]\}, \\
    ~\\
    \mathbb{E}_q[\log\Pr(\pi_1|a_0)]
    = & \mathbb{E}_q[ \log\mbox{Beta}(\pi_1 | a_0, 1)] \\
    = & \mathbb{E}_q[(a_0-1)\log(\pi_1)] + \mbox{constant} \\
    = & (a_0-1)[\psi(a)-\psi(a+b)] + \mbox{constant},\\
    ~\\
    \mathbb{E}_q[\log q(\beta)]
    = & \mathbb{E}_q\left[\log\mathcal{N}(\beta | \mu_{\beta}, \sigma_{\beta}^2)\right] \\
    = & \mathbb{E}_q\left[-\frac12\log(\sigma_{\beta}^2) - \frac{1}{2\sigma_{\beta}^2}(\beta-\mu_{\beta})^2\right] + \mbox{constant} \\
    = & - \frac12 \log(\sigma_{\beta}^2) + \mbox{constant}, \\
    ~\\
    \mathbb{E}_q[\log q({\boldsymbol{\gamma}})]
    = & \mathbb{E}_q\left[\sum\limits_{j=1}^N \log\mathcal{N}(\gamma_j | \mu_{\gamma_j}, \sigma_{\gamma_j}^2)\right] \\
    = & \mathbb{E}_q\left\{ \sum\limits_{j=1}^N[-\frac12\log(\sigma_{\gamma_j}^2) - \frac{1}{2\sigma_{\gamma_j}^2}(\gamma_j-\mu_{\gamma_j})^2] \right\} + \mbox{constant} \\
    = & - \frac12\sum\limits_{j=1}^N \log(\sigma_{\gamma_j}^2) + \mbox{constant}, \\
    ~\\
  \end{aligned}
$$
$$
  \begin{aligned} 
    \mathbb{E}_q[\log q({\boldsymbol{w}})] 
    = & \mathbb{E}_q\left[\sum\limits_{j=1}^N \log\mbox{Bernoulli}(w_j | \pi_{w_j})\right] \\
    = & \mathbb{E}_q\left\{ \sum\limits_{j=1}^N [w_j\log(\pi_{w_j}) + (1-w_j)\log(1-\pi_{w_j})] \right\} \\
    = & \sum\limits_{j=1}^N \left[\pi_{w_j}\log(\pi_{w_j}) + (1-\pi_{w_j})\log(1-\pi_{w_j})\right], \\
    ~\\
    \mathbb{E}_q[\log q(\pi_1)]
    = & \mathbb{E}_q[\log\mbox{Beta}(\pi_1 | a,b)] \\
    = & \mathbb{E}_q\{(a-1)\log(\pi_1) + (b-1)\log(1-\pi_1) - \log[\mbox{B}(a, b)]\} \\
    = & (a-1)[\psi(a)-\psi(a+b)] + (b-1)[\psi(b)-\psi(a+b)] - \log[\mbox{B}(a, b)].
    \end{aligned}
$$

Now we derive the updating equations for parameters $\sigma^2$ and $\tau^2$.

For parameter $\sigma^2$, the ELBO $\mathcal{L}$ terms with $\sigma^2$ are given as
$$
\mathcal{L}_{[\sigma^2]} = -\frac{N}{2}\log(\sigma^2) - \sum\limits_{j=1}^N \left( \frac{\mu_{\gamma_j}^2+\sigma_{\gamma_j}^2}{2\sigma^2} \right).
$$
Then taking derivative of ELBO $\mathcal{L}$ w.r.t. $\sigma^2$
$$
\dfrac{\partial \mathcal{L}}{\partial \sigma^2} = -\dfrac{N}{2\sigma^2} + \dfrac{1}{2(\sigma^2)^2}\sum\limits_{j=1}^N (\mu_{\gamma_j}^2+\sigma_{\gamma_j}^2).
$$
and setting it to zero gives the updating equation for $\sigma^2$ as
$$
\sigma^2 = \dfrac{\sum\limits_{j=1}^N (\mu_{\gamma_j}^2+\sigma_{\gamma_j}^2)}{N}.
$$

Similarly, for parameter $\tau^2$, we calculate the ELBO $\mathcal{L}$ terms with $\tau^2$ as
$$
\mathcal{L}_{[\tau^2]} = \sum\limits_{j=1}^N \left\{ \pi_{w_j}\left[ -\frac12\log(\sigma_{Y_j}^2+\tau^2) - \frac{(\mu_{\beta}^2+\sigma_{\beta}^2)(\mu_{\gamma_j}^2+\sigma_{\gamma_j}^2)-2\mu_{\beta}\mu_{\gamma_j}\hat{\Gamma}_j+\hat{\Gamma}_j^2}{2(\sigma_{Y_j}^2+\tau^2)} \right] \right\}.
$$
However, directly taking derivative w.r.t. $\tau^2$ and setting it to zero will not give a closed-form
updating equation for $\tau^2$. Now we propose some tricks. Consider the function $f(x, y)=\dfrac{x^2}{y}$ which is jointly convex in $(x, y)$ for $y > 0$ (This is known
as `Quadratic over linear'). Now we
consider function $f(\sigma_{Y_j}^2+(\tau^{(t)})^2, \sigma_{Y_j}^2+\tau^2) = [\sigma_{Y_j}^2+(\tau^{(t)})^2](\sigma_{Y_j}^2+\tau^2)^{-1}[\sigma_{Y_j}^2+(\tau^{(t)})^2]$, and we are going to make use of the convexity of $f(x, y)$: $f(\lambda x_1 + (1-\lambda)x_2, \lambda y_1 + (1-\lambda)y_2) \leq \lambda f(x_1, y_1) + (1-\lambda)f(x_2, y_2)$.
Denote
$$
  \begin{aligned}
  \lambda & = \dfrac{\sigma_{Y_j}^2}{\sigma_{Y_j}^2+(\tau^{(\mathrm{old})})^2}, \quad\quad 1-\lambda = \dfrac{(\tau^{(\mathrm{old})})^2}{\sigma_{Y_j}^2+(\tau^{(\mathrm{old})})^2}, \\
  x_1 & = \dfrac{\sigma_{Y_j}^2+(\tau^{(\mathrm{old})})^2}{\sigma_{Y_j}^2}\sigma_{Y_j}^2, \quad\quad x_2 = \dfrac{\sigma_{Y_j}^2+(\tau^{(\mathrm{old})})^2}{(\tau^{(\mathrm{old})})^2}(\tau^{(\mathrm{old})})^2, \\
  y_1 & = \dfrac{\sigma_{Y_j}^2+(\tau^{(\mathrm{old})})^2}{\sigma_{Y_j}^2}\sigma_{Y_j}^2, \quad\quad y_2 = \dfrac{\sigma_{Y_j}^2+(\tau^{(\mathrm{old})})^2}{(\tau^{(\mathrm{old})})^2}\tau^2.
  \end{aligned}
$$
Then we have
$$
  \begin{aligned}
  & [\sigma_{Y_j}^2+(\tau^{(\mathrm{old})})^2](\sigma_{Y_j}^2+\tau^2)^{-1}[\sigma_{Y_j}^2+(\tau^{(\mathrm{old})})^2]\\
  = & [\lambda x_1 + (1-\lambda)x_2][\lambda y_1 + (1-\lambda)y_2]^{-1}[\lambda x_1 + (1-\lambda)x_2] \\
  = & f(\lambda x_1 + (1-\lambda)x_2, \lambda y_1 + (1-\lambda)y_2) \\
  \leq & \lambda f(x_1, y_1) + (1-\lambda)f(x_2, y_2) \\
  = & \lambda x_1 y_1^{-1} x_1 + (1-\lambda) x_2 y_2^{-1} x_2
  \end{aligned}
$$
It is easy to check
$$
  \begin{aligned}
  \lambda x_1 y_1^{-1} x_1 = & \sigma_{Y_j}^2, \\
  (1-\lambda)x_2 y_2^{-1} x_2 = & \dfrac{(\tau^{(\mathrm{old})})^4}{\tau^2}.
  \end{aligned}
$$
In summary, we derive the following inequality
$$
(\sigma_{Y_j}^2 + \tau^2)^{-1} \leq [\sigma_{Y_j}^2 + (\tau^{(\mathrm{old})})^2]^{-2} \left[\sigma_{Y_j}^2 + \dfrac{(\tau^{(\mathrm{old})})^4}{\tau^2}\right].
$$
For the log term in $\mathcal{L}_{[\tau^2]}$, we also have a bound (first-order approximation to a concave function)
$$
-\log(\sigma_{Y_j}^2+\tau^2) \geq -\log[\sigma_{Y_j}^2+(\tau^{(\mathrm{old})})^2] - [\sigma_{Y_j}^2+(\tau^{(\mathrm{old})})^2]^{-1}[\tau^2-(\tau^{(\mathrm{old})})^2].
$$
Using such two bounds, the ELBO $\mathcal{L}$ terms containing $\tau^2$ are bounded as
$$
  \begin{aligned}
  \mathcal{L}_{[\tau^2]} 
  & \geq\sum\limits_{j=1}^N \left\{ -\dfrac{\pi_{w_j}[\tau^2-(\tau^{(\mathrm{old})})^2]}{2[\sigma_{Y_j}^2+(\tau^{(\mathrm{old})})^2]} 
  -\dfrac{\pi_{w_j}}{2}[(\mu_{\beta}^2+\sigma_{\beta}^2)(\mu_{\gamma_j}^2+\sigma_{\gamma_j}^2)-2\mu_{\beta}\mu_{\gamma_j}\hat{\Gamma}_j+\hat{\Gamma}_j^2]\dfrac{\sigma_{Y_j}^2+\frac{[\tau^{(\mathrm{old})}]^4}{\tau^2}}{[\sigma_{Y_j}^2+(\tau^{(\mathrm{old})})^2]^2}\right\}+\mbox{constant}
\end{aligned}
$$
Taking derivative of such bound of $\mathcal{L}_{[\tau^2]}$ w.r.t. $\tau^2$ and setting it to zero
$$
\sum\limits_{j=1}^N \left\{ -\dfrac{\pi_{w_j}}{2[\sigma_{Y_j}^2+(\tau^{(\mathrm{old})})^2]} 
+ \dfrac{\pi_{w_j}[(\mu_{\beta}^2+\sigma_{\beta}^2)(\mu_{\gamma_j}^2+\sigma_{\gamma_j}^2)-2\mu_{\beta}\mu_{\gamma_j}\hat{\Gamma}_j+\hat{\Gamma}_j^2](\tau^{(\mathrm{old})})^4}{2[\sigma_{Y_j}^2+(\tau^{(\mathrm{old})})^2]^2\tau^4} \right\}=0,
$$
gives the following updating equation for $\tau^2$:
$$
\tau^2 \leftarrow \sqrt{\dfrac{\sum\limits_{j=1}^N \dfrac{\pi_{w_j}[(\mu_{\beta}^2+\sigma_{\beta}^2)(\mu_{\gamma_j}^2+\sigma_{\gamma_j}^2)-2\mu_{\beta}\mu_{\gamma_j}\hat{\Gamma}_j+\hat{\Gamma}_j^2](\tau^{(\mathrm{old})})^4}{[\sigma_{Y_j}^2+(\tau^{(\mathrm{old})})^2]^2}}
{\sum\limits_{j=1}^N \frac{\pi_{w_j}}{\sigma_{Y_j}^2 + (\tau^{(\mathrm{old})})^2}}}.
$$
where $(\tau^{(\mathrm{old})})^2$ is the estimate of $\tau^2$ at previous iteration.

\subsection{Inference}
With the mean field assumption, VEM algorithm can provide accurate posterior mean of $\beta$ \citep{Blei2017}. However, MFVB often underestimates the posterior variance of latent variables \citep{Blei2017}. As inspired by the linear response methods \citep{Giordano2015,Giordano2018}, in this section we propose an exact closed-from formula to correct the underestimated variance, yielding an accurate inference for $\beta$. 

\subsubsection{An example showing the properties of MFVB}
First we give a simple example to vividly show the properties of MFVB. Suppose the true posterior distribution is a multivariate normal distribution 
$$p(\mathbf{z})=\mathcal{N}(\boldsymbol{\mu}_0, \boldsymbol{\Sigma}_0),$$
where $\mathbf{z}\in\mathbb{R}^{M\times 1}$, $\boldsymbol{\mu}_0\in\mathbb{R}^{M\times 1}$, $\boldsymbol{\Sigma}_0\in\mathbb{R}^{M\times M}$. And we use the variational distribution $q(\mathbf{z})$ to approximate it. According to MFVB, we assume that $q(\mathbf{z})$ can be factorized as 
$$q(\mathbf{z})=\prod\limits_{j=1}^M q(z_j).$$
Note that in the framework of VEM and MFVB, the optimal approximation $q^*(\mathbf{z})$ of $p(\mathbf{z})$ is obtained by maximizing ELBO w.r.t. $q$, which is equivalent to minimizing the Kullback-Leibler (KL) divergence ${KL}(q(\mathbf{z})\|p(\mathbf{z}))$ w.r.t. $q$. We obtain
\begin{equation}
{KL}(q(\mathbf{z})\|p(\mathbf{z}))
= \mathbb{E}_{q(\mathbf{z})}\left[\log\frac{q(\mathbf{z})}{p(\mathbf{z})}\right]
= \int \prod\limits_{j=1}^M q(z_j)\left[-\log p(\mathbf{z})+\log\prod\limits_{j=1}^M q(z_j)\right]\mbox{d}\mathbf{z}.
\label{eq: MFVB-kl-div}
\end{equation}
We now seek the minimum of ${KL}(q(\mathbf{z})\|p(\mathbf{z}))$ by making optimization of ${KL}(q(\mathbf{z})\|p(\mathbf{z}))$ w.r.t. $q(z_j),\,j=1,2,...,M$, in turn.  Let $\mathbf{z}_{-j}:=\{z_i\}_{i\neq j}$ and $q_{-j}(\mathbf{z}_{-j}):=\prod\limits_{i\neq j}q(z_i)$, then we re-arrange Eq. (\ref{eq: MFVB-kl-div}) as
$$\begin{aligned}
{KL}(q(\mathbf{z})\|p(\mathbf{z}))
=& \int q(z_j)\int q_{-j}(\mathbf{z}_{-j})[-\log p(\mathbf{z})+\log q_{-j}(\mathbf{z}_{-j})+\log q(z_j)]\mbox{d}\mathbf{z}_{-j}\mbox{d}z_j\\
=& \int q(z_j)\int q_{-j}(\mathbf{z}_{-j})[-\log p(\mathbf{z})+\log q(z_j)]\mbox{d}\mathbf{z}_{-j}\mbox{d}z_j
+\int q_{-j}(\mathbf{z}_{-j})\log q_{-j}(\mathbf{z}_{-j})\mbox{d}\mathbf{z}_{-j}\\
=& \int q(z_j)\left[-\int q_{-j}(\mathbf{z}_{-j})\log p(\mathbf{z})\mbox{d}\mathbf{z}_{-j} + \log q(z_j)\right]\mbox{d}z_j
+\int q_{-j}(\mathbf{z}_{-j})\log q_{-j}(\mathbf{z}_{-j})\mbox{d}\mathbf{z}_{-j}.
\end{aligned}$$
By defining a new distribution $\tilde{p}(z_j)$
\begin{equation}
\log\tilde{p}(z_j):=\int q_{-j}(\mathbf{z}_{-j})\log p(\mathbf{z})\mbox{d}\mathbf{z}_{-j}+\mbox{constant},
\label{eq: MFVB-new-distribution}
\end{equation}
we further derive
$$\begin{aligned}
{KL}(q(\mathbf{z})\|p(\mathbf{z}))
=& \int q(z_j)[-\log\tilde{p}(z_j)+\log q(z_j)]\mbox{d}z_j+\mbox{constant}\\
=& {KL}(q(z_j)\|\tilde{p}(z_j))+\mbox{constant}.
\end{aligned}$$
Given that the Kullback-Leibler (KL) divergence is nonnegative, we know that the optimal $q(z_j)$ is given as
\begin{equation}
q(z_j)=\tilde{p}(z_j).
\label{eq: MFVB-optimal}
\end{equation}
Thus, as $p(\mathbf{z})$ is a multivariate normal distribution, it is reasonable to write the optimal $q(z_j)$ as
\begin{equation}
q(z_j)=\mathcal{N}(\mu_j,\sigma_j^2).
\label{eq: MFVB-assum}
\end{equation}
Finally, from Eqs. (\ref{eq: MFVB-new-distribution},\ref{eq: MFVB-optimal},\ref{eq: MFVB-assum}), we derive
$$\mu_j=\boldsymbol{\mu}_0^{(j)},\,
\sigma_j^2=1/[\boldsymbol{\Sigma}_0^{-1}]^{(j,j)},$$
where $\boldsymbol{\mu}_0^{(j)}$ denotes the $j$-th element of $\boldsymbol{\mu}_0$ and $[\boldsymbol{\Sigma}_0^{-1}]^{(j,j)}$ denotes the element located in the $j$-th row and the $j$-th column of $\boldsymbol{\Sigma}_0^{-1}$. 
Therefore,
\begin{equation}
\begin{aligned}
& \hat{\boldsymbol{\mu}}_0^{(MFVB)} = \boldsymbol{\mu}_0,\\
& \hat{\boldsymbol{\Sigma}}_0^{(MFVB)} = \mbox{diag}\left(1/[\boldsymbol{\Sigma}_0^{-1}]^{(1,1)}, 1/[\boldsymbol{\Sigma}_0^{-1}]^{(2,2)},...,1/[\boldsymbol{\Sigma}_0^{-1}]^{(M,M)}\right).
\label{eq: MFVB-conclusion}
\end{aligned}
\end{equation}
As can be seen from Eq. (\ref{eq: MFVB-conclusion}), MFVB gives accurate mean estimations while it provides no information of covariances between different variables and often gives underestimated variances.

\subsubsection{An example showing the intuition for linear response variational Bayes (LRVB)}
We provides intuition for linear response variational Bayes \citep{Giordano2015} by continuing the above example. For the sake of simplicity, we assume 
$$\begin{aligned}
& p(\mathbf{z})=\mathcal{N}(\boldsymbol{\mu}_0, \boldsymbol{\Sigma}_0),\\
& \mathbf{z}=(z_1,z_2),\\
& \boldsymbol{\mu}_0=(\mu_{0,1},\mu_{0,2}),\\
& \boldsymbol{\Sigma}_0 =
  \begin{bmatrix}
      a_{11} & a_{12}\\
      a_{21} & a_{22}
  \end{bmatrix},\,
  \boldsymbol{\Sigma}_0^{-1} =
  \begin{bmatrix}
      r_{11} & r_{12}\\
      r_{21} & r_{22}
  \end{bmatrix}.\\
& q(\mathbf{z})=q(z_1)q(z_2),\,q(z_1)=\mathcal{N}(\mu_1,\sigma_1^2),\,q(z_2)=\mathcal{N}(\mu_2,\sigma_2^2).
\end{aligned}$$
and let $\boldeta$ be the collection of variational parameters, i.e.,
$$\boldeta = \{\mu_1,\sigma_1^2,\mu_2,\sigma_2^2\}.$$

Here we denote $p_0(\mathbf{z})$ as the true posterior of interest, i.e, $p_0(\mathbf{z})=p(\mathbf{z})$, and we define a perturbation of the posterior as
\begin{equation}
p_{\boldsymbol{t}}(\mathbf{z}):=\frac{p_0(\mathbf{z})\exp(\boldsymbol{t}^T\mathbf{z})}{\int p_0(\mathbf{z})\exp(\boldsymbol{t}^T\mathbf{z})\mbox{d}\mathbf{z}}.
\label{eq: LRVB-pert}
\end{equation}
Recall that $p_0(\mathbf{z})=\mathcal{N}(\boldsymbol{\mu}_0,\boldsymbol{\Sigma}_0)$, Eq. (\ref{eq: LRVB-pert}) implies 
\begin{equation}
\begin{aligned}
\log p_{\boldsymbol{t}}(\mathbf{z})
=& -\frac12(\mathbf{z}-\boldsymbol{\mu}_0)^T\boldsymbol{\Sigma}_0^{-1}(\mathbf{z}-\boldsymbol{\mu}_0)+\boldsymbol{t}^T\mathbf{z}+\mbox{constant}\\
=& -\frac12\mathbf{z}^T\boldsymbol{\Sigma}_0^{-1}\mathbf{z} + (\boldsymbol{\mu}_0^T\boldsymbol{\Sigma}_0^{-1}+\boldsymbol{t}^T)\mathbf{z}+\mbox{constant}\\
=& -\frac12[\mathbf{z}-(\boldsymbol{\mu}_0+\boldsymbol{\Sigma}_0\boldsymbol{t})]^T\boldsymbol{\Sigma}_0^{-1}[\mathbf{z}-(\boldsymbol{\mu}_0+\boldsymbol{\Sigma}_0\boldsymbol{t})]+\mbox{constant}.
\label{eq: LRVB-quadratic-norm}
\end{aligned}
\end{equation}
The quadratic form in Eq. (\ref{eq: LRVB-quadratic-norm}) yields 
\begin{equation}
p_{\boldsymbol{t}}(\mathbf{z})
=\mathcal{N}(\boldsymbol{\mu}_0+\boldsymbol{\Sigma}_0\boldsymbol{t},\boldsymbol{\Sigma}_0),
\label{eq: LRVB-pert-norm}
\end{equation}
implying 
\begin{equation}
\mbox{Cov}_{p(\mathbf{z})}[\mathbf{z}]=\boldsymbol{\Sigma}_0=\left.\frac{\partial\mathbb{E}_{p_{\boldsymbol{t}(\mathbf{z})}}[\mathbf{z}]}{\partial\boldsymbol{t}^T}\right|_{\boldt=\boldsymbol{0}}.
\label{eq: LRVB-norm-intuition}
\end{equation}
Eq. (\ref{eq: LRVB-norm-intuition}) shows that the sensitivity of posterior mean at $\boldt=\boldsymbol{0}$ can provide information about the true posterior variance.

We now introduce the key idea to approximate the posterior covariance matrix in Eq. (\ref{eq: LRVB-norm-intuition}) from MFVB. Let $q_\boldt(\mathbf{z})$ be the mean field approximation to $p_\boldt(\mathbf{z})$, i.e., 
$$q_\boldt(\mathbf{z}):=q(\mathbf{z};\boldeta^*_\boldt)=\mbox{arg}\,\mbox{min}_q KL(q(\mathbf{z};\boldeta)\|p_\boldt(\mathbf{z})).$$
Since MFVB often provides accurate inference on posterior mean \citep{Blei2017}, we assume the following conditions hold:
\begin{equation}
\begin{aligned}
&\mbox{Condition 1: }
\mathbb{E}_{q_\boldt}[\mathbf{z}]\approx\mathbb{E}_{p_\boldt}[\mathbf{z}]\mbox{ for all }\boldt,\mbox{ and}\\
&\mbox{Condition 2: }
\left.\frac{\partial\mathbb{E}_{q_{\boldsymbol{t}}}[\mathbf{z}]}{\partial\boldsymbol{t}^T}\right|_{\boldsymbol{t}=\boldsymbol{0}}
\approx\left.\frac{\partial\mathbb{E}_{p_{\boldsymbol{t}}}[\mathbf{z}]}{\partial\boldsymbol{t}^T}\right|_{\boldsymbol{t}=\boldsymbol{0}}.
\label{eq: conditions}
\end{aligned}
\end{equation}
Condition 1 says that MFVB can provide good approximations to posterior means of all the perturbations. Condition 2 further requires the accuracy of the first order approximation at $\boldt=\boldsymbol{0}$. Consequently, from Eq. (\ref{eq: LRVB-norm-intuition}) and Condition 2 in Eq. (\ref{eq: conditions}), we know that the posterior covariance matrix $\boldsymbol{\Sigma}_0$ can be approximated by
\begin{equation}
\boldsymbol{\Sigma}_0
\approx\left.\frac{\partial\mathbb{E}_{q_\boldt}[\mathbf{z}]}{\partial\boldt}\right|_{\boldt=\boldsymbol{0}}
= \left.\frac{\mbox{d}\mathbb{E}_{q_\boldt}[\mathbf{z}]}{\mbox{d}\boldeta}\right|_{\boldeta=\boldeta_0^*}
\left.\frac{\mbox{d}\boldeta^*_{\boldt}}{\mbox{d}\boldt}\right|_{\boldt=\boldsymbol{0}},
\label{eq: LRVB-estimator}
\end{equation}
where $\boldeta_0^*$ is exactly the MFVB solution obtained in Section 2.1.

We now use properties of MFVB solution to calculate $\left.\frac{\mbox{d}\boldeta^*_{\boldt}}{\mbox{d}\boldt}\right|_{\boldt=\boldsymbol{0}}$. Importantly, note that $KL(q(\mathbf{z;\boldeta})\|p_{\boldsymbol{t}}(\mathbf{z}))$ can be viewed as a function of $\boldsymbol{t}$, $\boldeta$, and using $q(\mathbf{z};\boldeta)$ to approximate $p_{\boldsymbol{t}}(\mathbf{z})$ in the framework of MFVB is equivalent to minimizing ${KL}(q(\mathbf{z;\boldeta})\|p_{\boldsymbol{t}}(\mathbf{z}))$ w.r.t. $\boldeta$. The minimization yeilds
$$\left.\frac{\partial{KL}(q(\mathbf{z;\boldeta})\|p_{\boldsymbol{t}}(\mathbf{z}))}{\partial\boldeta}\right|_{\boldeta=\boldeta^*_{\boldsymbol{t}}}
=\boldsymbol{0}.$$
By taking derivative w.r.t. $\boldsymbol{t}$, we then derive
$$\frac{\mbox{d}}{\mbox{d}\boldsymbol{t}}
\left(\left.\frac{\partial{KL}(q(\mathbf{z;\boldeta})\|p_{\boldsymbol{t}}(\mathbf{z}))}{\partial\boldeta}\right|_{\boldeta=\boldeta^*_{\boldsymbol{t}}}\right)
=\boldsymbol{0},$$
yeilding 
\begin{equation}
\left.\frac{\partial^2{KL}(q(\mathbf{z;\boldeta})\|p_{\boldsymbol{t}}(\mathbf{z}))}{\partial\boldeta\partial\boldeta^T}\right|_{\boldeta=\boldeta^*_{\boldsymbol{t}}}
\frac{\mbox{d}\boldeta^*_{\boldsymbol{t}}}{\mbox{d}\boldsymbol{t}}
+\left.\frac{\partial^2{KL}(q(\mathbf{z;\boldeta})\|p_{\boldsymbol{t}}(\mathbf{z}))}{\partial\boldt\partial\boldeta^T}\right|_{\boldeta=\boldeta^*_{\boldsymbol{t}}}
=\boldsymbol{0}.
\label{eq: LRVB-chain-rule}
\end{equation}
The strict minimization of ${KL}(q(\mathbf{z;\boldeta})\|p_{\boldsymbol{t}}(\mathbf{z}))$ requires that $\left.\frac{\partial^2{KL}(q(\mathbf{z;\boldeta})\|p_{\boldsymbol{t}}(\mathbf{z}))}{\partial\boldeta\partial\boldeta^T}\right|_{\boldeta=\boldeta^*_{\boldsymbol{t}}}$ is a positive definite matrix. So we can evaluate Eq. (\ref{eq: LRVB-chain-rule}) at $\boldsymbol{t}=\boldsymbol{0}$ and solve it to find that 
\begin{equation}
\left.\frac{\mbox{d}\boldeta^*_{\boldsymbol{t}}}{\mbox{d}\boldsymbol{t}^T}\right|_{\boldt=\boldsymbol{0}}
=-\left.\left(\left.\frac{\partial^2{KL}(q(\mathbf{z;\boldeta})\|p_{\boldsymbol{t}}(\mathbf{z}))}{\partial\boldeta\partial\boldeta^T}\right|_{\boldeta=\boldeta^*_0}\right)^{-1}
\frac{\partial^2{KL}(q(\mathbf{z;\boldeta})\|p_{\boldsymbol{t}}(\mathbf{z}))}{\partial\boldt\partial\boldeta^T}
\right|_{\boldeta=\boldeta^*_0,\boldt=\boldsymbol{0}}.
\label{eq: LRVB-eta-t}
\end{equation}
As the KL divergence is written as
$${KL}(q(\mathbf{z;\boldeta})\|p_{\boldsymbol{t}}(\mathbf{z}))
=\mathbb{E}_{q(\mathbf{z};\boldeta)}[\log q(\mathbf{z};\boldeta)-\log p(\mathbf{z})-\boldt^T\mathbf{z}]+\mbox{constant},$$
where the constant is not relevant to $\boldeta$, the term $\left.\frac{\partial^2{KL}(q(\mathbf{z;\boldeta})\|p_{\boldsymbol{t}}(\mathbf{z}))}{\partial\boldt\partial\boldeta^T}
\right|_{\boldeta=\boldeta^*,\boldt=\boldsymbol{0}}$ can be further summarized as
\begin{equation}
\left.\frac{\partial^2{KL}(q(\mathbf{z;\boldeta})\|p_{\boldsymbol{t}}(\mathbf{z}))}{\partial\boldt\partial\boldeta^T}
\right|_{\boldeta=\boldeta_0^*,\boldt=\boldsymbol{0}}
=-\left.\frac{\partial\mathbb{E}_{q(\mathbf{z};\boldeta)}[\mathbf{z}]}{\partial\boldeta^T}\right|_{\boldeta=\boldeta_0^*}.
\label{eq: LRVB-kl}
\end{equation}

Finally, Eqs. (\ref{eq: LRVB-estimator},\ref{eq: LRVB-eta-t},\ref{eq: LRVB-kl}) indicate an estimator of the posterior covariance matrix as
$$\hat{\boldsymbol{\Sigma}}_0^{(LRVB)}:=
\left(\left.\frac{\partial\mathbb{E}_{q(\mathbf{z};\boldeta)}[\mathbf{z}]}{\partial\boldeta^T}\right|_{\boldeta=\boldeta^*_0}\right)^T
\left(\left.\frac{\partial^2{KL}(q(\mathbf{z;\boldeta})\|p(\mathbf{z}))}{\partial\boldeta\partial\boldeta^T}\right|_{\boldeta=\boldeta^*_0}\right)^{-1}
\left(\left.\frac{\partial\mathbb{E}_{q(\mathbf{z};\boldeta)}[\mathbf{z}]}{\partial\boldeta^T}\right|_{\boldeta=\boldeta^*_0}\right).$$

In the above Gaussian example, ${KL}(q(\mathbf{z};\boldeta)\|p(\mathbf{z}))$ is given as 
$$\begin{aligned}
{KL}(q(\mathbf{z};\boldeta)\|p(\mathbf{z}))
=& \mathbb{E}_{q(\mathbf{z};\boldeta)}[\log q(\mathbf{z};\boldeta)]-\mathbb{E}_{q(\mathbf{z};\boldeta)}[\log p(\mathbf{z})]\\
=& -\frac12\log(\sigma_1^2)-\frac12\log(\sigma_2^2)\\
&+ \frac12 r_{11}[(\mu_1-\mu_{0,1})^2+\sigma_1^2]
+ \frac12 r_{22}[(\mu_2-\mu_{0,2})^2+\sigma_2^2]
+ r_{12}(\mu_1-\mu_{0,1})(\mu_2-\mu_{0,2})\\
&+ \mbox{constant}.
\end{aligned}$$
Thus, in this case,
$$\left.\frac{\partial^2{KL}(q(\mathbf{z;\boldeta})\|p(\mathbf{z}))}{\partial\boldeta\partial\boldeta^T}\right|_{\boldeta=\boldeta_0^*}
=\begin{bmatrix}
    r_{11} & 0 & r_{12} & 0 \\
    0 & \frac{1}{2\sigma_1^4} & 0 & 0 \\
    r_{21} & 0 & r_{22} & 0 \\
    0 & 0 & 0 & \frac{1}{2\sigma_2^4}.
\end{bmatrix}$$
Note that we already have
$$\sigma_1^2=1/r_{11},\,\sigma_2^2=1/r_{22},\,
\begin{bmatrix}
    a_{11} & a_{12} \\
    a_{21} & a_{22} 
\end{bmatrix}=
\begin{bmatrix}
    r_{11} & r_{12} \\
    r_{21} & r_{22} 
\end{bmatrix}^{-1}
=\frac{1}{r_{11}r_{22}-r_{12}r_{21}}
\begin{bmatrix}
    r_{22} & -r_{12} \\
    -r_{21} & r_{11} 
\end{bmatrix}.$$
With 
$$\left.\frac{\partial\mathbb{E}_{q(\mathbf{z};\boldeta)}[\mathbf{z}]}{\partial\boldeta^T}\right|_{\boldeta=\boldeta_0^*}
=\begin{bmatrix}
    1 & 0 & 0 & 0 \\
    0 & 0 & 1 & 0
\end{bmatrix}^T,$$
We finally calculate $\hat{\boldsymbol{\Sigma}}_0^{(LRVB)}$ as
$$\hat{\boldsymbol{\Sigma}}_0^{(LRVB)}
=\begin{bmatrix}
    1 & 0 & 0 & 0 \\
    0 & 0 & 1 & 0
\end{bmatrix}
\begin{bmatrix}
    r_{11} & 0 & r_{12} & 0 \\
    0 & \frac{1}{2\sigma_1^4} & 0 & 0 \\
    r_{21} & 0 & r_{22} & 0 \\
    0 & 0 & 0 & \frac{1}{2\sigma_2^4}.
\end{bmatrix}^{-1}
\begin{bmatrix}
    1 & 0 & 0 & 0 \\
    0 & 0 & 1 & 0
\end{bmatrix}^T
=\begin{bmatrix}
    a_{11} & a_{12} \\
    a_{21} & a_{22} 
\end{bmatrix}=\boldsymbol{\Sigma}_0.
$$
This Gaussian example not only gives the intuition of LRVB, but also helps to show that LRVB can indeed correct the covariance estimation provided by MFVB.

\subsubsection{Inference of posterior variance of one specific latent variable}
Note that, for BWMR model, in the inference part we only seek to derive the posterior variance of latent variable $\beta$, i.e., we only care about the inference of one specific latent variable among all latent variables. Before implementing inference for BWMR, we continue the Gaussian example to get more intuition. 

Let $\boldt=(t_1,0)$ in $p_{\boldsymbol{t}}(\mathbf{z})=\mathcal{N}(\boldsymbol{\mu}_0+\boldsymbol{\Sigma}_0\boldsymbol{t},\boldsymbol{\Sigma}_0)$ (Eq. (\ref{eq: LRVB-pert-norm})), we then have
\begin{equation}
p_{t_1}(\mathbf{z})=\mathcal{N}((\mu_{0,1}+a_{11}t_1,\mu_{0,2}+a_{12}t_1),\boldsymbol{\Sigma}_0).
\label{eq: LRVB-one-new-posterior}
\end{equation}
It is not hard to find out that Eq. (\ref{eq: LRVB-one-new-posterior}) corresponds to the perturbation 
$$p_{t_1}(\mathbf{z})=\frac{p(\mathbf{z})\exp(t_1 z_1)}{\int p(\mathbf{z})\exp(t_1 z_1)\mbox{d}\mathbf{z}}.$$
Eq. (\ref{eq: LRVB-one-new-posterior}) indicates an approximation to $\mbox{Var}(z_1)$:
$$\mbox{Var}(z_1)=a_{11}=\left.\frac{d\mathbb{E}_{p_{t_1}}[z_1]}{d t_1}\right|_{t_1=0}
\approx\left.\frac{d\mathbb{E}_{q_{t_1}}[z_1]}{d t_1}\right|_{t_1=0}=\widehat{\mbox{Var}}(z_1),$$
where $q_{t_1}(\mathbf{z})$ is the optimal solution of variational distribution used to approximate $p_{t_1}(\mathbf{z})$ in the framework of MFVB. Then
\begin{equation}
\widehat{\mbox{Var}}(z_1)
=\left.\frac{\partial\mathbb{E}_{q_{t_1}}[z_1]}{\partial\boldeta}\right|_{\boldeta=\boldeta_0^*}\left.\frac{\mbox{d}\boldeta_{t_1}^*}{\mbox{d}t_1}\right|_{t_1=0},
\label{eq: LRVB-one-1}
\end{equation}
where $\boldeta^*_{t_1}$ is the optimal solution of $\boldeta$ when approximating $p_{t_1}(\mathbf{z})$, and $\boldeta_0^*=\boldeta^*_{t_1}|_{t_1=0}$ still represents the optimal solution of $\boldeta$ when approximating true posterior $p_0(\mathbf{z})=p(\mathbf{z})$.

To conveniently derive $\left.\frac{\mbox{d}\boldeta_{t_1}^*}{\mbox{d}t_1}\right|_{t_1=0}$, similar as Eqs. (\ref{eq: LRVB-chain-rule},\ref{eq: LRVB-eta-t}), we now have 
$$\left.\frac{\partial^2{KL}(q(\mathbf{z;\boldeta})\|p_{t_1}(\mathbf{z}))}{\partial\boldeta\partial\boldeta^T}\right|_{\boldeta=\boldeta^*_{t_1}}
\frac{\mbox{d}\boldeta^*_{t_1}}{\mbox{d}t_1}
+\left.\frac{\partial^2{KL}(q(\mathbf{z;\boldeta})\|p_{t_1}(\mathbf{z}))}{\partial t_1\partial\boldeta^T}\right|_{\boldeta=\boldeta^*_{t_1}}
=\boldsymbol{0},$$
indicating
\begin{equation}
\left.\frac{\mbox{d}\boldeta^*_{t_1}}{\mbox{d}t_1}\right|_{t_1=0}
=-\left.\left(\left.\frac{\partial^2{KL}(q(\mathbf{z;\boldeta})\|p_{t_1}(\mathbf{z}))}{\partial\boldeta\partial\boldeta^T}\right|_{\boldeta=\boldeta^*_0,t_1=0}\right)^{-1}
\frac{\partial^2{KL}(q(\mathbf{z;\boldeta})\|p_{t_1}(\mathbf{z}))}{\partial t_1\partial\boldeta^T}
\right|_{\boldeta=\boldeta_0^*,\,t_1=0}.
\label{eq: LRVB-one-2}
\end{equation}
Then similar as the derivation of Eq. (\ref{eq: LRVB-kl}), given that now the KL divergence ${KL}(q(\mathbf{z;\boldeta})\|p_{t_1}(\mathbf{z}))$ is written as 
$${KL}(q(\mathbf{z;\boldeta})\|p_{t_1}(\mathbf{z}))
=\mathbb{E}_{q(\mathbf{z};\boldeta)}[\log q(\mathbf{z};\boldeta)-\log p(\mathbf{z})-t_1 z_1]+\mbox{constant},$$
where the constant is not relevant to $\boldeta$, we summary the term $\left.\frac{\partial^2{KL}(q(\mathbf{z;\boldeta})\|p_{t_1}(\mathbf{z}))}{\partial t_1\partial\boldeta^T}
\right|_{\boldeta=\boldeta^*_0,t_1=0}$ as
\begin{equation}
\left.\frac{\partial^2{KL}(q(\mathbf{z;\boldeta})\|p_{t_1}(\mathbf{z}))}{\partial t_1\partial\boldeta^T}
\right|_{\boldeta=\boldeta^*_0,t_1=0}
=-\left.\frac{\partial\mathbb{E}_{q(\mathbf{z};\boldeta)}[z_1]}{\partial\boldeta^T}\right|_{\boldeta=\boldeta_0^*}.
\label{eq: LRVB-one-3}
\end{equation}

From Eq. (\ref{eq: LRVB-one-1},\ref{eq: LRVB-one-2},\ref{eq: LRVB-one-3}), we finally derive
$$\begin{aligned}
\widehat{\mbox{Var}}(z_1)
=& \left(\left.\frac{\partial\mathbb{E}_{q(\mathbf{z};\boldeta)}[z_1]}{\partial\boldeta^T}\right|_{\boldeta=\boldeta^*_0}\right)^T
\left(\left.\frac{\partial^2\mbox{D}_{KL}(q(\mathbf{z;\boldeta})\|p(\mathbf{z}))}{\partial\boldeta\partial\boldeta^T}\right|_{\boldeta=\boldeta^*_0}\right)^{-1}
\left(\left.\frac{\partial\mathbb{E}_{q(\mathbf{z};\boldeta)}[z_1]}{\partial\boldeta^T}\right|_{\boldeta=\boldeta^*_0}\right)\\
=& [1, 0, 0, 0]
\begin{bmatrix}
    r_{11} & 0 & r_{12} & 0 \\
    0 & \frac{1}{2\sigma_1^4} & 0 & 0 \\
    r_{21} & 0 & r_{22} & 0 \\
    0 & 0 & 0 & \frac{1}{2\sigma_2^4}.
\end{bmatrix}^{-1}
[1, 0, 0, 0]^T \\
=& a_{11} \\
=& \mbox{Var}(z_1).
\end{aligned}$$

\subsubsection{Detailed derivations for inference}
For BWMR, $\mathbf{z}$ denotes the latent variables and $D$ denotes the input data. For notation convenience, we denote $p_0(\mathbf{z})$ as the true posterior of interest, i.e., $p_0(\mathbf{z})=\Pr(\mathbf{z}|D)$. As we want to derive the accurate posterior variance of latent $\beta$, inspired by the inference of $z_1$ in the Gaussian example, here we define a perturbation of the posterior $p_0(\mathbf{z})=\Pr(\mathbf{z}|D)$ in BWMR as
$$p_{t}(\mathbf{z})
:=\frac{p_0(\mathbf{z})\exp(t\beta)}{\int p_0(\mathbf{z})\exp(t\beta)\mbox{d}\mathbf{z}}.$$
Let 
\begin{equation}
C(t):=\log\int p_0(\mathbf{z})\exp(t\beta)\mbox{d}\mathbf{z}.
\label{eq: infer-normalizer}
\end{equation}
Then the perturbation $p_t(\mathbf{z})$ is written as
$$
p_t(\mathbf{z}) = p_0(\mathbf{z})\exp(t\beta-C(t)),
$$
where $C(t)$ normalizes $p_t(\mathbf{z})$.
\begin{itemize}
  \item {\textbf{use the cumulant generating properties}}
\end{itemize}
We introduce a new variable $u$ and seek to derive the cumulant generating function $K_{p_t}(u)$ for $p_t(\mathbf{z})$. From the definition of the cumulant generating function as well as Eq. (\ref{eq: infer-normalizer}), we have
$$\begin{aligned}
K_{p_t}(u)
=& \log\mathbb{E}_{p_t}[\exp(u\beta)]\\
=& \log\int\exp(u\beta)p_0(\mathbf{z})\exp(t\beta-C(t))\mbox{d}\mathbf{z}\\
=& \log\int p_0(\mathbf{z})\exp((u+t)\beta)\mbox{d}\mathbf{z}
-C(t)\\
=& C(u+t)-C(t).
\end{aligned}$$
The properties of the cumulant generating function implies
$$\begin{aligned}
& \mathbb{E}_{p_t}[\beta]
=\left.\frac{\partial K_{p_t}(u)}{\partial u}\right|_{u=0}
=\frac{\mbox{d}C(t)}{\mbox{d}t},\\
& \mbox{Var}_{p_t}[\beta]
= \left.\frac{\partial^2 K_{p_t}(u)}{\partial u^2}\right|_{u=0}
= \frac{\mbox{d}^2C(t)}{\mbox{d}t^2}.
\end{aligned}$$
Therefore we get
\begin{equation}
\mbox{Var}_{\Pr(\mathbf{z}|D)}[\beta]
=\mbox{Var}_{p_0}[\beta]
=\left.\mbox{Var}_{p_t}[\beta]\right|_{t=0}
=\left.\frac{\mbox{d}\mathbb{E}_{p_t}[\beta]}{\mbox{d}t}\right|_{t=0}.
\label{eq: LRVB-insight}
\end{equation}
\begin{itemize}
  \item {\textbf{use the properties of MFVB solution}}
\end{itemize}
Eq. (\ref{eq: LRVB-insight}) provides insight of using the accurate mean estimation from MFVB to correct the posterior variance inference. However, in practice we seek to only implement VEM and MFVB once without any perturbations of the posterior. The usage of MFVB solution properties helps to address this problem. 

The variational distributions from MFVB is
$$q(\mathbf{z})=q(\beta)q(\pi_1)\prod\limits_{j=1}^N q(\gamma_j)\prod\limits_{j=1}^N q(w_j),$$
with
\begin{align*}
& q(\beta) = \mathcal{N}(\mu_{\beta},\sigma_{\beta}^2),\,q(\gamma_j) = \mathcal{N}(\mu_{\gamma_j},\sigma_{\gamma_j}^2),\\
& q(w_j) = \mbox{Bernoulli}(\pi_{w_j}),\,q(\pi_1) = \mbox{Beta}(a, b).
\end{align*}
We now use the variational distribution 
$$q(\mathbf{z};\boldeta)=q(\mathbf{z}),$$
with the collection of variational parameters 
$$\boldeta=\{\mu_\beta,\sigma^2_\beta,\mu_{\gamma_1},\sigma^2_{\gamma_1},\pi_{w_1},...,\mu_{\gamma_j},\sigma^2_{\gamma_j},\pi_{w_j},...,\mu_{\gamma_N},\sigma^2_{\gamma_N},\pi_{w_N},a,b\}$$
to approximate the perturbation $p_{t}(\mathbf{z})$ of the posterior $p_0(\mathbf{z})=\Pr(\mathbf{z}|D)$. Let $q_t(\mathbf{z})=q(\mathbf{z};\boldeta^*_t)$ denotes the optimal solution of variational distribution in MFVB when inferring $p_{t}(\mathbf{z}|D)$, with $\boldeta^*_t$ denoting the corresponding optimal solution of $\boldeta$. Note that $\Pr(\mathbf{z}|D)=p_{t}(\mathbf{z})|_{t=0}$. Let $q_0(\mathbf{z})=q_t(\mathbf{z})|_{t=0}$ and $\boldeta_0^*=\boldeta^*_t|_{t=0}$ refer to the optimal solution in the VEM algorithm for BWMR. Since MFVB often provides accurate inference on posterior mean, similar as in the Gaussian example, here we assume the following conditions hold:
$$\begin{aligned}
&\mbox{Condition 1: }
\mathbb{E}_{q_t}[\beta]\approx\mathbb{E}_{p_t}[\beta]
\mbox{ for all }t,\mbox{ and}\\
&\mbox{Condition 2: }
\left.\frac{\mbox{d}\mathbb{E}_{q_t}[\beta]}{\mbox{d}t}\right|_{t=0}
\approx\left.\frac{\mbox{d}\mathbb{E}_{p_t}[\beta]}{\mbox{d}t}\right|_{t=0}.
\end{aligned}$$
Then we now have
\begin{equation}
\begin{aligned}
\mbox{Var}_{\Pr(\mathbf{z}|D)}[\beta]
=& \left.\frac{\mbox{d}\mathbb{E}_{p_t}[\beta]}{\mbox{d}t}\right|_{t=0}
\approx\left.\frac{\mbox{d}\mathbb{E}_{q_t}[\beta]}{\mbox{d}t}\right|_{t=0}
=:\widehat{\mbox{Var}}_{\Pr(\mathbf{z}|D)}[\beta],\\
\widehat{\mbox{Var}}_{\Pr(\mathbf{z}|D)}[\beta]
=& \left.\frac{\partial\mathbb{E}_{q(\mathbf{z};\boldeta)}[\beta]}{\partial\boldeta}\right|_{\boldeta=\boldeta_0^*}
\left.\frac{\mbox{d}\boldeta^*_t}{\mbox{d}t}\right|_{t=0}.
\end{aligned}
\label{eq: LRVB-BWMR-1}
\end{equation}

Similar as Eq. (\ref{eq: LRVB-one-2}) and Eq. (\ref{eq: LRVB-one-3}), we can derive
\begin{equation}
\begin{aligned}
& \left.\frac{\mbox{d}\boldeta^*_t}{\mbox{d}t}\right|_{t=0}
=-\left.\left(\left.\frac{\partial^2{KL}(q(\mathbf{z;\boldeta})\|p_t(\mathbf{z}|D))}{\partial\boldeta\partial\boldeta^T}\right|_{\boldeta=\boldeta_0^*,\,t=0}\right)^{-1}
\frac{\partial^2{KL}(q(\mathbf{z;\boldeta})\|p_t(\mathbf{z}|D))}{\partial t\partial\boldeta^T}
\right|_{\boldeta=\boldeta_0^*,t=0}, \\
& \left.\frac{\partial^2{KL}(q(\mathbf{z;\boldeta})\|p_t(\mathbf{z}|D))}{\partial t\partial\boldeta^T}
\right|_{\boldeta=\boldeta^*_0,t=0} = 
-\left.\frac{\partial\mathbb{E}_{q(\mathbf{z};\boldeta)}[\beta]}{\partial\boldeta}\right|_{\boldeta=\boldeta_0^*}.
\end{aligned}
\label{eq: LRVB-BWMR-2}
\end{equation}
Recall that the ELBO $\mathcal{L}$ in VEM algorithm can be written as 
$$
\mathcal{L}=-{KL}(q(\mathbf{z};\boldeta)\|\Pr(\mathbf{z}|D))+\mbox{constant},
$$
where the constant is not relevant to $\boldeta$, indicating
\begin{equation}
\frac{\partial^2{KL}(q(\mathbf{z;\boldeta})\|\Pr(\mathbf{z}|D))}{\partial\boldeta\partial\boldeta^T}
=-\frac{\partial^2\mathcal{L}}{\partial\boldeta\partial\boldeta^T}.
\label{eq: LRVB-BWMR-3}
\end{equation}

Combining Eqs. (\ref{eq: LRVB-BWMR-1}, \ref{eq: LRVB-BWMR-2}, \ref{eq: LRVB-BWMR-3}), we finally derive the estimator of posterior variance of $\beta$ as
\begin{equation}
\widehat{\mbox{Var}}_{\Pr(\mathbf{z}|D)}(\beta)
=-\mathbf{g}^T\mathbf{H}^{-1}\mathbf{g},
\label{eq: LRVB-result-eta-1}
\end{equation}
where
\begin{equation}
\mathbf{g}=\left.\frac{\partial\mathbb{E}_{q(\mathbf{z};\boldeta)}[\beta]}{\partial\boldeta^T}\right|_{\boldeta=\boldeta_0^*},\,
\mathbf{H}=\left.\frac{\partial^2\mathcal{L}}{\partial\boldeta\partial\boldeta^T}\right|_{\boldeta=\boldeta_0^*}.
\label{eq: LRVB-result-eta-2}
\end{equation}

\begin{itemize}
  \item {\textbf{calculate the matrix $\mathbf{H}$}}
\end{itemize}
The first derivatives of ELBO $\mathcal{L}$ is given as
$$\begin{aligned}
& \frac{\partial\mathcal{L}}{\partial\mu_\beta}
=-\left(\sum\limits_{j=1}^N
\pi_{w_j}\frac{\mu_\beta(\mu_{\gamma_j}^2+\sigma^2_{\gamma_j})-\mu_{\gamma_j}\hat{\Gamma}_j}{\sigma^2_{Y_j}+\tau^2}
\right)-\frac{\mu_\beta}{\sigma_0^2},\\
& \frac{\partial\mathcal{L}}{\partial\sigma^2_\beta}
=-\left(\sum\limits_{j=1}^N
\frac{\pi_{w_j}(\mu_{\gamma_j}^2+\sigma^2_{\gamma_j})}{2(\sigma^2_{Y_j}+\tau^2)}\right)
-\frac{1}{2\sigma^2_0} + \frac{1}{2\sigma^2_\beta},\\
& \frac{\partial\mathcal{L}}{\partial\mu_{\gamma_j}}
=-\frac{\mu_{\gamma_j}-\hat{\gamma}_j}{\sigma^2_{X_j}}
-\pi_{w_j}\frac{\mu_{\gamma_j}(\mu_\beta^2+\sigma^2_\beta)-\mu_\beta\hat{\Gamma}_j}{\sigma^2_{Y_j}+\tau^2}
-\frac{\mu_{\gamma_j}}{\sigma^2},\\
& \frac{\partial\mathcal{L}}{\partial\sigma^2_{\gamma_j}}
=-\frac{1}{2\sigma^2_{X_j}}
-\frac{\pi_{w_j}(\mu_\beta^2+\sigma^2_\beta)}{2(\sigma^2_{Y_j}+\tau^2)}
-\frac{1}{2\sigma^2}+\frac{1}{2\sigma^2_{\gamma_j}},
\end{aligned}$$
$$\begin{aligned}
\frac{\partial\mathcal{L}}{\partial\pi_{w_j}}
=&-\frac{\log(2\pi)}{2}
-\frac{\log(\sigma^2_{Y_j}+\tau^2)}{2}
-\frac{(\mu_\beta^2+\sigma^2_\beta)(\mu^2_{\gamma_j}+\sigma^2_{\gamma_j})-2\mu_\beta\mu_{\gamma_j}\hat{\Gamma}_j+\hat{\Gamma}_j^2}{2(\sigma^2_{Y_j}+\tau^2)}\\
&+\psi(a)-\psi(b)
+\log\left(\frac{1-\pi_{w_j}}{\pi_{w_j}}\right),\\
\frac{\partial\mathcal{L}}{\partial a}
=&\psi'(a)(a_0-a+\sum\limits_{j=1}^N\pi_{w_j})
+\psi'(a+b)(a+b-1-N-a_0),\\
\frac{\partial\mathcal{L}}{\partial b}
=&\psi'(b)(N+1-b-\sum\limits_{j=1}^N\pi_{w_j})
+\psi'(a+b)(a+b-1-N-a_0).
\end{aligned}$$

Among the second derivatives of ELBO $\mathcal{L}$, those do not have zero-values, are written as
$$\begin{aligned}
& \frac{\partial^2\mathcal{L}}{\partial\mu_\beta\partial\mu_\beta}=-\left(\sum\limits_{j=1}^N\frac{\pi_{w_j}(\mu^2_{\gamma_j}+\sigma^2_{\gamma_j})}{\sigma^2_{Y_j}+\tau^2}\right)-\frac{1}{\sigma^2_0},\\
& \frac{\partial^2\mathcal{L}}{\partial\mu_\beta\partial\mu_{\gamma_j}}
=\frac{\partial^2\mathcal{L}}{\partial\mu_{\gamma_j}\partial\mu_\beta}
=-\frac{\pi_{w_j}(2\mu_\beta\mu_{\gamma_j}-\hat{\Gamma}_j)}{\sigma^2_{Y_j}+\tau^2},\\
& \frac{\partial^2\mathcal{L}}{\partial\mu_\beta\partial\sigma^2_{\gamma_j}}
=\frac{\partial^2\mathcal{L}}{\partial\sigma^2_{\gamma_j}\partial\mu_\beta}
=-\frac{\pi_{w_j}\mu_\beta}{\sigma^2_{Y_j}+\tau^2},\\
& \frac{\partial^2\mathcal{L}}{\partial\mu_\beta\partial\pi_{w_j}}
=\frac{\partial^2\mathcal{L}}{\partial\pi_{w_j}\partial\mu_\beta}
=-\frac{\mu_\beta(\mu^2_{\gamma_j}+\sigma^2_{\gamma_j})-\mu_{\gamma_j}\hat{\Gamma}_j}{\sigma^2_{Y_j}+\tau^2},\\
& \frac{\partial^2\mathcal{L}}{\partial\sigma^2_\beta\partial\sigma^2_\beta}
=-\frac{1}{2(\sigma^2_\beta)^2},\\
& \frac{\partial^2\mathcal{L}}{\partial\sigma^2_\beta\partial\mu_{\gamma_j}}
=\frac{\partial^2\mathcal{L}}{\partial\mu_{\gamma_j}\partial\sigma^2_\beta}
=-\frac{\pi_{w_j}\mu_{\gamma_j}}{\sigma^2_{Y_j}+\tau^2}, \\
& \frac{\partial^2\mathcal{L}}{\partial\sigma^2_\beta\partial\sigma^2_{\gamma_j}}
=\frac{\partial^2\mathcal{L}}{\partial\sigma^2_{\gamma_j}\partial\sigma^2_\beta}
=-\frac{\pi_{w_j}}{2(\sigma^2_{Y_j}+\tau^2)},\,
\frac{\partial^2\mathcal{L}}{\partial\sigma^2_\beta\partial\pi_{w_j}}
=\frac{\partial^2\mathcal{L}}{\partial\pi_{w_j}\partial\sigma^2_\beta}
=-\frac{\mu^2_{\gamma_j}+\sigma^2_{\gamma_j}}{2(\sigma^2_{Y_j}+\tau^2)},\\
& \frac{\partial^2\mathcal{L}}{\partial\mu_{\gamma_j}\partial\mu_{\gamma_j}}
=-\frac{1}{\sigma^2_{X_j}}
-\frac{\pi_{w_j}(\mu_\beta^2+\sigma^2_\beta)}{\sigma^2_{Y_j}+\tau^2}
-\frac{1}{\sigma^2},\\
& \frac{\partial^2\mathcal{L}}{\partial\mu_{\gamma_j}\partial\pi_{w_j}}
=\frac{\partial^2\mathcal{L}}{\partial\pi_{w_j}\partial\mu_{\gamma_j}}
=-\frac{\mu_{\gamma_j}(\mu_\beta^2+\sigma^2_\beta)-\mu_\beta\hat{\Gamma}_j}{\sigma^2_{Y_j}+\tau^2},\\
& \frac{\partial^2\mathcal{L}}{\partial\sigma^2_{\gamma_j}\partial\sigma^2_{\gamma_j}}
=-\frac{1}{2(\sigma^2_{\gamma_j})^2},\\
& \frac{\partial^2\mathcal{L}}{\partial\sigma^2_{\gamma_j}\partial\pi_{w_j}}
=\frac{\partial^2\mathcal{L}}{\partial\pi_{w_j}\partial\sigma^2_{\gamma_j}}=-\frac{\mu^2_\beta+\sigma^2_\beta}{2(\sigma^2_{Y_j}+\tau^2)},\\
& \frac{\partial^2\mathcal{L}}{\partial\pi_{w_j}\partial\pi_{w_j}}=
-\frac{1}{\pi_{w_j}}-\frac{1}{1-\pi_{w_j}},\\
\end{aligned}$$
$$\begin{aligned}
& \frac{\partial^2\mathcal{L}}{\partial\pi_{w_j}\partial a}
=\frac{\partial^2\mathcal{L}}{\partial a\partial\pi_{w_j}}
=\psi'(a),\\
& \frac{\partial^2\mathcal{L}}{\partial\pi_{w_j}\partial b}
=\frac{\partial^2\mathcal{L}}{\partial b\partial\pi_{w_j}}
=-\psi'(b),\\
& \frac{\partial^2\mathcal{L}}{\partial a\partial a}
=\psi''(a)(a_0-a+\sum\limits_{j=1}^N\pi_{w_j})
-\psi'(a)
+\psi''(a+b)(a+b-a_0-1-N)
+\psi'(a+b),\\
& \frac{\partial^2\mathcal{L}}{\partial a\partial b}
=\frac{\partial^2\mathcal{L}}{\partial b\partial a}
=\psi''(a+b)(a+b-a_0-1-N)
+\psi'(a+b),\\
& \frac{\partial^2\mathcal{L}}{\partial b\partial b}
=\psi''(b)(N+1-b-\sum\limits_{j=1}^N\pi_{w_j})
-\psi'(b)
+\psi''(a+b)(a+b-a_0-1-N)
+\psi'(a+b).
\end{aligned}$$

\begin{itemize}
  \item {\textbf{calculate the vector $\mathbf{g}$}}
\end{itemize}
$$\mathbf{g}=\left.\frac{\partial\mathbb{E}_{q(\mathbf{z};\boldeta)[\beta]}}{\partial\boldeta}\right|_{\boldeta=\boldeta_0^*}
=\left.\frac{\partial\mu_\beta}{\partial\boldeta}\right|_{\boldeta=\boldeta_0^*}
=[1,0,0,...,0,0]^T.$$

\begin{itemize}
  \item {\textbf{Numerical problem of the matrix $\mathbf{H}$}}
\end{itemize}
When the variational parameter $\pi_{w_j}$ is close to its boundary 0, 1, the term $\frac{\partial^2\mathcal{L}}{\partial\pi_{w_j}\partial\pi_{w_j}}=-\frac{1}{\pi_{w_j}}-\frac{1}{1-\pi_{w_j}}$ makes the matrix $\mathbf{H}$ numerically not invertible.

\begin{itemize}
  \item {\textbf{Address the numerical problem by a reparameterized trick}}
\end{itemize}
Note that
$$\frac{\partial^2\mathcal{L}}{\partial\pi_{w_j}\partial\pi_{w_j}}
=-\frac{1}{\pi_{w_j}}-\frac{1}{1-\pi_{w_j}}
=\frac{\partial^2\mathbb{E}_{q(\mathbf{z};\boldeta)}[-\log q(\mathbf{z};\boldeta)]}{\partial\pi_{w_j}\partial\pi_{w_j}}.$$
The numerical problem is deal with the term $\left.\frac{\partial^2\mathbb{E}_{q(\mathbf{z};\boldeta)}[-\log q(\mathbf{z};\boldeta)]}{\partial\boldeta\partial\boldeta^T}\right|_{\boldeta=\boldeta_0^*}$. Next we introduce a reparameterized trick to address the numerical problem.

The variational distribution $q(\mathbf{z};\boldeta)$ 
$$q(\mathbf{z};\boldeta)=q(\beta;\boldeta)q(\pi_1;\boldeta)\prod\limits_{j=1}^N q(\gamma_j;\boldeta)\prod\limits_{j=1}^N q(w_j;\boldeta),$$
with
\begin{align*}
& q(\beta;\boldeta) = \mathcal{N}(\mu_{\beta},\sigma_{\beta}^2),\,q(\gamma_j;\boldeta) = \mathcal{N}(\mu_{\gamma_j},\sigma_{\gamma_j}^2),\\
& q(w_j;\boldeta) = \mbox{Bernoulli}(\pi_{w_j}),\,q(\pi_1;\boldeta) = \mbox{Beta}(a, b),\\
& \boldeta=\{\mu_\beta,\sigma^2_\beta,\mu_{\gamma_1},\sigma^2_{\gamma_1},\pi_{w_1},...,\mu_{\gamma_j},\sigma^2_{\gamma_j},\pi_{w_j},...,\mu_{\gamma_N},\sigma^2_{\gamma_N},\pi_{w_N},a,b\},
\end{align*}
is in an exponential family, i.e., the variational distribution can be written in the following form:
$$q(\mathbf{z};\boldeta)=\exp(\boldsymbol{\zeta}^T\mathbf{z}_s-A(\boldsymbol{\zeta})),$$
where $\mathbf{z}_s=T(\mathbf{z})$ are the sufficient statistics, $\boldsymbol{\zeta}=h(\boldeta)$, $h(\cdot)$ and $A(\cdot)$ are known functions. Then the entropy $\mathbb{E}_{q(\mathbf{z};\boldeta)}[-\log q(\mathbf{z};\boldeta)]$ can be written as
\begin{equation}
\mathbb{E}_{q(\mathbf{z};\boldeta)}[-\log q(\mathbf{z};\boldeta)]=-\boldsymbol{\zeta}^T\mathbb{E}_{q(\mathbf{z};\boldeta)}[\mathbf{z}_s]+A(\boldsymbol{\zeta}).
\label{eq: m-1}
\end{equation}
Moreover, using the cumulant generating properties again yeilds 
$$\begin{aligned}
& \mathbb{E}_{q(\mathbf{z};\boldeta)}[\mathbf{z}_s]
=\frac{\partial A(\boldsymbol{\zeta})}{\partial\boldsymbol{\zeta}},\\
& \mbox{Cov}_{q(\mathbf{z};\boldeta)}[\mathbf{z}_s]
=\frac{\partial^2A(\boldsymbol{\zeta})}{\partial\boldsymbol{\zeta}^T\partial\boldsymbol{\zeta}},
\end{aligned}$$
indicating 
\begin{equation}
\mbox{Cov}_{q(\mathbf{z};\boldeta)}[\mathbf{z}_s]=\frac{\partial\mathbb{E}_{q(\mathbf{z};\boldeta)}[\mathbf{z}_s]}{\partial\boldsymbol{\zeta}}.
\label{eq: m-2}
\end{equation}
For notation convenience, let 
\begin{equation}
\mathbf{m}:=\mathbb{E}_{q(\mathbf{z};\boldeta)}[\mathbf{z}_s].
\label{eq: m-3}
\end{equation}
Then combining Eqs. (\ref{eq: m-1},\ref{eq: m-2},\ref{eq: m-3}), we derive
$$\begin{aligned}
\frac{\partial\mathbb{E}_{q(\mathbf{z};\boldeta)}[-\log q(\mathbf{z};\boldeta)]}{\partial\mathbf{m}}
=& \frac{\partial[-\boldsymbol{\zeta}^T\mathbf{m}+A(\boldsymbol{\zeta})]}{\partial\boldsymbol{\zeta}^T}\frac{\partial\boldsymbol{\zeta}}{\partial\mathbf{m}}
+\frac{\partial[-\boldsymbol{\zeta}^T\mathbf{m}+A(\boldsymbol{\zeta})]}{\partial\mathbf{m}}\\
=& \left(-\mathbf{m}+\frac{\partial A(\boldsymbol{\zeta})}{\partial\boldsymbol{\zeta}}\right)\frac{\partial\boldsymbol{\zeta}}{\partial\mathbf{m}}-\boldsymbol{\zeta}\\
=& -\boldsymbol{\zeta},
\end{aligned}$$
and
\begin{equation}
\frac{\partial^2\mathbb{E}_{q(\mathbf{z};\boldeta)}[-\log q(\mathbf{z};\boldeta)]}{\partial\mathbf{m}\partial\mathbf{m}^T}
=-\frac{\partial\boldsymbol{\zeta}}{\partial\mathbf{m}}
=-\left(\frac{\partial\mathbf{m}}{\partial\boldsymbol{\zeta}}\right)^{-1}
=-(\mbox{Cov}_{q(\mathbf{z};\boldeta)}[\mathbf{z}_s])^{-1}.
\label{eq: LRVB-entropy-secondderiv}
\end{equation}
Note that $q(\beta),q(\gamma_j),q(w_j),q(\pi_1)$ can be written in exponential family forms as
$$\begin{aligned}
& q(\beta) = \frac{1}{\sqrt{2\pi\sigma_{\beta}^2}}\exp\left(-\frac{1}{2\sigma_{\beta}^2}(\beta-\mu_{\beta})^2\right) = \frac{1}{\sqrt{2\pi\sigma_{\beta}^2}}\exp\left( -\frac{\mu_{\beta}}{2\sigma_{\beta}^2} \right)\exp\left( \frac{\mu_{\beta}}{\sigma_{\beta}^2}\beta - \frac{1}{2\sigma_{\beta}^2}\beta^2 \right),\\
& q(\gamma_j) =  \frac{1}{\sqrt{2\pi\sigma_{\gamma_j}^2}}\exp\left(-\frac{1}{2\sigma_{\gamma_j}^2}(\gamma_j-\mu_{\gamma_j})^2\right)
= \frac{1}{\sqrt{2\pi\sigma_{\gamma_j}^2}}\exp\left( -\frac{\mu_{\gamma_j}}{2\sigma_{\gamma_j}^2} \right)\exp\left( \frac{\mu_{\gamma_j}}{\sigma_{\gamma_j}^2}\gamma_j - \frac{1}{2\sigma_{\gamma_j}^2}\gamma_j^2 \right),\\
& q(w_j) = \pi_{w_j}^{w_j}(1-\pi_{w_j})^{1-w_j}
=(1-\pi_{w_j})\exp\left[(\log\pi_{w_j}-\log (1-\pi_{w_j}))w_j \right],\\
& q(\pi_1) = \frac{\pi_1^{a-1}(1-\pi_1)^{b-1}}{\mbox{B}(a, b)}
= \frac{1}{\mbox{B}(a,b)}\exp[(a-1)\log\pi_1+(b-1)\log(1-\pi_1)],
\end{aligned}$$
where $\mbox{B}(\cdot,\cdot)$ denotes the beta function. Clearly, we get the sufficient statistics of latent variables as
$$\begin{aligned}
& T(\beta)=(\beta,\beta^2),\,T(\gamma_j)=(\gamma_j,\gamma_j^2)\\
& T(w_j)=w_j,\,T(\pi_1)=(\log\pi_1,\log(1-\pi_1)),
\end{aligned}$$
implying
$$\begin{aligned}
\mathbf{z}_s
&= T(\mathbf{z})\\
&= (T(\beta),T(\gamma_1),T(w_1),...,T(\gamma_j),T(w_j),...,T(\gamma_N),T(w_N),T(\pi_1))\\
&= (\beta,\beta^2,\gamma_1,\gamma_1^2,w_1,...,\gamma_j,\gamma_j^2,w_j,...,\gamma_N,\gamma_N^2,w_N,\log\pi_1,\log(1-\pi_1)).
\end{aligned}$$
Then we derive $\mathbf{m}$ as
$$\begin{aligned}
{\mathbf{m}} 
=& \mathbb{E}_q[\mathbf{z}_s] \\
=& (m_{\beta,1}, m_{\beta,2}, m_{\gamma_{1,1}}, m_{\gamma_{1,2}}, m_{w_1}, ..., m_{\gamma_{j,1}}, m_{\gamma_{j,2}}, m_{w_j}, ..., m_{\gamma_{N,1}}, m_{\gamma_{N,1}}, m_{w_N}, m_{\pi_{1,1}}, m_{\pi_{1,2}}),
\end{aligned}$$
with
$$\begin{aligned}
& m_{\beta,1}:=\mathbb{E}_q[\beta]=\mu_{\beta},\,m_{\beta,2}:=\mathbb{E}_q[\beta^2]=\mu_{\beta}^2+\sigma_{\beta}^2,\\
& m_{\gamma_j,1}:=\mathbb{E}_q[\gamma_j]=\mu_{\gamma_j},\,m_{\gamma_j,2}:=\mathbb{E}_q[\gamma_j^2]=\mu_{\gamma_j}^2+\sigma_{\gamma_j}^2\\
& m_{w_j}:=\mathbb{E}_q[w_j]=\pi_{w_j},\\
& m_{\pi_1,1}:=\mathbb{E}_q[\log\pi_1]=\psi(a)-\psi(a+b),\,
m_{\pi_1,2}:=\mathbb{E}_q[\log(1-\pi_1)]=\psi(b)-\psi(a+b).
\end{aligned}$$

Importantly, note that the variational distribution $q(\mathbf{z})=q(\mathbf{z};\boldeta)$ can also be parameterized by the vector $\mathbf{m}$, here adopt this reparamterized trick and let $q(\mathbf{z})=q(\mathbf{z};\mathbf{m})$ represents the variational distribution paramterized by $\mathbf{m}$. We assume that the parameters $\boldeta_0^*$ (optimal solution of $\boldeta$) and $\mathbf{m}_0^*$ (optimal solution of $\mathbf{m}$) in MFVB from the VEM algorithm are in the interior of their feasible space. Then under the same conditions (Condition 1, 2), similar as the result in Eqs. (\ref{eq: LRVB-result-eta-1}, \ref{eq: LRVB-result-eta-2}), i.e.,
$$\widehat{\mbox{Var}}_{\Pr(\mathbf{z}|D)}(\beta)
=-\mathbf{g}^T\mathbf{H}^{-1}\mathbf{g},$$
where
$$\mathbf{g}=\left.\frac{\partial\mathbb{E}_{q(\mathbf{z};\boldeta)}[\beta]}{\partial\boldeta^T}\right|_{\boldeta=\boldeta_0^*},\,
\mathbf{H}=\left.\frac{\partial^2\mathcal{L}}{\partial\boldeta\partial\boldeta^T}\right|_{\boldeta=\boldeta_0^*},$$
with the reparameterization we now have
$$\widehat{\mbox{Var}}_{\Pr(\mathbf{z}|D)}(\beta)
=-\mathbf{g}_{\mathbf{m}}^T(\mathbf{H}_{\mathbf{m}\mathbf{m}})^{-1}\mathbf{g}_{\mathbf{m}},$$
where
$$\mathbf{g}_{\mathbf{m}}=\left.\frac{\partial\mathbb{E}_{q(\mathbf{z};\mathbf{m})}[\beta]}{\partial\mathbf{m}^T}\right|_{\mathbf{m}=\mathbf{m}_0^*},\,
\mathbf{H}_{\mathbf{m}\mathbf{m}}=\left.\frac{\partial^2\mathcal{L}}{\partial\mathbf{m}\partial\mathbf{m}^T}\right|_{\mathbf{m}=\mathbf{m}_0^*}.$$

Note that the 
$$\mathcal{L}=
\mathbb{E}_{q(\mathbf{z};\mathbf{m})}[\log\Pr(\hat{\boldsymbol\gamma}, {\hat{\boldsymbol{\Gamma}}}, \mathbf{z} | {\boldsymbol{\sigma}}_X^2, {\boldsymbol{\sigma}}_Y^2; \boldsymbol{\theta}; \boldsymbol{h}_0)]
+\mathbb{E}_{q(\mathbf{z};\mathbf{m})}[-\log q(\mathbf{z};\mathbf{m})],$$
and that the results given in Eq. (\ref{eq: LRVB-entropy-secondderiv}) as
$$\frac{\partial^2\mathbb{E}_{q(\mathbf{z};\mathbf{m})}[-\log q(\mathbf{z};\mathbf{m})]}{\partial\mathbf{m}\partial\mathbf{m}^T}
=-(\mbox{Cov}_{q(\mathbf{z};\mathbf{m})}[\mathbf{z}_s])^{-1}.$$
Let 
$$\mathbf{H}^1_{\mathbf{m}\mathbf{m}}:=\left.\frac{\partial^2\mathbb{E}_{q(\mathbf{z};\mathbf{m})}[\log\Pr(\hat{\boldsymbol\gamma}, {\hat{\boldsymbol{\Gamma}}}, \mathbf{z} | {\boldsymbol{\sigma}}_X^2, {\boldsymbol{\sigma}}_Y^2; \boldsymbol{\theta}; \boldsymbol{h}_0)]}{\partial\mathbf{m}\partial\mathbf{m}^T}\right|_{\mathbf{m}=\mathbf{m}_0^*},\,
\mathbf{V}:=\left.\mbox{Cov}_{q(\mathbf{z};\mathbf{m})}[\mathbf{z}_s]\right|_{\mathbf{m}=\mathbf{m}_0^*}.$$
Then we can write $(\mathbf{H}_{\mathbf{m}\mathbf{m}})^{-1}$ as
$$(\mathbf{H}_{\mathbf{m}\mathbf{m}})^{-1}
=(\mathbf{H}^1_{\mathbf{m}\mathbf{m}}-\mathbf{V}^{-1})^{-1}
=(\mathbf{V}\mathbf{H}^1_{\mathbf{m}\mathbf{m}}-\mathbf{I})^{-1}\mathbf{V}.$$
Calculating $(\mathbf{H}_{\mathbf{m}\mathbf{m}})^{-1}$ in this way can prevent the numerical unstability in practice, because the term corresponding to $-\frac{1}{w_j}-\frac{1}{1-w_j}=-\frac{1}{w_j(1-w_j)}$ becomes $w_j(1-w_j)$ in matrix $\mathbf{V}$.

\begin{itemize}
  \item {\textbf{calculate matrix $\mathbf{V}$}}
\end{itemize}
Recall that $\mathbf{V}=\left.\mbox{Cov}_{q(\mathbf{z};\mathbf{m})}[\mathbf{z}_s]\right|_{\mathbf{m}=\mathbf{m}_0^*}$. As MFVB provides no information of covariances between different latent variables , we only need to calculate the covriance matrices $\mbox{Cov}_{q(\mathbf{z})}[T(\beta)]$, $\mbox{Cov}_{q(\mathbf{z})}[T(\gamma_j)]$, $\mbox{Cov}_{q(\mathbf{z})}[T(w_j)]$, $\mbox{Cov}_{q(\mathbf{z})}[T(\pi_1)]$. Intuitively, we give a illustrative diagram for matrix $\mathbf{V}$ in the following Figure.
\begin{figure}[!htbp]
\begin{centering}
\includegraphics[scale=0.45]{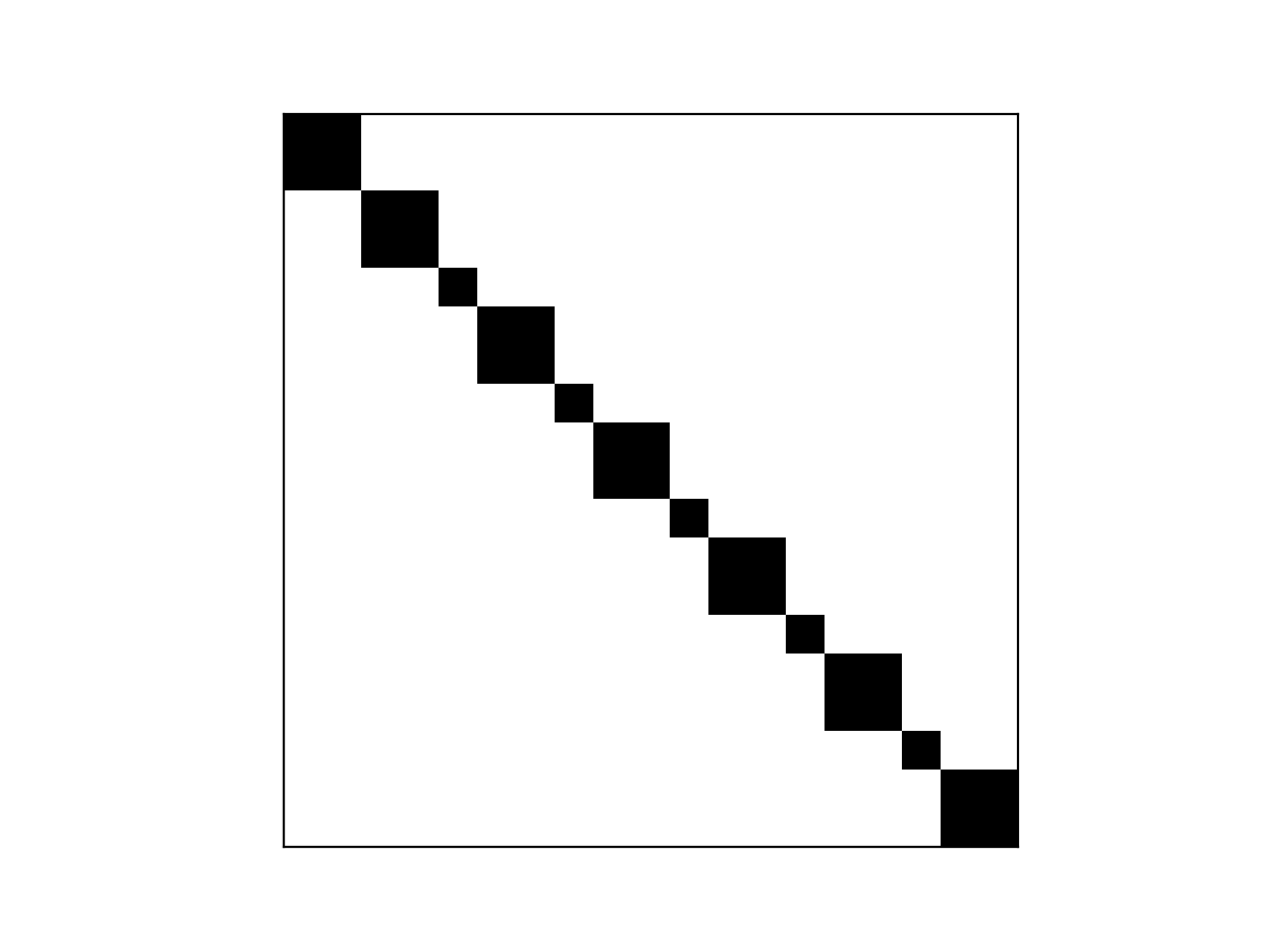}
\label{fig: V}
\par\end{centering}
\caption{Sparsity pattern for the $(3N+4)\times(3N+4)$ matrix $\mathbf{V}$. For simplicity, here we choose $N=5$ as an example.}
\end{figure}

As $q(\beta)$ and $q(\gamma_j)$ are normal distributions, the covariance matrices $\mbox{Cov}_{q(\mathbf{z})}[T(\beta)],\mbox{Cov}_{q(\mathbf{z})}[T(\gamma_j)]$ are given as 
$$\begin{aligned}
  & \mbox{Cov}_{q(\mathbf{z})}[T(\beta)] =
  \begin{bmatrix}
      \mathrm{Var}_q(\beta) & \mathrm{Cov}_q(\beta, \beta^2)\\
      \mathrm{Cov}_q(\beta^2, \beta) & \mathrm{Var}_q(\beta^2)
  \end{bmatrix},\\
  & \mbox{Cov}_{q(\mathbf{z})}[T(\gamma_j)] =
  \begin{bmatrix}
      \mathrm{Var}_q(\gamma_j) &\mathrm{Cov}_q(\gamma_j, \gamma_j^2)\\
      \mathrm{Cov}_q(\gamma_j^2,\gamma_j) &\mathrm{Var}_q(\gamma_j^2)
  \end{bmatrix}.
\end{aligned}$$
Recall that $q(\beta)=\mathcal{N}(\mu_{\beta},\sigma_{\beta}^2)$, and that if $\xi\sim\mathcal{N}(\xi|0, 1)$ then
$$
\mathbb{E}(\xi^k) = \left\{
\begin{aligned}
  &(k-1)!! ,\quad\quad k=2n, n\in\mathbb{N}_+, \\
  &0,\quad\quad\quad\quad\quad k=2n-1, n\in\mathbb{N}_+.
\end{aligned}
\right.
$$
To use this property, we rewrite $\beta$ as $\beta=\sigma_{\beta}\left( \frac{\beta-\mu_{\beta}}{\sigma_{\beta}} + \frac{\mu_{\beta}}{\sigma_{\beta}} \right) = \sigma_{\beta} \left( \xi + \frac{\mu_{\beta}}{\sigma_{\beta}} \right)$.
Therefore, we have
$$\begin{aligned}
\mathbb{E}_q(\beta^2) &= \mu_{\beta}^2 + \sigma_{\beta}^2,\\
\mathbb{E}_q(\beta^3) 
&= \mathbb{E}\left[\sigma_{\beta}^3 \left( \xi + \frac{\mu_{\beta}}{\sigma_{\beta}}\right)^3 \right]
=\sigma_{\beta}^3\mathbb{E}\left[\xi^3 + 3\frac{\mu_{\beta}}{\sigma_{\beta}}\xi^2 + 3\left(\frac{\mu_{\beta}}{\sigma_{\beta}}\right)^2\xi + \left(\frac{\mu_{\beta}}{\sigma_{\beta}}\right)^3\right]
= 3\mu_{\beta}\sigma_{\beta}^2 + \mu_{\beta}^3,\\
\mathbb{E}_q(\beta^4) 
&= \mathbb{E}\left[ \sigma_{\beta}^4 \left( \xi + \frac{\mu_{\beta}}{\sigma_{\beta}}\right)^4 \right]
= \sigma_{\beta}^4\mathbb{E}\left[ \xi^4 + 4\frac{\mu_{\beta}}{\sigma_{\beta}}\xi^3 + 6\left(\frac{\mu_{\beta}}{\sigma_{\beta}}\right)^2\xi^2 + 4\left(\frac{\mu_{\beta}}{\sigma_{\beta}}\right)^3\xi + \left(\frac{\mu_{\beta}}{\sigma_{\beta}}\right)^4\right] = 3\sigma_{\beta}^4 + 6\mu_{\beta}^2\sigma_{\beta}^2 + \mu_{\beta}^4,
\end{aligned}$$
Then, we derive
$$\begin{aligned}
& \mbox{Var}_q(\beta^2)
=\mathbb{E}(\beta^4) - \mathbb{E}^2(\beta^2)
=3\sigma_{\beta}^4 + 6\mu_{\beta}^2\sigma_{\beta}^2 + \mu_{\beta}^4 - (\sigma_{\beta}^2 + \mu_{\beta}^2)^2
=2\sigma_{\beta}^4 + 4\mu_{\beta}^2\sigma_{\beta}^2,\\
& \mathrm{Cov}(\beta^2, \beta) 
= \mathbb{E}(\beta^3) - \mathbb{E}(\beta^2)\mathbb{E}(\beta)
=3\mu_{\beta}\sigma_{\beta}^2 + \mu_{\beta}^3 - (\sigma_{\beta}^2 + \mu_{\beta}^2)\mu_{\beta}
= 2\mu_{\beta}\sigma_{\beta}^2,
\end{aligned}$$
yielding
$$
\mbox{Cov}_{q(\mathbf{z})}[T(\beta)] =
  \begin{bmatrix}
      \sigma_{\beta}^2 & 2\mu_{\beta}\sigma_{\beta}^2\\
      2\mu_{\beta}\sigma_{\beta}^2 & 2\sigma_{\beta}^4 + 4\mu_{\beta}^2\sigma_{\beta}^2
  \end{bmatrix}.
  \label{eq: Vj equation}
$$
Similarly, we have
$$
  \mbox{Cov}_{q(\mathbf{z})}[T(\gamma_j)]=
  \begin{bmatrix}
      \sigma_{\gamma_j}^2 & 2\mu_{\gamma_j}\sigma_{\gamma_j}^2\\
      2\mu_{\gamma_j}\sigma_{\gamma_j}^2 & 2\sigma_{\gamma_j}^4 + 4\mu_{\gamma_j}^2\sigma_{\gamma_j}^2
  \end{bmatrix}.
  \label{eq: Vj equation}
$$
Given that $q(w_j)=\mbox{Bernoulli}(\pi_{w_j})$, the variance of $w_j$ is given as
$$
\mbox{Var}_{q(\mathbf{z})}[T(w_j)]
=\mathbb{E}_q(w_j^2)-\mathbb{E}_q^2(w_j)=\pi_{w_j}-\pi_{w_j}^2.
$$
Recall that $q(\pi_1)=\mbox{Beta}(a,b)$ and $T(\pi_1)=(\log\pi_1,\log(1-\pi_1))$, then we know
\begin{equation}
  \mbox{Cov}_{q(\mathbf{z})}[T(\pi_1)] =
  \begin{bmatrix}
      \mathrm{Var}_q(\log\pi_1) & \mathrm{Cov}_q(\log\pi_1, \log(1-\pi_1))\\
      \mathrm{Cov}_q(\log(1-\pi_1), \log\pi_1) & \mathrm{Var}_q(\log(1-\pi_1))
  \end{bmatrix}.
  \label{eq: V0}
\end{equation}
It is not hard to derive
$$
\mathrm{Var}_q(\log\pi_1) = \psi_1(a) - \psi_1(a+b),
$$
where $\psi_1(\cdot)$ represents the trigamma function. Note that $(1-\pi_1)\sim \mbox{Beta}(b,a)$, then we know
$$
\mathrm{Var}_q(\log(1-\pi_1)) = \psi_1(b) - \psi_1(a+b).
$$
To calculate $\mathrm{Cov}_q(\log(1-\pi_1), \log\pi_1)$, we need to find $\mathbb{E}_q[\log\pi_1\log(1-\pi_1)]$ firstly:
$$
\mathbb{E}_q[\log\pi_1\log(1-\pi_1)] = \int_0^1 \log x\log(1-x)\frac{x^{a-1}(1-x)^{b-1}}{\mbox{B}(a,b)}\mbox{d}x.
$$
Note that, 
$$
\frac{\partial^2[x^{a-1}(1-x)^{b-1}]}{\partial a\partial b} = \frac{\partial^2[x^{a-1}(1-x)^{b-1}]}{\partial b\partial a} = \log x\log(1-x)x^{a-1}(1-x)^{b-1}.
$$
Therefore, 
$$
  \begin{aligned}
  \mathbb{E}_q[\log\pi_1\log(1-\pi_1)]
  = & \frac{1}{\mbox{B}(a, b)}\int_0^1 \frac{\partial^2[x^{a-1}(1-x)^{b-1}]}{\partial a\partial b} dx \\
  = & \frac{1}{\mbox{B}(a, b)}\frac{\partial^2}{\partial a\partial b}\int_0^1 x^{a-1}(1-x)^{b-1} dx \\
  = & \frac{1}{\mbox{B}(a, b)}\frac{\partial^2\mbox{B}(a, b)}{\partial a\partial b}.
  \end{aligned}
$$
With detailed derivations as
$$
  \begin{aligned}
  \frac{\partial\mbox{B}(a, b)}{\partial a}
  = & \frac{\partial}{\partial a}\left[ \frac{\Gamma(a)\Gamma(b)}{\Gamma(a+b)} \right] \\
  = & \Gamma(b)\frac{\Gamma'(a)\Gamma(a+b)-\Gamma(a)\Gamma'(a+b)}{\Gamma^2(a+b)} \\
  = & \frac{\Gamma(a)\Gamma(b)}{\Gamma(a+b)}\left[ \frac{\Gamma'(a)}{\Gamma(a)}-\frac{\Gamma'(a+b)}{\Gamma(a+b)} \right] \\
  = & \mbox{B}(a, b)[\psi(a)-\psi(a+b)]
  \end{aligned}
$$
and 
$$
  \begin{aligned}
  \frac{\partial^2\mbox{B}(a, b)}{\partial b\partial a}
  = & \frac{\partial\mbox{B}(a, b)}{\partial b}[\psi(a)-\psi(a+b)]+\mbox{B}(a, b)\frac{\partial}{\partial b}[\psi(a)-\psi(a+b)] \\
  = &\mbox{B}(a, b)[\psi(b)-\psi(a+b)][\psi(a)-\psi(a+b)]-\mbox{B}(a, b)\psi_1(a+b) \\
  = & B(a, b)\{[\psi(b)-\psi(a+b)][\psi(a)-\psi(a+b)]-\psi_1(a+b)\},
  \end{aligned}
$$
we finally derive
\begin{equation}
  \mathbb{E}_q[\log\pi_1\log(1-\pi_1)] 
  = [\psi(a)-\psi(a+b)][\psi(b)-\psi(a+b)]-\psi_1(a+b).
  \label{eq: infer-E-hard}
\end{equation}
Further considering that $\mathbb{E}_q[\log \pi_1]=\psi(a)-\psi(a+b)$, $\mathbb{E}_q[\log (1-\pi_1)]=\psi(b)-\psi(a+b)$ and Eq. (\ref{eq: infer-E-hard}), we have
$$\begin{aligned}
\mathrm{Cov}_q(\log\pi_1, \log(1-\pi_1)) 
= & \mathbb{E}_q[\log p(\pi_1)\log p(1-\pi_1)]-\mathbb{E}_q[\log p(\pi_1)]\mathbb{E}_q[\log p(1-\pi_1)] \\
= & [\psi(a)-\psi(a+b)][\psi(b)-\psi(a+b)]-\psi_1(a+b)\\
& -[\psi(a)-\psi(a+b)][\psi(b)-\psi(a+b)] \\
= & -\psi_1(a+b),
\end{aligned}$$
i.e.,
$$
\mathrm{Cov}_{q(\mathbf{z})}(\log\pi_1, \log(1-\pi_1)) 
= \mathrm{Cov}_{q(\mathbf{z})}(\log(1-\pi_1), \log\pi_1) = -\psi_1(a+b).
$$
Thus, we derive
$$
\mathrm{Cov}_{q(\mathbf{z})}[T(\pi_1)] =
\begin{bmatrix}
\psi_1(a)-\psi_1(a+b) & -\psi_1(a+b)\\
-\psi_1(a+b) & \psi_1(b)-\psi_1(a+b)
\end{bmatrix}.
$$
\begin{itemize}
  \item {\textbf{calculate matrix $\mathbf{H}^1_{\mathbf{m}\mathbf{m}}$}}
\end{itemize}
Recall that $\mathbf{H}^1_{\mathbf{m}\mathbf{m}} = \left.\frac{\partial^2 \mathbb{E}_q[\log\Pr(\hat{\boldsymbol\gamma}, {\hat{\boldsymbol{\Gamma}}}, \mathbf{z} | {\boldsymbol{\sigma}}_X^2, {\boldsymbol{\sigma}}_Y^2; \boldsymbol{\theta}; \boldsymbol{h}_0)]}{\partial {\mathbf{m}}^T \partial {\mathbf{m}}} \right|_{{\mathbf{m}} = {\mathbf{m}}_0^*}$. To calculate matrix $\mathbf{H}^1_{\mathbf{m}\mathbf{m}}$, we summarize the terms with $\mathbf{m}$ in $\mathbb{E}_q[\log\Pr(\hat{\boldsymbol\gamma}, {\hat{\boldsymbol{\Gamma}}}, \mathbf{z} | {\boldsymbol{\sigma}}_X^2, {\boldsymbol{\sigma}}_Y^2; \boldsymbol{\theta}; \boldsymbol{h}_0)]$ as:
$$
  \begin{aligned}
  & \mathbb{E}_q[\log\Pr(\hat{\boldsymbol\gamma}, {\hat{\boldsymbol{\Gamma}}}, \mathbf{z} | {\boldsymbol{\sigma}}_X^2, {\boldsymbol{\sigma}}_Y^2; \boldsymbol{\theta}; \boldsymbol{h}_0)]\\
  =& \sum\limits_{j=1}^N \left(- \frac{m_{\gamma_{j,2}}-2\hat{\gamma}_j m_{\gamma_{j,1}}}{2\sigma_{X_j}^2}\right) \\
  &+ \sum\limits_{j=1}^N m_{w_j}\left[ -\frac12\log(2\pi) - \frac12\log(\sigma_{Y_j}^2+\tau^2) - \frac{m_{\beta,2} m_{\gamma_{j,2}}-2m_{\beta,1}\hat{\Gamma}_j m_{\gamma_{j,1}} + \hat{\Gamma}_j^2}{2(\sigma_{Y_j}^2+\tau^2)} \right] \\
  &- \frac{m_{\beta,2}}{2\sigma_0^2} \\
  &+ \sum\limits_{j=1}^N \left( - \frac{m_{\gamma_{j,2}}}{2\sigma^2} \right) \\
  & + \sum\limits_{j=1}^N \left[ m_{w_j} m_{\pi_1,1} + (1-m_{w_j}) m_{\pi_1,2} \right] \\
  & + (a_0-1)m_{\pi_1,1}\\
  & + \mbox{constant}.
  \end{aligned}
$$
Therefore we can calculate $\frac{\partial^2 \mathbb{E}_q[\log\Pr(\hat{\boldsymbol\gamma}, {\hat{\boldsymbol{\Gamma}}}, \mathbf{z} | {\boldsymbol{\sigma}}_X^2, {\boldsymbol{\sigma}}_Y^2; \boldsymbol{\theta}; \boldsymbol{h}_0)]}{\partial {\mathbf{m}}^T \partial {\mathbf{m}}}$ as follows. The first derivatives are given as
$$\begin{aligned}
\frac{\partial \mathbb{E}_q[\log\Pr(\hat{\boldsymbol\gamma}, {\hat{\boldsymbol{\Gamma}}}, \mathbf{z} | {\boldsymbol{\sigma}}_X^2, {\boldsymbol{\sigma}}_Y^2; \boldsymbol{\theta}; \boldsymbol{h}_0)]}{\partial m_{\beta,1}}
  =& \sum\limits_{j=1}^N \frac{m_{w_j}\hat{\Gamma}_j m_{\gamma_j,1}}{\sigma_{Y_j}^2+\tau^2},\\
\frac{\partial \mathbb{E}_q[\log\Pr(\hat{\boldsymbol\gamma}, {\hat{\boldsymbol{\Gamma}}}, \mathbf{z} | {\boldsymbol{\sigma}}_X^2, {\boldsymbol{\sigma}}_Y^2; \boldsymbol{\theta}; \boldsymbol{h}_0)]}{\partial m_{\beta,2}}
  =& -\sum\limits_{j=1}^N \left[\frac{m_{w_j}m_{\gamma_j,2}}{2(\sigma_{Y_j}^2+\tau^2)}\right] - \frac{1}{2\sigma_0^2},\\
\frac{\partial \mathbb{E}_q[\log\Pr(\hat{\boldsymbol\gamma}, {\hat{\boldsymbol{\Gamma}}}, \mathbf{z} | {\boldsymbol{\sigma}}_X^2, {\boldsymbol{\sigma}}_Y^2; \boldsymbol{\theta}; \boldsymbol{h}_0)]}{\partial m_{\gamma_j,1}}
  =& \frac{\hat{\gamma}_j}{\sigma_{X_j}^2} + \frac{m_{w_j}m_{\beta,1}\hat{\Gamma}_j}{\sigma_{Y_j}^2+\tau^2},\\
\frac{\partial \mathbb{E}_q[\log\Pr(\hat{\boldsymbol\gamma}, {\hat{\boldsymbol{\Gamma}}}, \mathbf{z} | {\boldsymbol{\sigma}}_X^2, {\boldsymbol{\sigma}}_Y^2; \boldsymbol{\theta}; \boldsymbol{h}_0)]}{\partial m_{\gamma_j,2}}
  =& - \frac{1}{2\sigma_{X_j}^2} - \frac{m_{w_j}m_{\beta,2}}{2(\sigma_{Y_j}^2+\tau^2)} - \frac{1}{2\sigma^2},\\
\frac{\partial \mathbb{E}_q[\log\Pr(\hat{\boldsymbol\gamma}, {\hat{\boldsymbol{\Gamma}}}, \mathbf{z} | {\boldsymbol{\sigma}}_X^2, {\boldsymbol{\sigma}}_Y^2; \boldsymbol{\theta}; \boldsymbol{h}_0)]}{\partial m_{w_j}}
  =& -\frac12\log(2\pi) - \frac12\log(\sigma_{Y_j}^2+\tau^2) \\
  -& \frac{m_{\beta,2} m_{\gamma_{j,2}}-2m_{\beta,1}\hat{\Gamma}_j m_{\gamma_{j,1}} + \hat{\Gamma}_j^2}{2(\sigma_{Y_j}^2+\tau^2)} + m_{\pi,1}-m_{\pi,2},\\
\end{aligned}$$
$$\begin{aligned}
\frac{\partial \mathbb{E}_q[\log\Pr(\hat{\boldsymbol\gamma}, {\hat{\boldsymbol{\Gamma}}}, \mathbf{z} | {\boldsymbol{\sigma}}_X^2, {\boldsymbol{\sigma}}_Y^2; \boldsymbol{\theta}; \boldsymbol{h}_0)]}{\partial m_{\pi,1}}
  =& (a_0-1) + \sum\limits_{j=1}^N m_{w_j},\\
\frac{\partial \mathbb{E}_q[\log\Pr(\hat{\boldsymbol\gamma}, {\hat{\boldsymbol{\Gamma}}}, \mathbf{z} | {\boldsymbol{\sigma}}_X^2, {\boldsymbol{\sigma}}_Y^2; \boldsymbol{\theta}; \boldsymbol{h}_0)]}{\partial m_{\pi,2}}
  =& N - \sum\limits_{j=1}^N m_{w_j}.
\end{aligned}$$
Among the second derivatives, those which do not have zero values are:
$$\begin{aligned}
& \frac{\partial^2 \mathbb{E}_q[\log\Pr(\hat{\boldsymbol\gamma}, {\hat{\boldsymbol{\Gamma}}}, \mathbf{z} | {\boldsymbol{\sigma}}_X^2, {\boldsymbol{\sigma}}_Y^2; \boldsymbol{\theta}; \boldsymbol{h}_0)]}{\partial m_{\beta,1}\partial m_{\gamma_j,1}} 
= \frac{\partial^2 \mathbb{E}_q[\log\Pr(\hat{\boldsymbol\gamma}, {\hat{\boldsymbol{\Gamma}}}, \mathbf{z} | {\boldsymbol{\sigma}}_X^2, {\boldsymbol{\sigma}}_Y^2; \boldsymbol{\theta}; \boldsymbol{h}_0)]}{\partial m_{\gamma_j,1}\partial m_{\beta,1}} 
= \frac{m_{w_j}\hat{\Gamma}_j}{\sigma_{Y_j}^2+\tau^2},\\
&\frac{\partial^2 \mathbb{E}_q[\log\Pr(\hat{\boldsymbol\gamma}, {\hat{\boldsymbol{\Gamma}}}, \mathbf{z} | {\boldsymbol{\sigma}}_X^2, {\boldsymbol{\sigma}}_Y^2; \boldsymbol{\theta}; \boldsymbol{h}_0)]}{\partial m_{\beta,1}\partial m_{w_j}} 
= \frac{\partial^2 \mathbb{E}_q[\log\Pr(\hat{\boldsymbol\gamma}, {\hat{\boldsymbol{\Gamma}}}, \mathbf{z} | {\boldsymbol{\sigma}}_X^2, {\boldsymbol{\sigma}}_Y^2; \boldsymbol{\theta}; \boldsymbol{h}_0)]}{\partial m_{w_j}\partial m_{\beta,1}} = \frac{m_{\gamma_j,1}\hat{\Gamma}_j}{\sigma_{Y_j}^2+\tau^2},\\
& \frac{\partial^2 \mathbb{E}_q[\log\Pr(\hat{\boldsymbol\gamma}, {\hat{\boldsymbol{\Gamma}}}, \mathbf{z} | {\boldsymbol{\sigma}}_X^2, {\boldsymbol{\sigma}}_Y^2; \boldsymbol{\theta}; \boldsymbol{h}_0)]}{\partial m_{\beta,2}\partial m_{\gamma_j,2}} 
= \frac{\partial^2 \mathbb{E}_q[\log\Pr(\hat{\boldsymbol\gamma}, {\hat{\boldsymbol{\Gamma}}}, \mathbf{z} | {\boldsymbol{\sigma}}_X^2, {\boldsymbol{\sigma}}_Y^2; \boldsymbol{\theta}; \boldsymbol{h}_0)]}{\partial m_{\gamma_j,2}\partial m_{\beta,2}} 
= -\frac{m_{w_j}}{2(\sigma_{Y_j}^2+\tau^2)},\\
& \frac{\partial^2 \mathbb{E}_q[\log\Pr(\hat{\boldsymbol\gamma}, {\hat{\boldsymbol{\Gamma}}}, \mathbf{z} | {\boldsymbol{\sigma}}_X^2, {\boldsymbol{\sigma}}_Y^2; \boldsymbol{\theta}; \boldsymbol{h}_0)]}{\partial m_{\beta,2}\partial m_{w_j}} 
= \frac{\partial^2 \mathbb{E}_q[\log\Pr(\hat{\boldsymbol\gamma}, {\hat{\boldsymbol{\Gamma}}}, \mathbf{z} | {\boldsymbol{\sigma}}_X^2, {\boldsymbol{\sigma}}_Y^2; \boldsymbol{\theta}; \boldsymbol{h}_0)]}{\partial m_{w_j}\partial m_{\beta,2}} = -\frac{m_{\gamma_j,2}}{2(\sigma_{Y_j}^2+\tau^2)},\\
& \frac{\partial^2 \mathbb{E}_q[\log\Pr(\hat{\boldsymbol\gamma}, {\hat{\boldsymbol{\Gamma}}}, \mathbf{z} | {\boldsymbol{\sigma}}_X^2, {\boldsymbol{\sigma}}_Y^2; \boldsymbol{\theta}; \boldsymbol{h}_0)]}{\partial m_{\gamma_j,1}\partial m_{w_j}} 
= \frac{\partial^2 \mathbb{E}_q[\log\Pr(\hat{\boldsymbol\gamma}, {\hat{\boldsymbol{\Gamma}}}, \mathbf{z} | {\boldsymbol{\sigma}}_X^2, {\boldsymbol{\sigma}}_Y^2; \boldsymbol{\theta}; \boldsymbol{h}_0)]}{\partial m_{w_j}\partial m_{\gamma_j,1}} 
= \frac{m_{\beta,1}\hat{\Gamma}_j}{\sigma_{Y_j}^2+\tau^2},\\
& \frac{\partial^2 \mathbb{E}_q[\log\Pr(\hat{\boldsymbol\gamma}, {\hat{\boldsymbol{\Gamma}}}, \mathbf{z} | {\boldsymbol{\sigma}}_X^2, {\boldsymbol{\sigma}}_Y^2; \boldsymbol{\theta}; \boldsymbol{h}_0)]}{\partial m_{\gamma_j,2}\partial m_{w_j}} 
= \frac{\partial^2 \mathbb{E}_q[\log\Pr(\hat{\boldsymbol\gamma}, {\hat{\boldsymbol{\Gamma}}}, \mathbf{z} | {\boldsymbol{\sigma}}_X^2, {\boldsymbol{\sigma}}_Y^2; \boldsymbol{\theta}; \boldsymbol{h}_0)]}{\partial m_{w_j}\partial m_{\gamma_j,2}} 
= - \frac{m_{\beta,2}}{2(\sigma_{Y_j}^2+\tau^2)},\\
& \frac{\partial^2 \mathbb{E}_q[\log\Pr(\hat{\boldsymbol\gamma}, {\hat{\boldsymbol{\Gamma}}}, \mathbf{z} | {\boldsymbol{\sigma}}_X^2, {\boldsymbol{\sigma}}_Y^2; \boldsymbol{\theta}; \boldsymbol{h}_0)]}{\partial m_{\pi,1}\partial m_{w_j}} 
= \frac{\partial^2 \mathbb{E}_q[\log\Pr(\hat{\boldsymbol\gamma}, {\hat{\boldsymbol{\Gamma}}}, \mathbf{z} | {\boldsymbol{\sigma}}_X^2, {\boldsymbol{\sigma}}_Y^2; \boldsymbol{\theta}; \boldsymbol{h}_0)]}{\partial m_{w_j}\partial m_{\pi,1}} 
= 1,\\
& \frac{\partial^2 \mathbb{E}_q[\log\Pr(\hat{\boldsymbol\gamma}, {\hat{\boldsymbol{\Gamma}}}, \mathbf{z} | {\boldsymbol{\sigma}}_X^2, {\boldsymbol{\sigma}}_Y^2; \boldsymbol{\theta}; \boldsymbol{h}_0)]}{\partial m_{\pi,2}\partial m_{w_j}} 
= \frac{\partial^2 \mathbb{E}_q[\log\Pr(\hat{\boldsymbol\gamma}, {\hat{\boldsymbol{\Gamma}}}, \mathbf{z} | {\boldsymbol{\sigma}}_X^2, {\boldsymbol{\sigma}}_Y^2; \boldsymbol{\theta}; \boldsymbol{h}_0)]}{\partial m_{w_j}\partial m_{\pi,2}} 
= -1.
\end{aligned}$$
We finally derive the matrix ${\mathbf{H}^1_{\mathbf{m}\mathbf{m}}}$. Intuitively, we give a illustrative diagram for matrix ${\mathbf{H}^1_{\mathbf{m}\mathbf{m}}}$ in the following figure:
\begin{figure}[!htbp]
\begin{centering}
\includegraphics[scale=0.45]{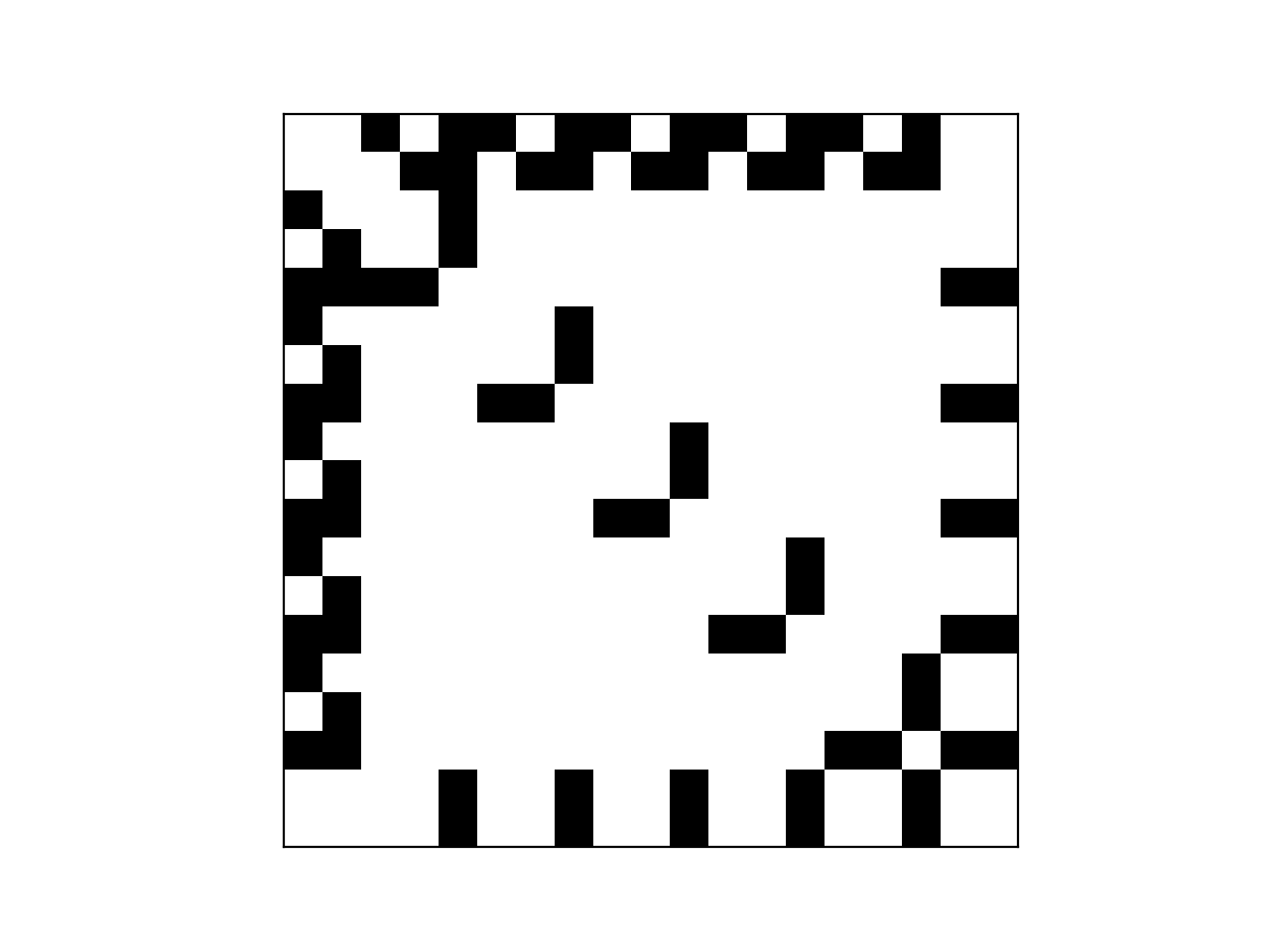}
\label{fig: H}
\par\end{centering}
\caption{Sparsity pattern for the $(3N+4)\times(3N+4)$ matrix ${\mathbf{H}^1_{\mathbf{m}\mathbf{m}}}$. For simplicity, here we choose $N=5$ as an example.}
\end{figure}

\begin{itemize}
  \item {\textbf{calculate the vector $\mathbf{g}_{\mathbf{m}}$}}
\end{itemize}
$$\mathbf{g_\mathbf{m}}=\left.\frac{\partial\mathbb{E}_{q(\mathbf{z};\mathbf{m})[\beta]}}{\partial\mathbf{m}}\right|_{\boldeta=\boldeta_0^*}
=\left.\frac{\partial m_{\beta,1}}{\partial\mathbf{m}}\right|_{\mathbf{m}=\mathbf{m}_0^*}
=[1,0,0,...,0,0]^T.$$

\begin{itemize}
  \item {\textbf{calculate the standard error of $\beta$}}
\end{itemize}
Finally, the standard error of $\beta$ is provided as the element in the first row and the first column of the matrix $\mathbf{H}_{\mathbf{m}\mathbf{m}}=(\mathbf{V}\mathbf{H}^1_{\mathbf{m}\mathbf{m}}-\mathbf{I})^{-1}\mathbf{V}$.

\subsection{Comparison of different MR Methods}
Methods like Egger and GSMR are closely related to standard Inverse Variance Weighted (IVW) meta analysis approach. The development of these MR methods put more efforts to address the issue of horizontal pleiotropy. To connect these methods, we first introduce the IVW method here.

Assuming that the involved genetic variants satisfy the conditions of instrumental variables (see Methods section in main text), the IVW approach serves as a standard approach for Mendelian Randomization analysis, which can also be viewed as a meta analysis of single causal estimates \citep{Bowden2015}. Let $\{G_j\}_{j=1}^N$ be $N$ independent variants which serve as valid instrumental variables (IVs). For a single variant $G_j$, the causal effect of exposure on the outcome can be estimated as $\hat\beta_j=\frac{\hat\Gamma_j}{\hat\gamma_j}$, where $\hat\gamma_j$ and $\hat\Gamma_j$ are the estimated SNP-exposure and SNP-outcome effects, respectively \citep{Lawlor2008}.
Since the variants are selected to be strongly associated with exposure, the variance of $\hat\beta_j$ can be given as: $\frac{\sigma_{Y_j}^2}{\hat\gamma_j^2}$, where the estimation errors $\sigma^2_{X_j}$ are ignored.

Therefore, the IVW estimator based on the $N$ IVs is given as,
\begin{equation}\label{1}
\hat\beta_{IVW} = \frac{\sum_{j=1}^{N}\hat\Gamma_j \hat\gamma_j {\sigma_{Y_j}^{-2}}}{\sum_{j=1}^{N}\hat\gamma_j^2\sigma_{Y_j}^{-2}}.
\end{equation}
In fact, this can be obtained by solving the following weighted least square problem where $\sigma^{-2}_{Y_j},j=1,\cdots, N$ are the weights,

\begin{equation}\label{2}
\hat\beta_{IVW} = \arg\min_{\beta} \sum_{j=1}^{N}\frac{(\hat\Gamma_j - \beta \hat\gamma_j)^2}{{\sigma_{Y_j}^2}}.
\end{equation}
Then the standard error of $\hat\beta_{IVW}$ is given as,
\begin{equation}\label{3}
SE(\hat\beta_{IVW}) = \sqrt{\frac{\hat\sigma^2}{\sum_{j=1}^{N}\hat\gamma_j^2\sigma_{Y_j}^{-2}}}.
\end{equation}
where $\hat\sigma^2 $ is the variance of residuals from the weighed regression.

The Egger method extends the IVW method by introducing an intercept term $\beta_0$ to address the influence of horizontal pleiotropy \citep{Bowden2015},
\begin{equation}\label{4}
\hat\beta_{Egger} = \arg\min_{\beta} \sum_{j=1}^{N}\frac{(\hat\Gamma_j - \beta_0 - \beta \hat\gamma_j)^2}{{\sigma_{Y_j}^2}}.
\end{equation}
Then the standard error of the Egger estimate can be obtained by
\begin{equation}\label{5}
SE(\hat\beta_{Egger}) = \sqrt{\frac{\hat\sigma^2}{\sum_{j=1}^{N}(\hat\gamma_j-\bar{\hat{\gamma}}_w)^2 {\sigma_{Y_j}^{2}}}}.
\end{equation}
where $\bar{\hat{\gamma}}_w$ is the weighted average of the estimated exposure effects that $\bar{\hat{\gamma}}_w =  \frac{\sum_{i=1}^{N} {\hat\gamma_j\sigma_{Y_j}^{-2}}}{\sum_{i=1}^{N}{\sigma_{Y_j}^{-2}}}$.

From the model assumption of Egger, we can see that Egger may not have a satisfactory performance in the presence of weak but non-constant pleiotropic effects. In addition, Egger may not be robust if there exist a few strong pleiotropic effects (i.e., outliers).

Unlike Egger, GSMR introduces an outlier detection procedure (called  HEIDI-outlier) to reduce the strong effects of horizontal pleiotropy. It further takes into account the measurement error of the SNP-exposure effect and weak LD between SNPs \citep{Zhu2018}. Let $\hat{\boldsymbol{\beta}} = (\hat\beta_1, \hat\beta_2,\cdots,\hat\beta_N)$ be the vector of single variant MR estimates and $\mathbf{U}$ be the variance covariance matrix of $\hat{\boldsymbol{\beta}}$, then we have $\hat{\boldsymbol{\beta}} \sim N(\beta\mathbf{1}, \mathbf{U})$. Notice that, if variants are correlated, the non-diagonal elements in matrix $\mathbf{U}$ is non zero. The generalized least square approach is then applied to estimate the causal effect, yielding $\hat\beta_{GSMR} = (\mathbf{1}^T\mathbf{U}^{-1}\mathbf1)^{-1}\mathbf{1}^T\mathbf{U}^{-1}\hat{\boldsymbol{\beta}}$.
In the special case that uncorrelated and strong variants are used as IVs (now $\mathbf{U}$ is a diagonal matrix), the GSMR estimator can be written as follows:

\begin{equation}\label{6}
\hat\beta_{GSMR} = \arg\min_{\beta}\sum_{j=1}^{N}\frac{(\hat\Gamma_j - \beta\gamma_j)^2}{\sigma_{Yj}^2 + \hat\beta_j^2\sigma_{X_j}^2},
\end{equation}
Then the corresponding standard error is
\begin{equation}\label{7}
SE(\hat\beta_{GSMR})  = \frac{1}{\sum_{j=1}^{N}\hat\gamma_j^2/(\sigma_{Y_j}^2 + \hat\beta_j^2\sigma_{X_j}^2)}.
\end{equation}

To deal with horizontal pleiotropy, the HEIDI-outlier method firstly chooses a variant as the target, and then tests the estimated causal effect of the target variant and each of other variants. The variants that are significantly different from the target variant will be considered as outliers and then be removed in model fitting (\ref{6}, \ref{7}). To avoid choosing a potential pleiotropic outlier as the target variant, the HEIDI-outlier method adopts an ad-hoc way: it chooses the top exposure-associated variant among those with the corresponding causal effect estimates in the third quintile (41\% to 60\%) of the single causal variant estimates ($\hat{\beta_1}, \cdots, \hat{\beta_N}$). The HEIDI-outlier method has the risk of selecting an impropriate variant as the target variant leading to biased estimate or inflated type I errors.

RAPS uses a random effects model to model the weak Horizontal pleiotropy that $\alpha_j \sim N(0, \tau^2)$ with a variance component $\tau^2$ and the measurement error of $\hat\gamma_j$ is also taken into account:

 \begin{equation}\label{8}
 \hat\Gamma_j \sim N(\beta\gamma_j, \tau^2+\sigma_{Y_j}^2),\ \   \hat\gamma_j \sim N(\gamma_j, \sigma_{X_j}^2).
 \end{equation}

Parameter estimation of RAPS is obtained by maximizing the profile likelihood and in which $\gamma_1,\dots,\gamma_N$ are considered as nuisance parameters being profiled out. The profile likelihood function is given as follows:
 \begin{equation}\label{9}
 l(\beta,\tau) = -\frac{1}{2}\sum_{j=1}^{N}\frac{(\hat\Gamma_j - \beta\gamma_j)^2}{\sigma_{Y_j}^2 + \tau^2 + \beta^2\sigma_{X_j}^2} + \log(\sigma_{Y_j}^2 + \tau^2).
\end{equation}
An adjusted profile score (APS) \citep{mccullagh1990simple} is used in RAPS to obtain an consistent estimator of $\beta$ and $\tau$. To reduce the influence of the large pleiotropic effects ( i.e. outliers) , robust loss functions, like the Huber loss and Tukey's biweight loss functions, are suggested to replace the $l_2$ loss in (\ref{9}).

Clearly, if we set $\tau=0$ , the optimum obtained by maximizing (\ref{9}) w.r.t. $\beta$ equals to an IVW  estimate with weights: $w(\beta) = 1/(\sigma_{Y_j}^2  + \beta^2\sigma_{X_j}^2)$, which is a function of the true causal effect $\beta$.  From this perspective, the estimates from  IVW (\ref{2}) and Egger (\ref{4}) estimates  correspond to setting $\sigma_{X_j}=0$ in $w(\beta)$, while the estimates from GSMR (\ref{6}) are obtained by replacing the  unknown true effect size $\beta$ with $\hat{\beta}_j$.

Compared with RAPS in (\ref{9}), we conclude that GSMR ignores weak pleiotropic effects ($\tau^2 = 0$) which could lead to its inflated type I errors, as shown in simulation study and real data analysis.

\subsection{More simulation results}
\subsubsection{Summary-level simulations}
We simulated summary-level data $D=\{\hat{\boldsymbol{\gamma}},\hat{\boldsymbol{\Gamma}},\boldsymbol{\sigma}_X,\boldsymbol{\sigma}_Y\}$ to test the robustness of BWMR and its related methods. We first generated $\Gamma_j$ and $\gamma_j$ in the following five cases:
\begin{itemize}
  \item Case-1: $\gamma_j\overset{\text{i.i.d}}{\sim}\mathcal{N}(0,\sigma^2),\,\alpha_j\overset{\text{i.i.d}}{\sim}\mathcal{N}(0,\tau^2),\,\Gamma_j=\beta\gamma_j+\alpha_j$.
  \item Case-2: $\gamma_j\overset{\text{i.i.d}}{\sim}\mathcal{N}(0,\sigma^2),\,\alpha_j\overset{\text{i.i.d}}{\sim}\mathcal{N}(0,\tau^2),$
  $$\Gamma_j=\left\{
  \begin{aligned}
  & \beta\gamma_j+\alpha_j,\, j=1,2,...,(N\times C),\\
  & \beta_c\gamma_j+\alpha_j,\, j=(N\times C+1),...,N,
  \end{aligned}
  \right.$$
  with $\beta_c$ denoting corrupted value of $\beta$ and $C$ denoting the corresponding corrupted rate.
  \item Case-3: $\gamma_j\overset{\text{i.i.d.}}{\sim}\mathcal{N}(0,\sigma^2),\,\Gamma_j=\beta\gamma_j+\alpha_j$ and 
  $$\alpha_j\overset{\text{i.i.d.}}{\sim}\left\{
  \begin{aligned}
  & \mathcal{N}(0,\tau^2),\, j=1,2,...,(N\times C),\\
  & \mathcal{N}(0,\tau_c^2),\, j=(N\times C+1),...,N,
  \end{aligned}
  \right.$$
  with $\tau_c^2$ denoting corrupted value of $\tau^2$.
  \item Case-4: $\gamma_j\overset{\text{i.i.d}}{\sim}(1-R)\mathcal{N}(0,\sigma^2)+R\mathcal{N}(0, 10\sigma^2),\,\alpha_j\overset{\text{i.i.d.}}{\sim}\mathcal{N}(0,\tau^2)$ and $\Gamma_j=\beta\gamma_j+\alpha_j$, where $0<R<1$.
  \item Case-5: $\gamma_j\overset{\text{i.i.d}}{\sim}\mathcal{N}(0,\sigma^2),\,\Gamma_j=\beta\gamma_j+\alpha_j$ and $\alpha_j\overset{\text{i.i.d.}}{\sim}\tau\mbox{Laplace}(r)$, where $\mbox{Laplace}(r)$ represents Laplace (double exponential) distribution with the rate $r$.
\end{itemize} 
With $\Gamma_j$ and $\gamma_j$ generated above, we simulated $\hat{\Gamma}_j$ and $\hat{\gamma}_j$ from $\hat{\gamma}_j|\gamma_j \sim \mathcal{N}(\gamma_j, \sigma_{X_j}^2)$,$\hat{\Gamma}_j|\Gamma_j \sim \mathcal{N}(\Gamma_j, \sigma_{Y_j}^2)$, where $\sigma_{X_j}$ and $\sigma_{Y_j}$ were i.i.d. from the uniform distribution $U[c,d]$.

\newpage{}
\begin{itemize}
  \item Summary-level simulation results in Case-1
\end{itemize}

\begin{figure}[H]
\begin{centering}
\includegraphics[scale=0.34]{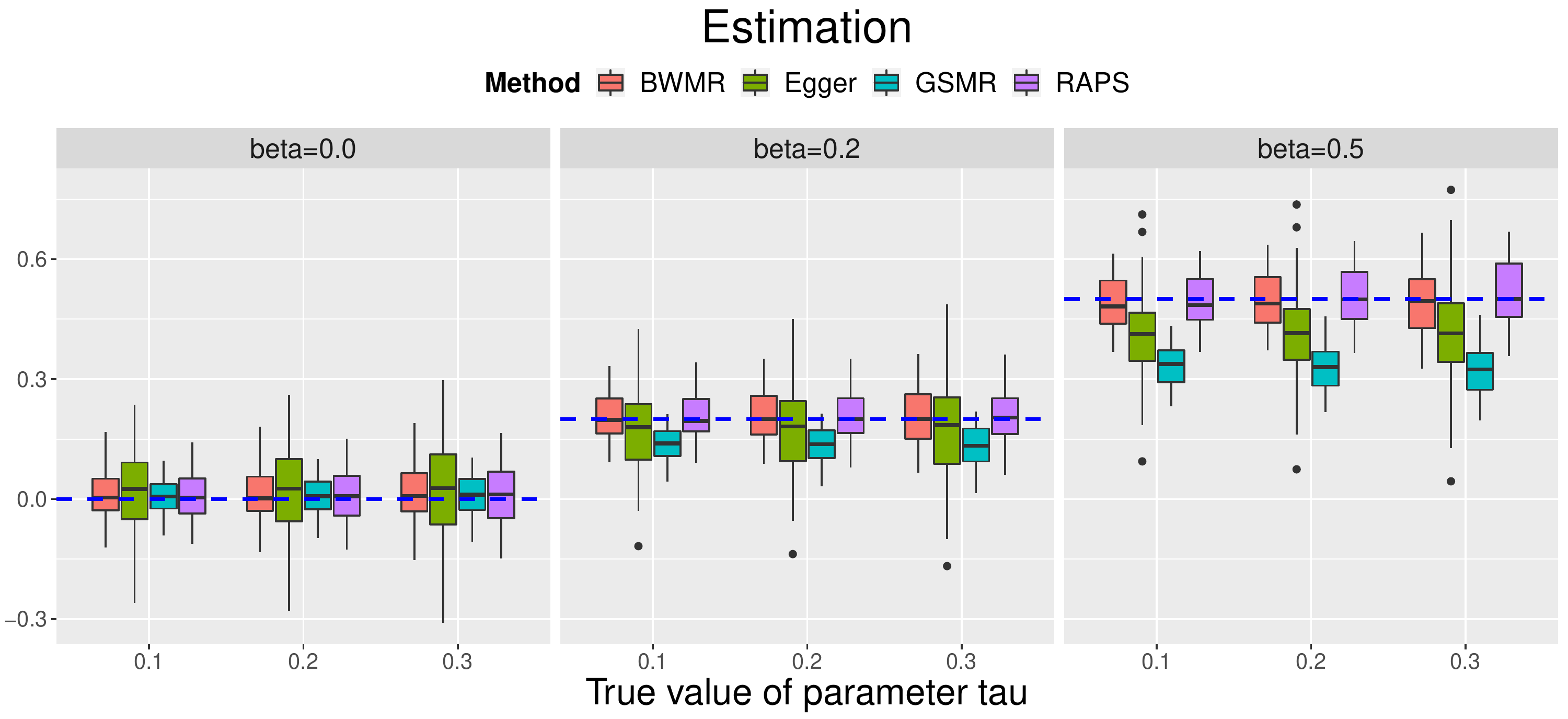}
\par\end{centering}
\caption{Comparison of estimation accuracy of BWMR, Egger, GSMR and RAPS in Case-1 of the summary-level data simulation. The simulation parameters were varied in the following range: $\beta\in\{0.0,0.2,0.5\}$, $\tau\in\{0.1,0.2,0.3\}$, $\sigma=0.8$, $[c,d]=[0.3,0.5]$. The results were summarized from 50 replications.}
\end{figure}

\begin{figure}[H]
\begin{centering}
\includegraphics[scale=0.34]{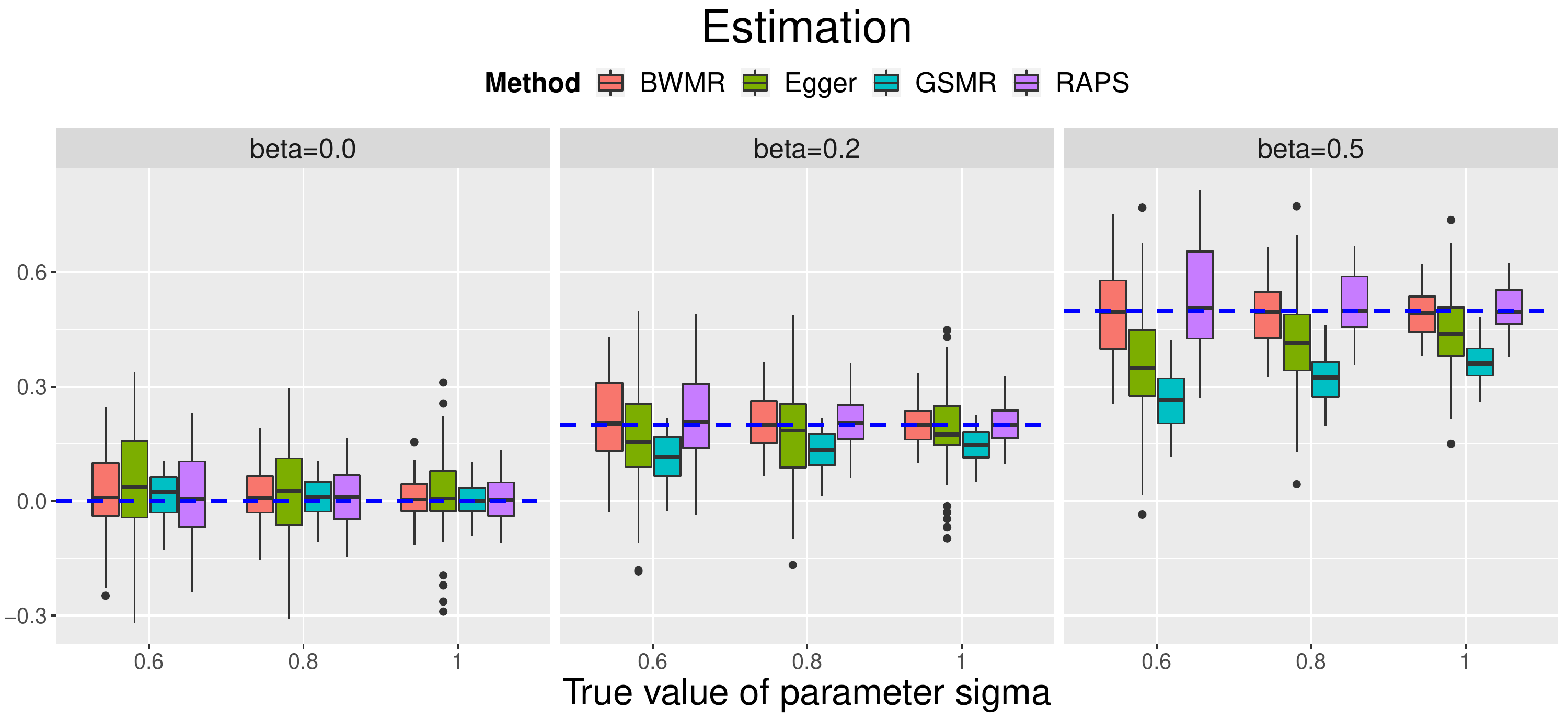}
\par\end{centering}
\caption{Comparison of estimation accuracy of BWMR, Egger, GSMR and RAPS in Case-1 of the summary-level data simulation. The simulation parameters were varied in the following range: $\beta\in\{0.0,0.2,0.5\}$, $\sigma\in\{0.6,0.8,1.0\}$, $\tau=0.2$, $[c,d]=[0.3,0.5]$. The results were summarized from 50 replications.}
\end{figure}

\begin{figure}[H]
\begin{centering}
\includegraphics[scale=0.375]{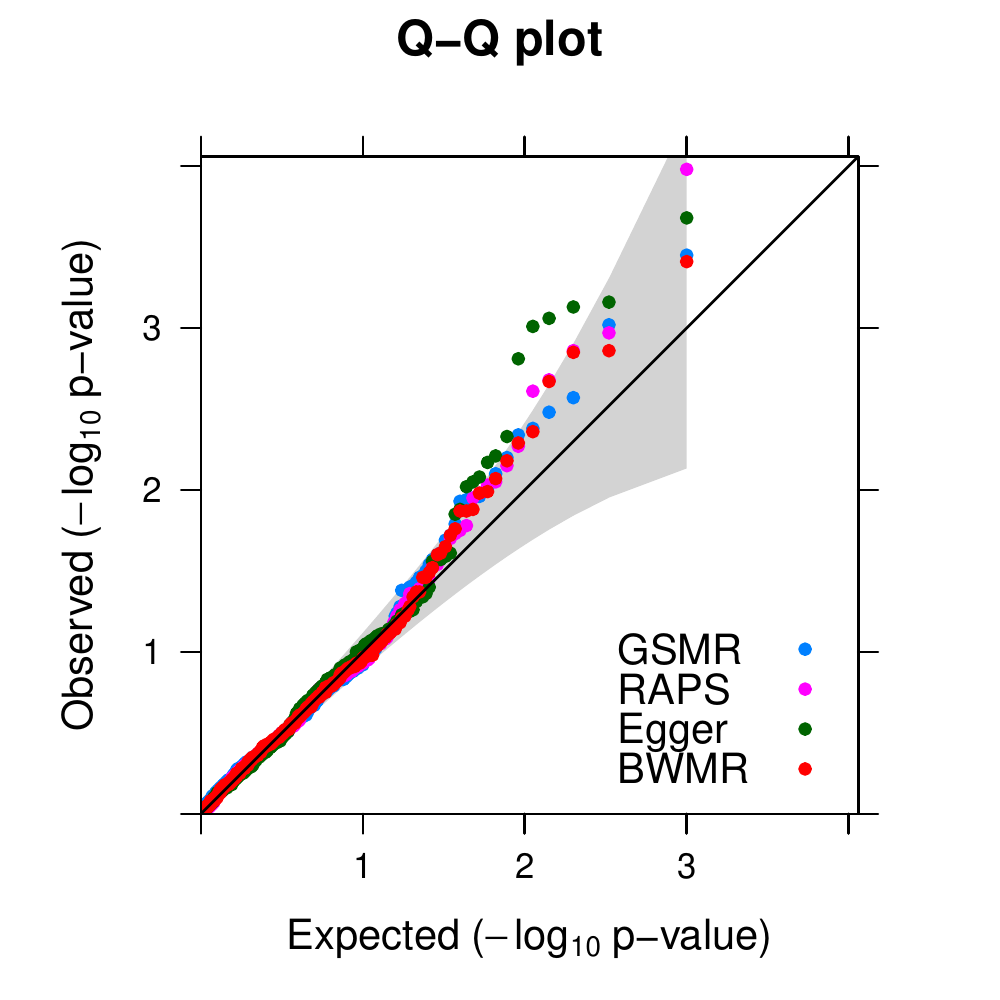}
\includegraphics[scale=0.225]{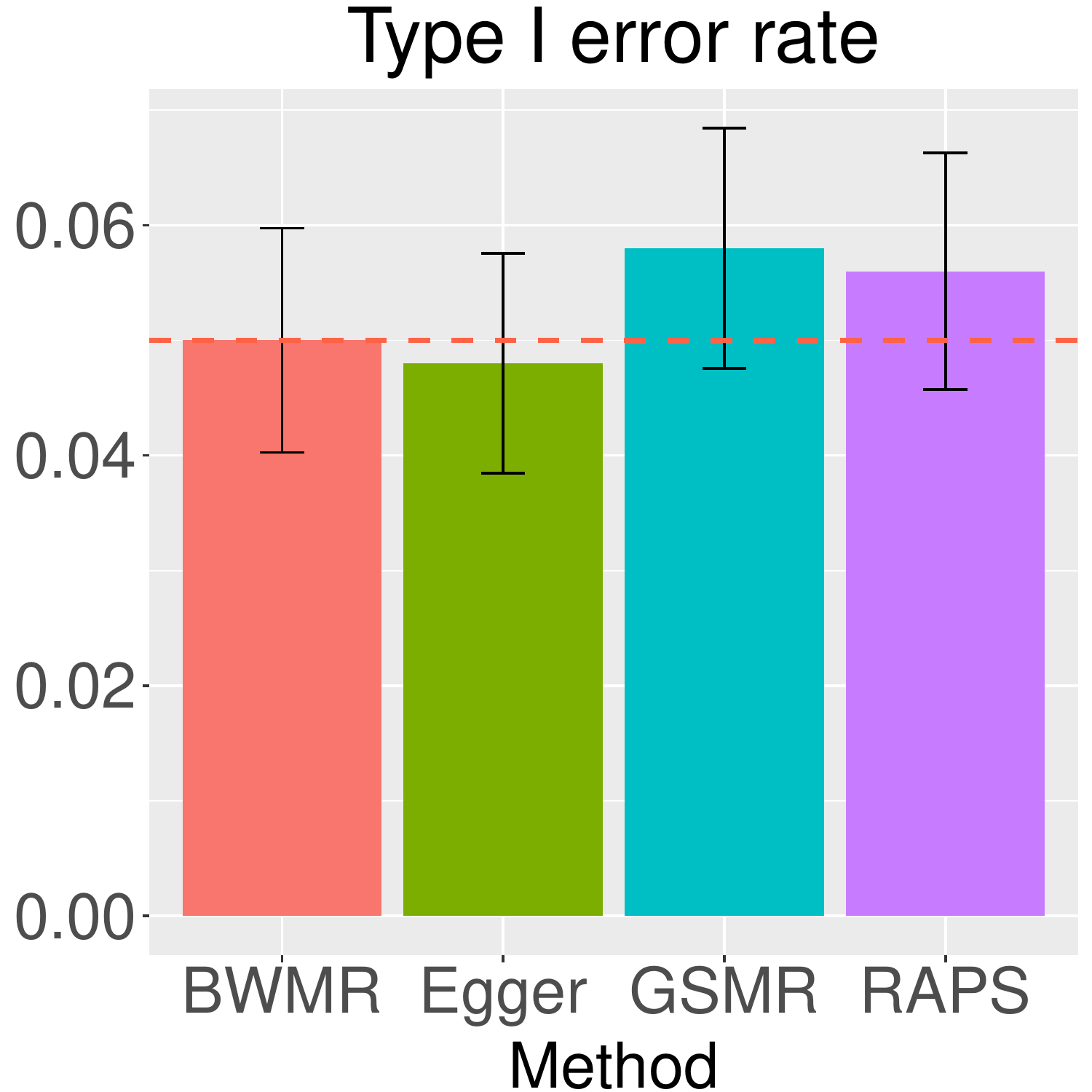}
\includegraphics[scale=0.225]{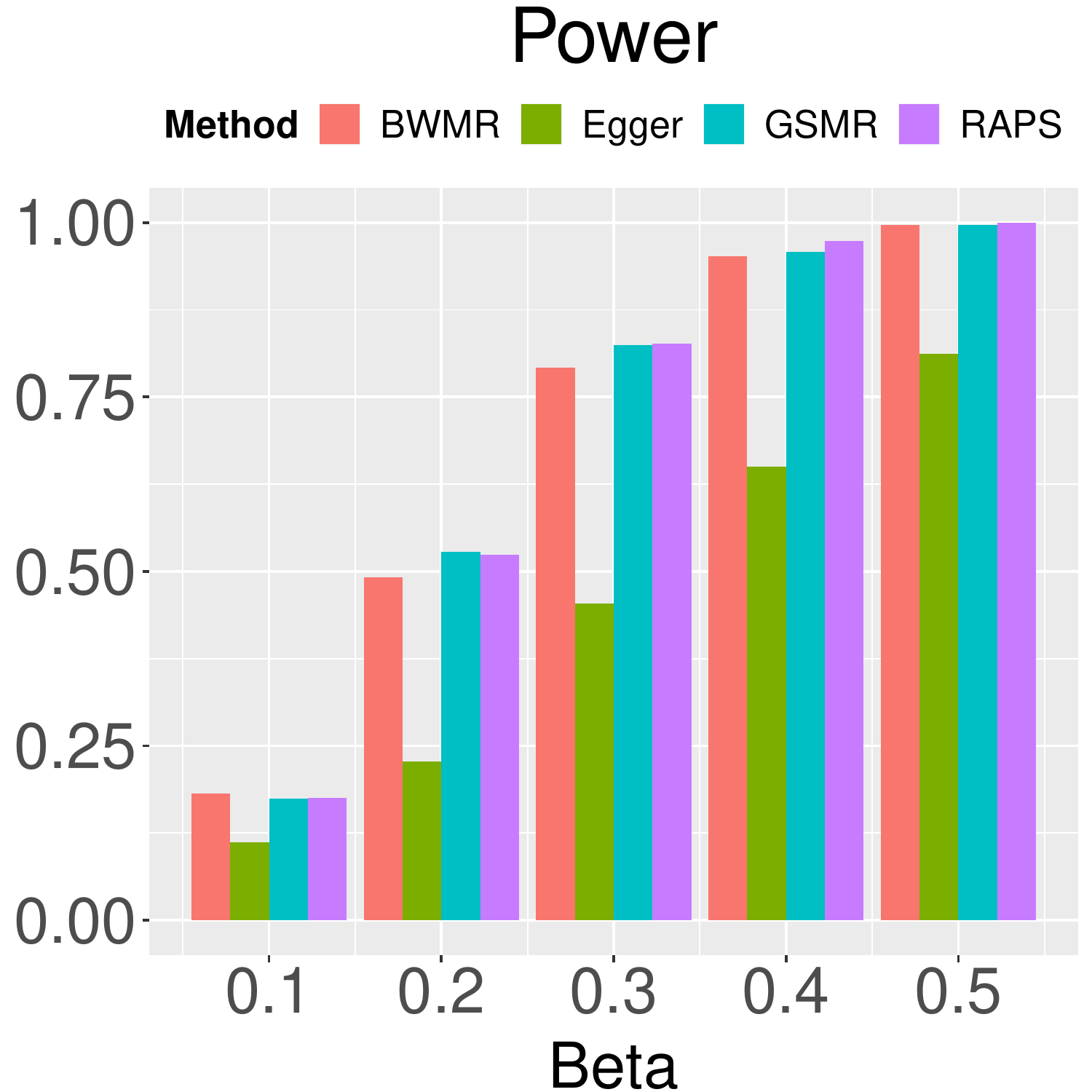}
\par\end{centering}
\caption{Comparison of type I error control and statistical power of BWMR, Egger, GSMR and RAPS in Case-1 of the summary-level data simulation. The simulation parameters were varied in the following range: $\beta\in\{0.0,0.1,0.2,0.3,0.4,0.5\}$, $\tau=0.3$, $\sigma=0.8$, $[c,d]=[0.3,0.5]$. We evaluated the empirical type I error rate and power by controlling type I error rates at the nomial level 0.05. The results were summarized from 500 replications.}
\end{figure}

\newpage{}
\begin{itemize}
  \item Summary-level simulation results in Case-2
\end{itemize}

\begin{figure}[H]
\begin{centering}
\includegraphics[scale=0.4]{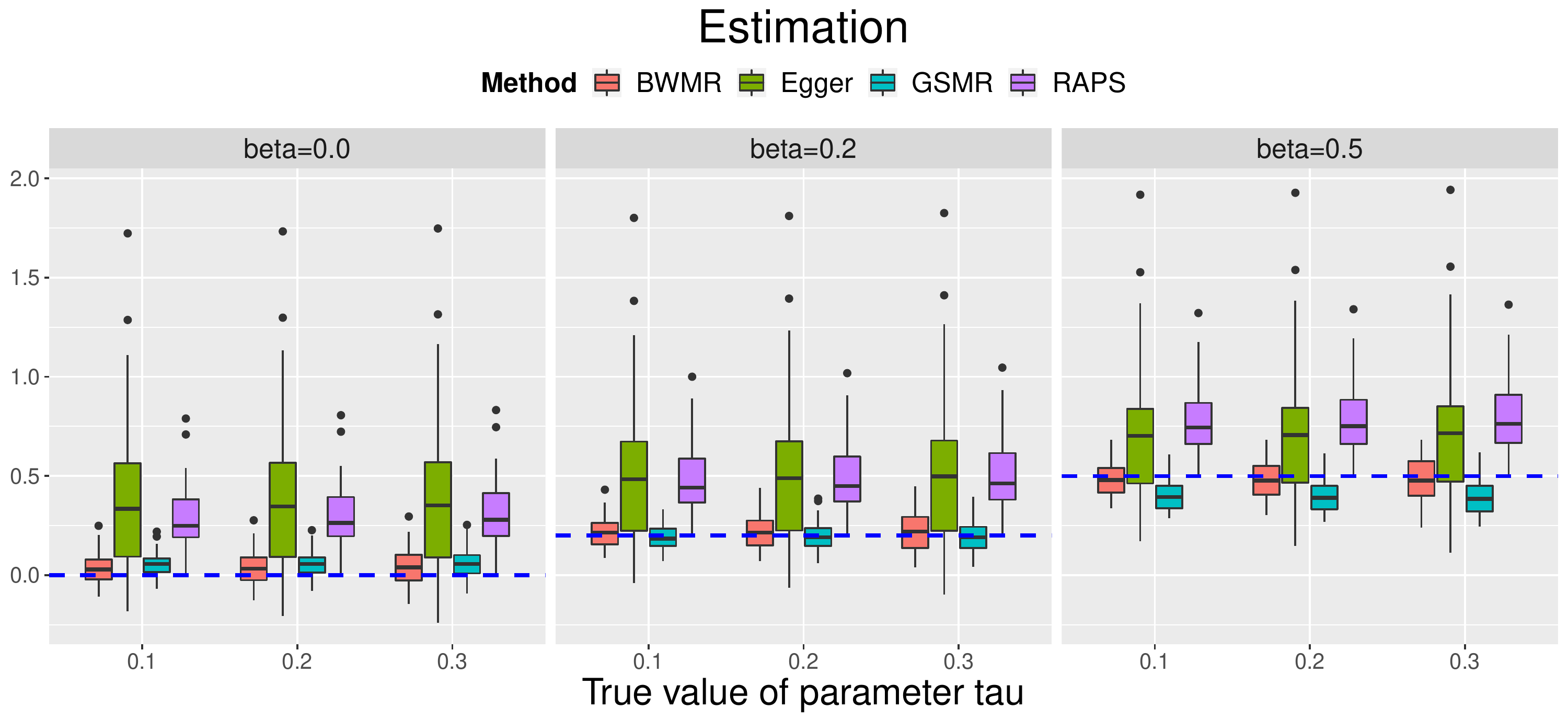}
\par\end{centering}
\caption{Comparison of estimation accuracy of BWMR, Egger, GSMR and RAPS in Case-2 of the summary-level data simulation. The simulation parameters were varied in the following range: $\beta\in\{0.0,0.2,0.5\}$, $\tau\in\{0.1,0.2,0.3\}$, $\sigma=0.8$, $[c,d]=[0.3,0.5]$, $\beta_c=3$, $C=0.2$. The results were summarized from 50 replications.}
\end{figure}

\begin{figure}[H]
\begin{centering}
\includegraphics[scale=0.4]{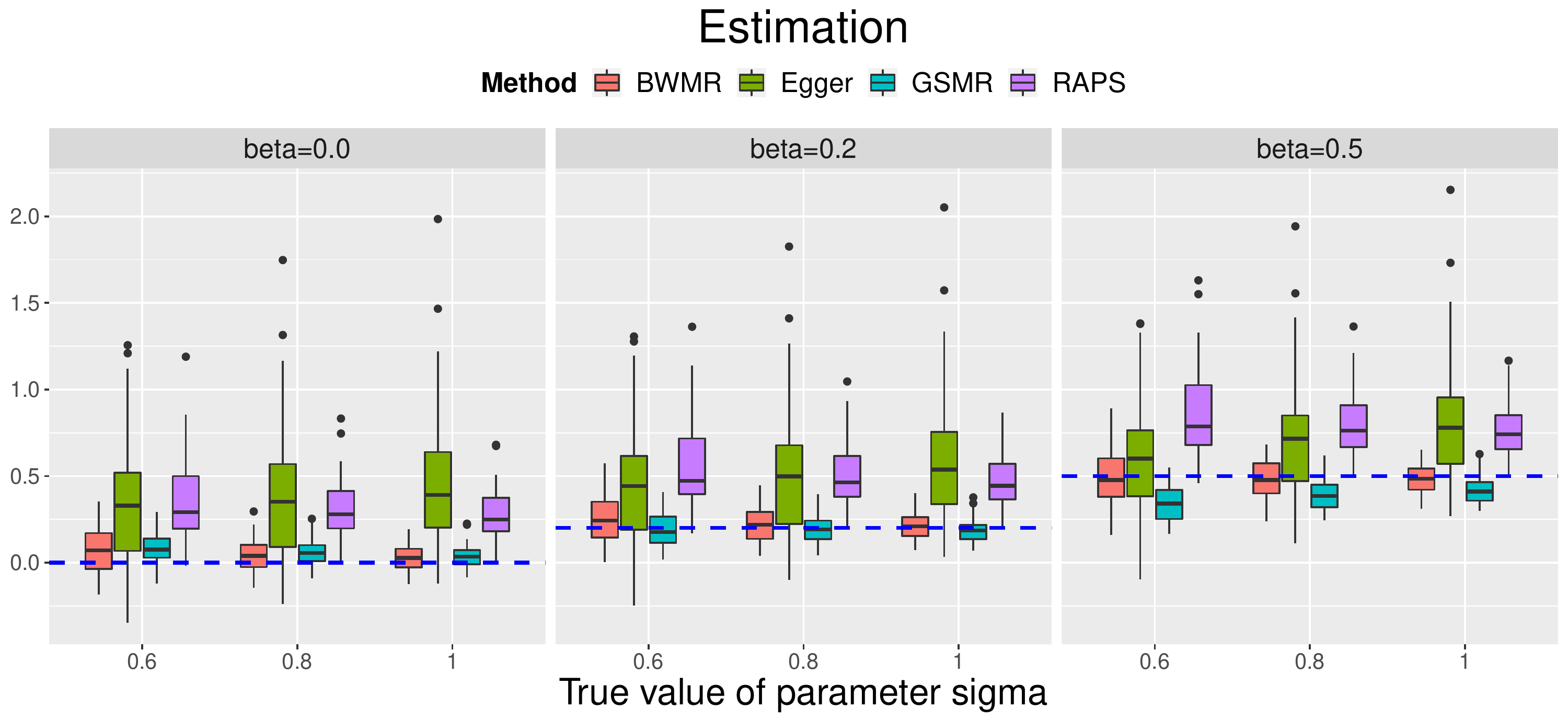}
\par\end{centering}
\caption{Comparison of estimation accuracy of BWMR, Egger, GSMR and RAPS in Case-2 of the summary-level data simulation. The simulation parameters were varied in the following range: $\beta\in\{0.0,0.2,0.5\}$, $\sigma\in\{0.6,0.8,1.0\}$, $\tau=0.2$, $[c,d]=[0.3,0.5]$, $\beta_c=3$, $C=0.2$. The results were summarized from 50 replications.}
\end{figure}

\begin{figure}[H]
\begin{centering}
\includegraphics[scale=0.4]{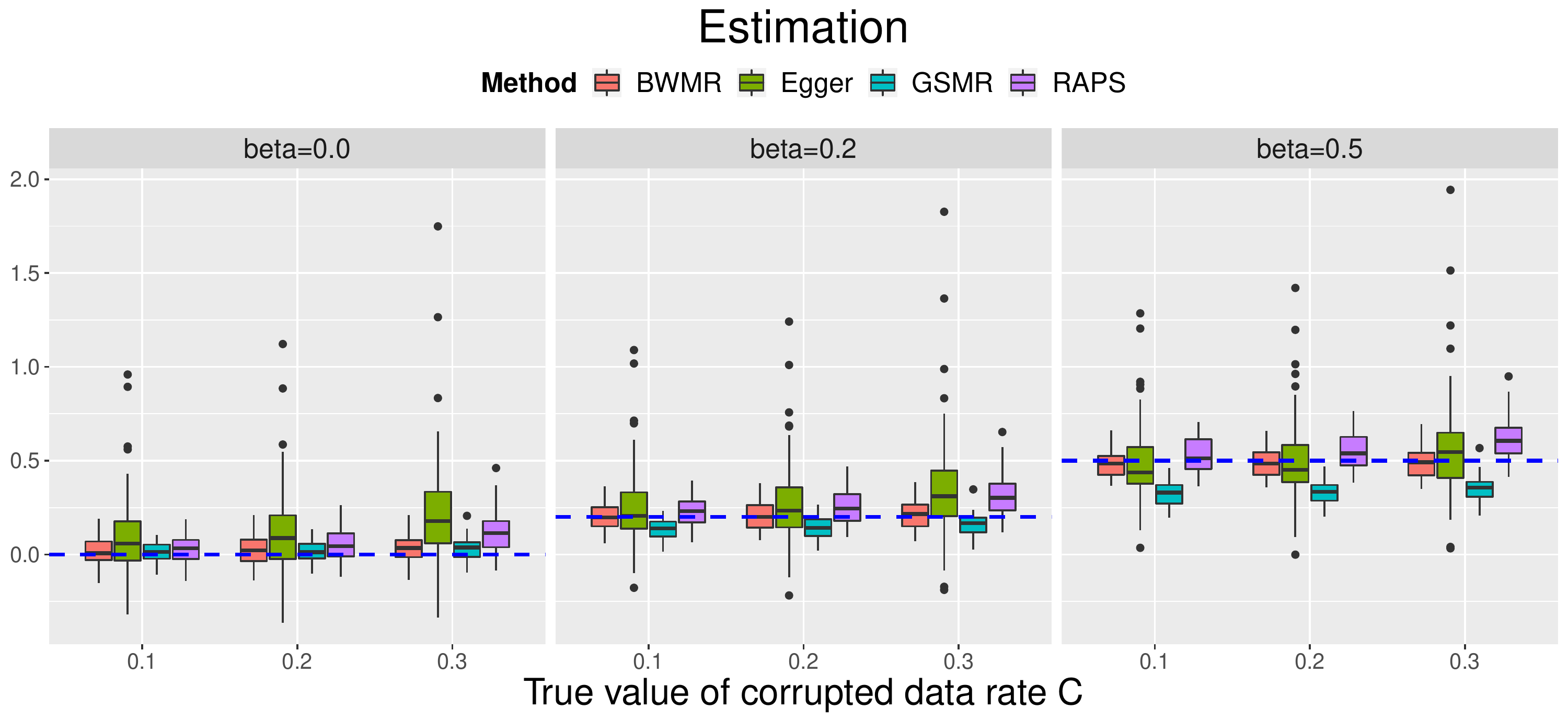}
\par\end{centering}
\caption{Comparison of estimation accuracy of BWMR, Egger, GSMR and RAPS in Case-2 of the summary-level data simulation. The simulation parameters were varied in the following range: $\beta\in\{0.0,0.2,0.5\}$, $C\in\{0.1,0.2,0.3\}$, $\tau=0.2$, $\sigma=0.8$, $[c,d]=[0.3,0.5]$, $\beta_c=3$. The results were summarized from 50 replications.}
\end{figure}

\begin{figure}[H]
\begin{centering}
\includegraphics[scale=0.45]{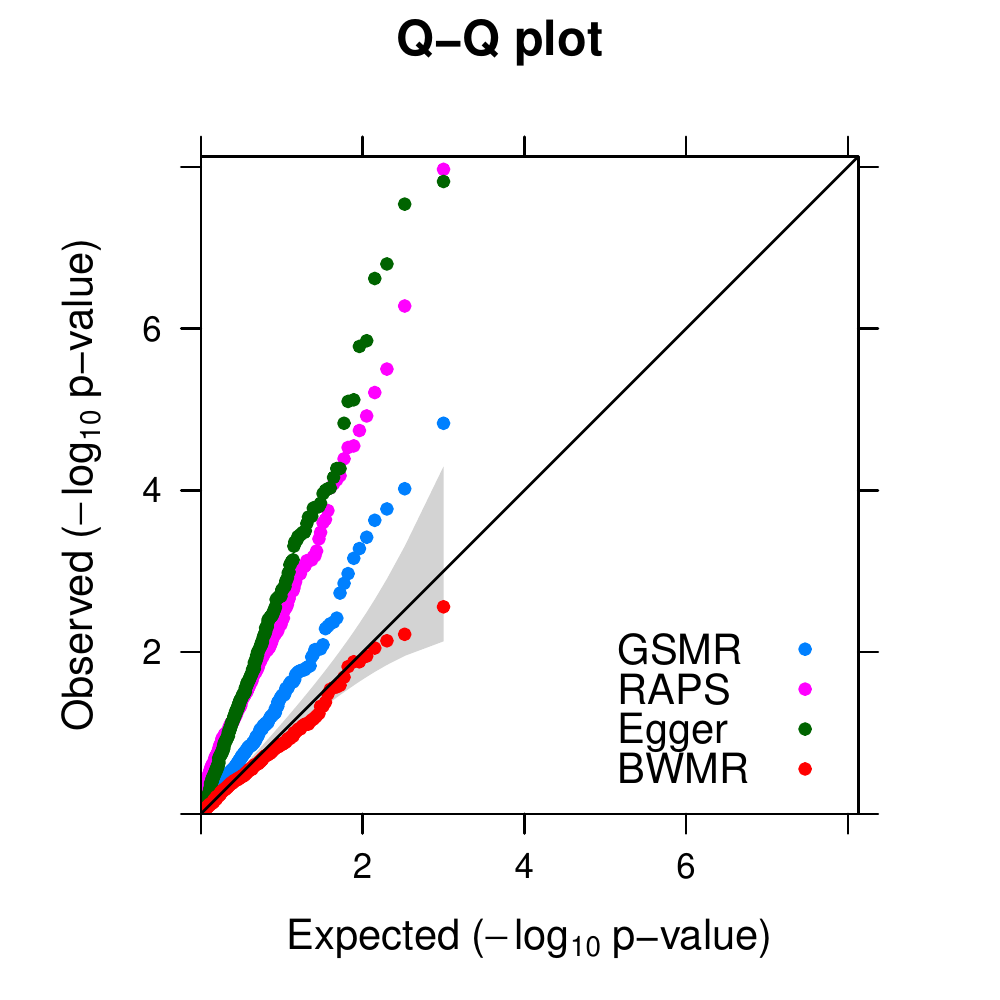}
\includegraphics[scale=0.27]{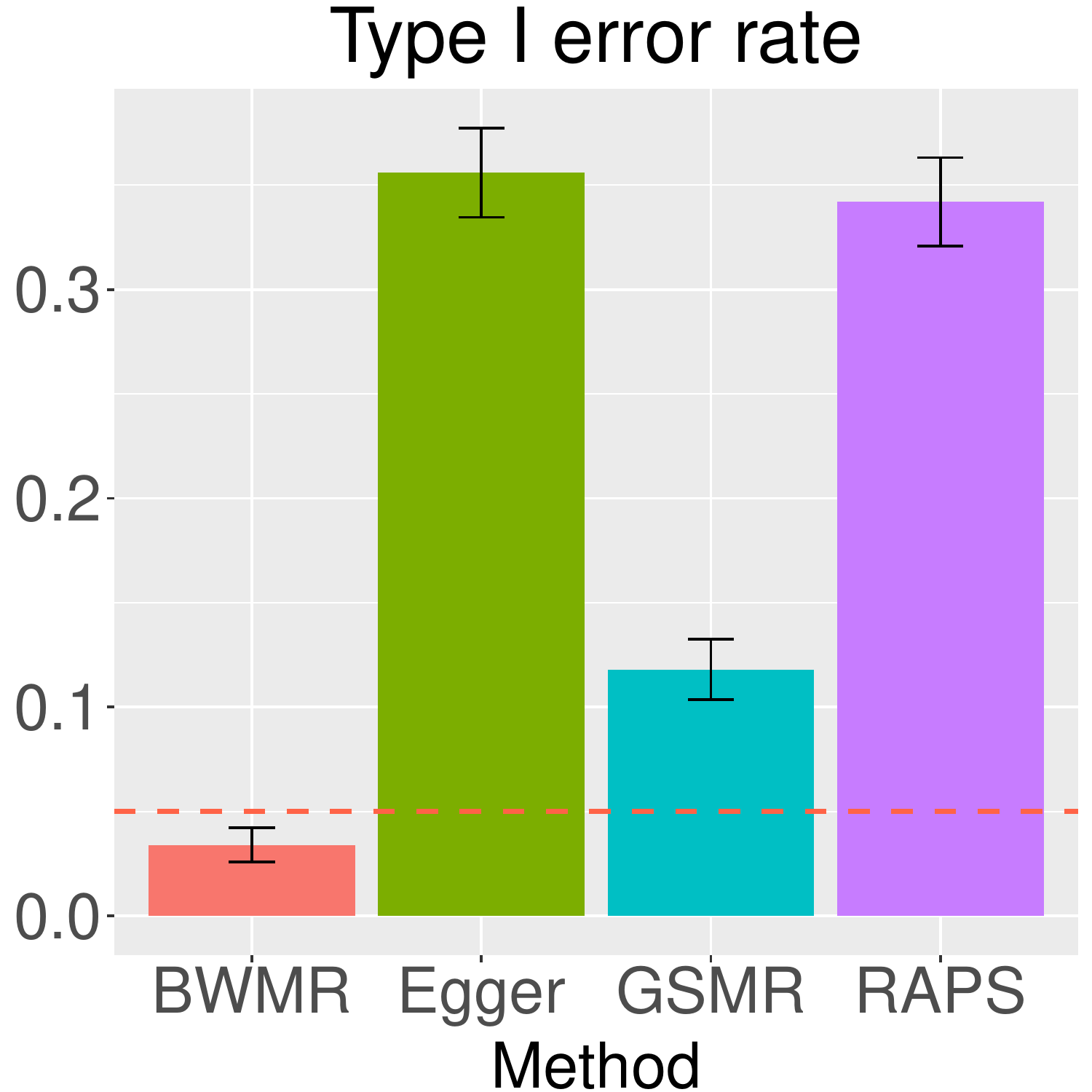}
\includegraphics[scale=0.27]{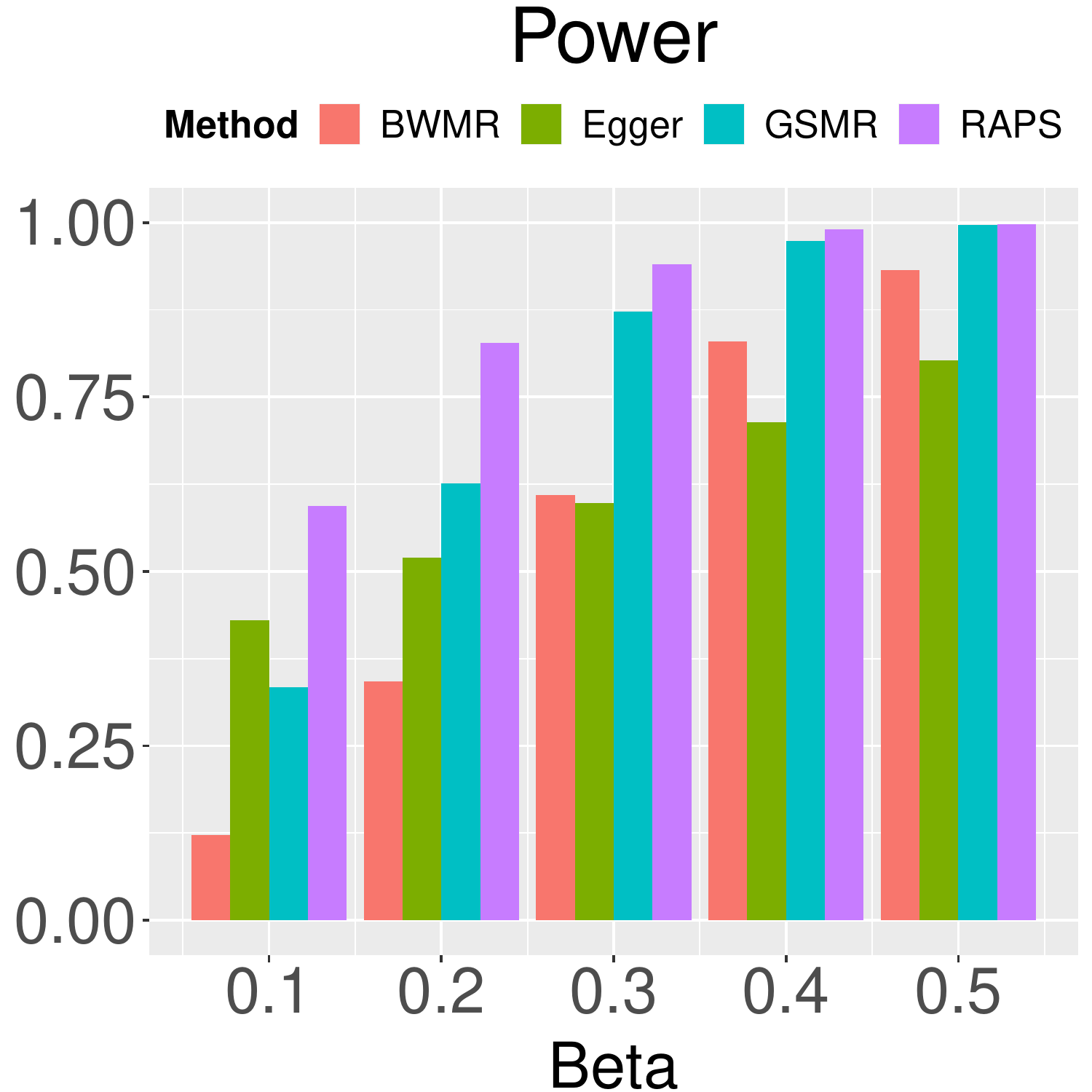}
\par\end{centering}
\caption{Comparison of type I error control and statistical power of BWMR, Egger, GSMR and RAPS in Case-2 of the summary-level data simulation. The simulation parameters were varied in the following range: $\beta\in\{0.0,0.1,0.2,0.3,0.4,0.5\}$, $\tau=0.3$, $\sigma=0.8$, $[c,d]=[0.3,0.5]$, $\beta_c=3$, $C=0.2$. We evaluated the empirical type I error rate and power by controlling type I error rates at the nomial level 0.05. The results were summarized from 500 replications.}
\end{figure}

\newpage{}
\begin{itemize}
  \item Summary-level simulation results in Case-3
\end{itemize}

\begin{figure}[H]
\begin{centering}
\includegraphics[scale=0.4]{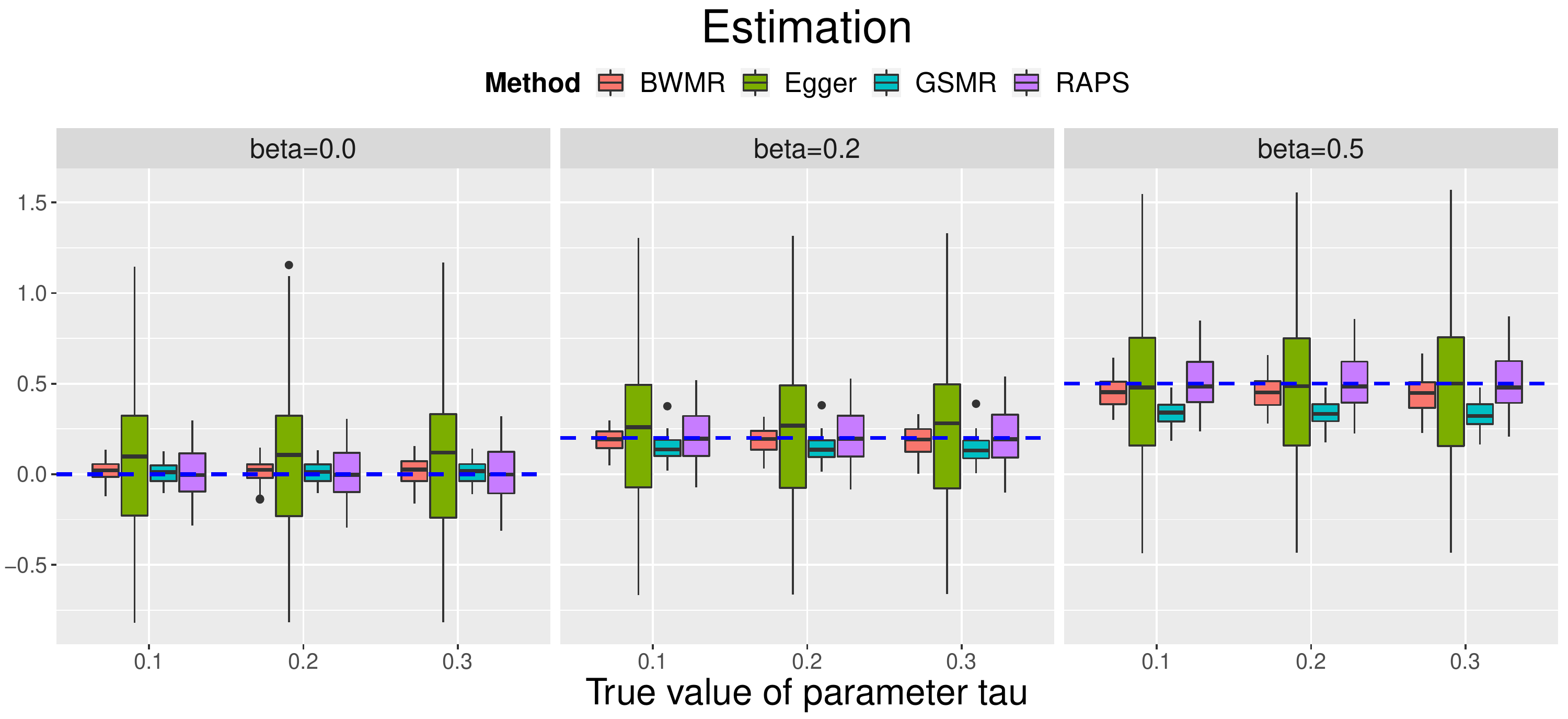}
\par\end{centering}
\caption{Comparison of estimation accuracy of BWMR, Egger, GSMR and RAPS in Case-3 of the summary-level data simulation. The simulation parameters were varied in the following range: $\beta\in\{0.0,0.2,0.5\}$, $\tau\in\{0.1,0.2,0.3\}$, $\sigma=0.8$, $[c,d]=[0.3,0.5]$, $\tau_c=3$, $C=0.2$. The results were summarized from 50 replications.}
\end{figure}

\begin{figure}[H]
\begin{centering}
\includegraphics[scale=0.4]{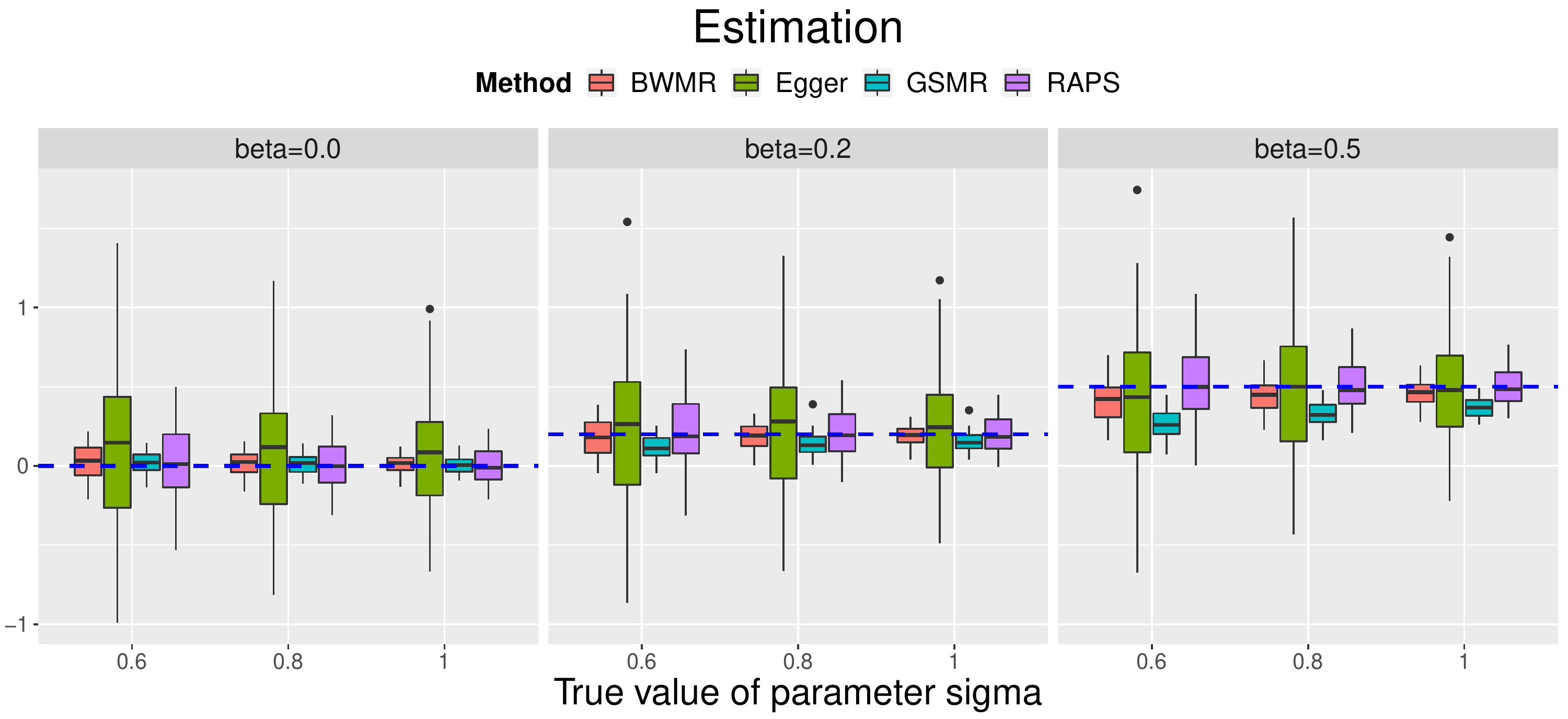}
\par\end{centering}
\caption{Comparison of estimation accuracy of BWMR, Egger, GSMR and RAPS in Case-3 of the summary-level data simulation. The simulation parameters were varied in the following range: $\beta\in\{0.0,0.2,0.5\}$, $\sigma\in\{0.6,0.8,1.0\}$, $\tau=0.2$, $[c,d]=[0.3,0.5]$, $\tau_c=3$, $C=0.2$. The results were summarized from 50 replications.}
\end{figure}

\begin{figure}[H]
\begin{centering}
\includegraphics[scale=0.4]{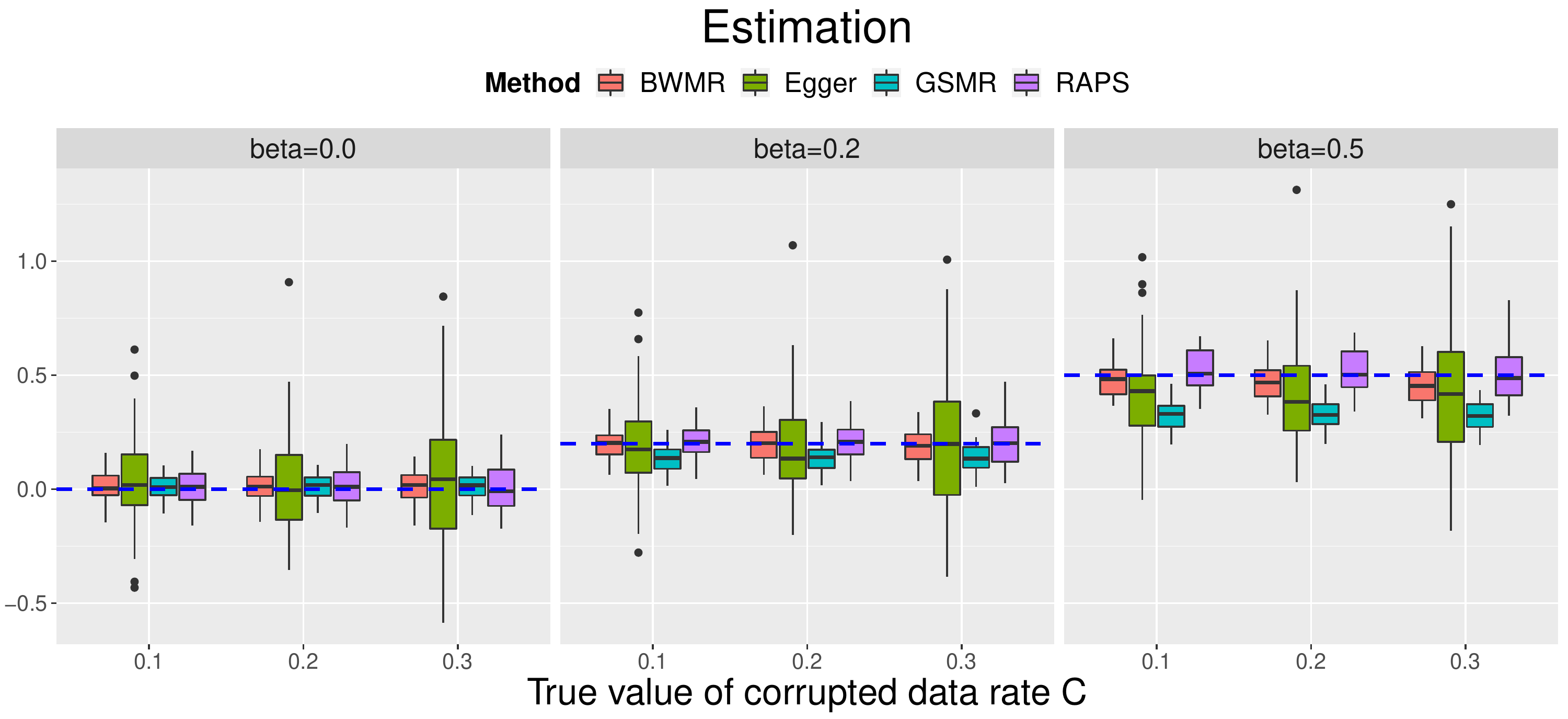}
\par\end{centering}
\caption{Comparison of estimation accuracy of BWMR, Egger, GSMR and RAPS in Case-3 of the summary-level data simulation. The simulation parameters were varied in the following range: $\beta\in\{0.0,0.2,0.5\}$, $C\in\{0.1,0.2,0.3\}$, $\tau=0.2$, $\sigma=0.8$, $[c,d]=[0.3,0.5]$, $\tau_c=3$. The results were summarized from 50 replications.}
\end{figure}

\begin{figure}[H]
\begin{centering}
\includegraphics[scale=0.45]{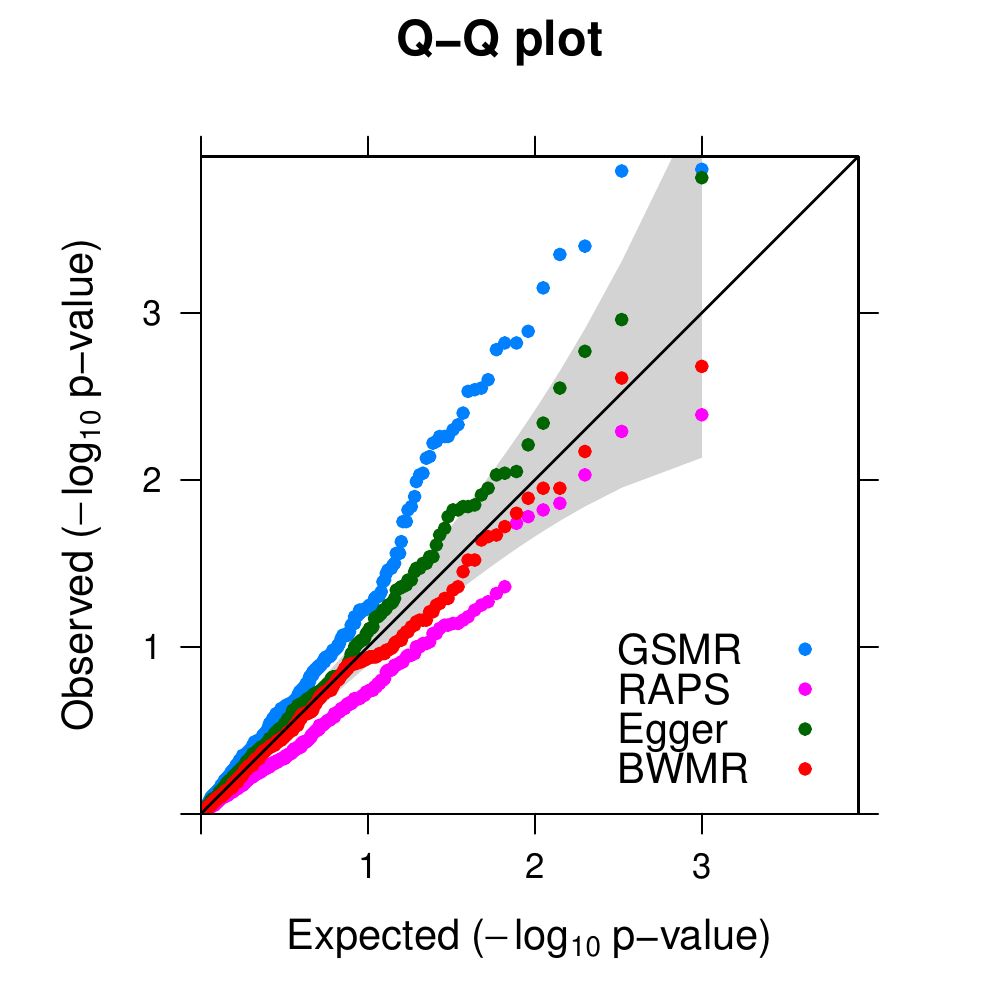}
\includegraphics[scale=0.27]{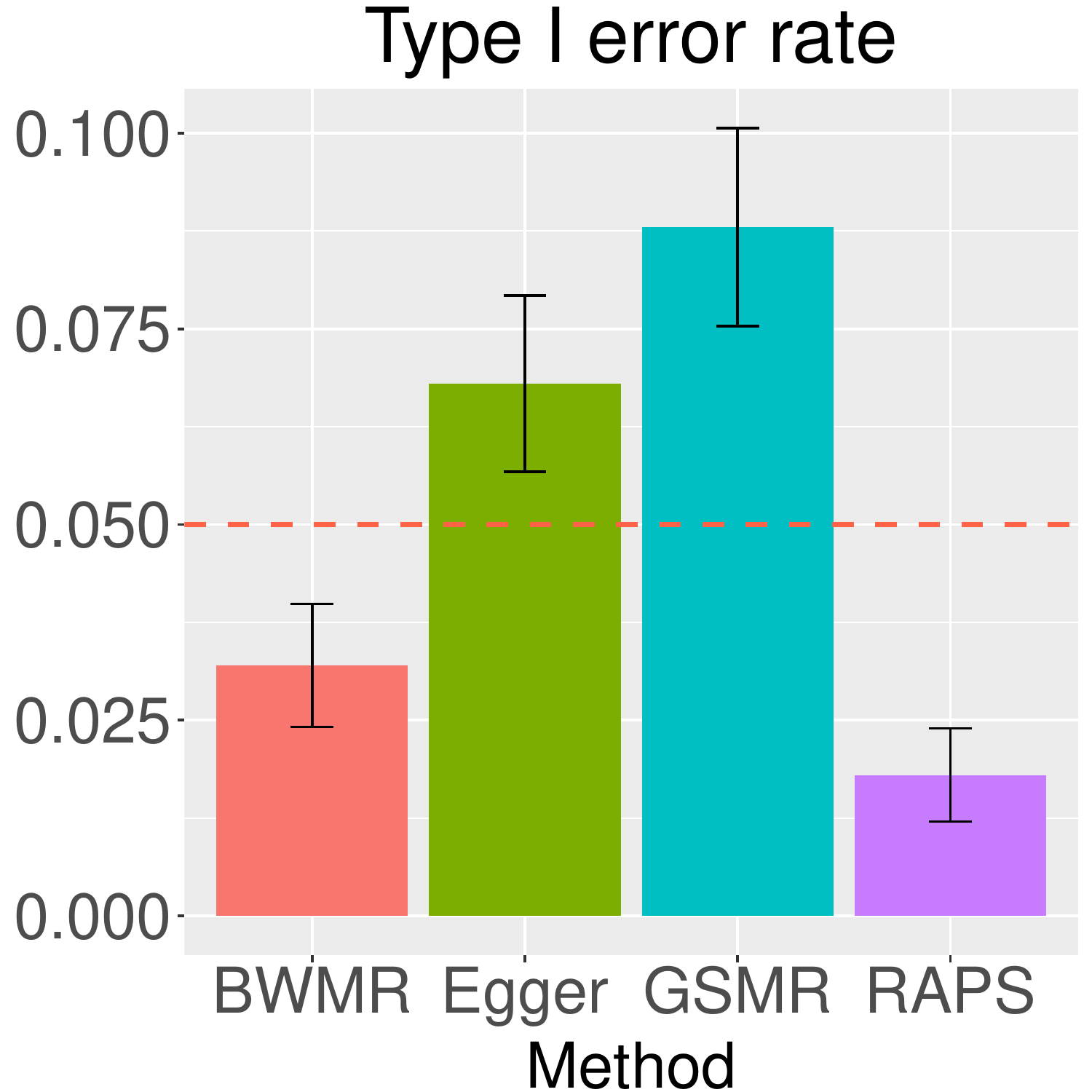}
\includegraphics[scale=0.27]{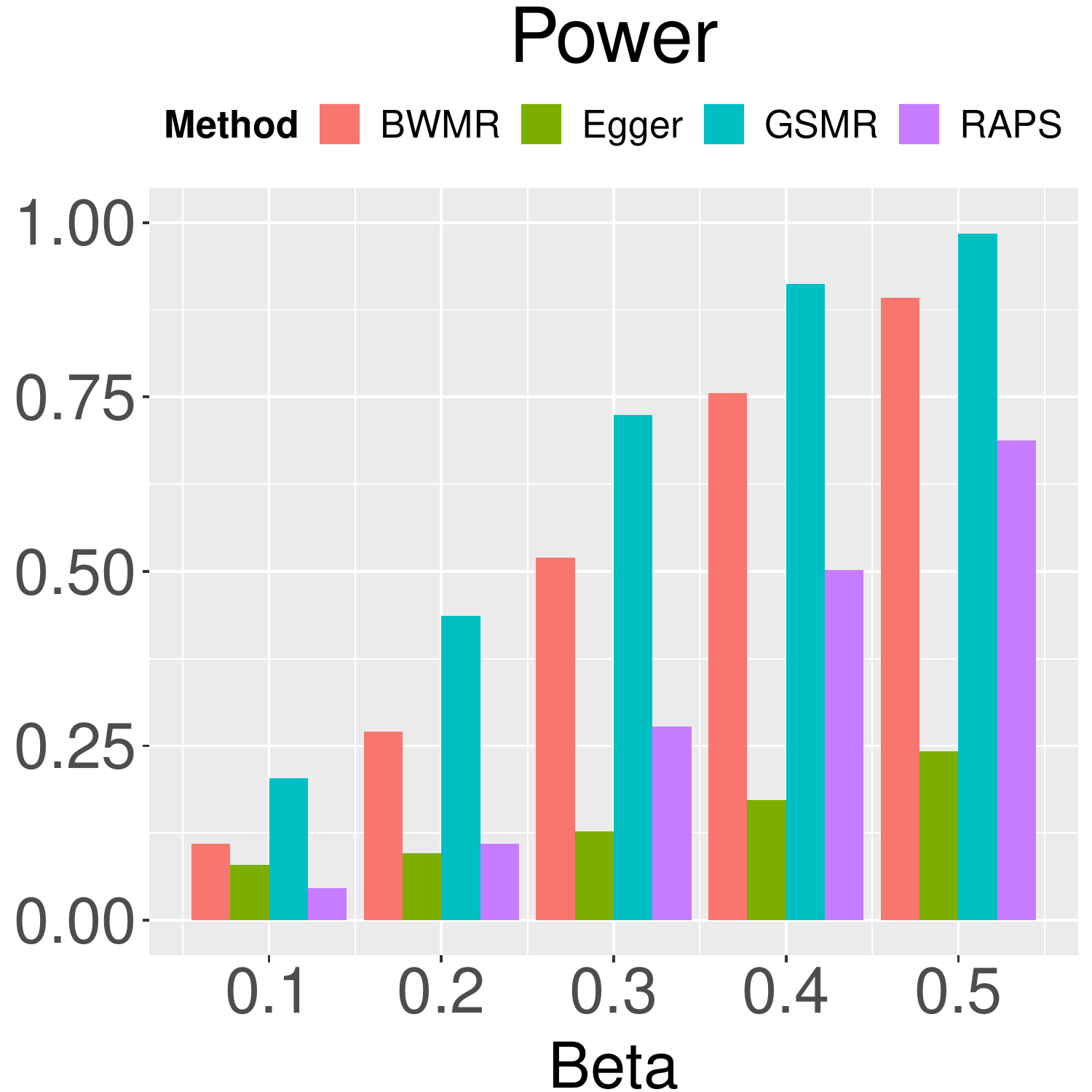}
\par\end{centering}
\caption{Comparison of type I error control and statistical power of BWMR, Egger, GSMR and RAPS in Case-3 of the summary-level data simulation. The simulation parameters were varied in the following range: $\beta\in\{0.0,0.1,0.2,0.3,0.4,0.5\}$, $\tau=0.3$, $\sigma=0.8$, $[c,d]=[0.3,0.5]$, $\tau_c=3$, $C=0.2$. We evaluated the empirical type I error rate and power by controlling type I error rates at the nomial level 0.05. The results were summarized from 500 replications.}
\end{figure}

\newpage{}
\begin{itemize}
  \item Summary-level simulation results in Case-4
\end{itemize}

\begin{figure}[H]
\begin{centering}
\includegraphics[scale=0.34]{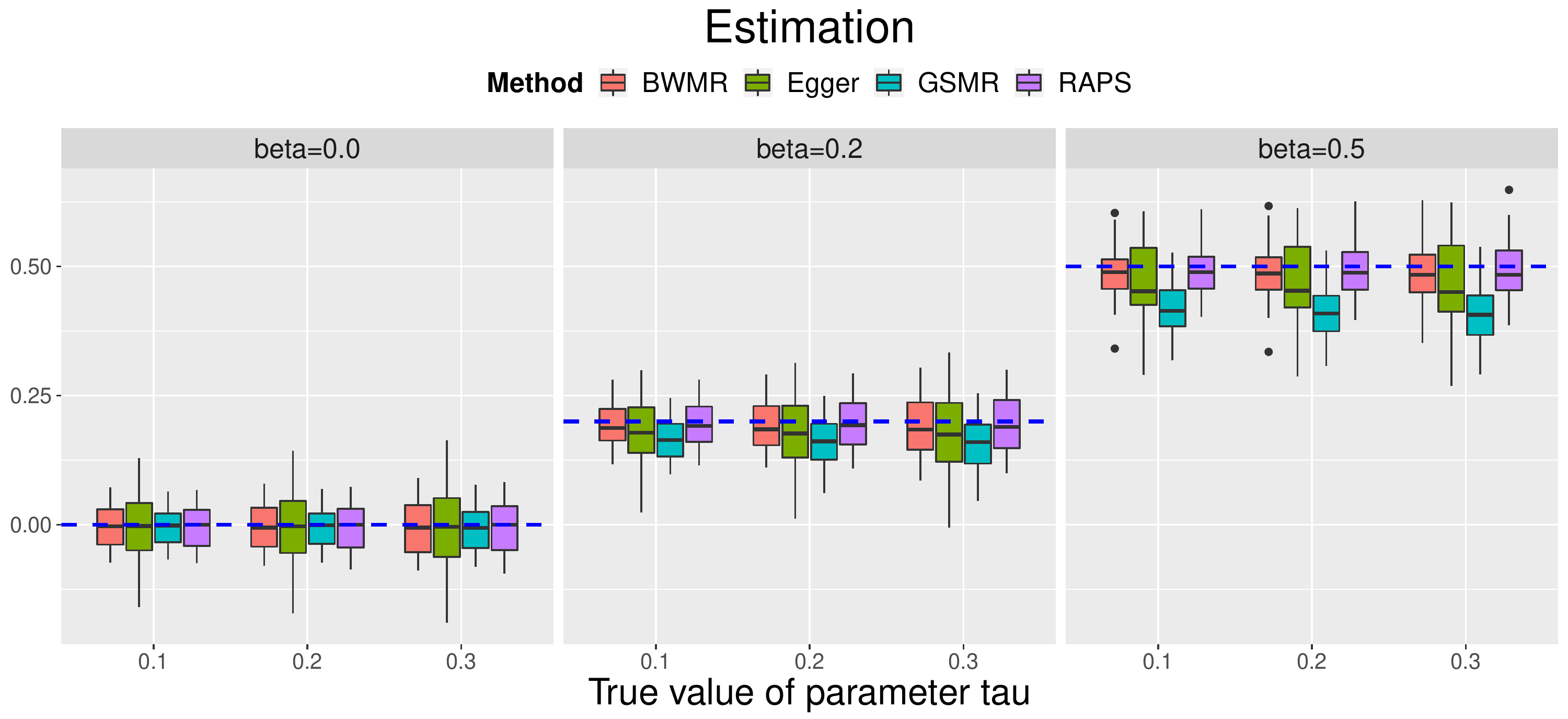}
\par\end{centering}
\caption{Comparison of estimation accuracy of BWMR, Egger, GSMR and RAPS in Case-4 of the summary-level data simulation. The simulation parameters were varied in the following range: $\beta\in\{0.0,0.2,0.5\}$, $\tau\in\{0.1,0.2,0.3\}$, $\sigma=0.8$, $[c,d]=[0.3,0.5]$, $R=0.2$. The results were summarized from 50 replications.}
\end{figure}

\begin{figure}[H]
\begin{centering}
\includegraphics[scale=0.34]{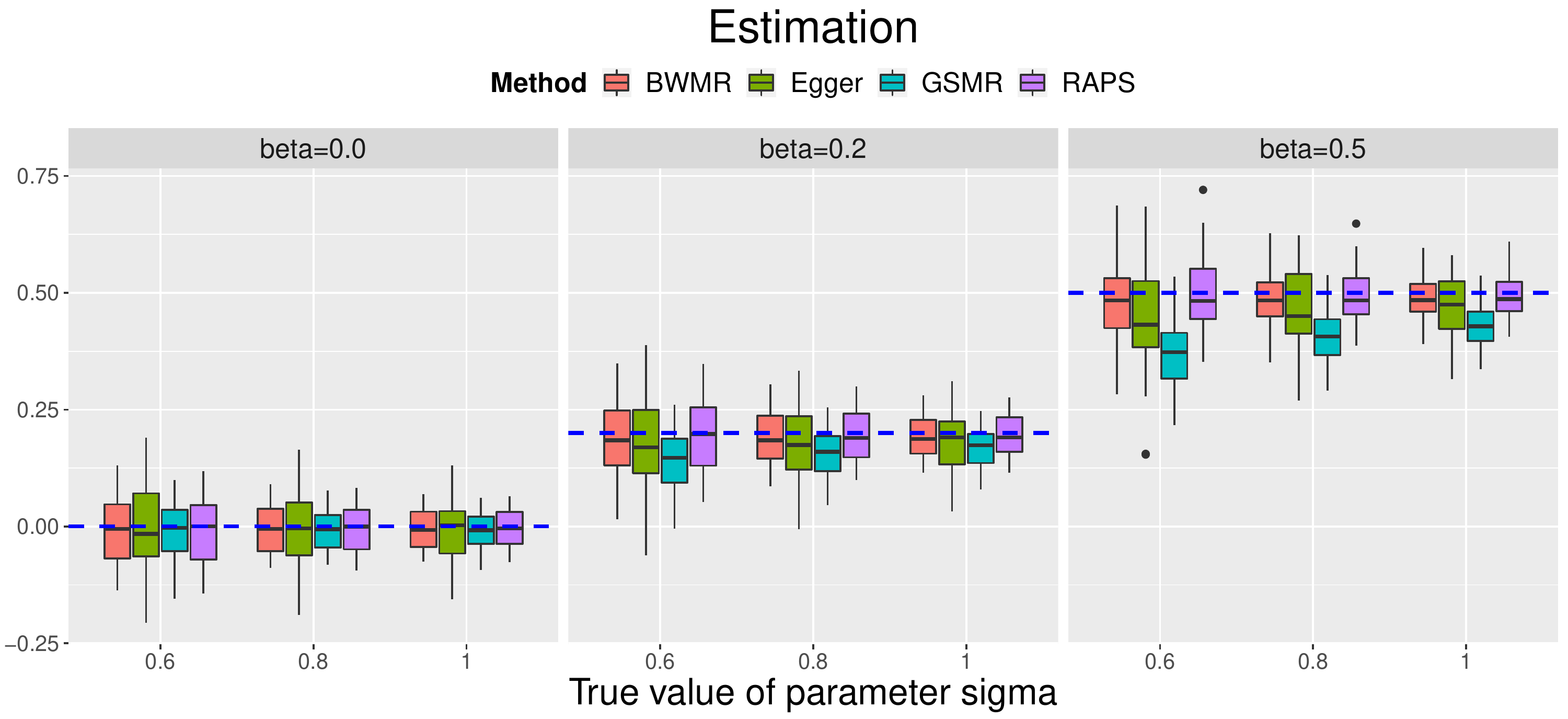}
\par\end{centering}
\caption{Comparison of estimation accuracy of BWMR, Egger, GSMR and RAPS in Case-4 of the summary-level data simulation. The simulation parameters were varied in the following range: $\beta\in\{0.0,0.2,0.5\}$, $\sigma\in\{0.6,0.8,1.0\}$, $\tau=0.2$, $[c,d]=[0.3,0.5]$, $R=0.2$. The results were summarized from 50 replications.}
\end{figure}

\begin{figure}[H]
\begin{centering}
\includegraphics[scale=0.375]{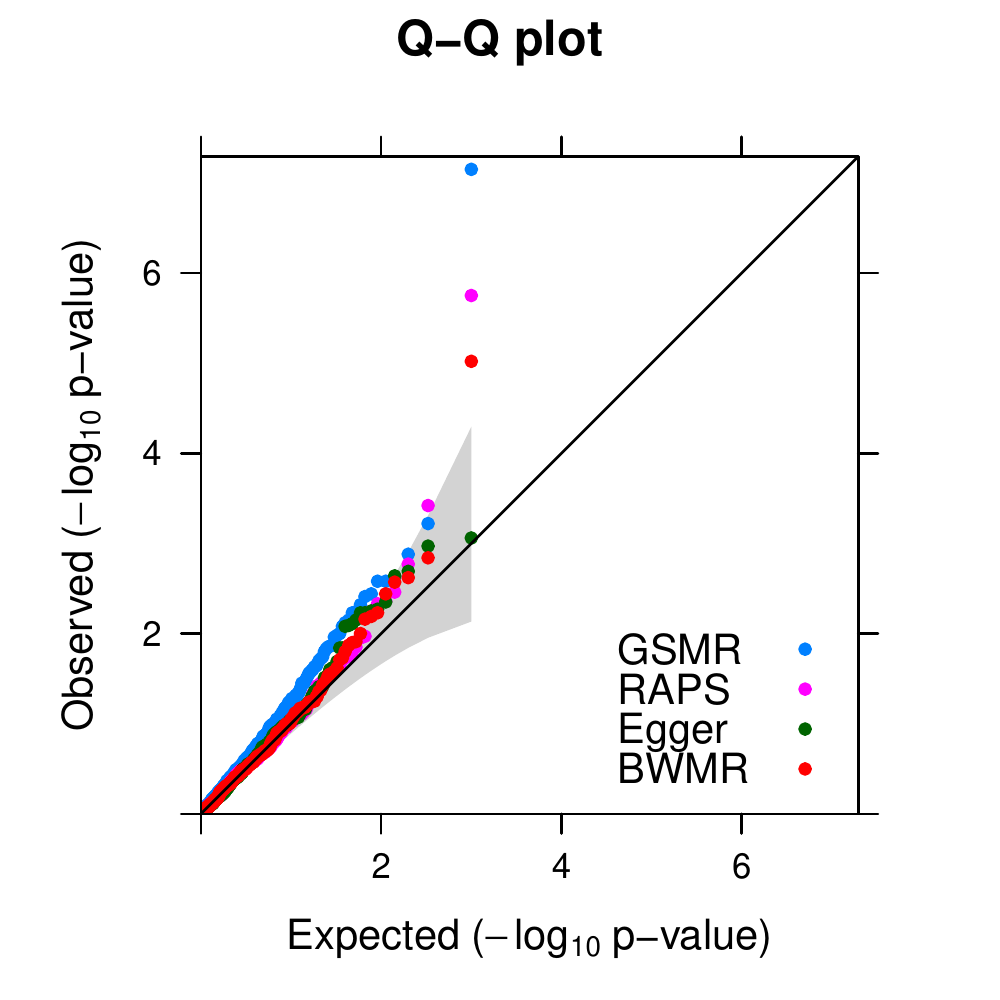}
\includegraphics[scale=0.225]{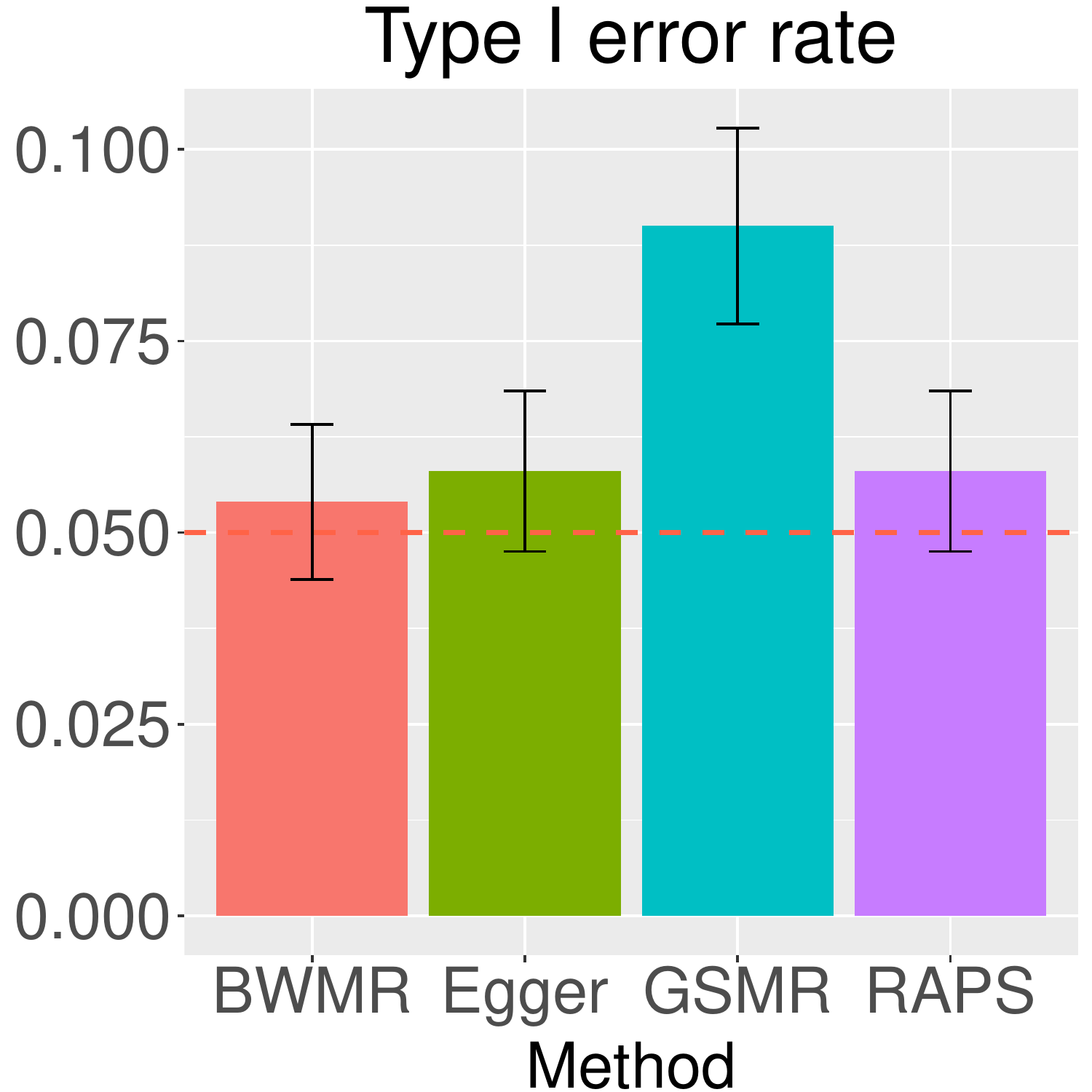}
\includegraphics[scale=0.225]{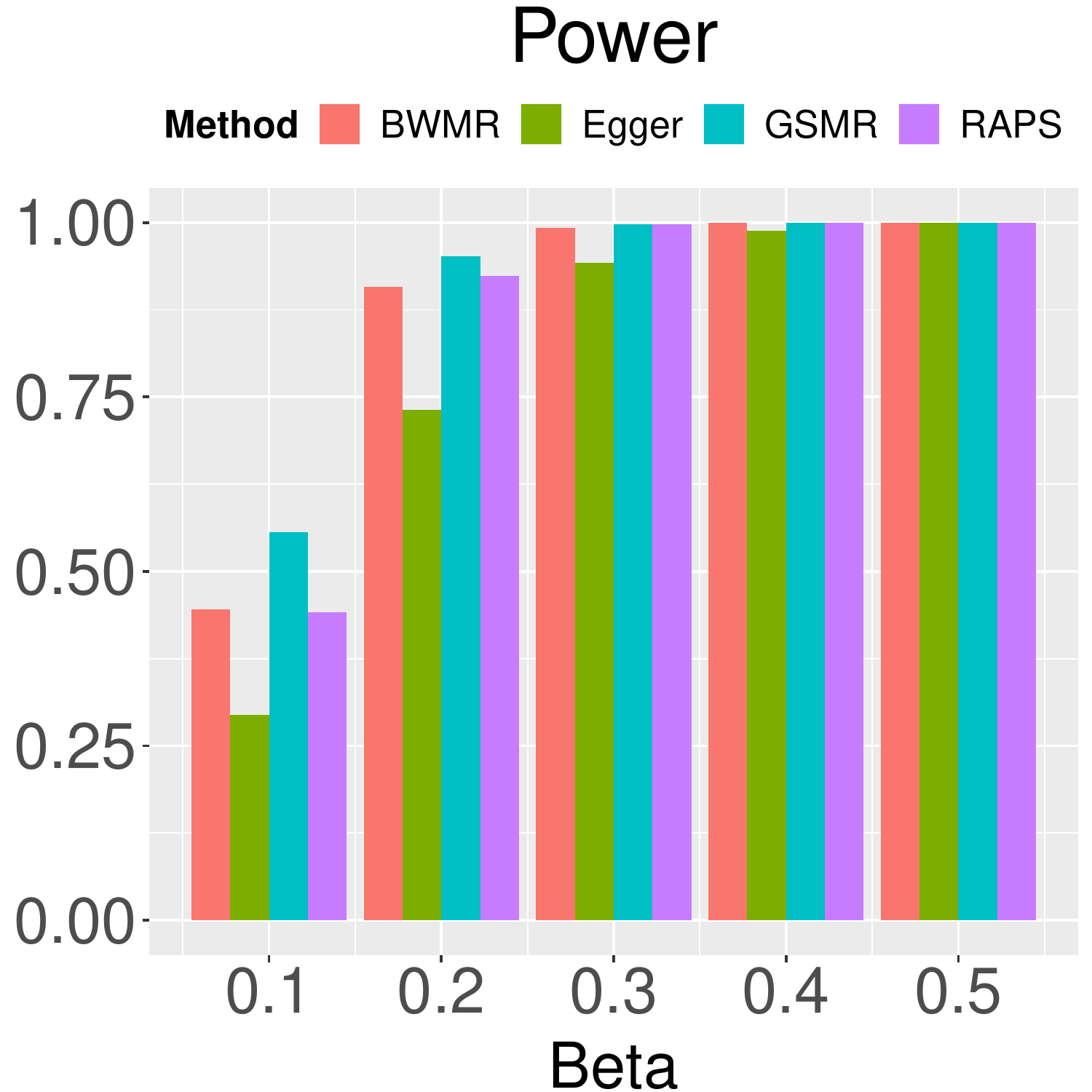}
\par\end{centering}
\caption{Comparison of type I error control and statistical power of BWMR, Egger, GSMR and RAPS in Case-4 of the summary-level data simulation. The simulation parameters were varied in the following range: $\beta\in\{0.0,0.1,0.2,0.3,0.4,0.5\}$, $\tau=0.3$, $\sigma=0.8$, $[c,d]=[0.3,0.5]$, $R=0.2$. We evaluated the empirical type I error rate and power by controlling type I error rates at the nomial level 0.05. The results were summarized from 500 replications.}
\end{figure}

\newpage{}
\begin{itemize}
  \item Summary-level simulation results in Case-5
\end{itemize}

\begin{figure}[H]
\begin{centering}
\includegraphics[scale=0.34]{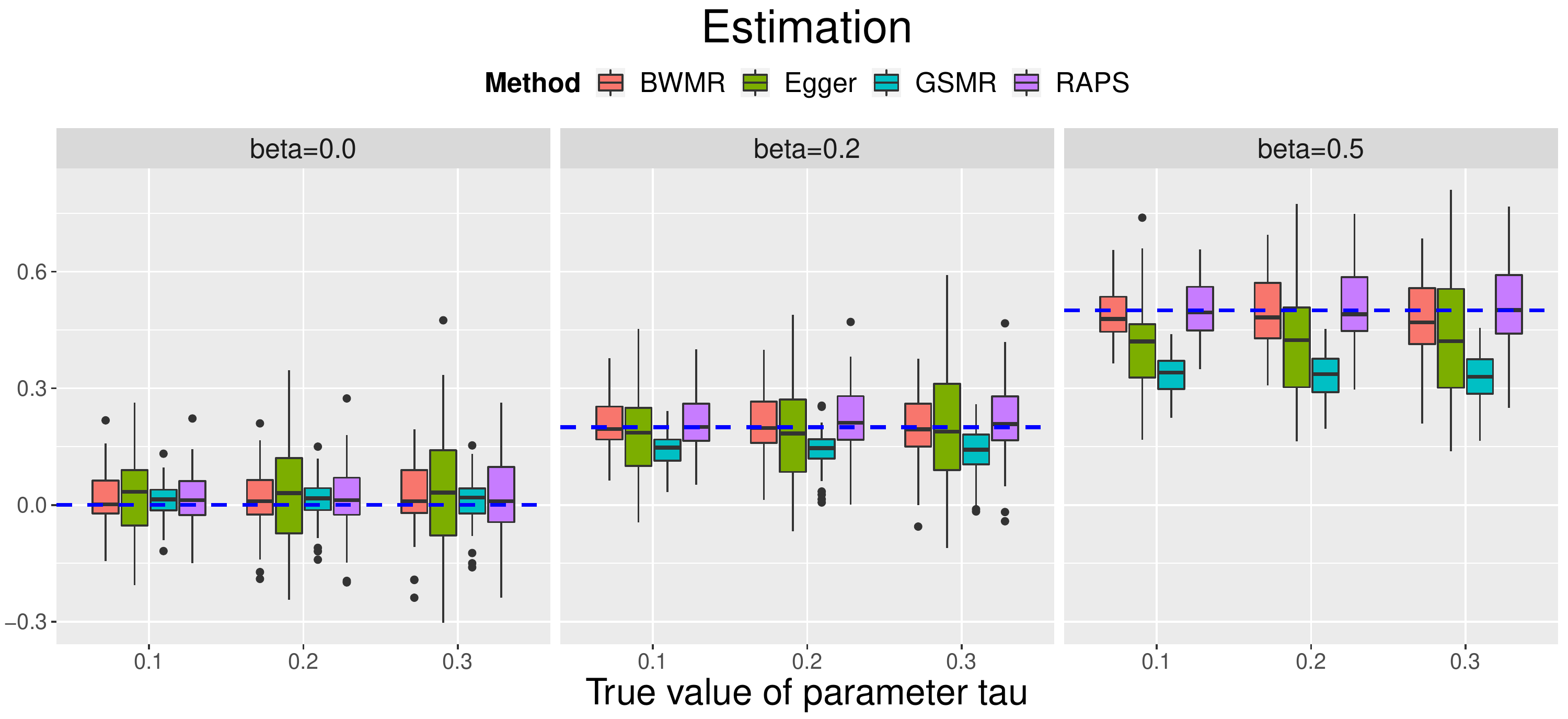}
\par\end{centering}
\caption{Comparison of estimation accuracy of BWMR, Egger, GSMR and RAPS in Case-5 of the summary-level data simulation. The simulation parameters were varied in the following range: $\beta\in\{0.0,0.2,0.5\}$, $\tau\in\{0.1,0.2,0.3\}$, $\sigma=0.8$, $[c,d]=[0.3,0.5]$, $r=1$. The results were summarized from 50 replications.}
\end{figure}

\begin{figure}[H]
\begin{centering}
\includegraphics[scale=0.34]{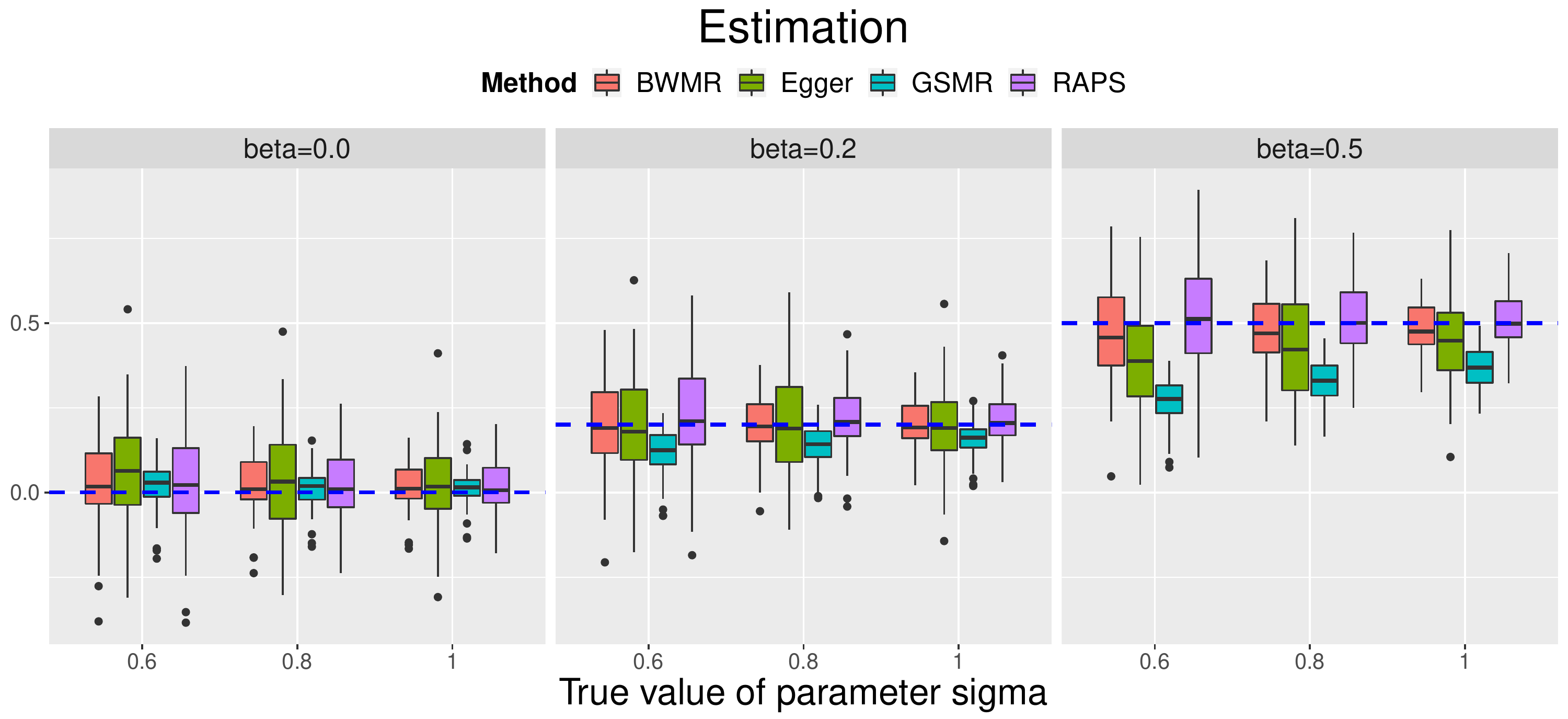}
\par\end{centering}
\caption{Comparison of estimation accuracy of BWMR, Egger, GSMR and RAPS in Case-5 of the summary-level data simulation. The simulation parameters were varied in the following range: $\beta\in\{0.0,0.2,0.5\}$, $\sigma\in\{0.6,0.8,1.0\}$, $\tau=0.2$, $[c,d]=[0.3,0.5]$, $r=1$. The results were summarized from 50 replications.}
\end{figure}

\begin{figure}[H]
\begin{centering}
\includegraphics[scale=0.375]{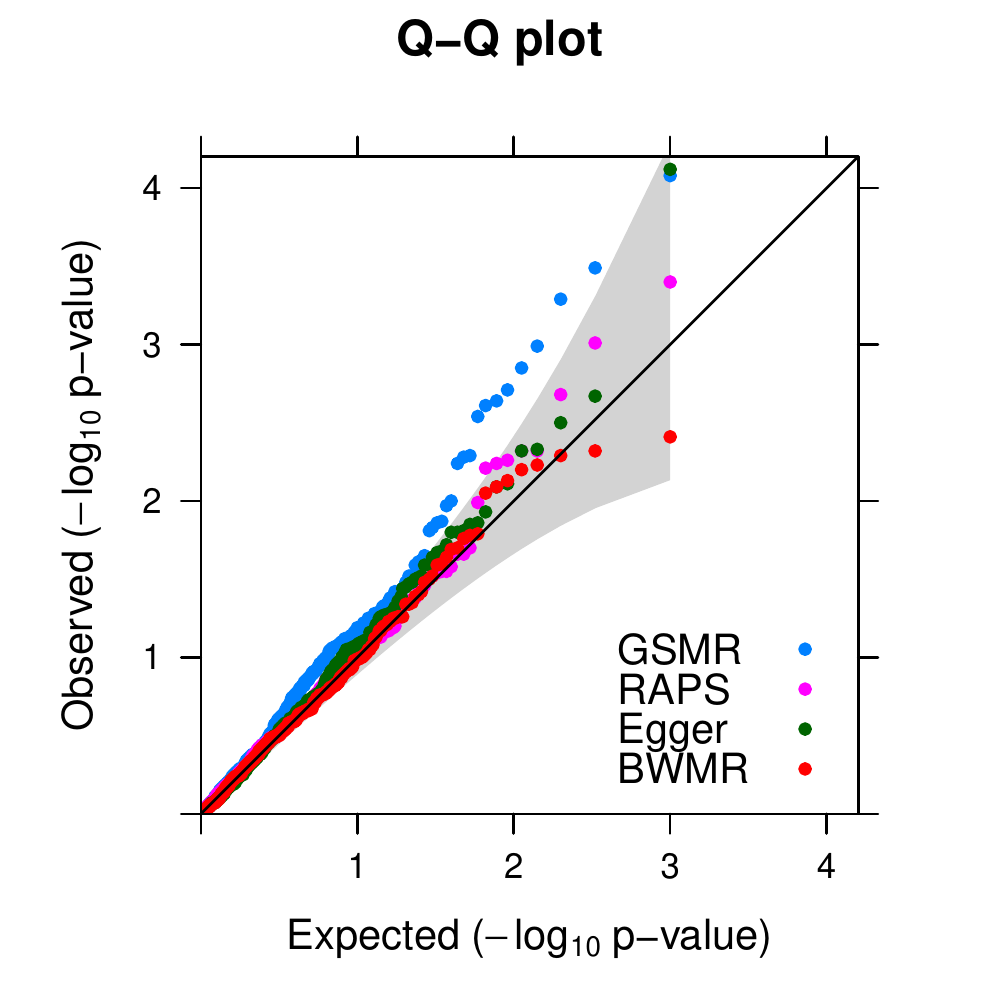}
\includegraphics[scale=0.225]{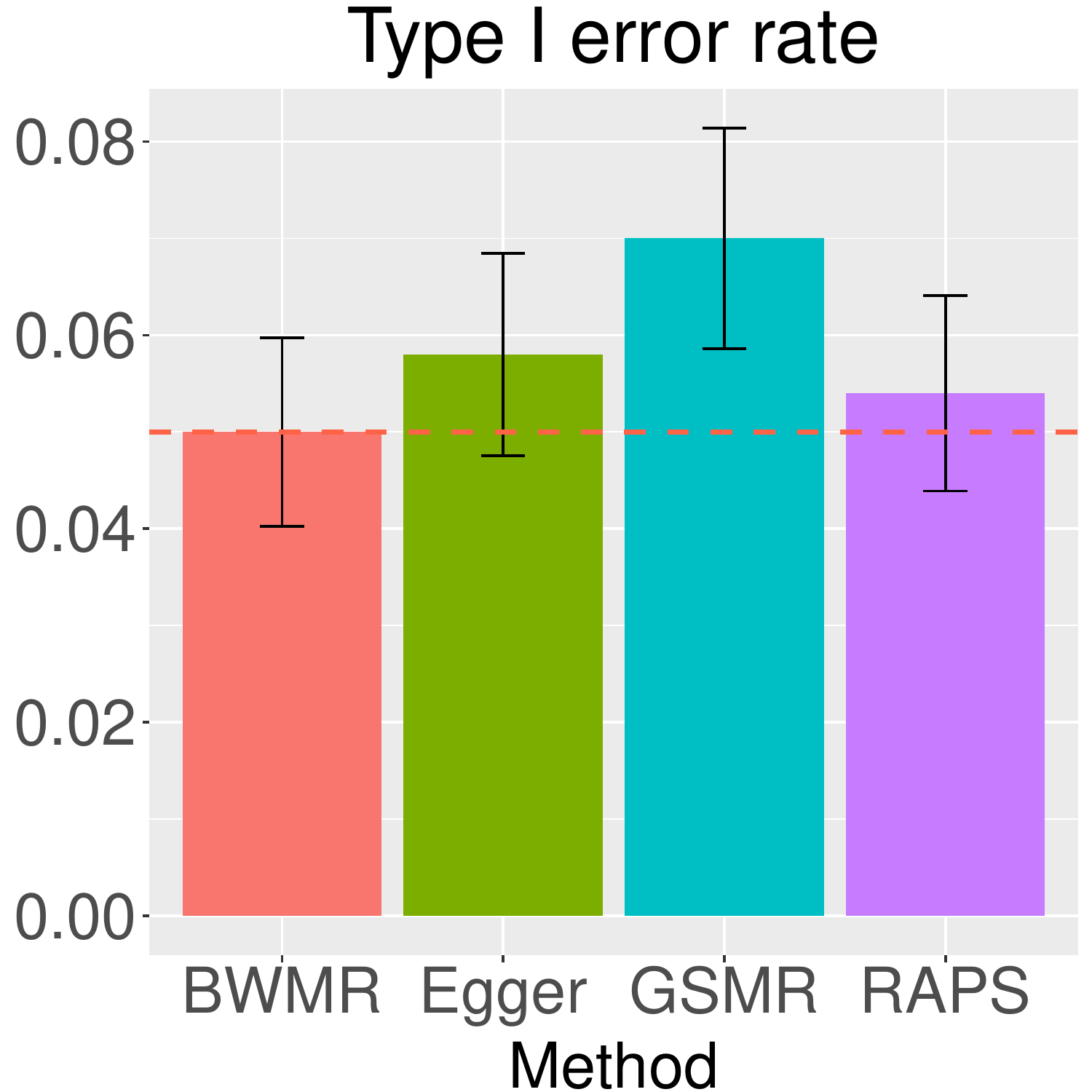}
\includegraphics[scale=0.225]{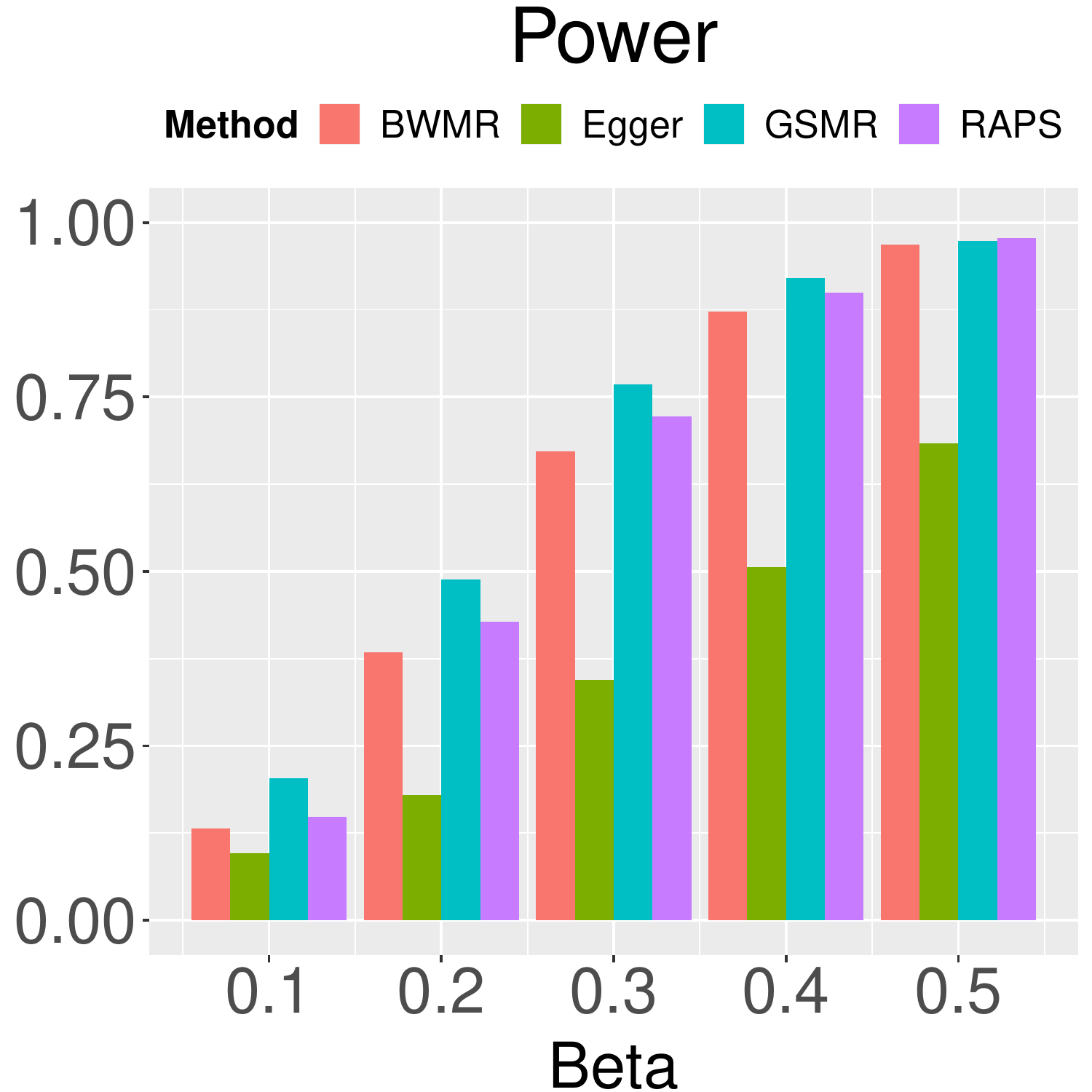}
\par\end{centering}
\caption{Comparison of type I error control and statistical power of BWMR, Egger, GSMR and RAPS in Case-5 of the summary-level data simulation. The simulation parameters were varied in the following range: $\beta\in\{0.0,0.1,0.2,0.3,0.4,0.5\}$, $\tau=0.3$, $\sigma=0.8$, $[c,d]=[0.3,0.5]$, $r=1$. We evaluated the empirical type I error rate and power by controlling type I error rates at the nomial level 0.05. The results were summarized from 500 replications.}
\end{figure}

\newpage{}
\subsubsection{Individul-level simulations}
\begin{itemize}
  \item Individual-level simulations discussing the selection bias in MR
\end{itemize}

\begin{figure}[H]
\begin{centering}
\includegraphics[scale=0.42]{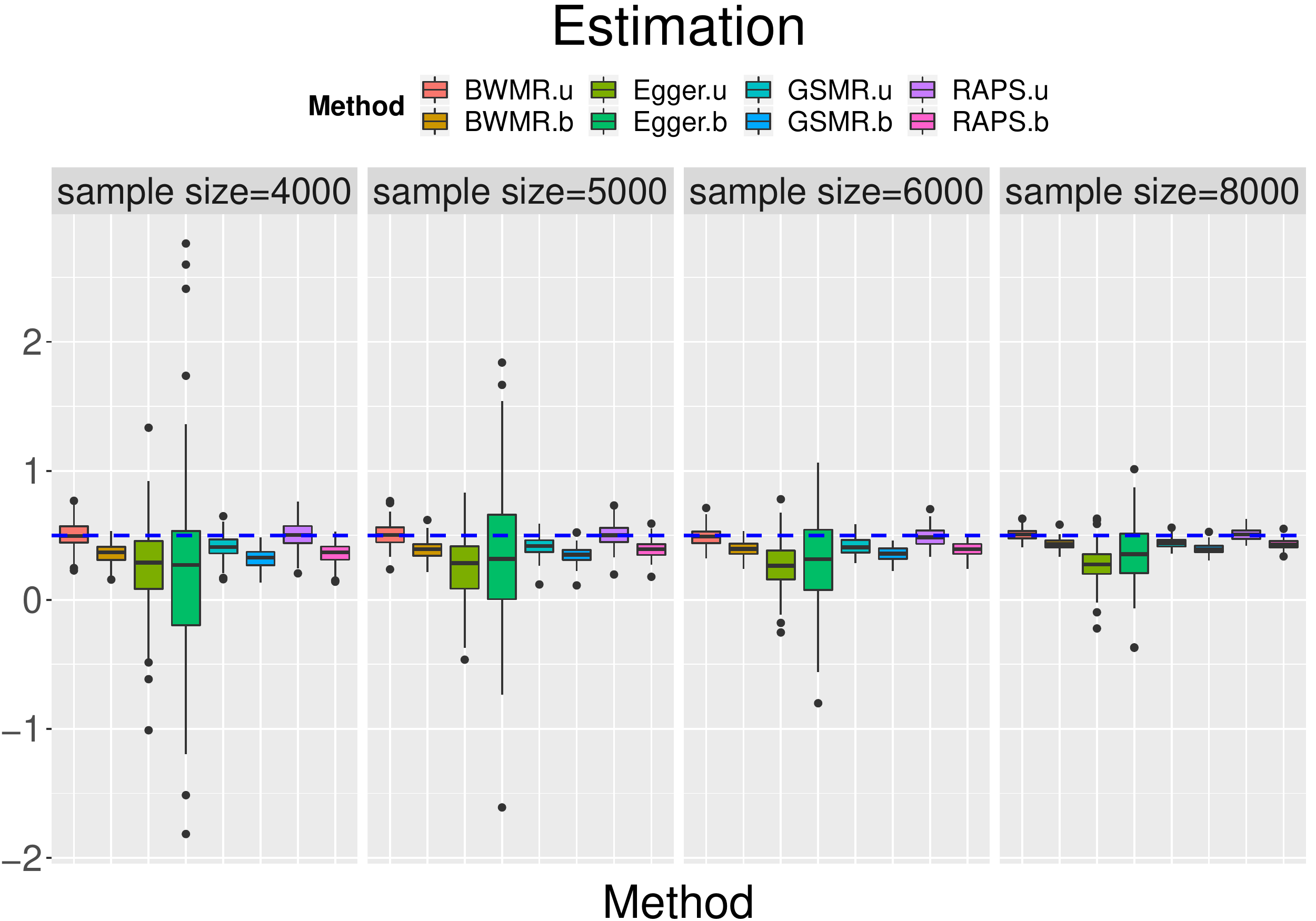}
\par\end{centering}
\caption{Comparison of BWMR, Egger, GSMR and RAPS in the individual-level data simulation, with / without selection bias, under the choices of $\beta\in\{0.0,0.1,0.2,0.3,0.4,0.5\}$, $N_0=10,000$, $n_1=n_2\in\{4,000,\,5,000,\,6,000,\,8,000\}$, $\pi_{11}=0.02$, $\pi_{10}=0.08$, $\pi_{01}=0.08$, $\pi_{00}=0.82$, $SNR_1=SNR_2=1:1$, and $p\mbox{-value threshold}=1\times 10^{-5}$. Let ``.u'' denote simulations without selection bias (unbias), let ``.b'' denote simulations with selection bias. The results were summarized from 100 replications.}
\end{figure}

\begin{figure}[H]
\begin{centering}
\includegraphics[scale=0.2]{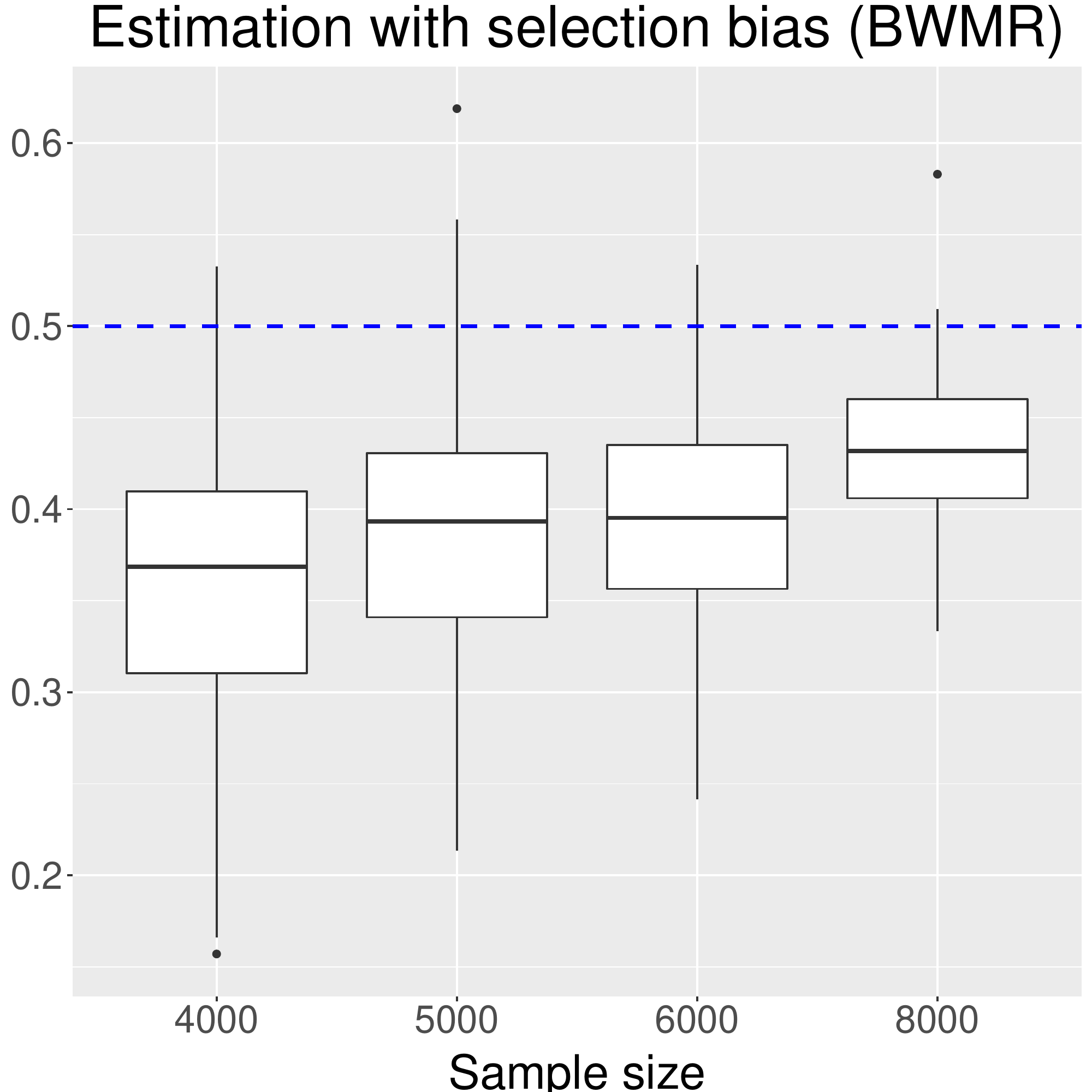}
\includegraphics[scale=0.2]{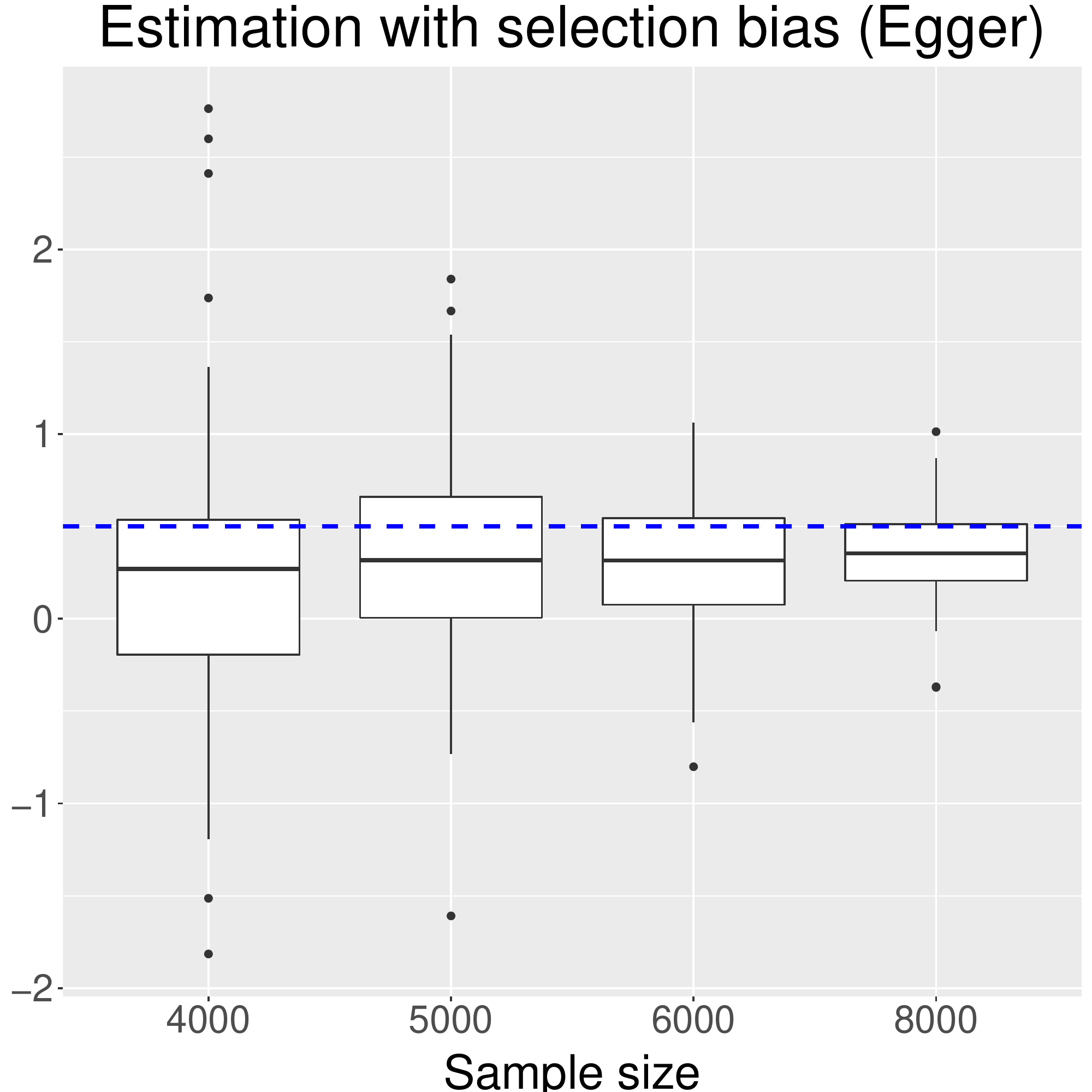}

\includegraphics[scale=0.2]{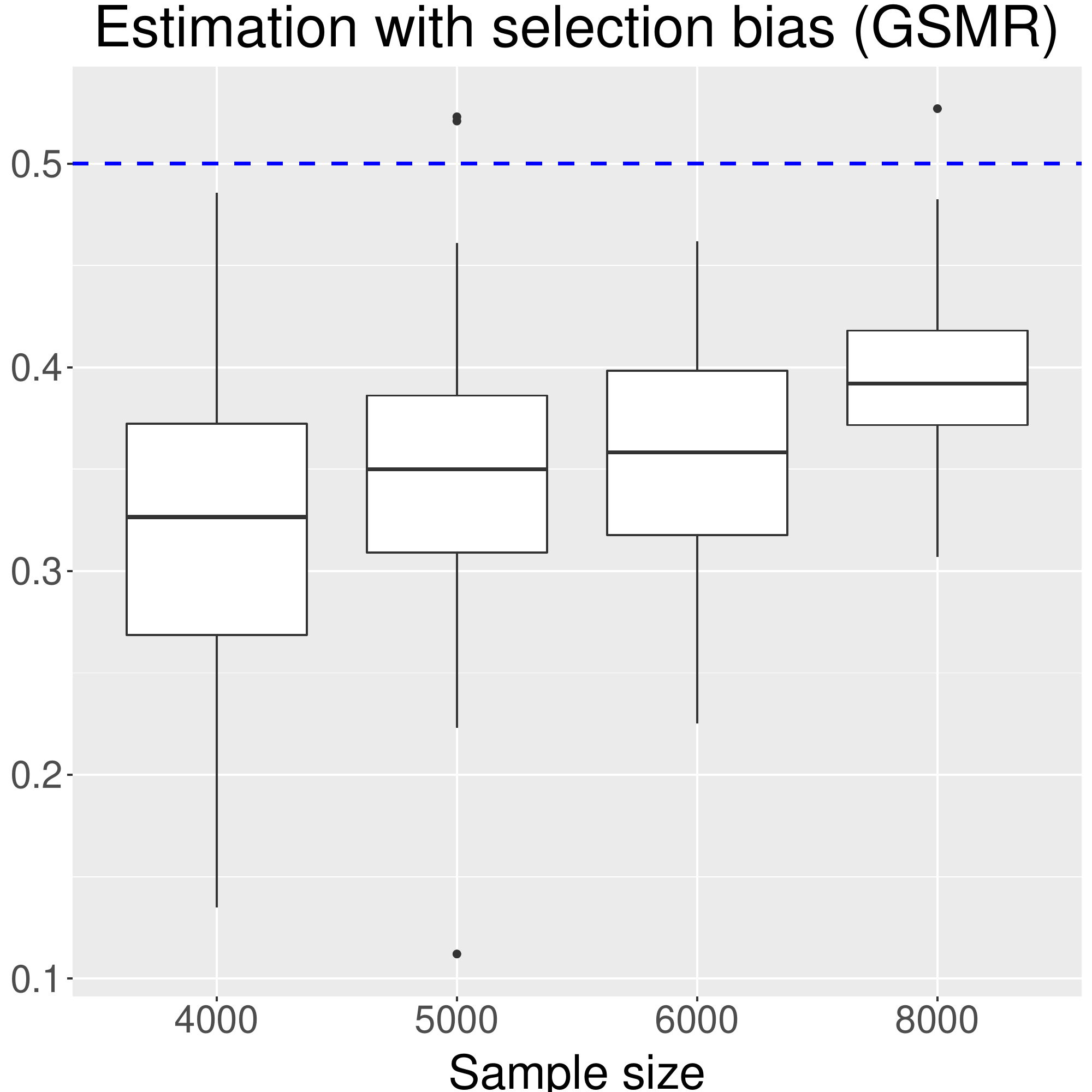}
\includegraphics[scale=0.2]{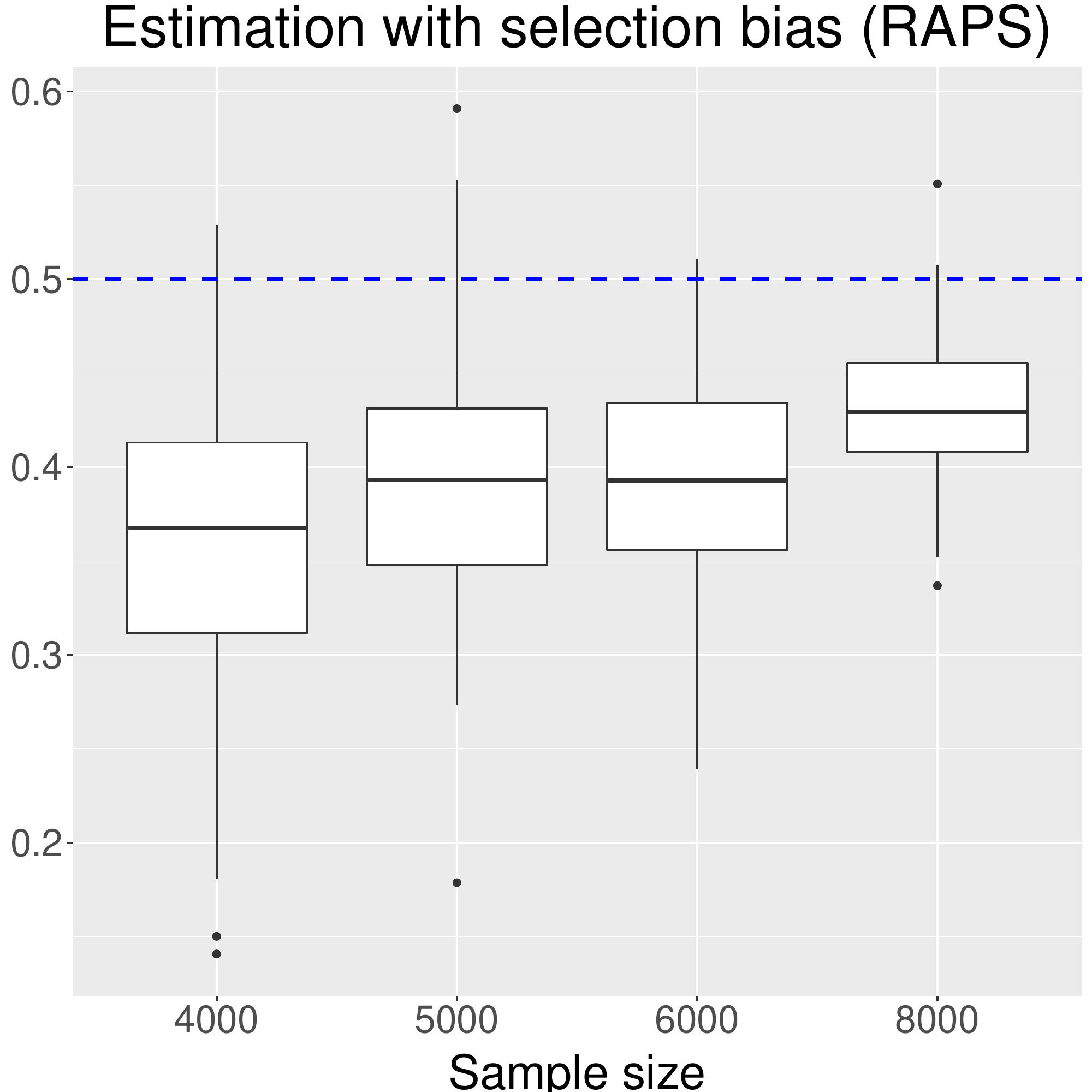}
\par\end{centering}
\caption{Comparison of individual-level data simulation with selection bias using different sample sizes, under the choices of $\beta\in\{0.0,0.1,0.2,0.3,0.4,0.5\}$, $N_0=10,000$, $n_1=n_2\in\{4,000,\,5,000,\,6,000,\,8,000\}$, $\pi_{11}=0.02$, $\pi_{10}=0.08$, $\pi_{01}=0.08$, $\pi_{00}=0.82$, $SNR_1=SNR_2=1:1$, and $p\mbox{-value threshold}=1\times 10^{-5}$. The results were summarized from 100 replications.}
\end{figure}

\newpage{}
\subsection{Real data analysis}
\subsubsection{Data sources}

\begin{table}[H]
\scriptsize
\centering
\setlength{\tabcolsep}{4mm}
\caption{Supplementary Table for GWAS sources of metabolites}\label{tab: S1}
\begin{tabular}{llll}
\hline
Metabolite& \# Sample Size& Reference\\
\hline
123 Blood metabolits from multiple metabolic pathways& up to 24925& \citep{Kettunen2016}\\
Fasting glucose& up to 122,743& \citep{Dupuis2010}\\
Vitamin D levels& up to 41,274& \citep{Manousaki2017}\\
Serum Urate& up to 14,000& \citep{Koettgen2013}\\
Total Cholesterol& up to 188,577& \citep{Willer2013}\\ 
Triglycerides& up to 188,577& \citep{Willer2013}\\ 
HDL cholesterol& up to 188,577& \citep{Willer2013}\\ 
LDL cholesterol& up to 188,577& \citep{Willer2013}\\ 
\end{tabular}
\end{table}

\begin{table}[H]
\scriptsize
\centering
\setlength{\tabcolsep}{3.5mm}
\caption{Supplementary Table for GWAS sources of complex human traits and diseases}\label{tab: Supplementary Table for GWAS sources of complex human traits and diseases}
\begin{tabular}{lllll}
\hline
Category& Name& \# Sample Size& Reference\\
\hline
Anthropometric Trait& Body mass index& 339,224& \citep{Locke2015}\\
Anthropometric Trait& Body fat percentage& 100,716& \citep{Lu2016}\\
Anthropometric Trait& Height& 253,288& \citep{Wood2014}\\
Anthropometric Trait& Hip circumference& 224,459& \citep{Shungin2015}\\
Anthropometric Trait& Waist circumference& 224,459& \citep{Shungin2015}\\
Anthropometric Trait& Waist hip ratio& 224,459& \citep{Shungin2015}\\
Anthropometric Trait& Birth length& 28,459& \citep{Valk2015}\\
Anthropometric Trait& Birth weight& 153,781& \citep{Horikoshi2016}\\
Anthropometric Trait& Childhood obesity& 13,848& \citep{Bradfield2012}\\
Anthropometric Trait& Infant head circumference& 10,768& \citep{Taal2012}\\
Cardiovascular measure& Diastolic blood pressure& 120,473& \citep{Liu2016}\\
Cardiovascular measure& Hypertension& 120,473& \citep{Liu2016}\\
Cardiovascular measure& Mean arterial pressure& 120,473& \citep{Liu2016}\\
Cardiovascular measure& Pulse pressure& 120,473& \citep{Liu2016}\\
Cardiovascular measure& Systolic blood pressure& 120,473& \citep{Liu2016}\\
Cardiovascular measure& Coronary artery disease& 184,305& \citep{Nikpay2015}\\
Cardiovascular measure& Heart rate& 181,171& \citep{DenHoed2013}\\
Cardiovascular measure& Heart rate variability pvRSA& 53,174& \citep{Nolte2017}\\
Cardiovascular measure& Heart rate variability RMSSD& 53,174& \citep{Nolte2017}\\
Cardiovascular measure& Heart rate variability SDNN& 53,174& \citep{Nolte2017}\\
Cardiovascular measure& Peripheral vascular disease& 53,991& \citep{Zhu2018}\\
\hline
\end{tabular}
\end{table}

\begin{table}[H]
\scriptsize
\centering
\setlength{\tabcolsep}{2mm}
\begin{tabular}{lllll}
\hline
Category& Name& \# Sample Size& Reference\\
\hline
Immune system disorder& Atopic dermatitis& 103,066& \citep{Paternoster2015}\\
Immune system disorder& Crohn's disease& 20,883& \citep{Liu2015}\\
Immune system disorder& Inflammary bowel disease& 34,652& \citep{Liu2015}\\
Immune system disorder& Ulcerative colitis& 34,652& \citep{Liu2015}\\
Immune system disorder& Celiac disease& 15,283& \citep{Dubois2010}\\
Immune system disorder& Eczema& 40,835& \citep{Paternoster2015}\\
Immune system disorder& Multiple sclerosis& 27,098& \citep{Sawcer2011}\\
Immune system disorder& Primary biliary cirrhosis& 13,239& \citep{Cordell2015}\\
Immune system disorder& Rheumatoid arthritis& 58,284& \citep{Okbay2016}\\
Immune system disorder& Systemic lupus erythematosus& 23,210& \citep{Bentham2015}\\
Immune system disorder& Type 1 diabetes& 14,741& \citep{Censin2017}\\
Metabolic Trait& Age at menarche& 182,416& \citep{Perry2014}\\
Metabolic Trait& Age at natural menopause& 69,360& \citep{Day2015}\\
Metabolic Trait& Dyslipidemia& 53,991& \citep{Zhu2018}\\
Metabolic Trait& Estimated glomerular filtration rate& 111,666& \citep{Li2017}\\
Metabolic Trait& Fasting insulin& 51,750& \citep{Manning2012}\\
Metabolic Trait& Fasting proinsulin& 10,701& \citep{Strawbridge2011}\\
Metabolic Trait& Glycated haemoglobin levels& 123,665& \citep{Wheeler2017}\\
Metabolic Trait& Gout& 69,374& \citep{Koettgen2013}\\
Metabolic Trait& Hemoglobin A1c& 123,665& \citep{Wheeler2017}\\
Metabolic Trait& Hemorrhoids& 53,991& \citep{Zhu2018}\\
Metabolic Trait& Iron deficiency& 53,991& \citep{Zhu2018}\\
Metabolic Trait& Type 2 diabetes& 69,033& \citep{Morris2012}\\
Metabolic Trait& Urinary albumin to creatinine ratio& 54,450& \citep{Teumer2016}\\
Metabolic Trait& Varicose veins& 53,991& \citep{Zhu2018}\\
Neurodegenerative disease& Alzheimer's disease& 54,162& \citep{Lambert2013}\\
Neurodegenerative disease& Amyotrophic lateral sclerosis& 36,052& \citep{Benyamin2017}\\
Neurodegenerative disease& Age related macular degeneration& 53,991& \citep{Zhu2018}\\
Neurodegenerative disease& Parkinson& 8,477& \citep{Pankratz2012}\\
Other complex trait& Asthma& 26,475& \citep{Moffatt2010}\\
Other complex trait& Breast cancer& 2,287& \citep{Hunter2007}\\
Other complex trait& Dermatophytosis& 53,991& \citep{Zhu2018}\\
Other complex trait& Leptin& 32,161& \citep{Kilpelaeinen2016}\\
Other complex trait& Leptin adjusted for BMI& 32,161& \citep{Kilpelaeinen2016}\\
Other complex trait& Osteoarthritis& 53,991& \citep{Zhu2018}\\
Other complex trait& Osteoporosis& 53,991& \citep{Zhu2018}\\
Psychiatric disorder& Angst& 18,000& \citep{Otowa2016}\\
Psychiatric disorder& Bipolar disorder& 16,731& \citep{Sklar2011}\\
Psychiatric disorder& Attention deficit hyperactivity disorder& 5,422& *\\
Psychiatric disorder& Autism spectrum disorder& 10,763& *\\
Psychiatric disorder& Major depressive disorder& 16,610& *\\
Psychiatric disorder& Schizophrenia& 17,115& *\\
Psychiatric disorder& Depress& 53,991& \citep{Zhu2018}\\
\hline
\end{tabular}
\end{table}
*: \tiny{\citep{PsychiatricGenomicsConsortium2013}}

\begin{table}[H]
\scriptsize
\centering
\setlength{\tabcolsep}{4mm}
\begin{tabular}{lllll}
\hline
Category& Name& \# Sample Size& Reference\\
\hline
Psychiatric disorder& Child aggressive behaviour& 18,988& \citep{Pappa2016}\\
Psychiatric disorder& Anorexia nervosa& 14,477& \citep{Duncan2017}\\
Psychiatric disorder& Loneliness& 10,760& \citep{Gao2017}\\
Psychiatric disorder& Obsessive compulsive disorder& 9,995& \citep{Stewart2013}\\
Psychiatric disorder& Post-traumatic stress disorder& 9,954& \citep{Duncan2017}\\
Psychiatric disorder& Stress& 53,991& \citep{Zhu2018}\\
Social Trait& Alcohol continuous& 70,460& \citep{Schumann2016}\\
Social Trait& Alcohol light heavy& 74,711& \citep{Schumann2016}\\
Social Trait& Cognitive performance& 106,736& \citep{Rietveld2014}\\
Social Trait& Chronotype& 127,898& \citep{Jones2016}\\
Social Trait& Oversleepers& 127,573& \citep{Jones2016}\\
Social Trait& Sleep duration& 127,573& \citep{Jones2016}\\
Social Trait& Undersleepers& 127,573& \citep{Jones2016}\\
Social Trait& Insomnia complaints& 113,006& \citep{Hammerschlag2017}\\
Social Trait& Educational attainment college& 95,429& \citep{Rietveld2013}\\
Social Trait& Educational attainment eduyears& 101,069& \citep{Rietveld2013}\\
Social Trait& Agreeableness& 17,375& \citep{DeMoor2012}\\
Social Trait& Conscientiousness& 17,375& \citep{DeMoor2012}\\
Social Trait& Extraversion& 17,375& \citep{DeMoor2012}\\
Social Trait& Neuroticism& 17,375& \citep{DeMoor2012}\\
Social Trait& Openness& 17,375& \citep{DeMoor2012}\\
Social Trait& Intelligence& 78,308& \citep{Sniekers2017}\\
Social Trait& Depressive symptoms& 161,460& \citep{Okbay2016}\\
Social Trait& Neuroticism& 170,911& \citep{Okbay2016}\\
Social Trait& Subjective well being& 298,420& \citep{Okbay2016}\\
Social Trait& Age onset& 47,961& \citep{Furberg2010}\\
Social Trait& Cigarette per day& 68,028& \citep{Furberg2010}\\
Social Trait& Ever smoked& 74,035& \citep{Furberg2010}\\
Social Trait& Former smoker& 41,969& \citep{Furberg2010}\\
\hline
\end{tabular}
\end{table}

\normalsize
\subsubsection{MR implementation}
The $p$-value threshold for selecting IVs was set to be $5\times 10^{-8}$. The LD clumping was implemented by using PLINK \citep{Purcell2007} with a $r^2$ threshold of $0.001$ and a window size of 1Mb. Before implementing MR methods, we harmonised the alleles for the exposure and the outcome by using the R package ``TwoSampleMR'' \citep{Hemani2018}, and standarded the SNP-exposure and SNP-outcome effects by using the R package ``gsmr'' \citep{Zhu2018}. Egger and RAPS were implemented with the R package ``TwoSampleMR'', GSMR was implemented with its R package ``gsmr'', and BWMR was implemented with our R package ``BWMR''.

\newpage{}
\subsubsection{Examination of selection bias}
\begin{figure}[H]
\begin{centering}
\includegraphics[scale=0.19]{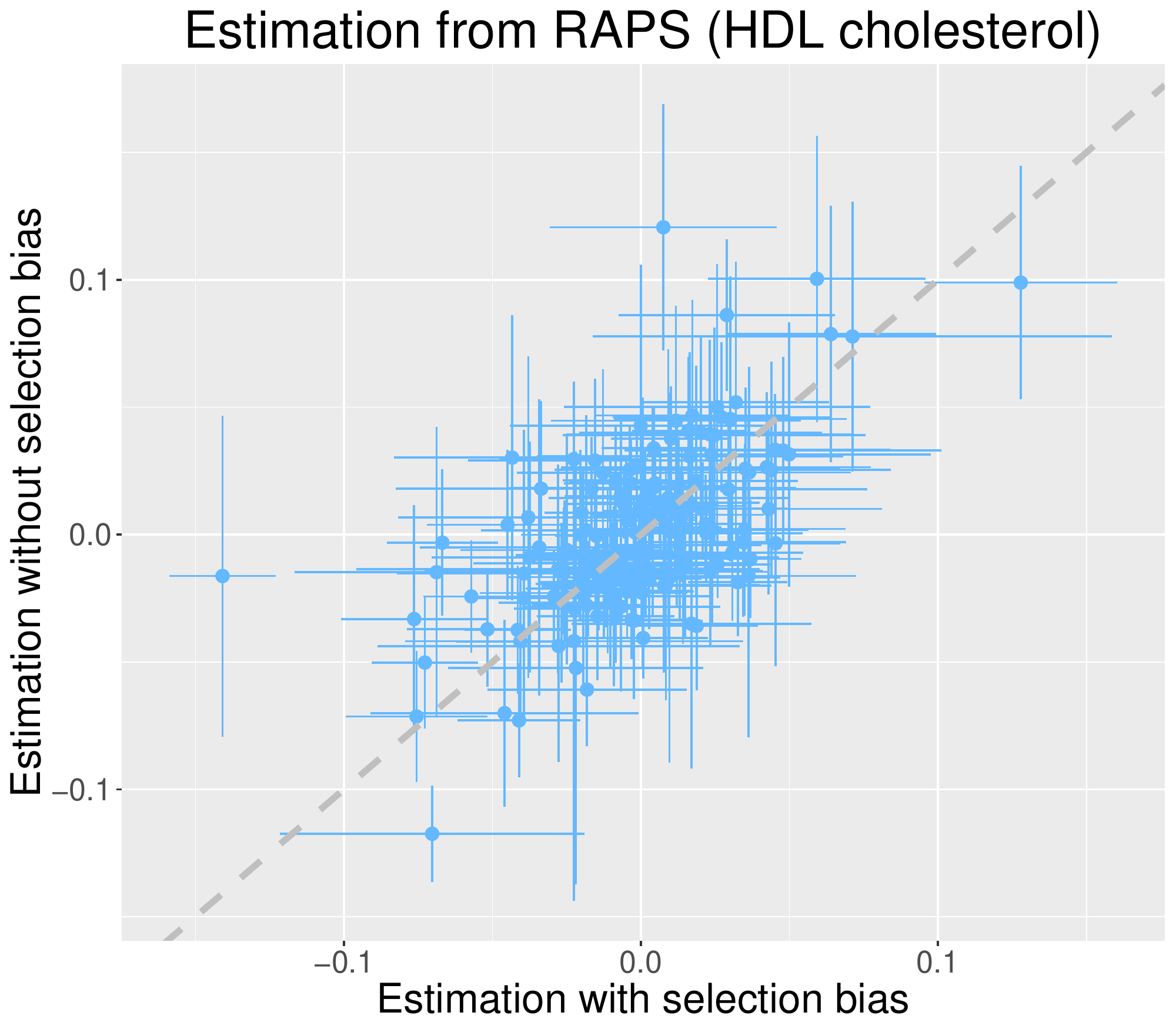}
\includegraphics[scale=0.19]{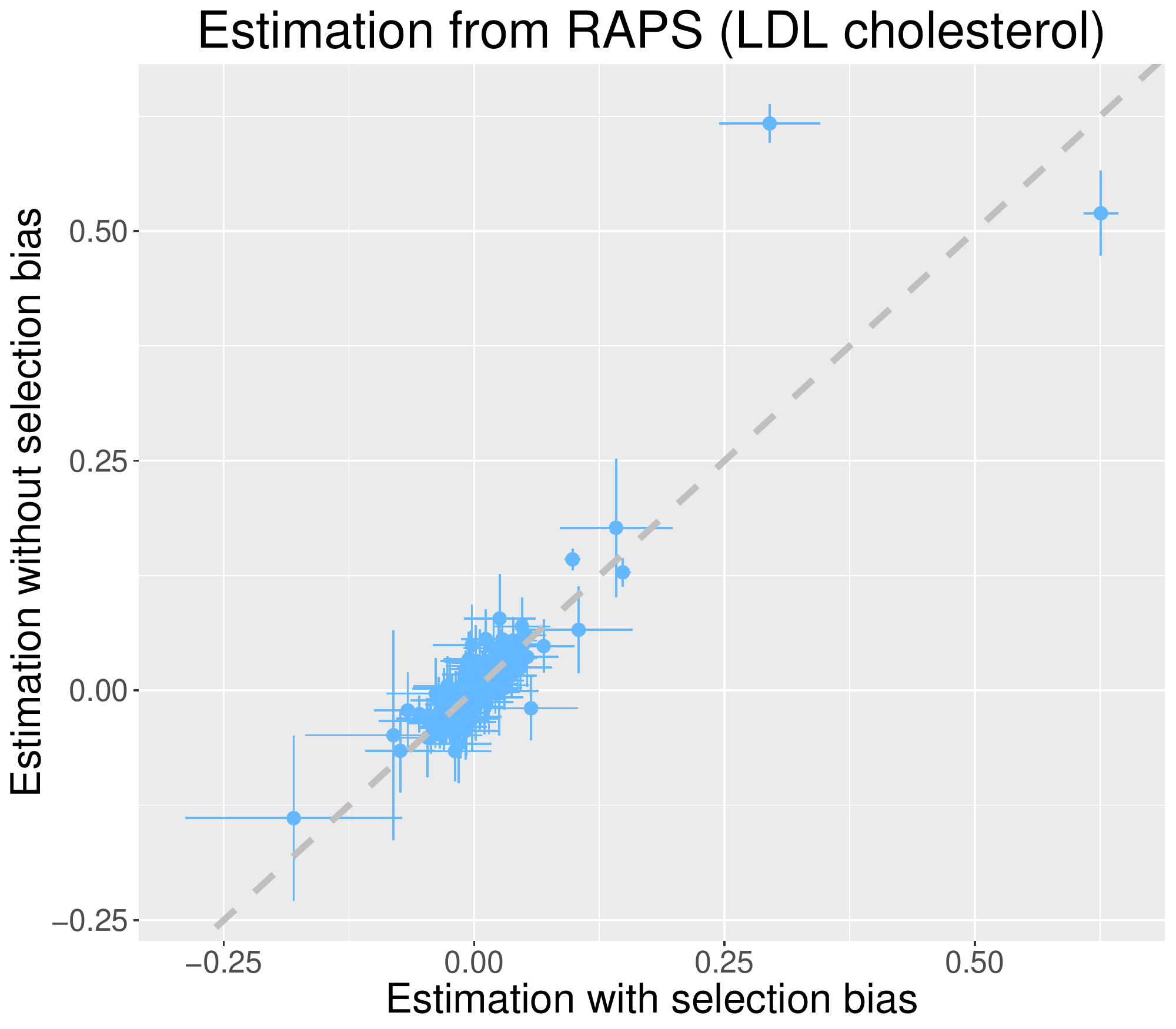}
\includegraphics[scale=0.19]{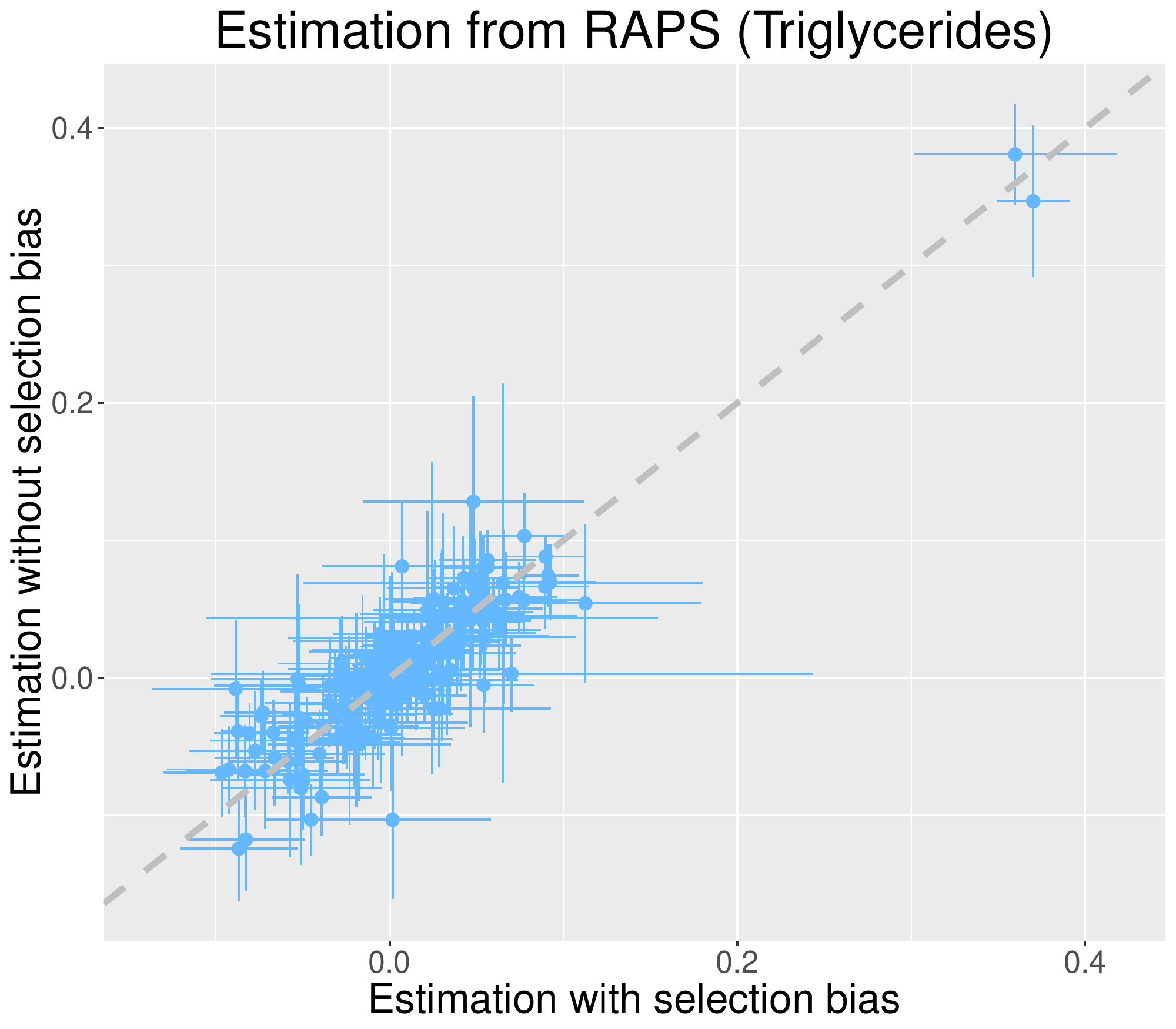}
\includegraphics[scale=0.19]{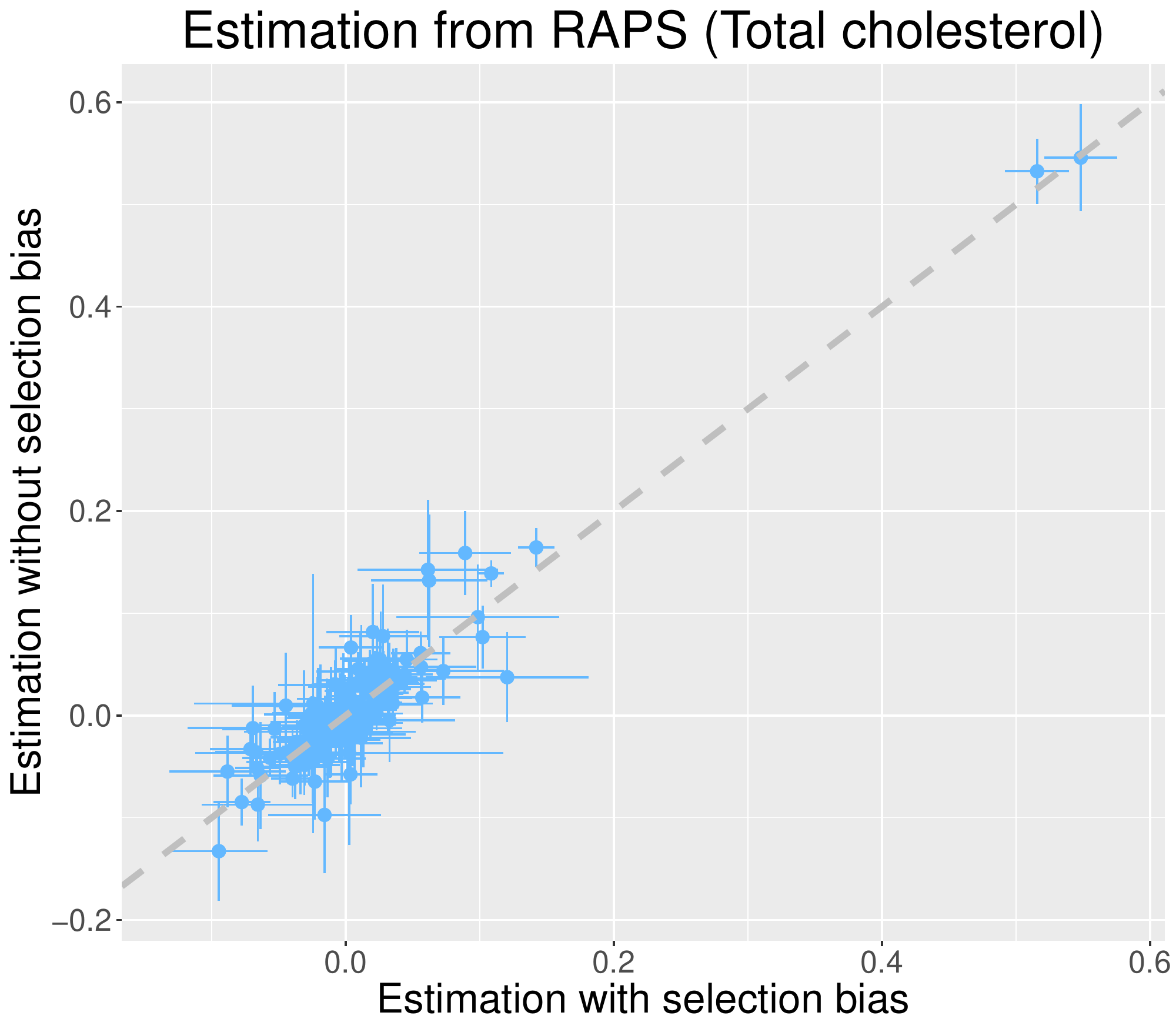}
\par\end{centering}
\caption{Comparisons of analysis results provided by RAPS with selection bias ($x$-axis) and without selection bias ($y$-axis). The dots represent estimated causal effect sizes $\hat{\beta}$ and the bars represent their standard errors $\mathrm{se}(\hat{\beta})$. The diagonal is indicated by the dashed line.}
\end{figure}
\begin{figure}[H]
\begin{centering}
\includegraphics[scale=0.19]{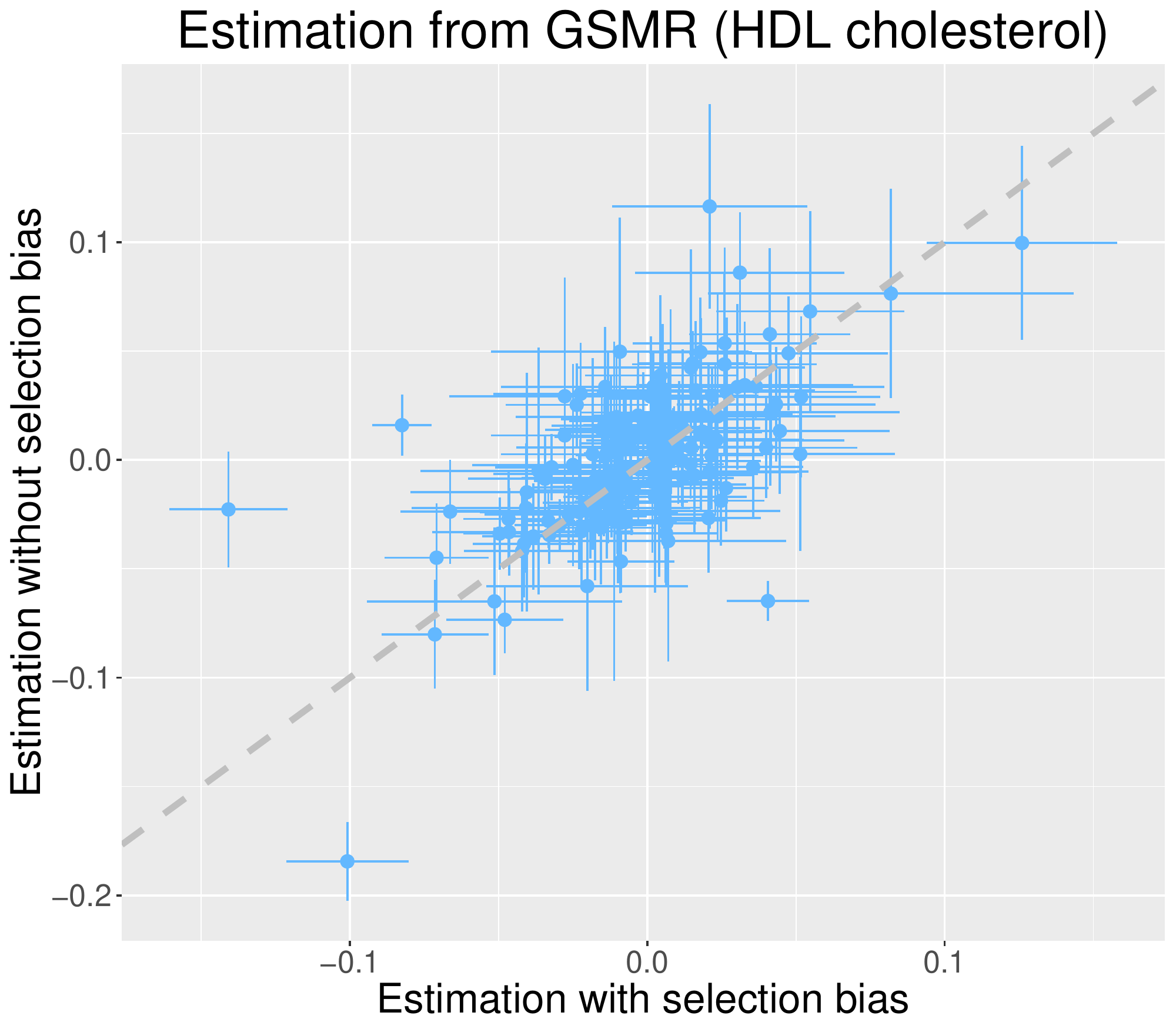}
\includegraphics[scale=0.19]{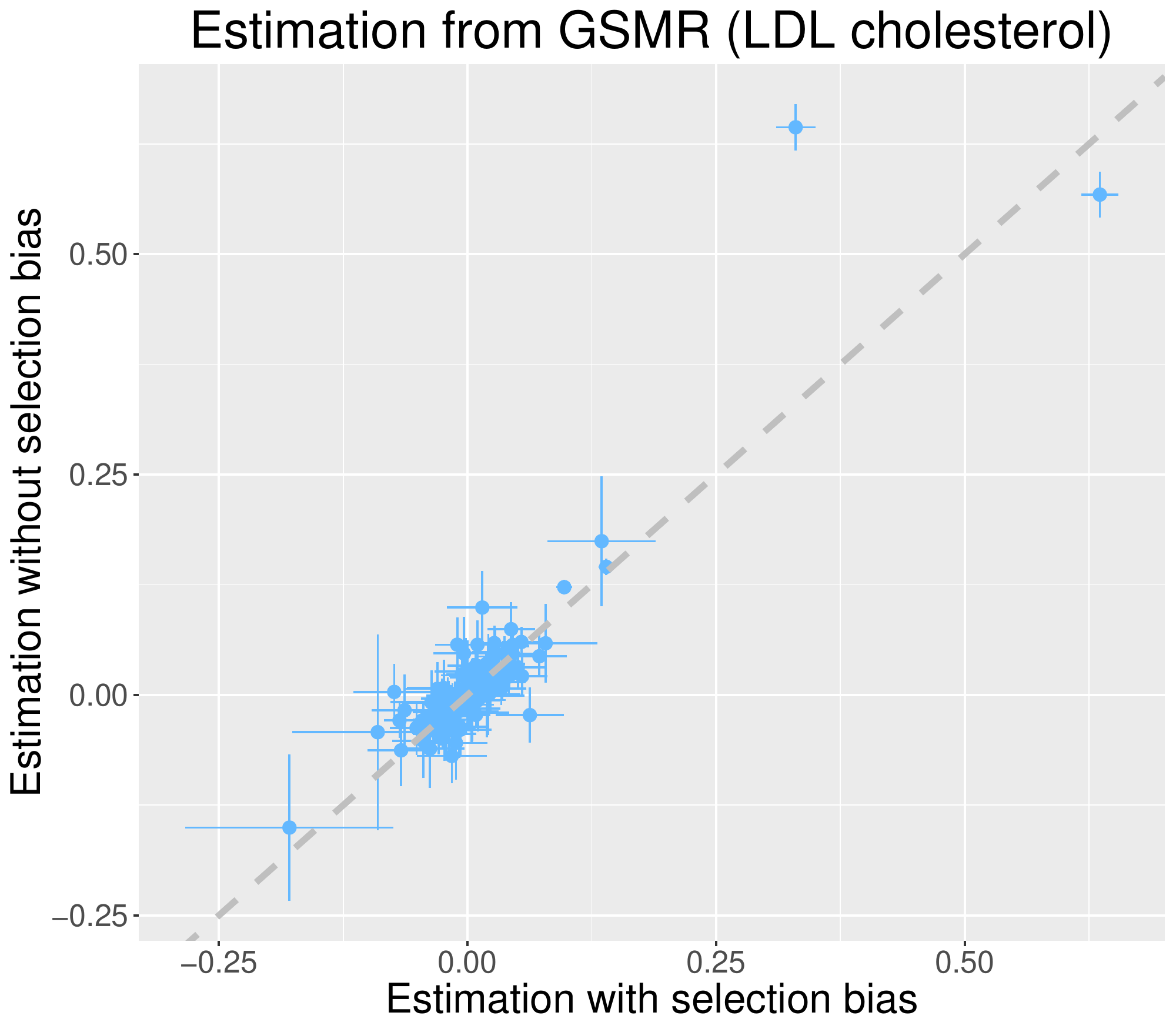}
\includegraphics[scale=0.19]{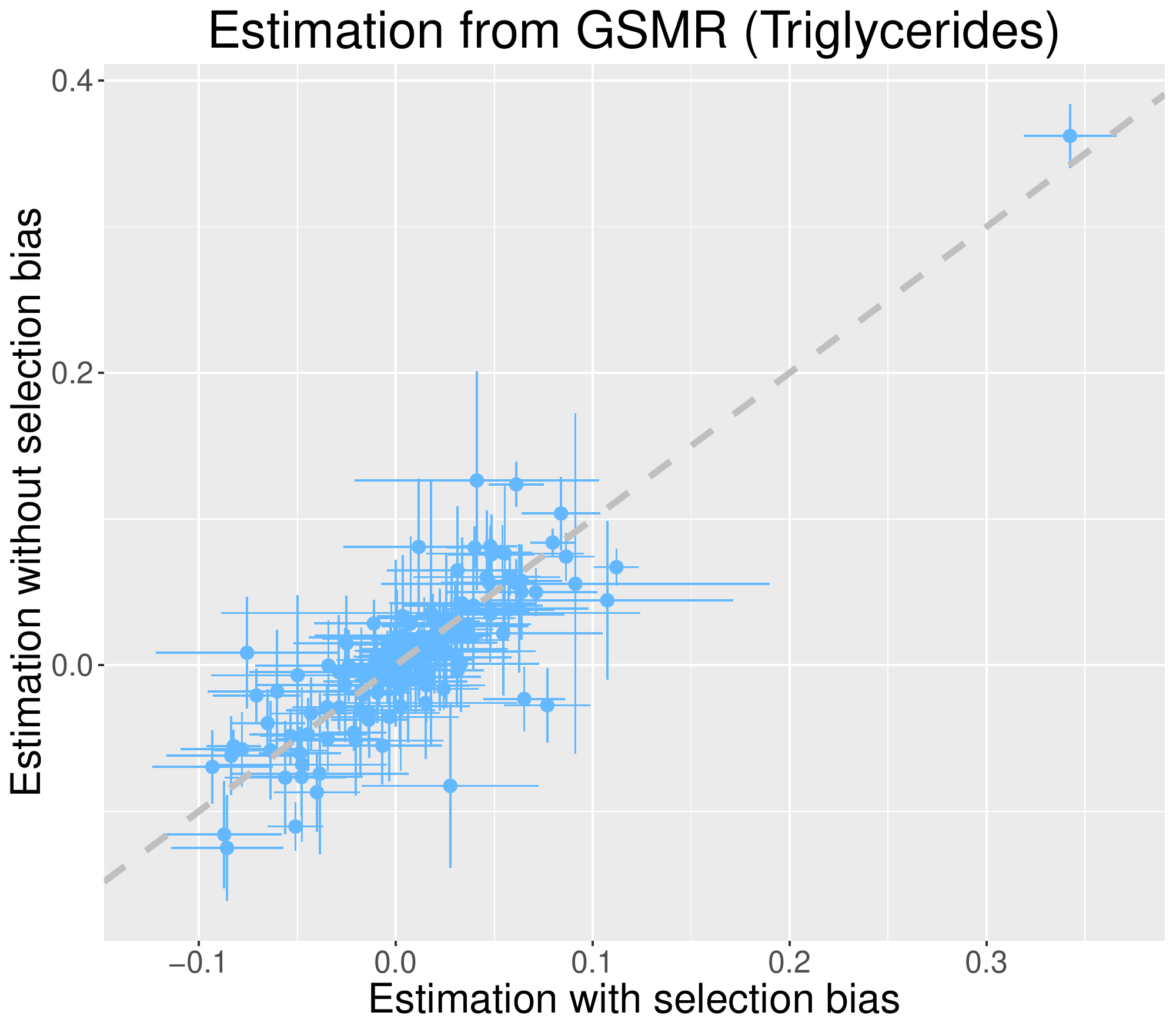}
\includegraphics[scale=0.19]{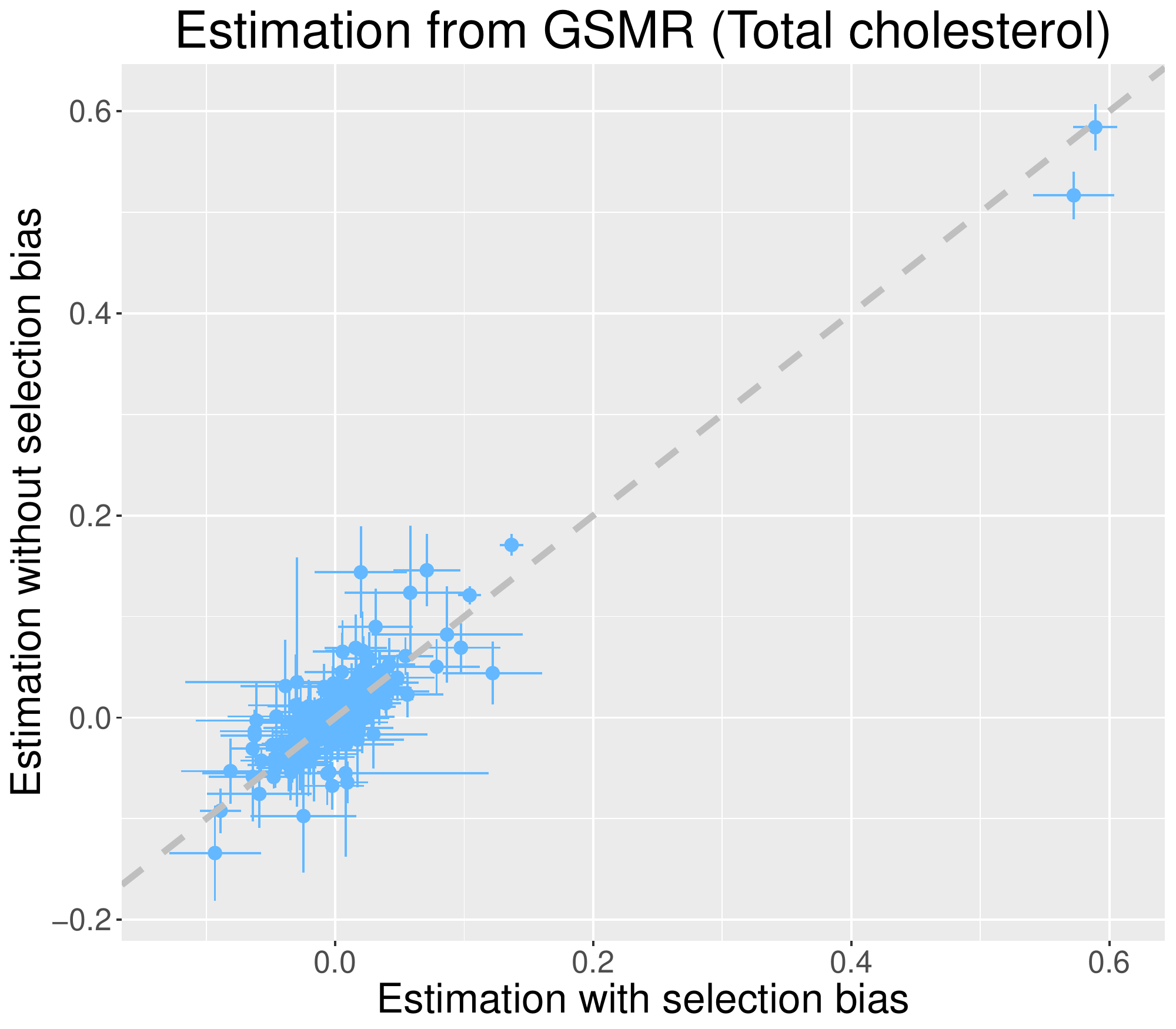}
\par\end{centering}
\caption{Comparisons of analysis results provided by GSMR with selection bias ($x$-axis) and without selection bias ($y$-axis). The dots represent estimated causal effect sizes $\hat{\beta}$ and the bars represent their standard errors $\mathrm{se}(\hat{\beta})$. The diagonal is indicated by the dashed line.}
\end{figure}
\begin{figure}[H]
\begin{centering}
\includegraphics[scale=0.19]{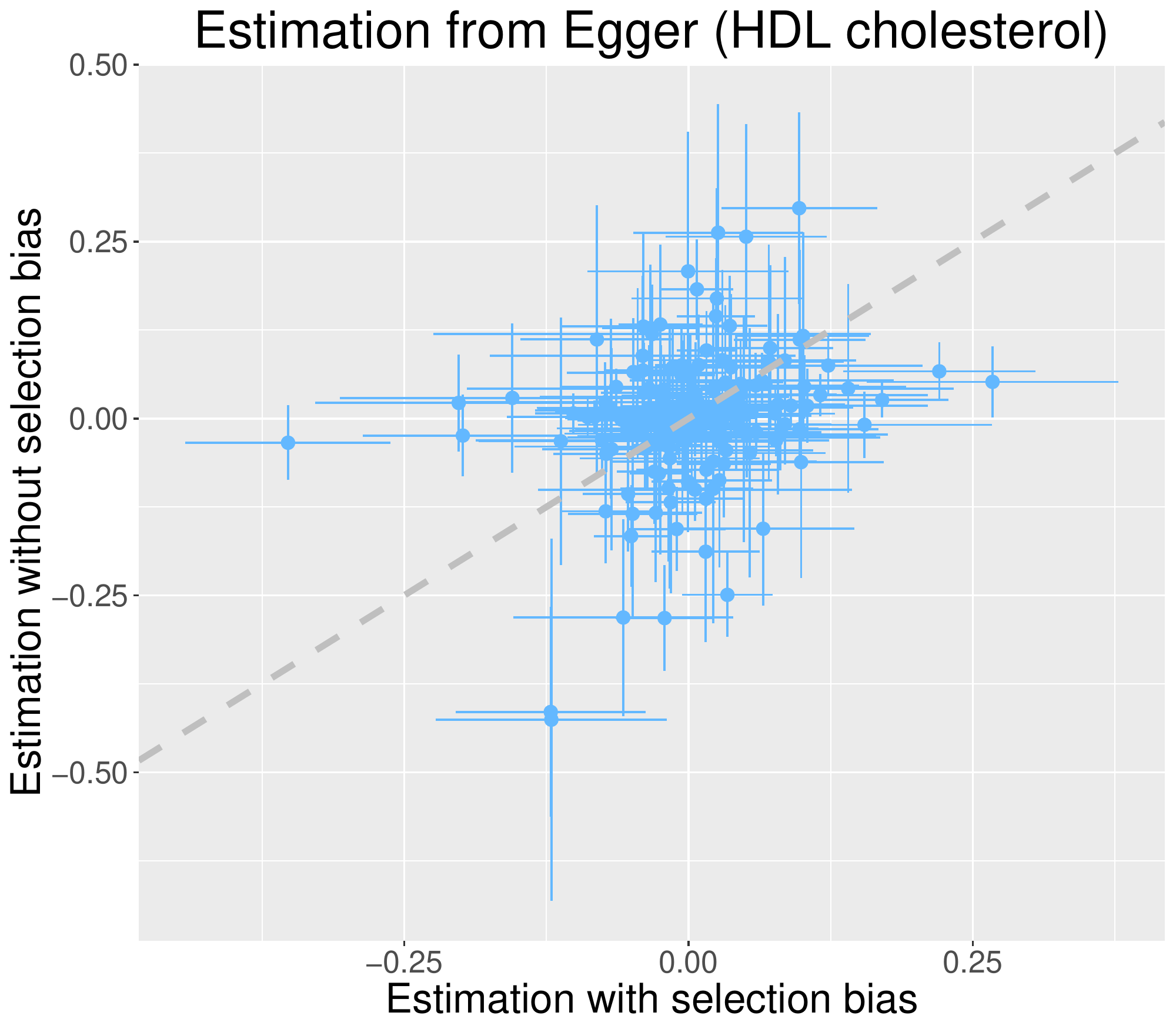}
\includegraphics[scale=0.19]{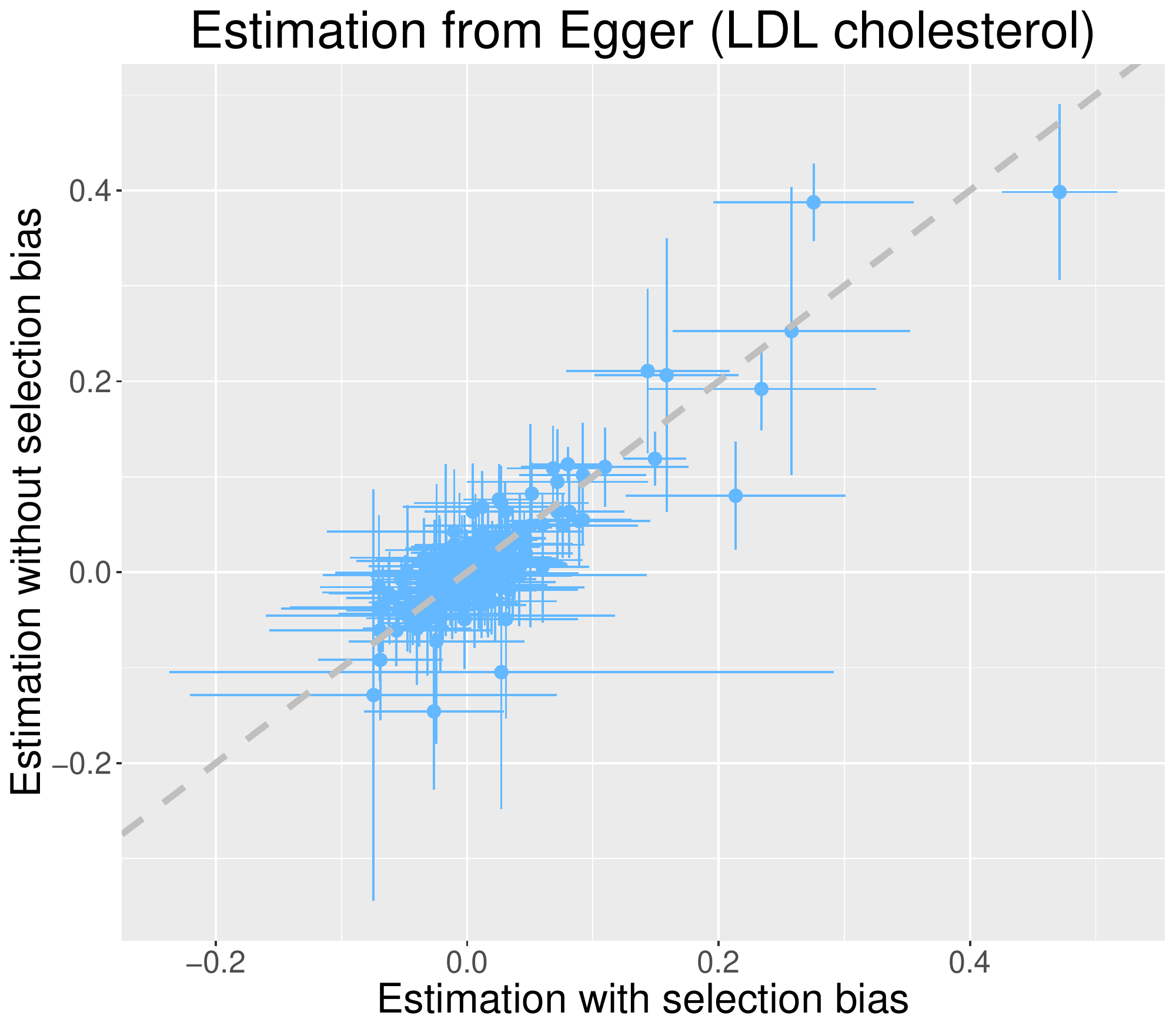}
\includegraphics[scale=0.19]{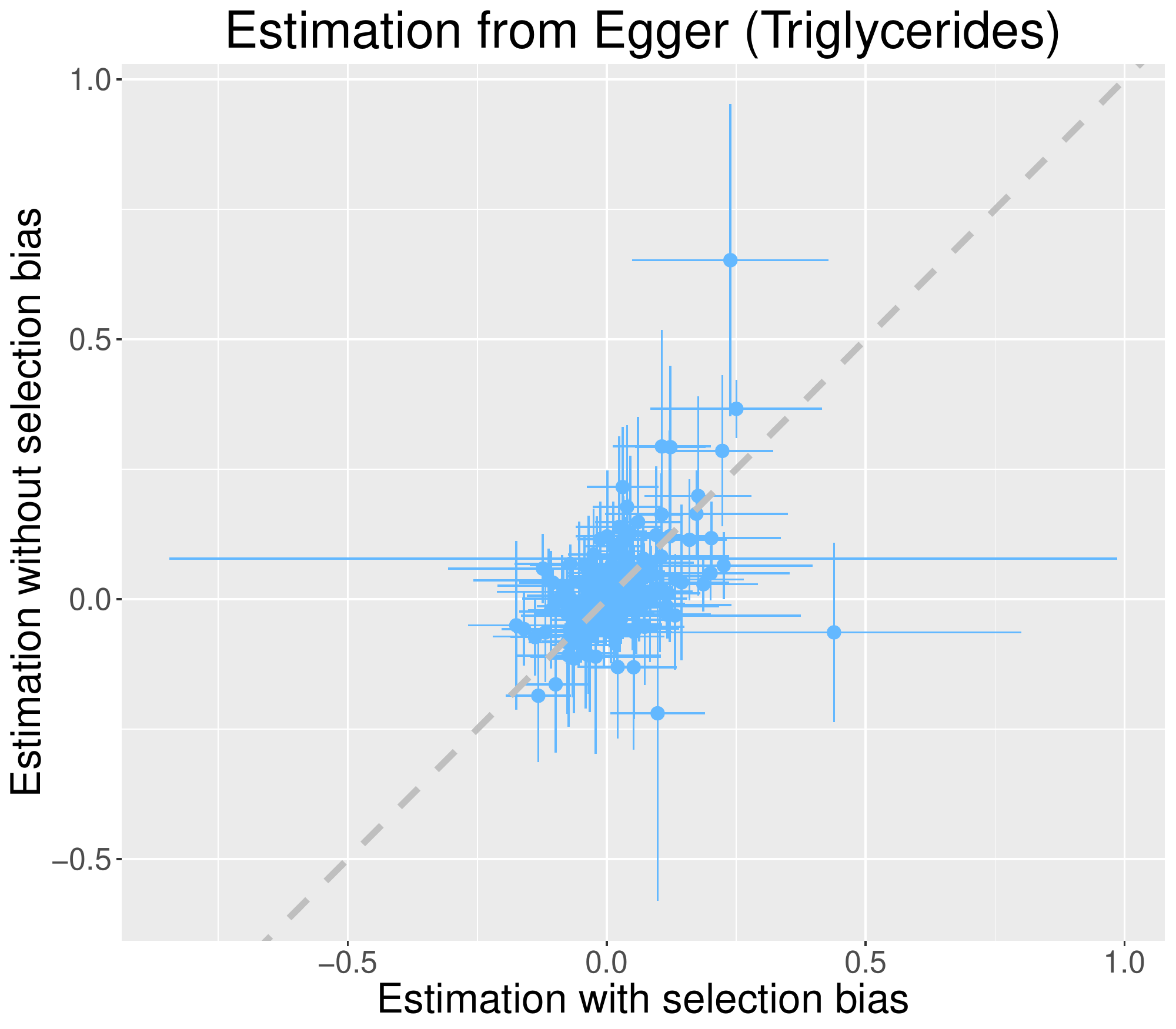}
\includegraphics[scale=0.19]{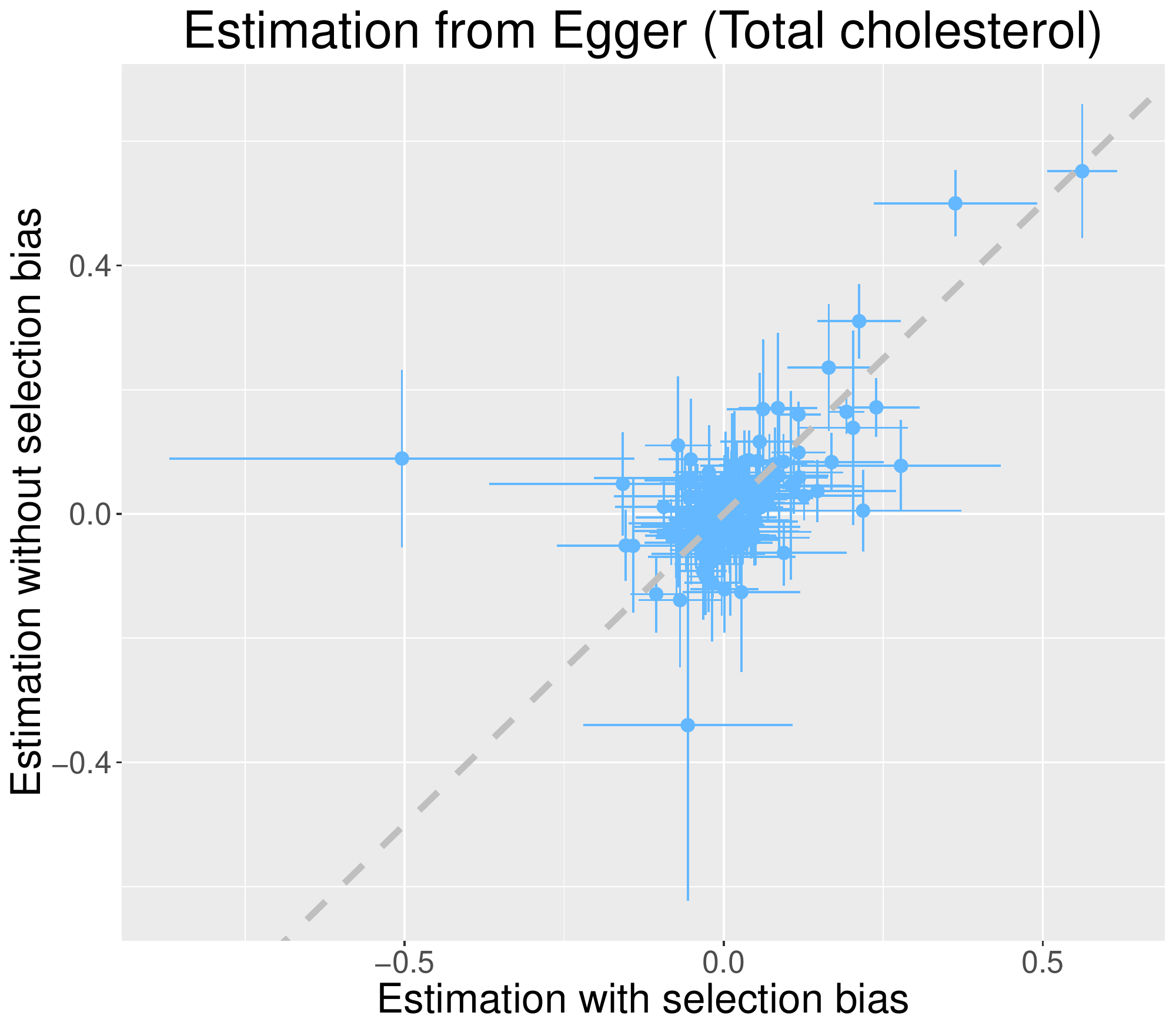}
\par\end{centering}
\caption{Comparisons of analysis results provided by Egger with selection bias ($x$-axis) and without selection bias ($y$-axis). The dots represent estimated causal effect sizes $\hat{\beta}$ and the bars represent their standard errors $\mathrm{se}(\hat{\beta})$. The diagonal is indicated by the dashed line.}
\end{figure}

\subsubsection{Consistency of analysis results}
\begin{figure}[H]
\begin{centering}
\includegraphics[scale=0.19]{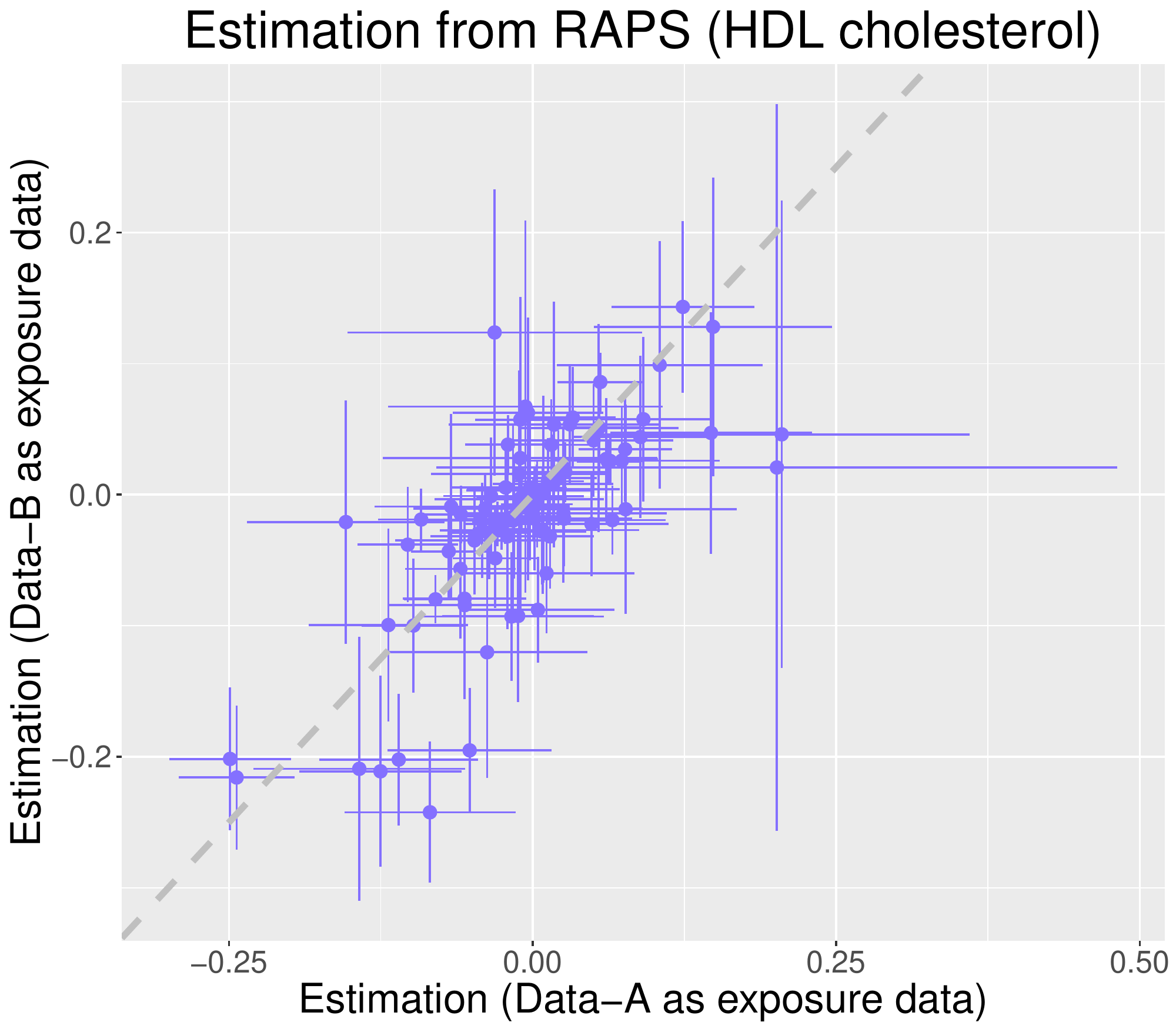}
\includegraphics[scale=0.19]{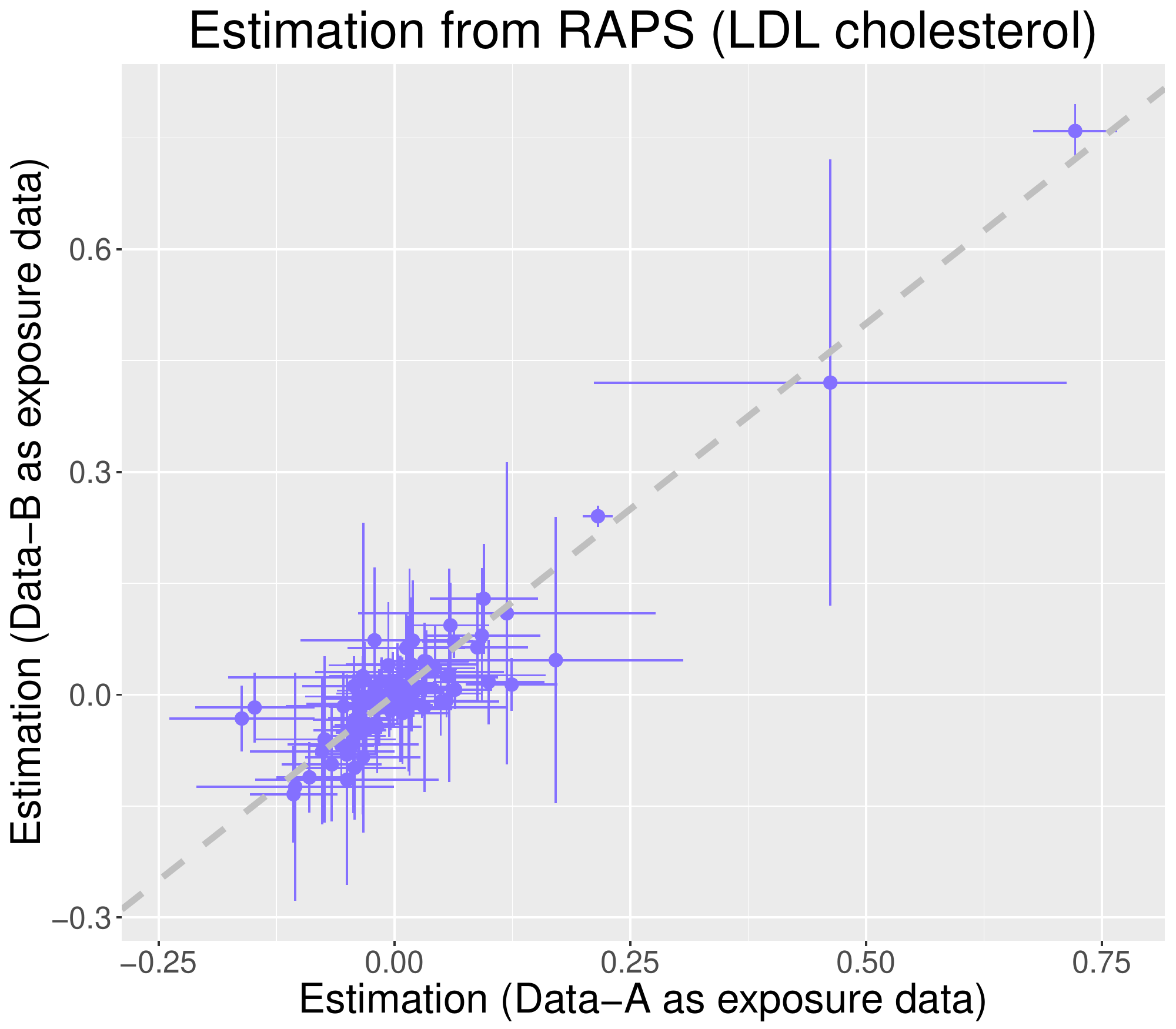}
\includegraphics[scale=0.19]{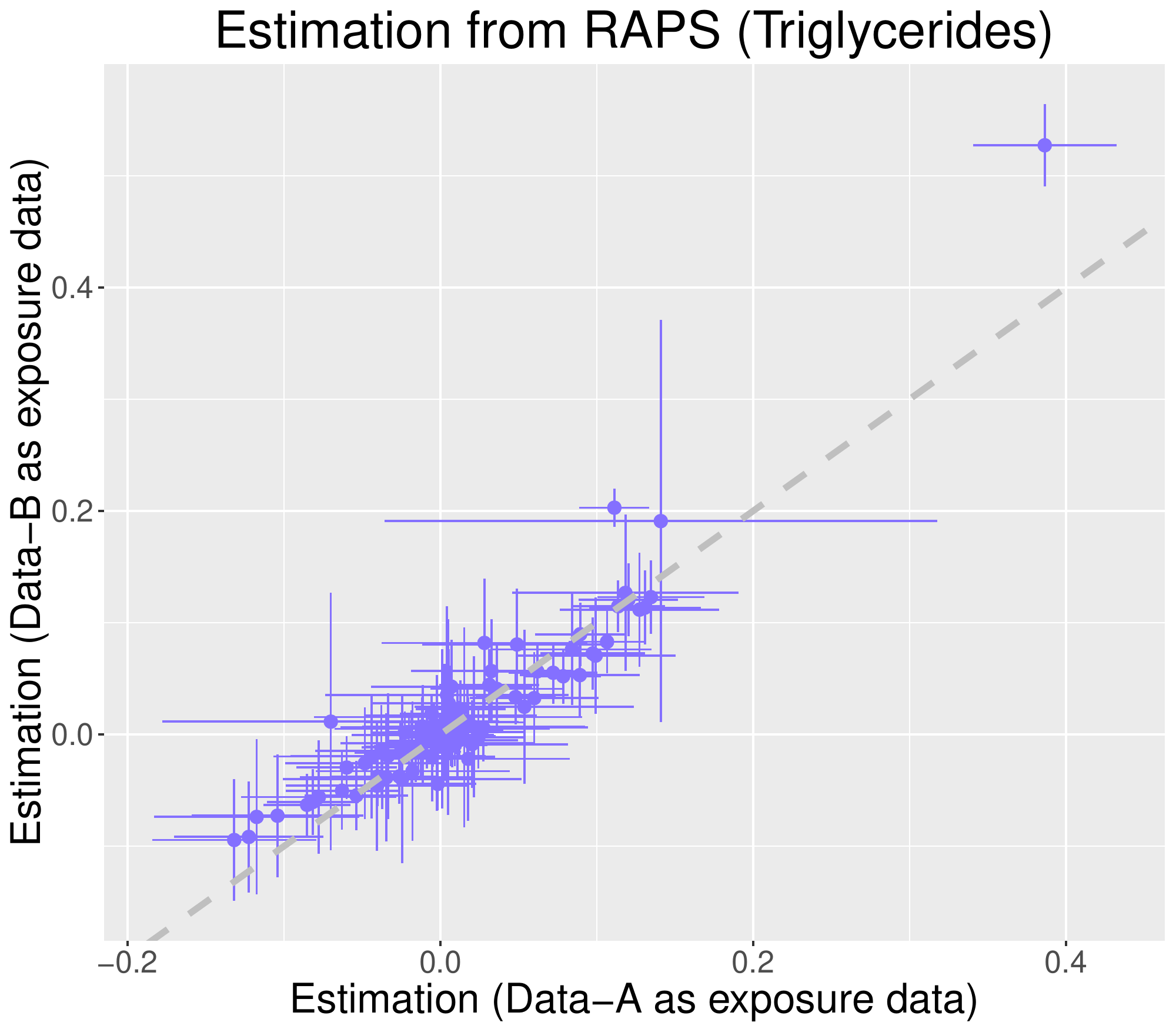}
\includegraphics[scale=0.19]{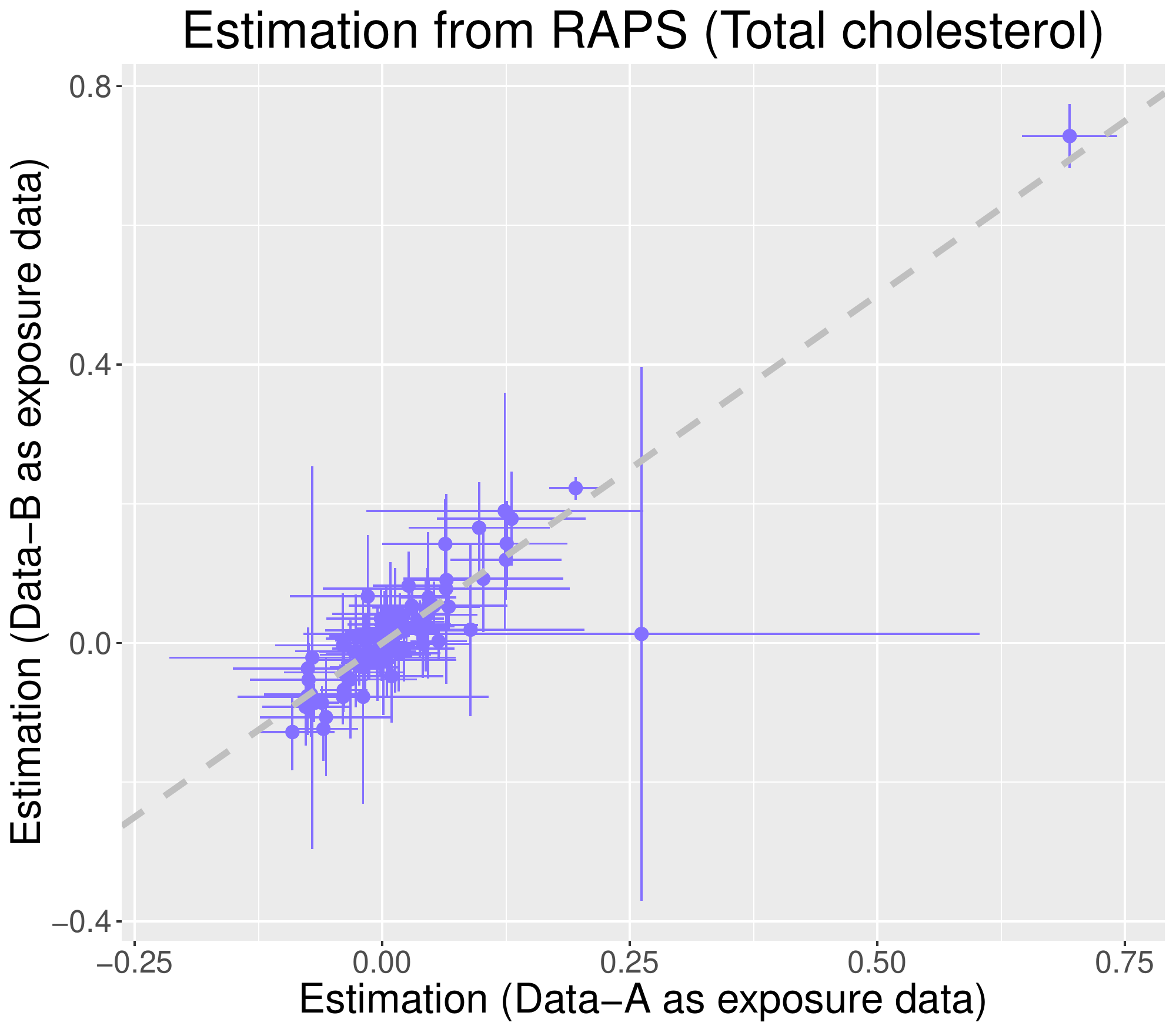}
\par\end{centering}
\caption{Comparisons of analysis results provided by RAPS using Data-A as exposure data ($x$-axis) and using Data-B as exposure data ($y$-axis). The dots represent estimated causal effect sizes $\hat{\beta}$ and the bars represent their standard errors $\mathrm{se}(\hat{\beta})$. The diagonal is indicated by the dashed line.}
\end{figure}
\begin{figure}[H]
\begin{centering}
\includegraphics[scale=0.19]{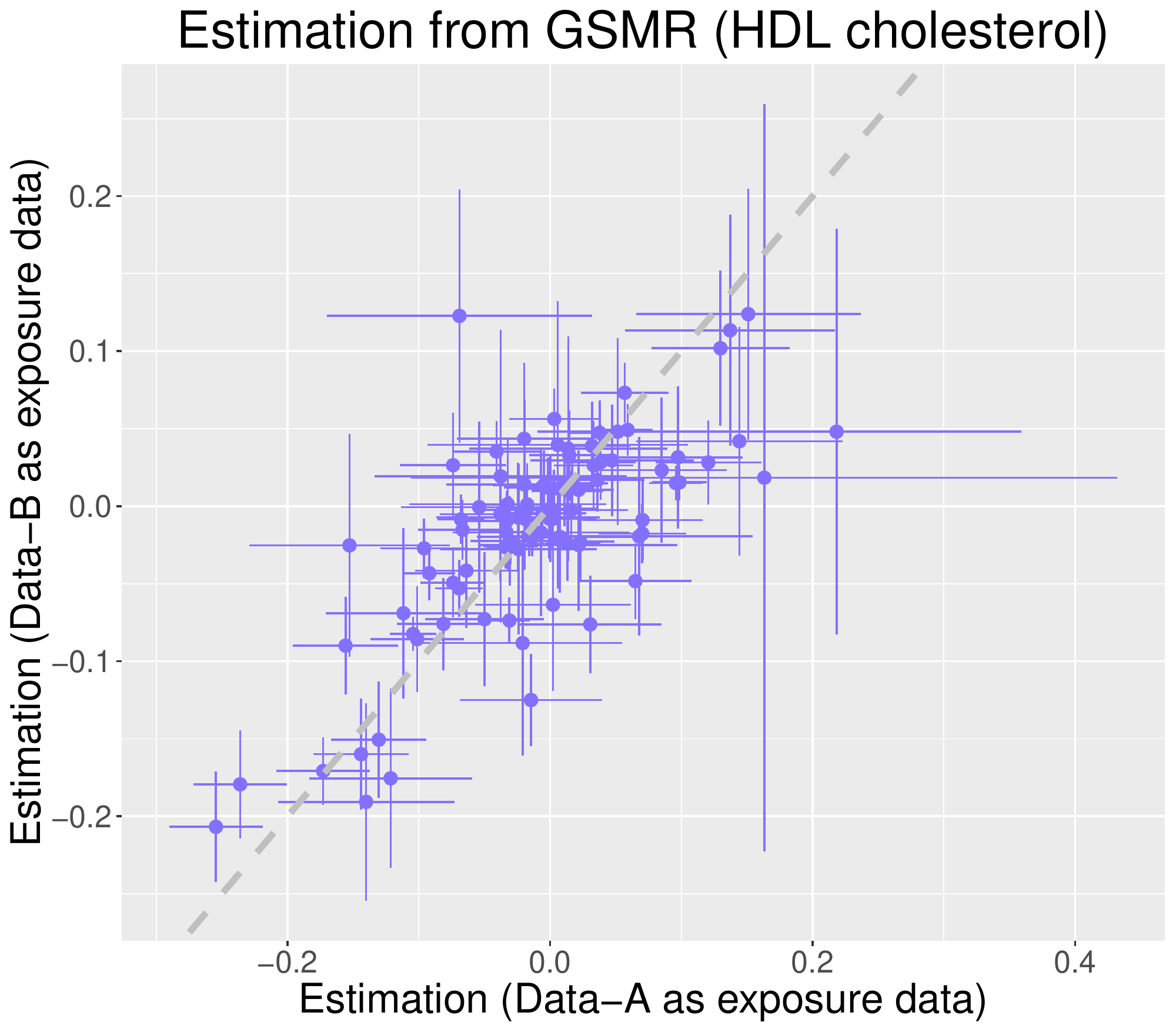}
\includegraphics[scale=0.19]{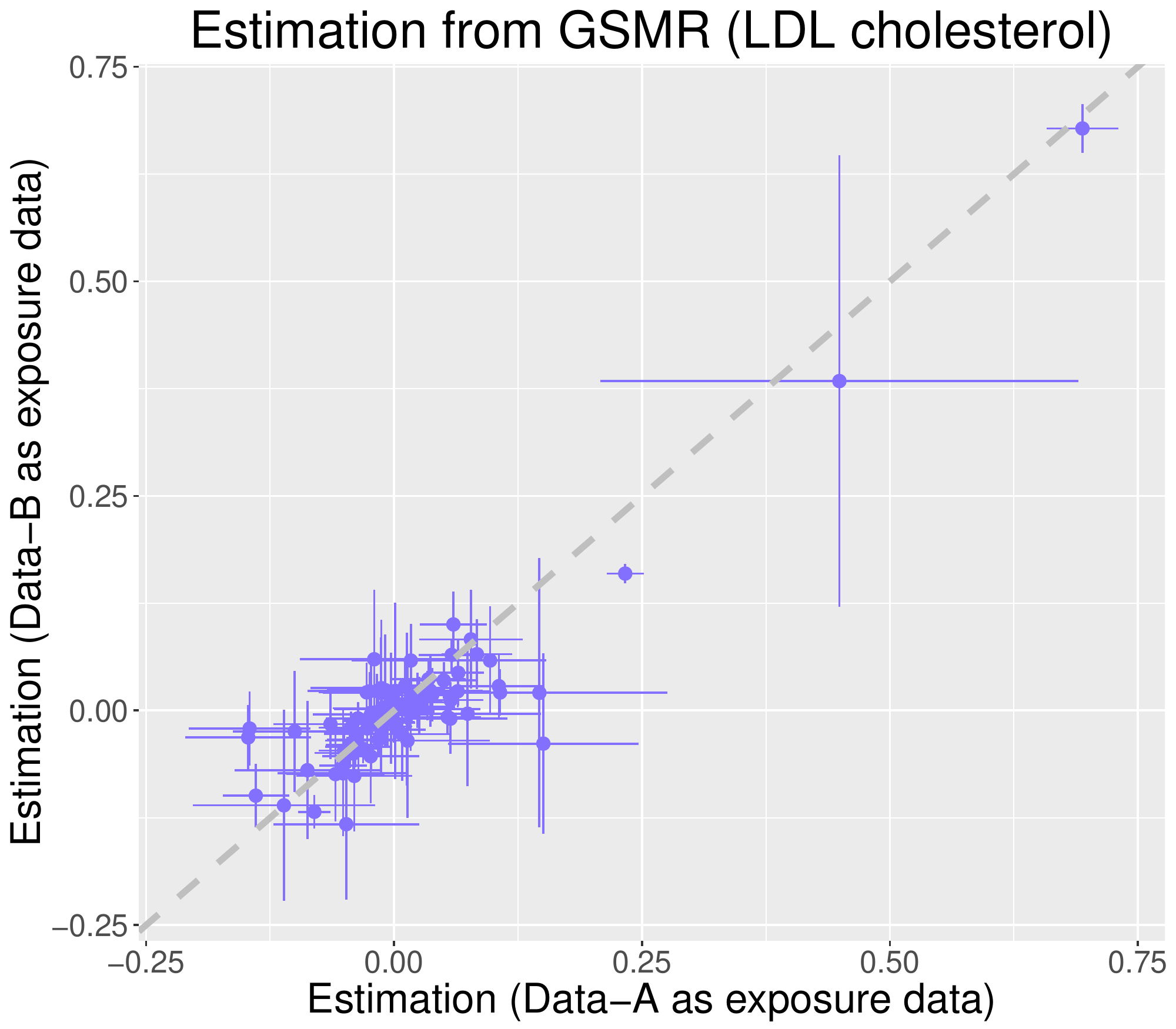}
\includegraphics[scale=0.19]{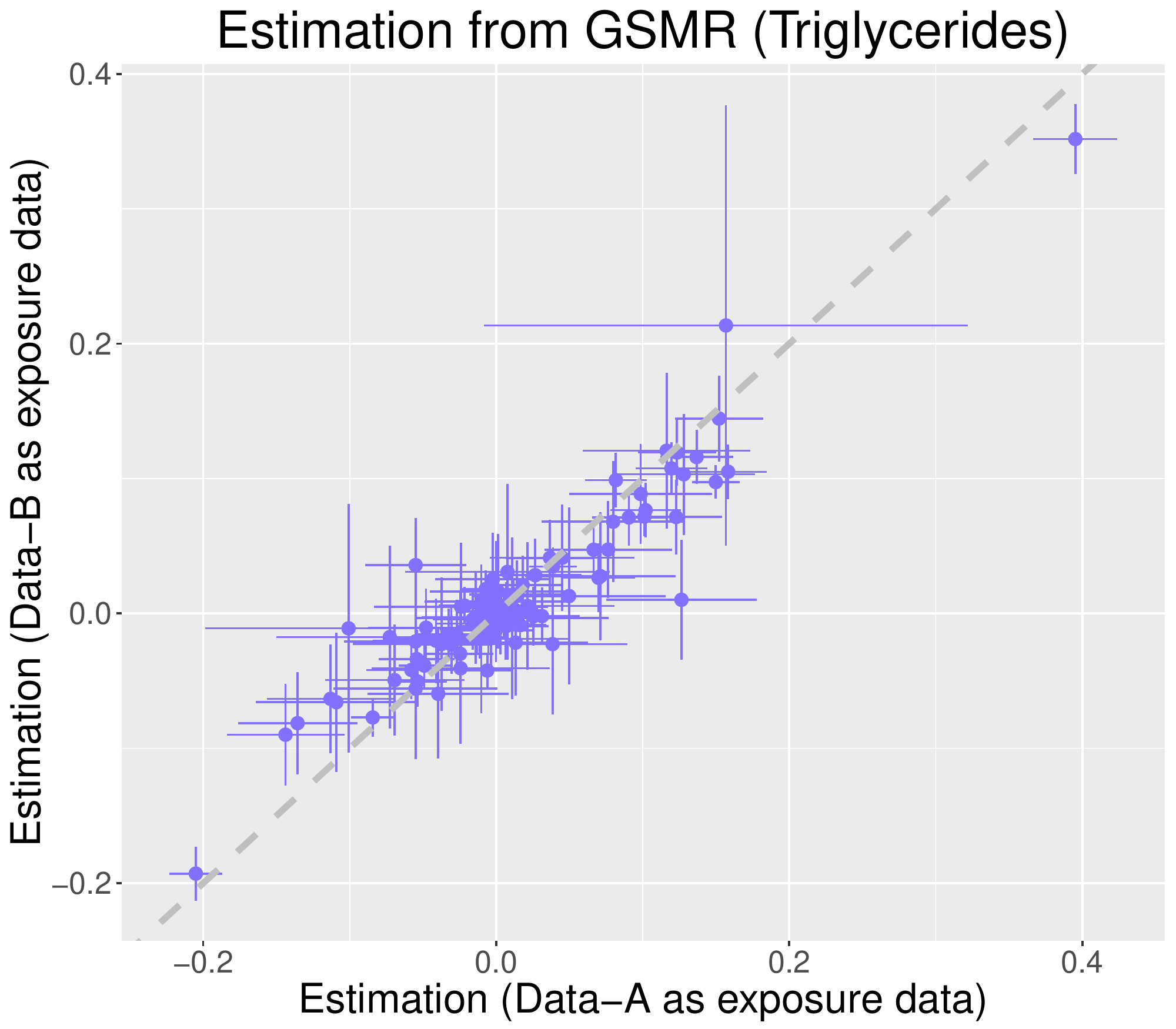}
\includegraphics[scale=0.19]{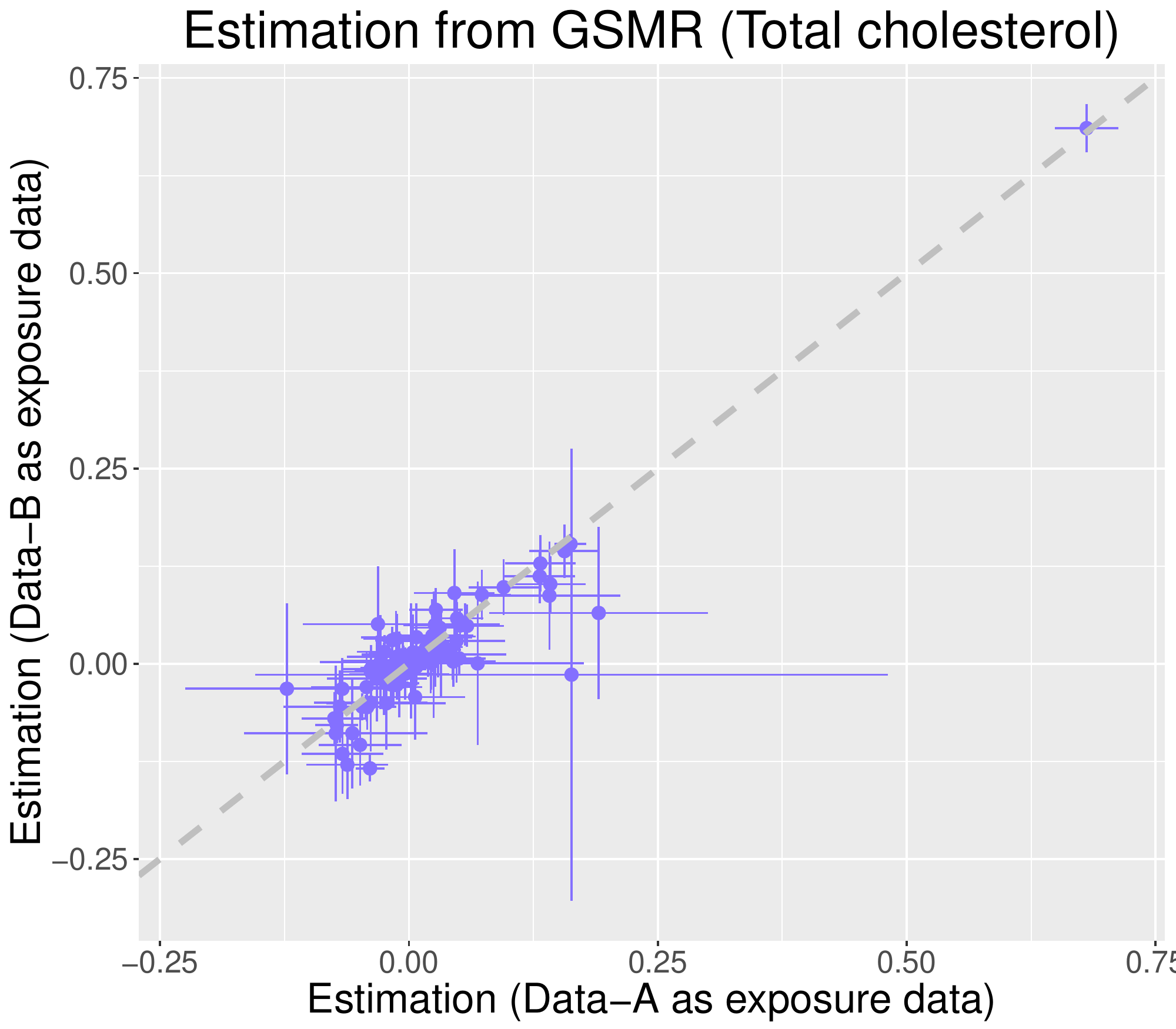}
\par\end{centering}
\caption{Comparisons of analysis results provided by GSMR using Data-A as exposure data ($x$-axis) and using Data-B as exposure data ($y$-axis). The dots represent estimated causal effect sizes $\hat{\beta}$ and the bars represent their standard errors $\mathrm{se}(\hat{\beta})$. The diagonal is indicated by the dashed line.}
\end{figure}
\begin{figure}[H]
\begin{centering}
\includegraphics[scale=0.19]{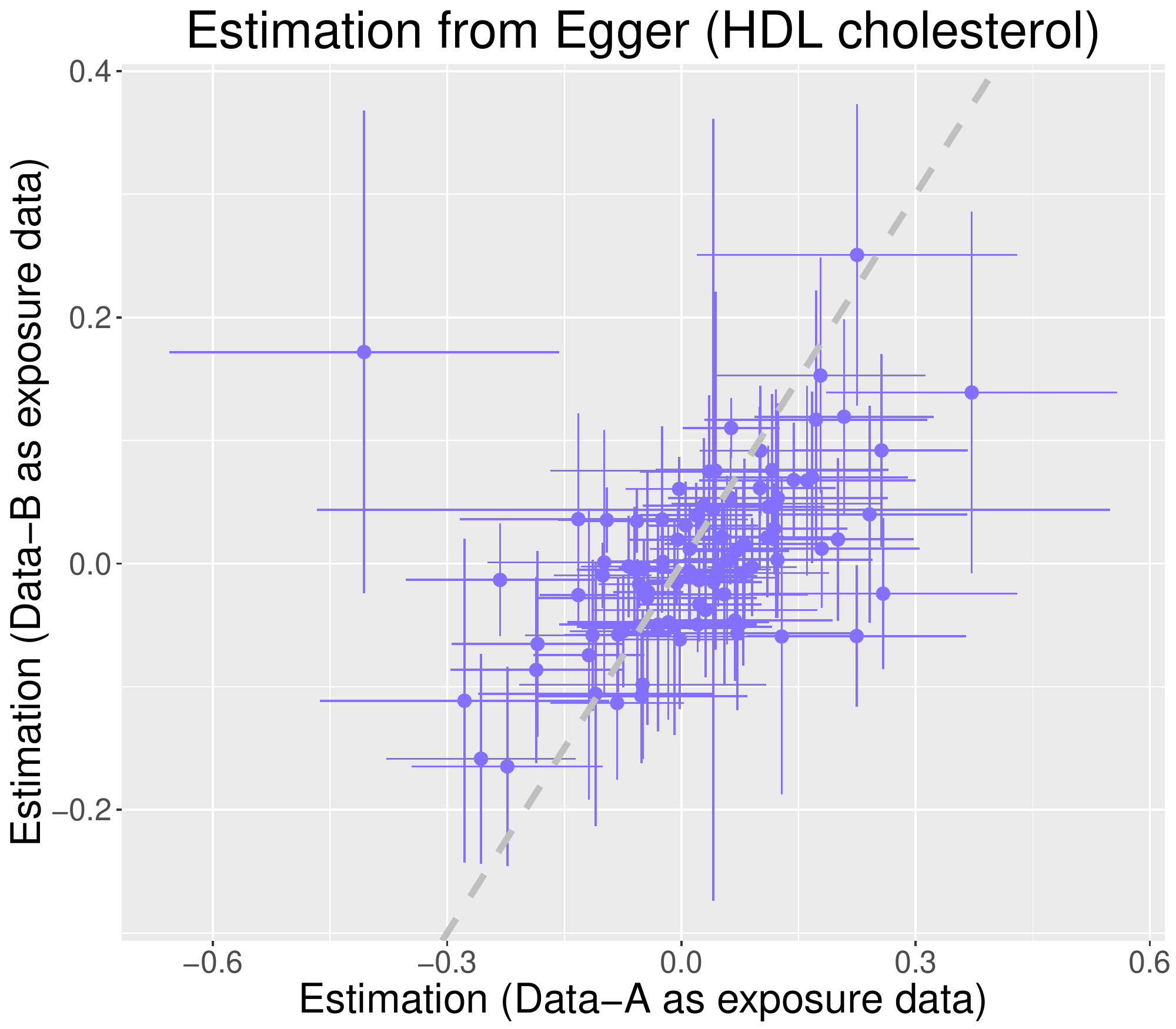}
\includegraphics[scale=0.19]{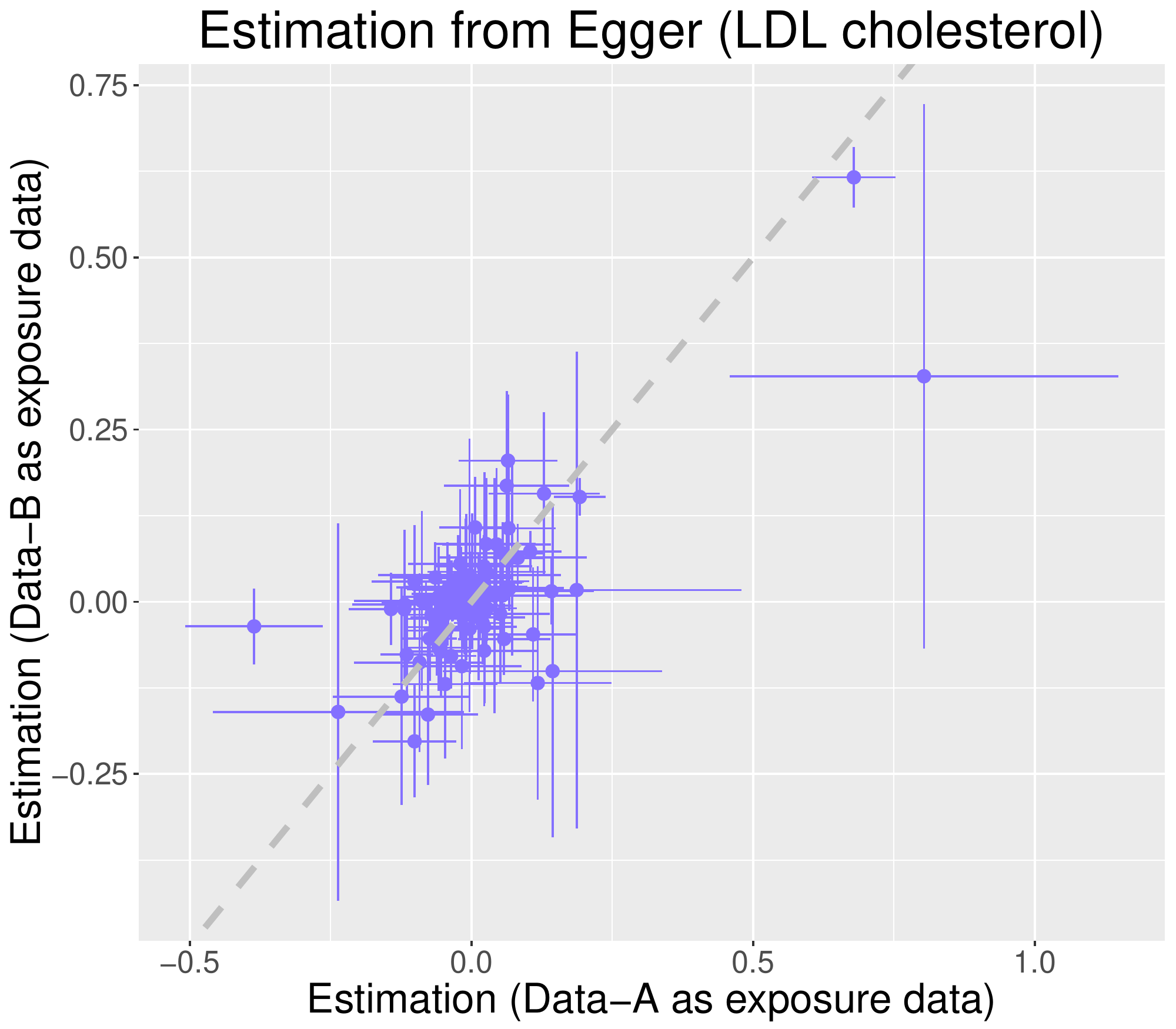}
\includegraphics[scale=0.19]{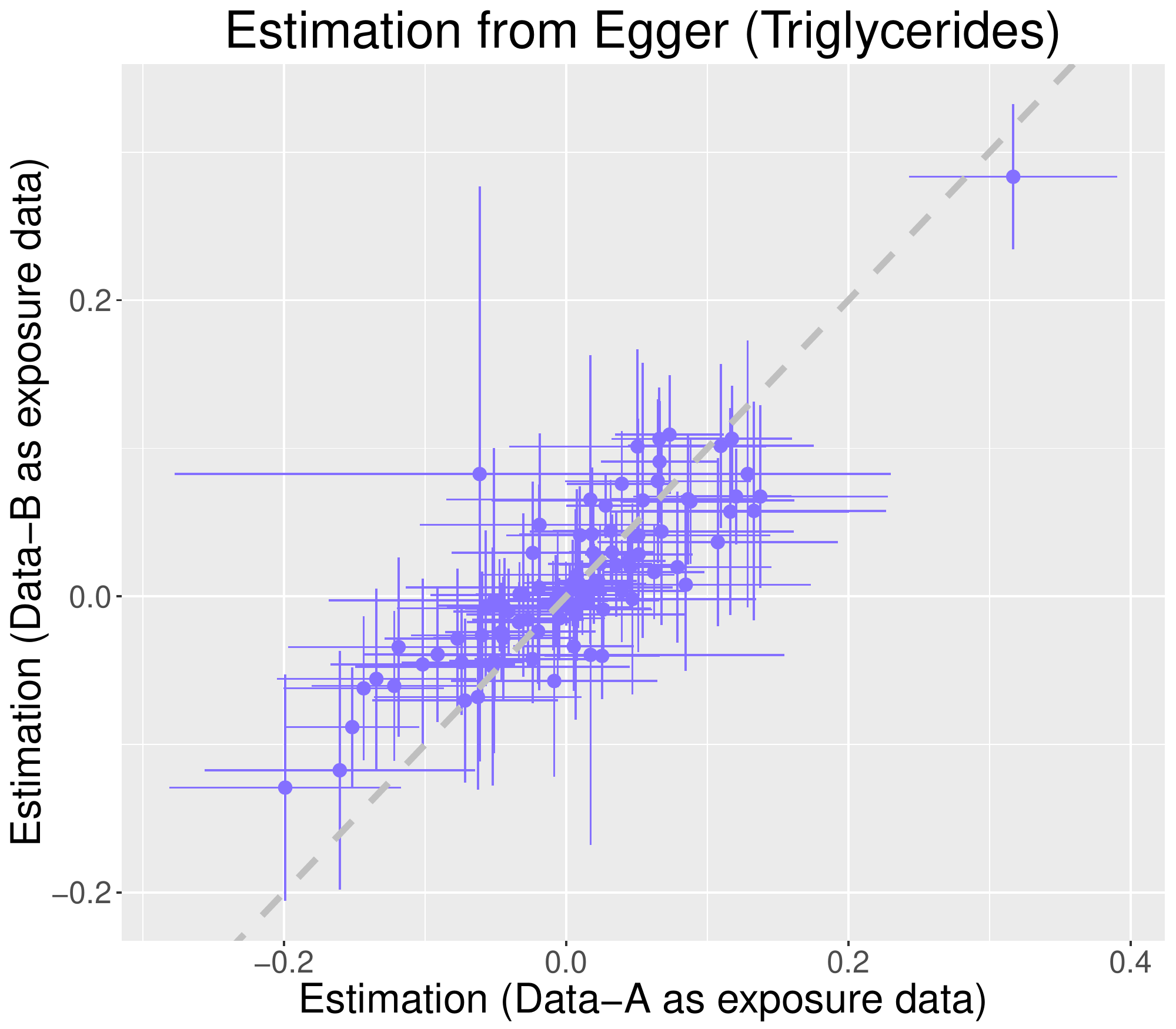}
\includegraphics[scale=0.19]{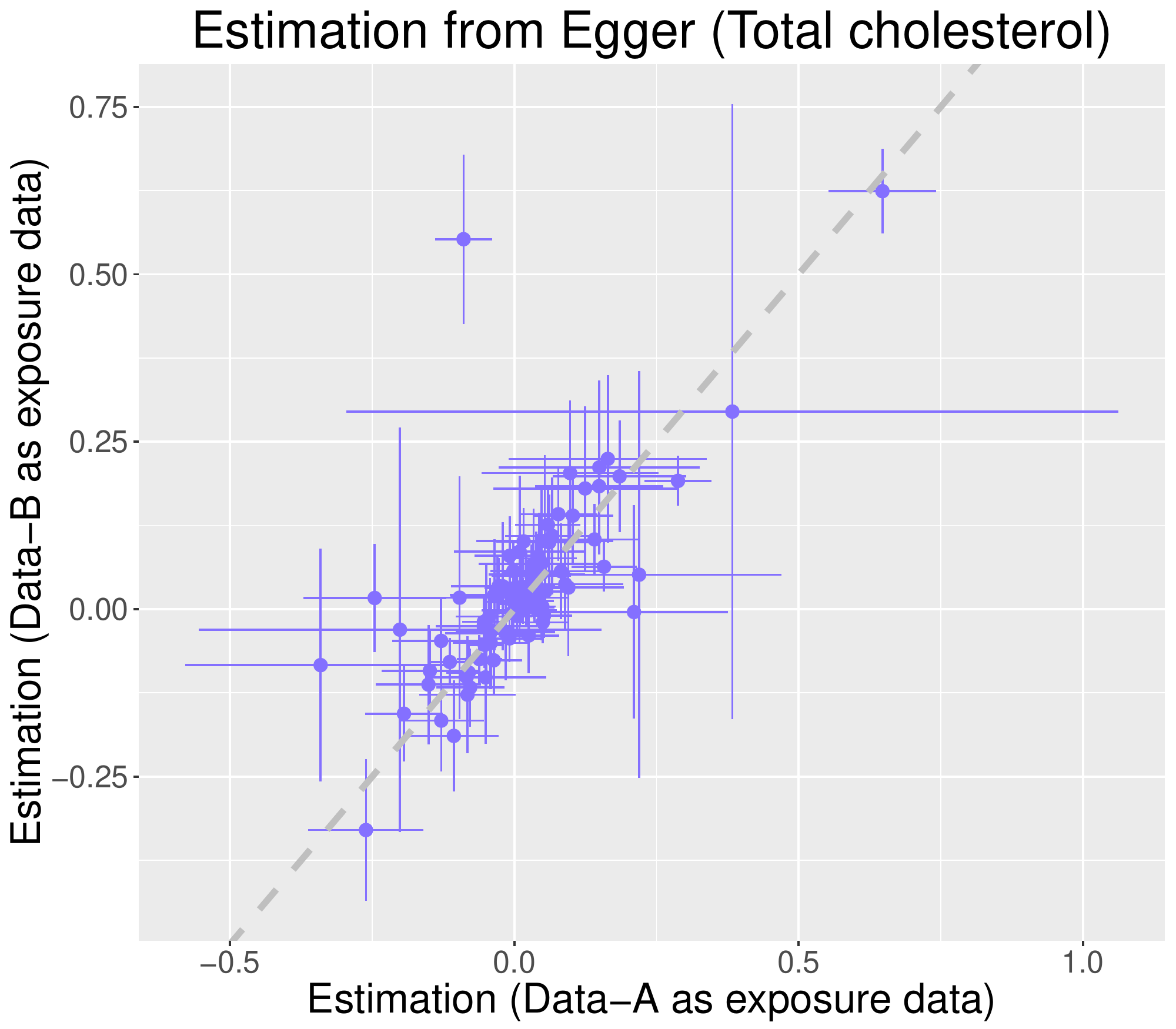}
\par\end{centering}
\caption{Comparisons of analysis results provided by Egger using Data-A as exposure data ($x$-axis) and using Data-B as exposure data ($y$-axis). The dots represent estimated causal effect sizes $\hat{\beta}$ and the bars represent their standard errors $\mathrm{se}(\hat{\beta})$. The diagonal is indicated by the dashed line.}
\end{figure}

\newpage{}
\subsubsection{Some illustrative examples in real data analysis}

\begin{itemize}
  \item exposure: LDL cholesterol; outcome: height.
\end{itemize}

\begin{table}[h]
\caption{MR results of ``LDL cholesterol - height''}
\footnotesize
\centering
\begin{tabular}{cccc}
\hline
method& $\hat{\beta}$& $\hat{se}$& $p$-value\\
\hline
BWMR& -0.0295& 0.0164& 0.0721\\
Egger& -0.0135& 0.0378& 0.7222\\
GSMR& -0.0655& 0.0088& 7.8146$\times 10^{-14}$\\
RAPS& -0.0290& 0.0169& 0.0856\\
\hline
\end{tabular}
\end{table}

\begin{figure}[H]
\begin{centering}
\includegraphics[scale=0.3]{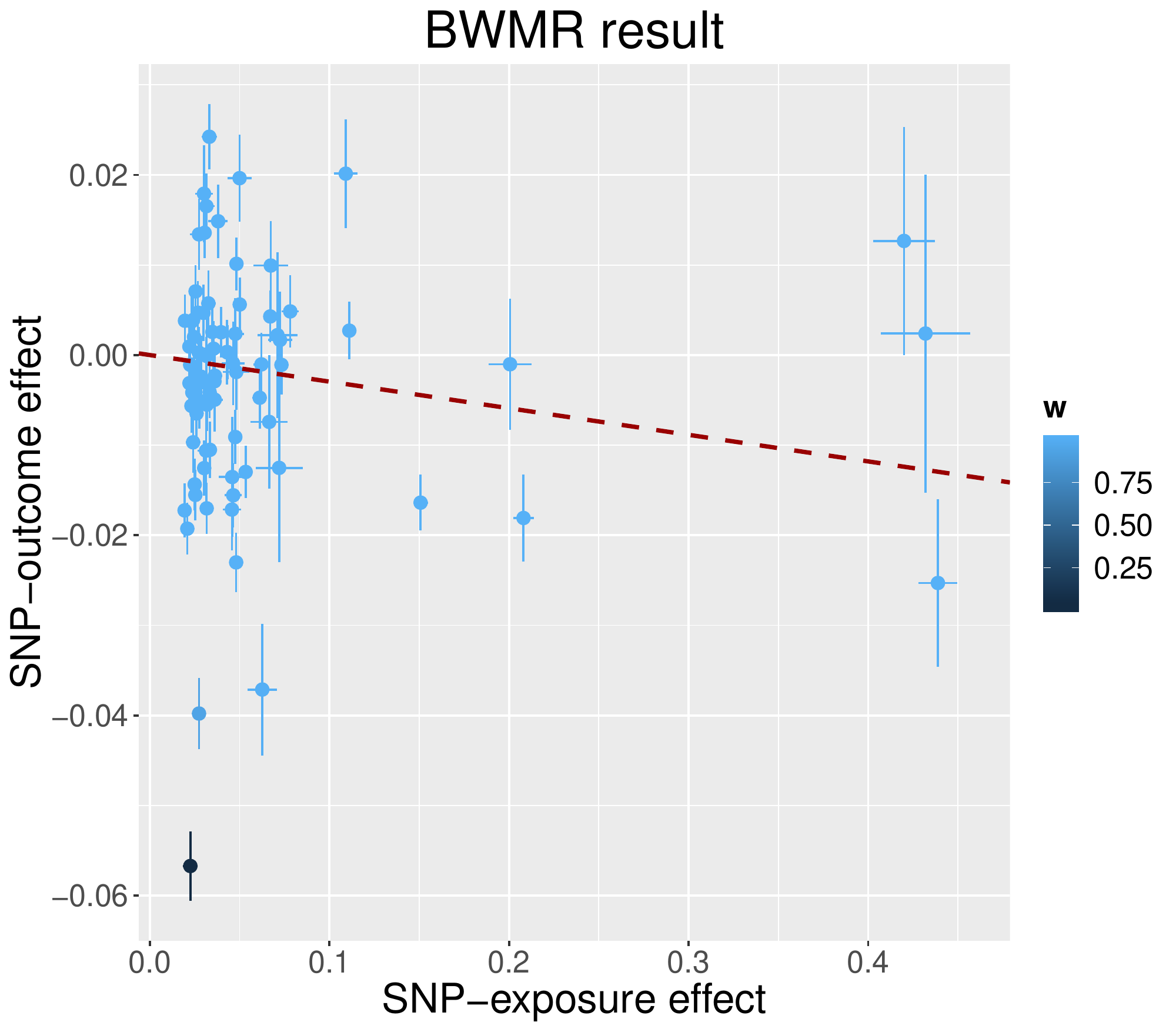}
\includegraphics[scale=0.3]{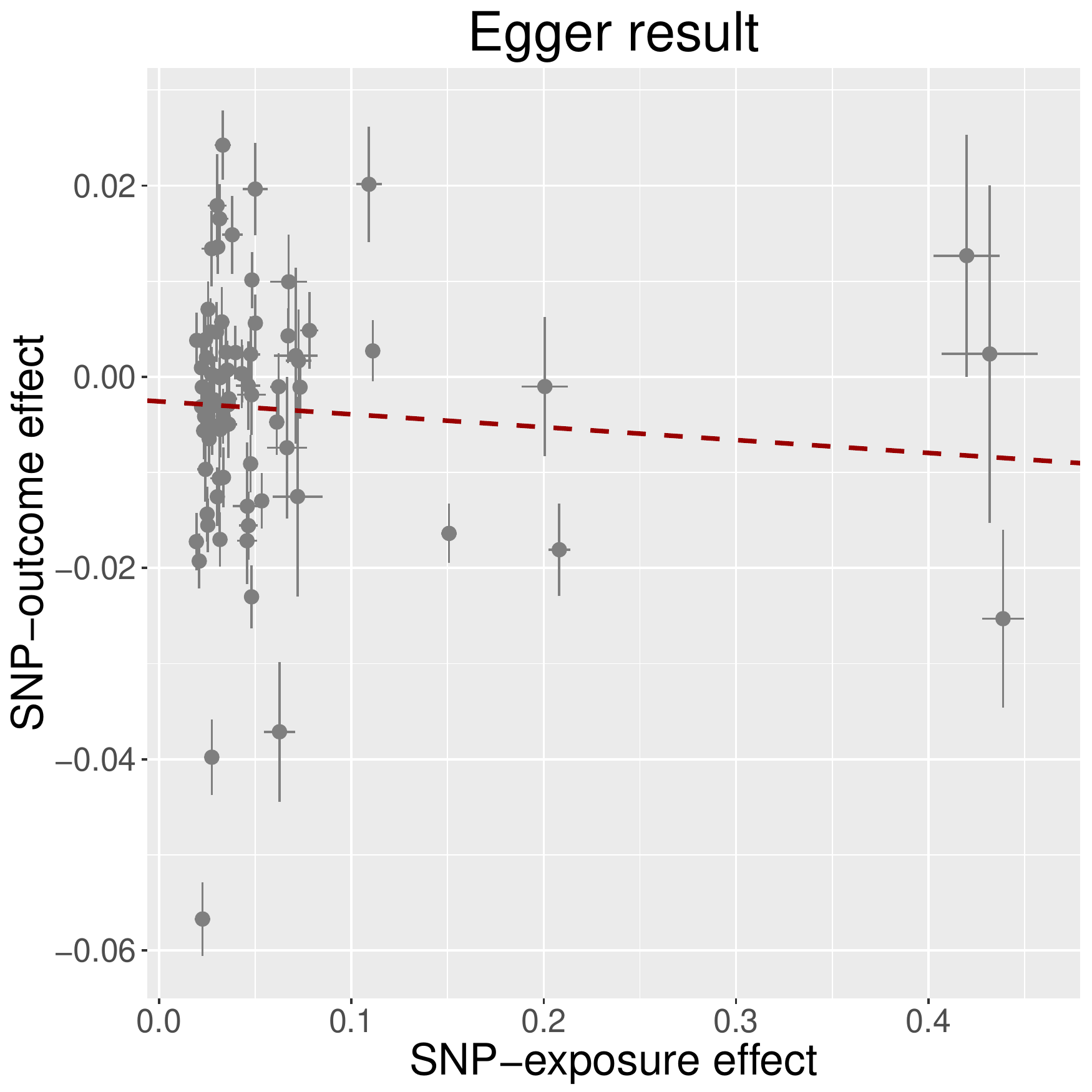}

\includegraphics[scale=0.32]{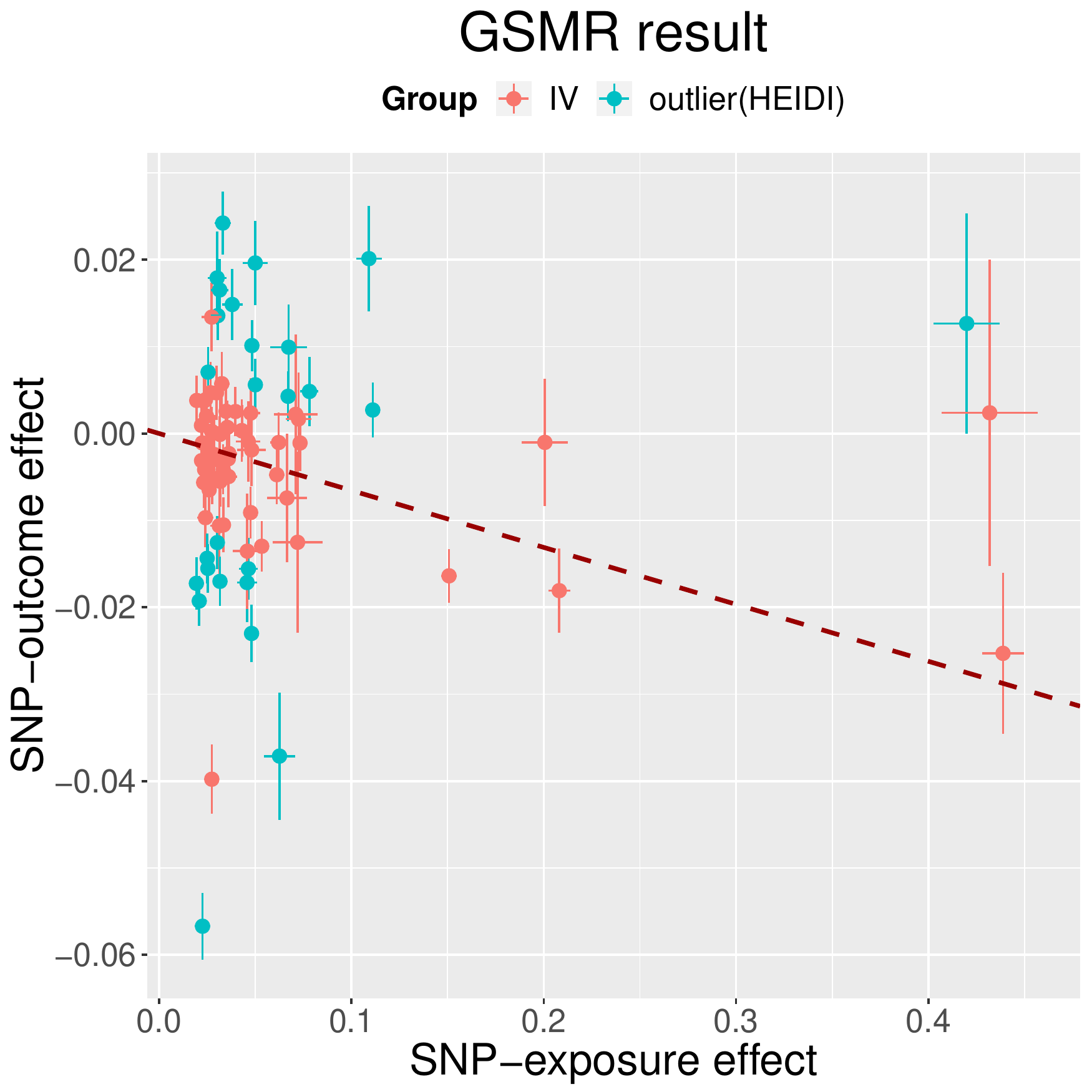}
\includegraphics[scale=0.3]{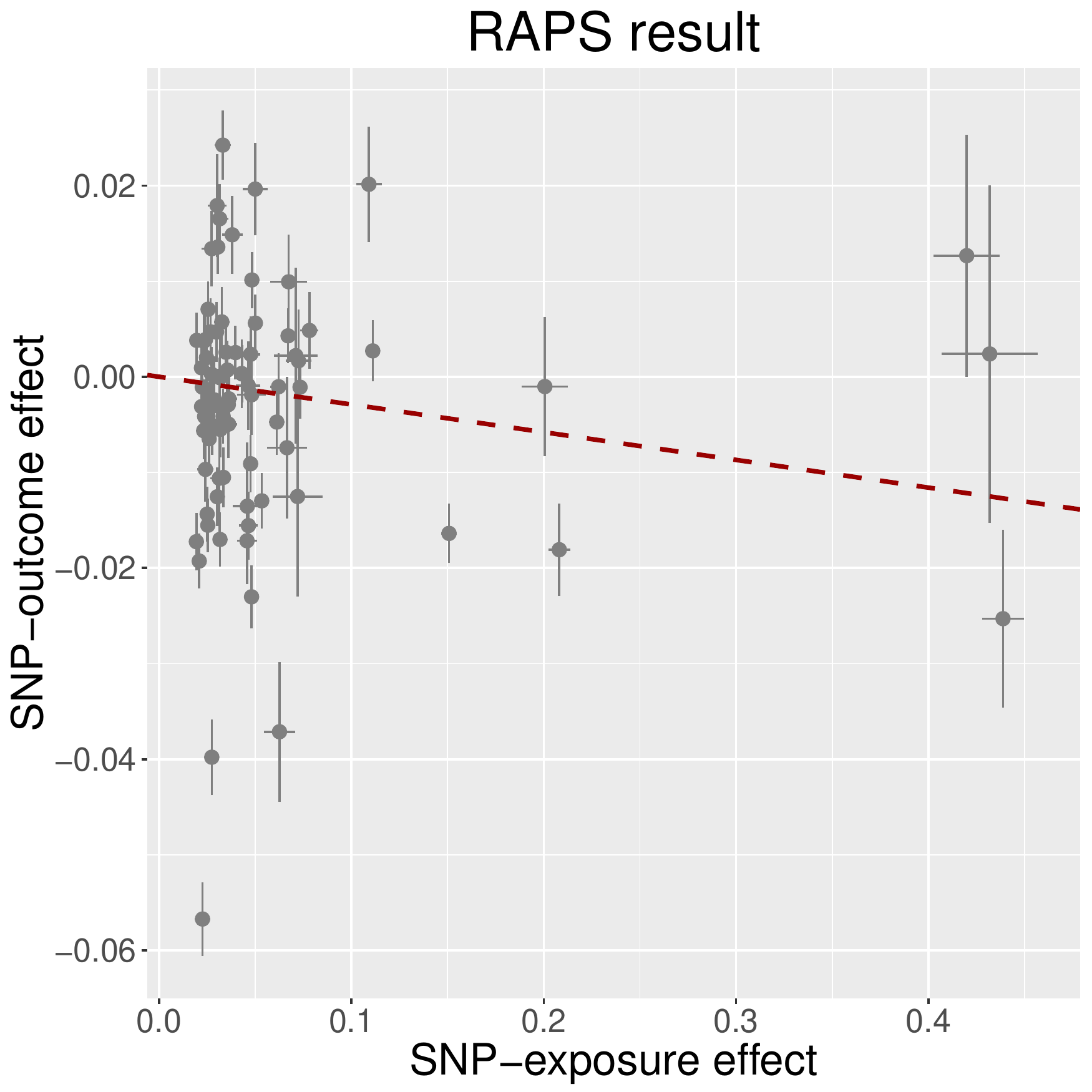}
\par\end{centering}
\caption{Comparisons of causal inference results when estimating the causal effect of LDL cholesterol on height. We use this example to show that GSMR may suffer from inflated type I error rate when IVs are affected by weak horizontal pleiotropic effects, and the HEIDI outlier method in GSMR has a risk of selecting an impropriate SNP (e.g., pleiotropic IV) as the top variant leading to false discovery.}
\end{figure}

\newpage{}
\begin{itemize}
  \item exposure: Gp; outcome: Crohn's disease.
\end{itemize}

\begin{table}[h]
\caption{MR results of ``Gp - Crohn's disease''}
\footnotesize
\centering
\begin{tabular}{cccc}
\hline
method& $\hat{\beta}$& $\hat{se}$& $p$-value\\
\hline
BWMR& 0.0711& 0.0678& 0.2946\\
Egger& -0.0781& 0.1668& 0.6521\\
GSMR& 0.1483& 0.0417& 0.0004\\
RAPS& 0.0691& 0.0700& 0.3242\\
\hline
\end{tabular}
\end{table}

\begin{figure}[H]
\begin{centering}
\includegraphics[scale=0.3]{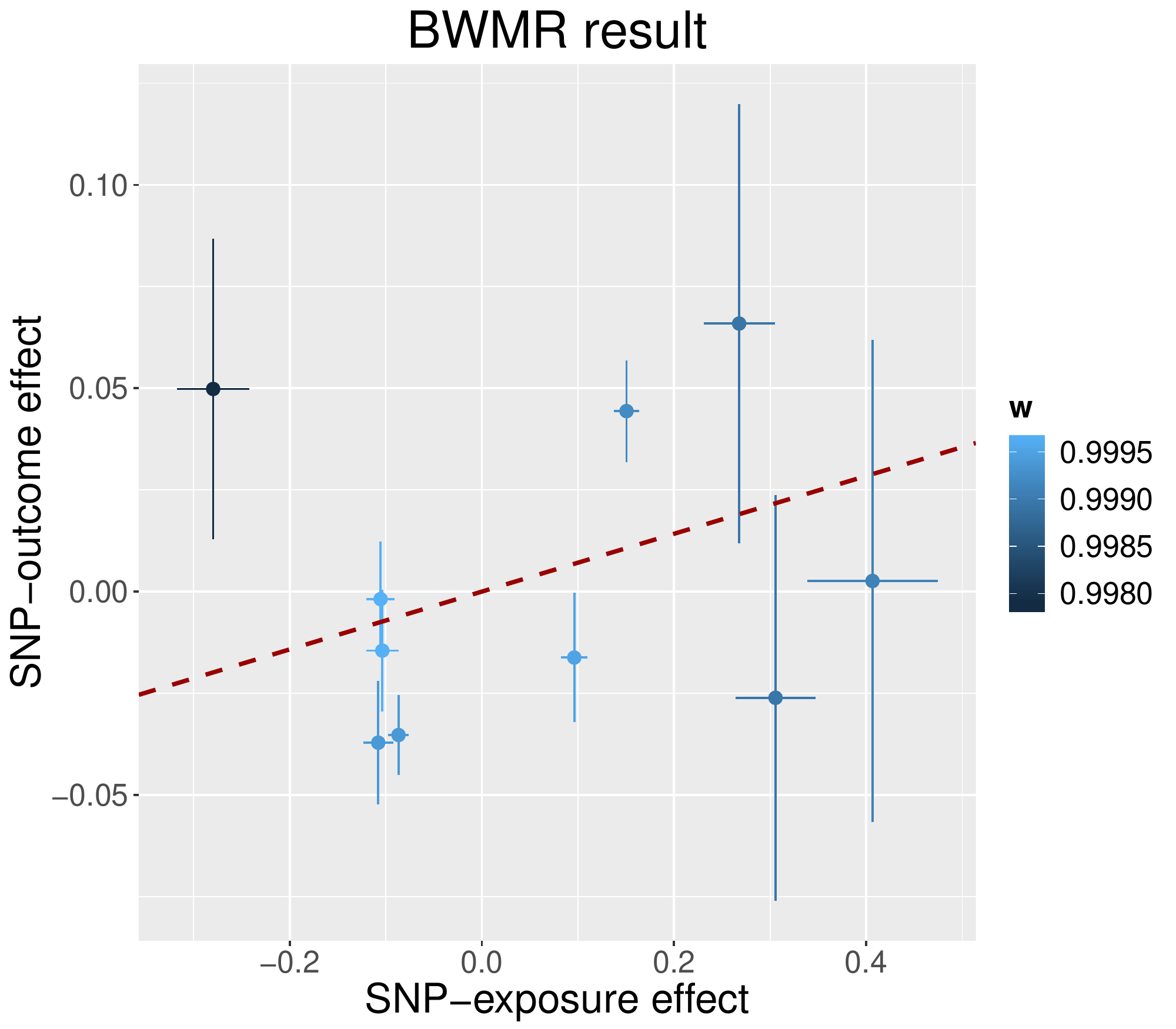}
\includegraphics[scale=0.3]{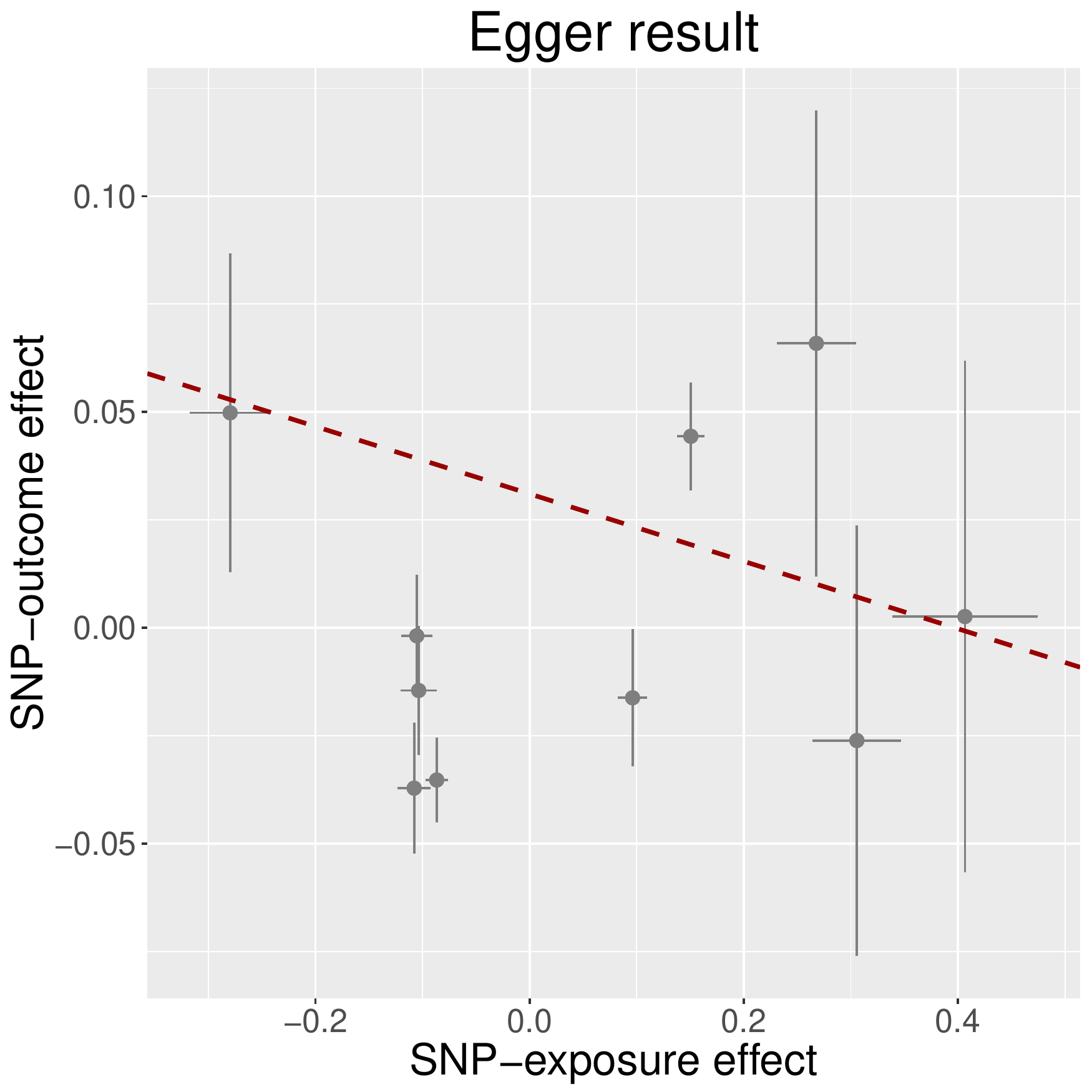}

\includegraphics[scale=0.32]{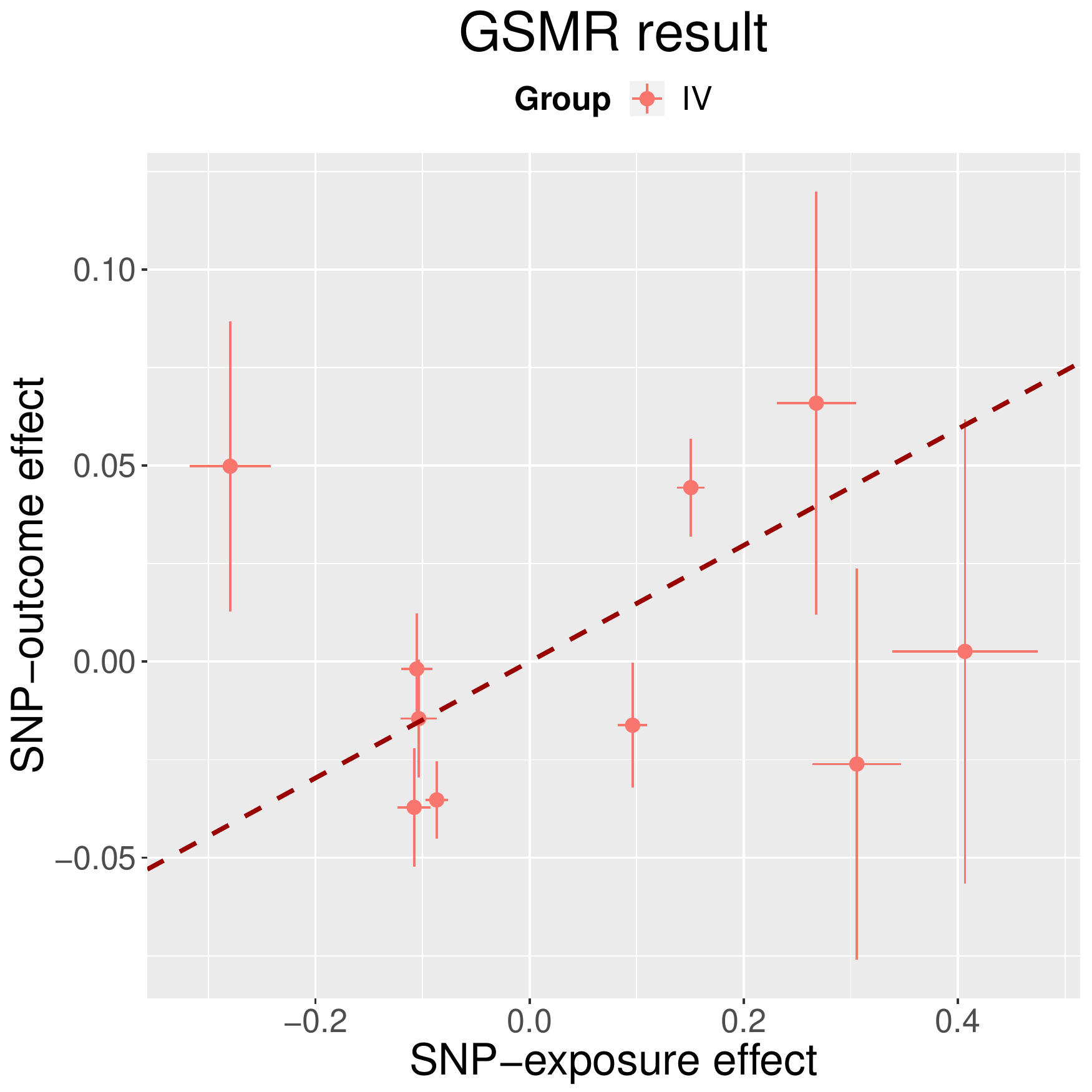}
\includegraphics[scale=0.3]{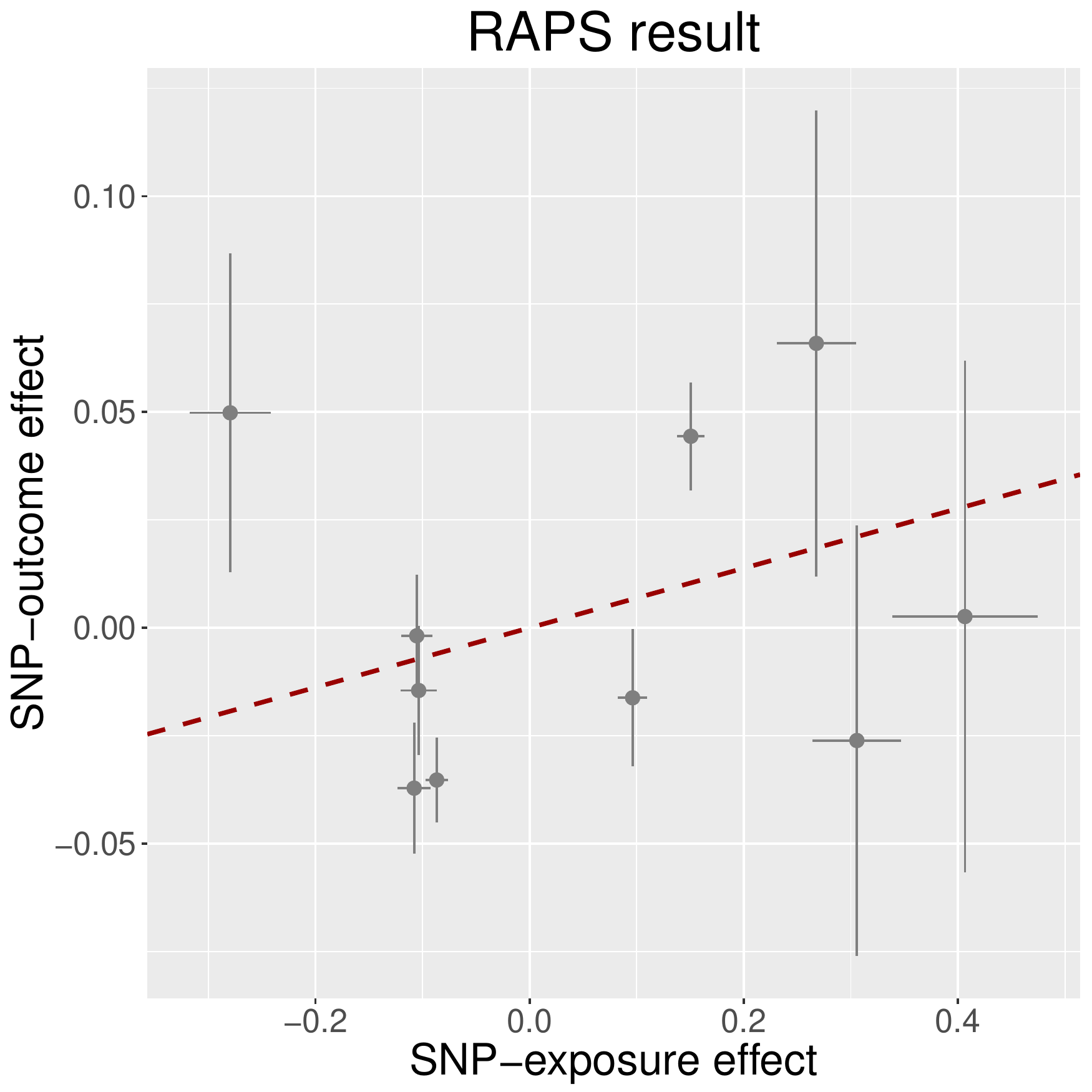}
\par\end{centering}
\caption{Comparisons of causal inference results when estimating the causal effect of Gp on Crohn's disease. We use this example to show that GSMR may suffer from inflated type I error rate when IVs are affected by weak horizontal pleiotropic effects, leading to false discovery.}
\end{figure}

\newpage{}
\begin{itemize}
  \item exposure: waist circumference; outcome: L.LDL.L.
\end{itemize}

\begin{table}[h]
\caption{MR results of ``waist circumference - L.LDL.L''}
\footnotesize
\centering
\begin{tabular}{cccc}
\hline
method& $\hat{\beta}$& $\hat{se}$& $p$-value\\
\hline
BWMR& -0.005& 0.0632& 0.9334\\
Egger& 0.2431& 0.2574& 0.3482\\
GSMR& 0.0002& 0.0587& 0.9968\\
RAPS& 0.0004& 0.0625& 0.9461\\
\hline
\end{tabular}
\end{table}

\begin{figure}[H]
\begin{centering}
\includegraphics[scale=0.3]{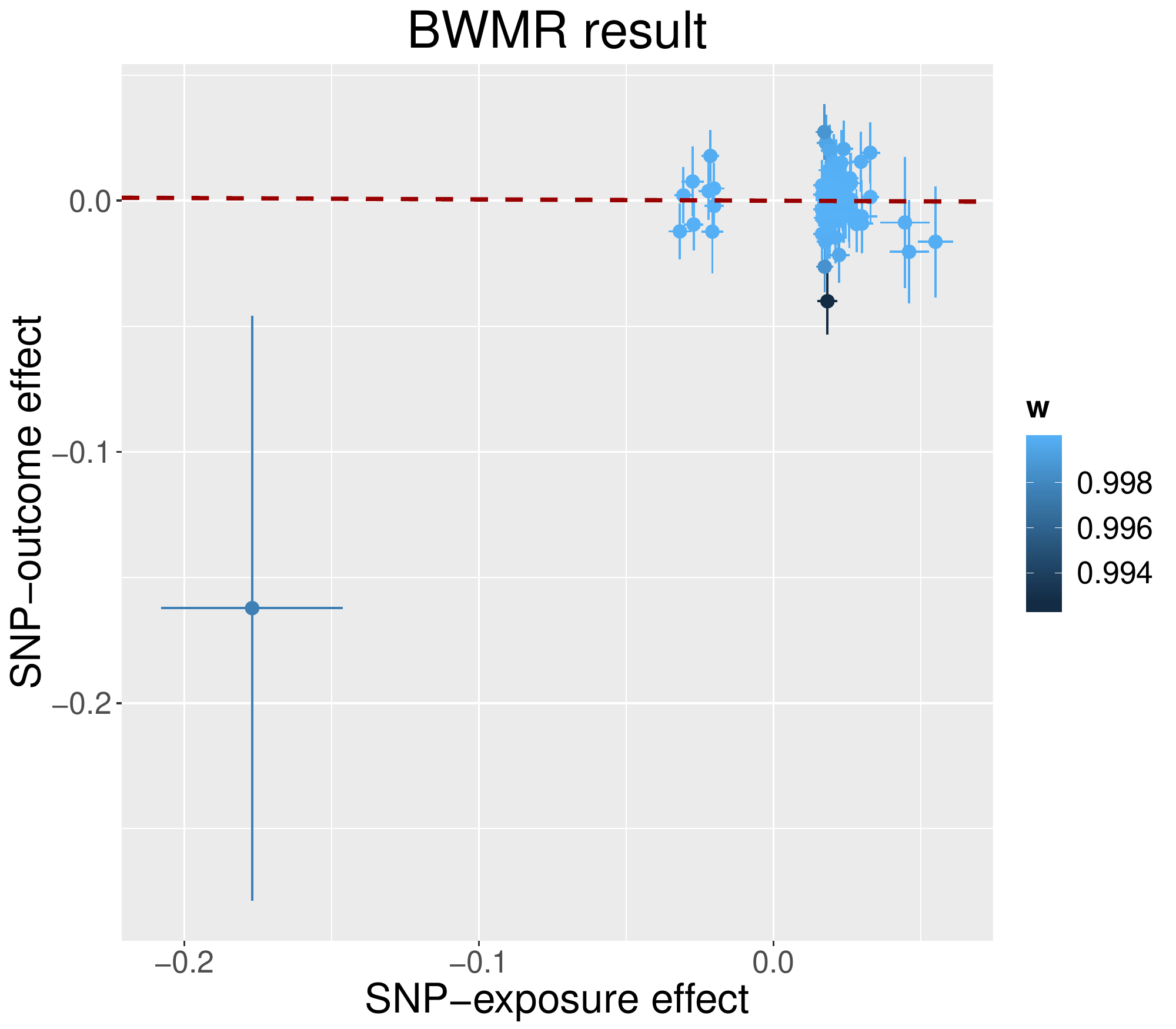}
\includegraphics[scale=0.3]{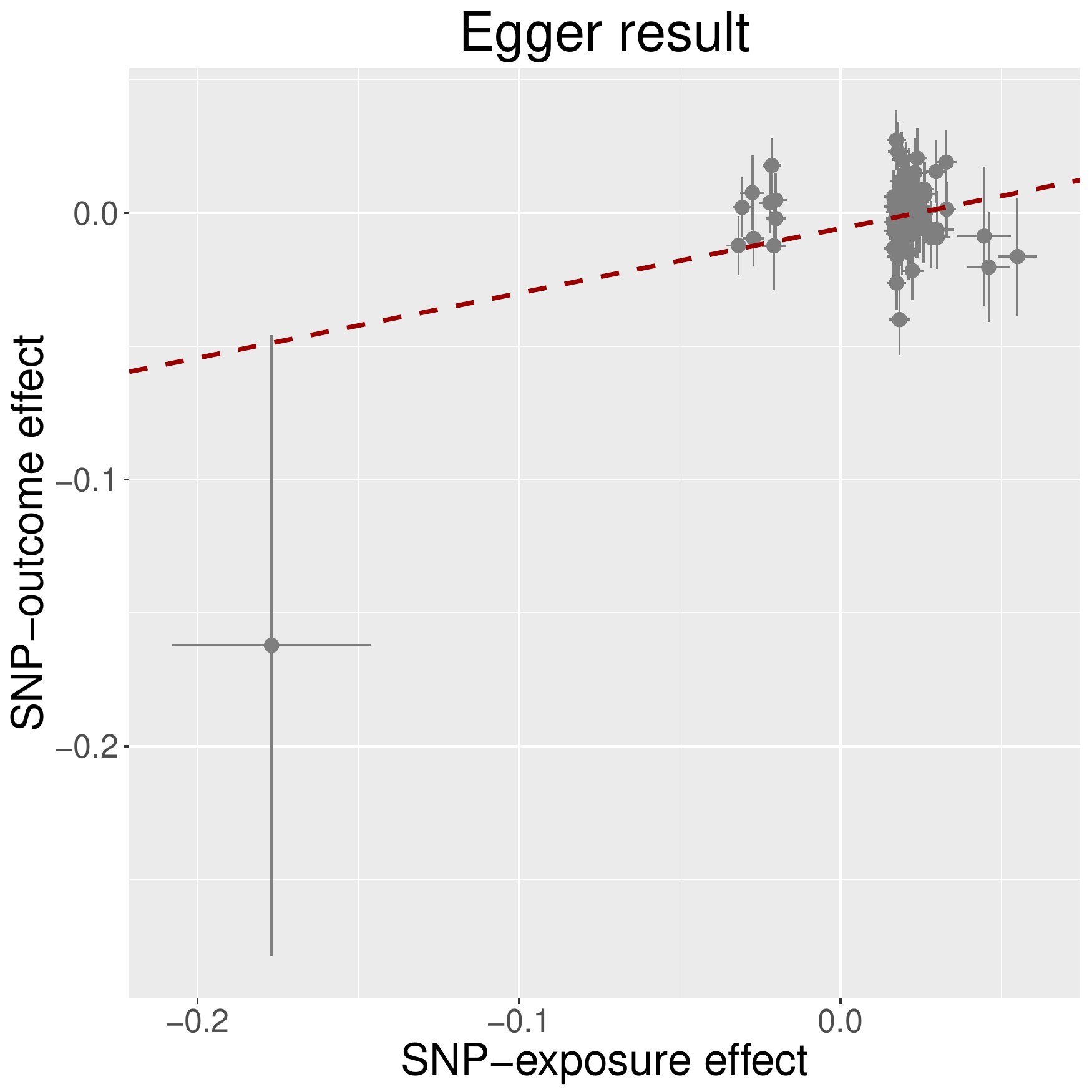}

\includegraphics[scale=0.32]{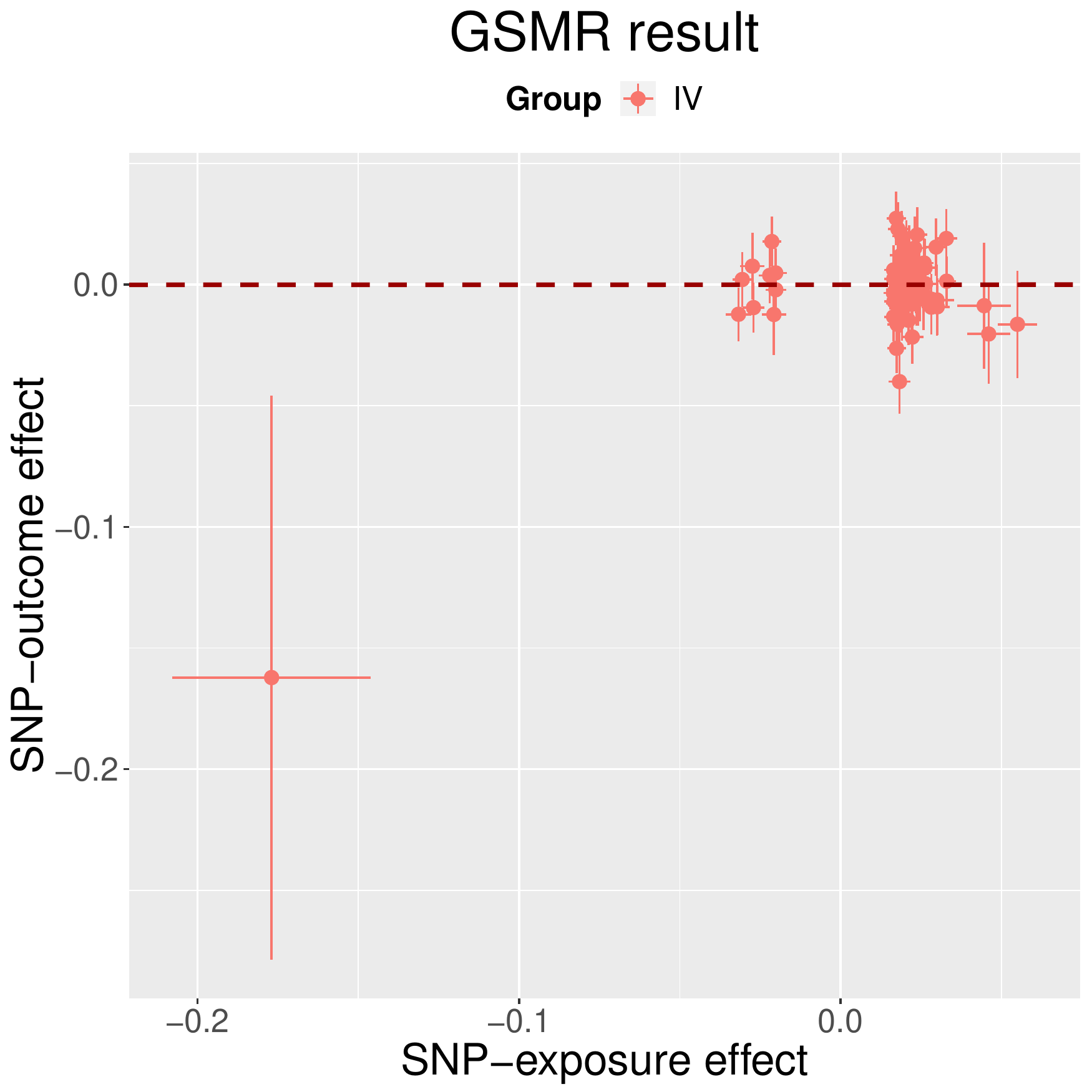}
\includegraphics[scale=0.3]{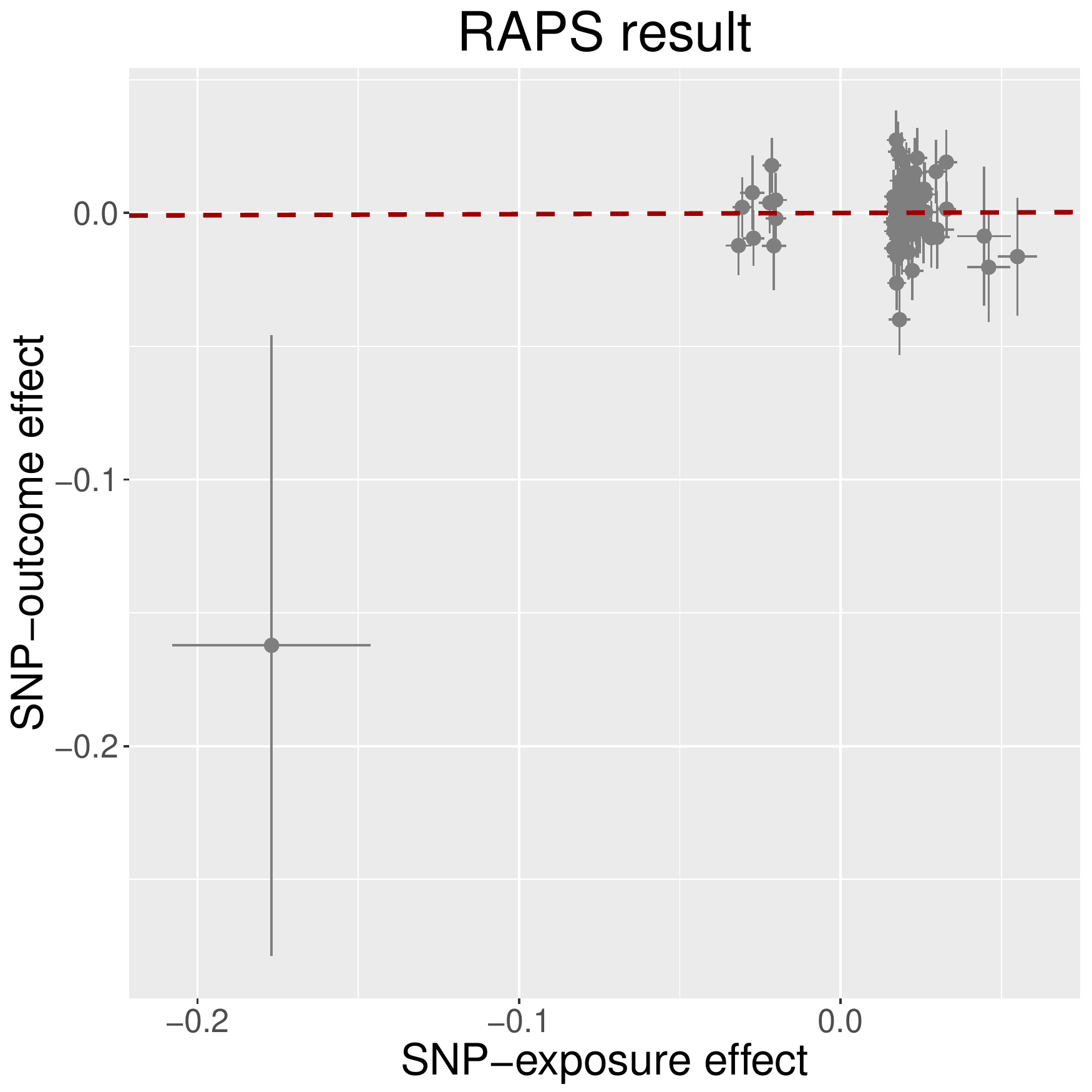}
\par\end{centering}
\caption{Comparisons of causal inference results when estimating the causal effect of waist circumference on L.LDL.L. We use this example to show that Egger is sensitive to outliers due to a few large pleiotropic effects.}
\end{figure}

\newpage{}
\begin{itemize}
  \item exposure: age at menarche; outcome: S.VLDL.C.
\end{itemize}

\begin{table}[h]
\caption{MR results of ``age at menarche - S.VLDL.C''}
\footnotesize
\centering
\begin{tabular}{cccc}
\hline
method& $\hat{\beta}$& $\hat{se}$& $p$-value\\
\hline
BWMR& -0.0163& 0.0463& 0.7251\\
Egger& -0.6315& 0.1917& 0.0015\\
GSMR& -0.0155& 0.0436& 0.7212\\
RAPS& -0.0069& 0.0472& 0.8838\\
\hline
\end{tabular}
\end{table}

\begin{figure}[H]
\begin{centering}
\includegraphics[scale=0.3]{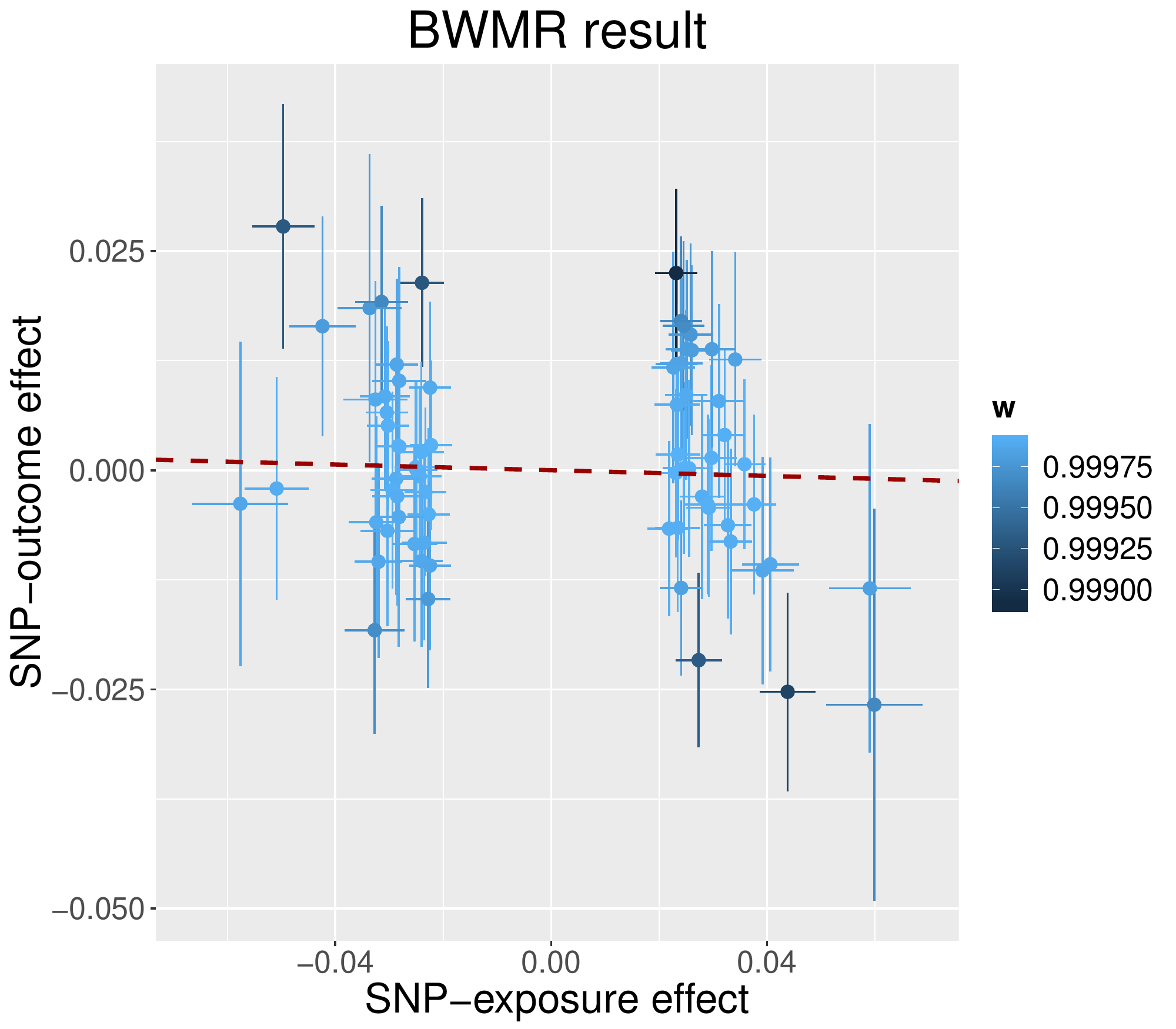}
\includegraphics[scale=0.3]{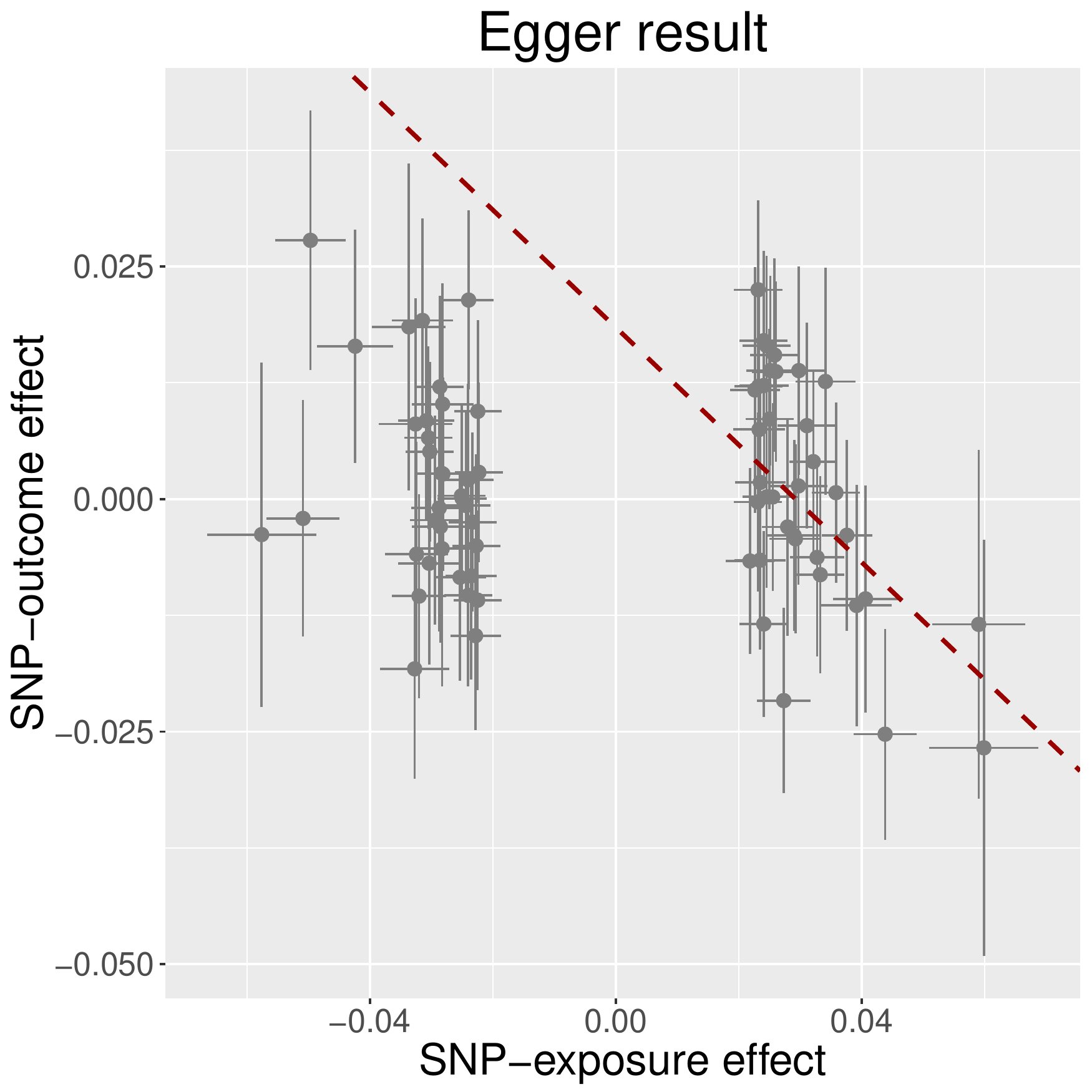}

\includegraphics[scale=0.32]{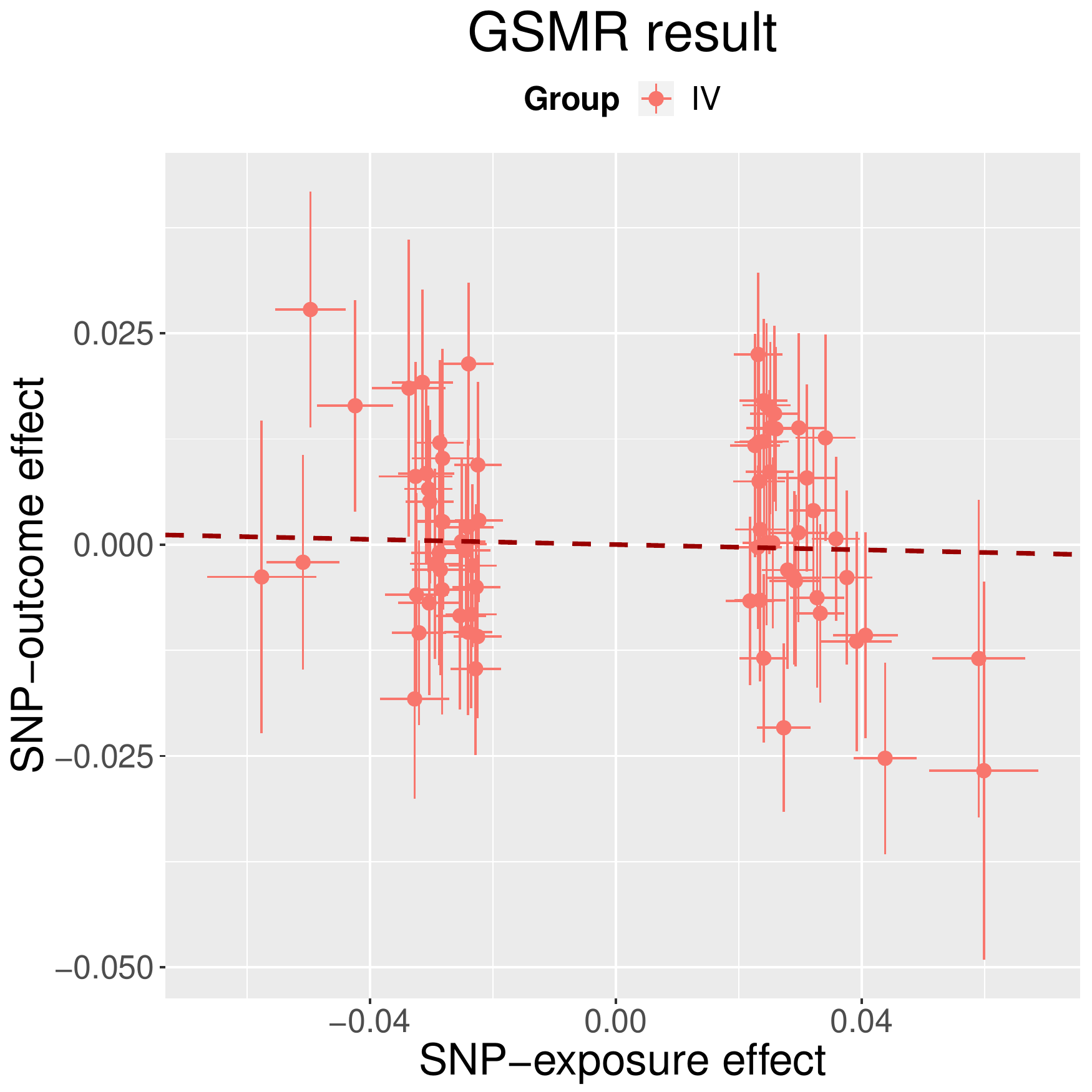}
\includegraphics[scale=0.3]{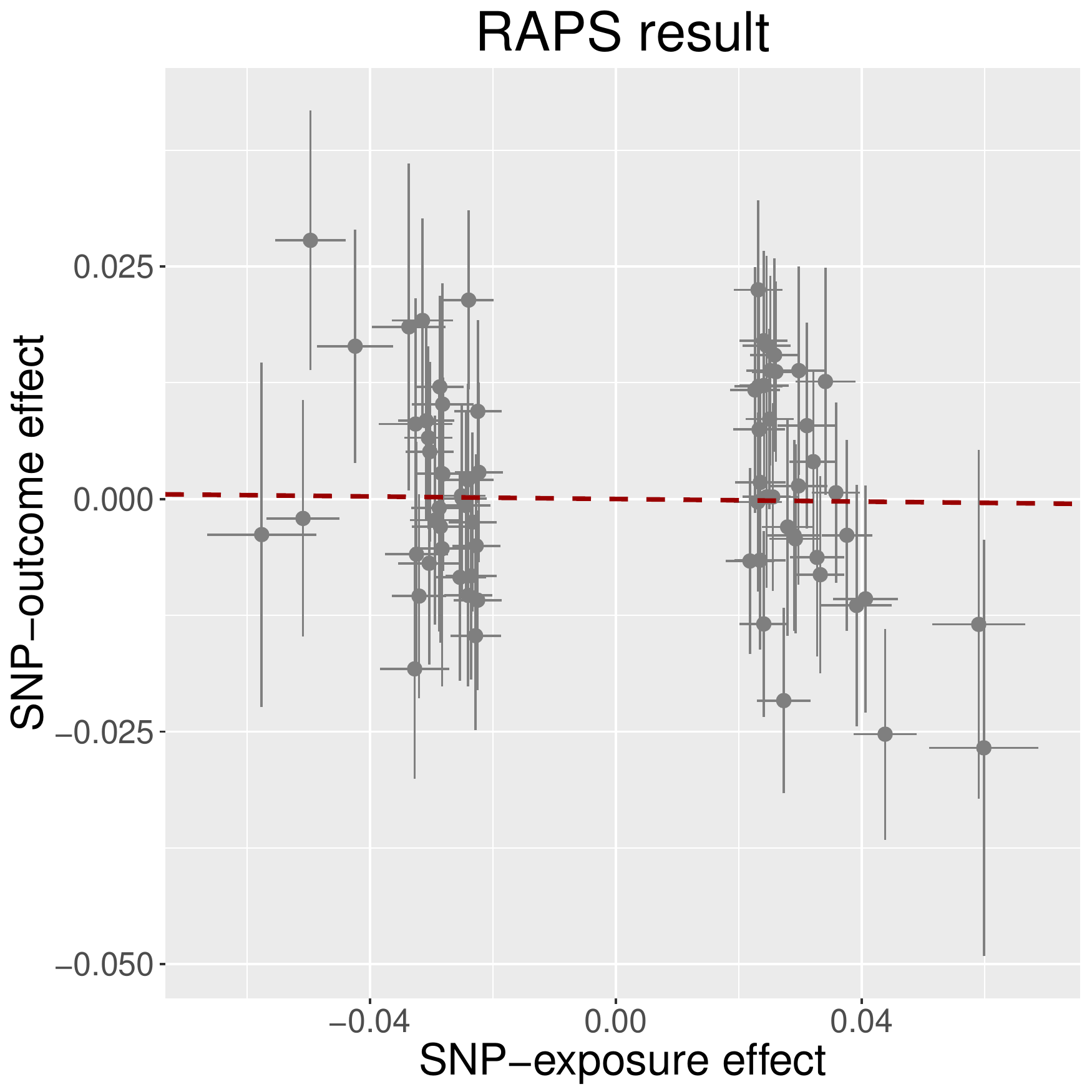}
\par\end{centering}
\caption{Comparisons of causal inference results when estimating the causal effect of age at menarche on S.VLDL.C. We use this example to show that Egger may mistakenly detect pleiotropic IVs or outliers, leading to biased estimates. Indeed, the two components in summary statistics are due to the phenomenon of selection bias.}
\end{figure}

\newpage{}
\begin{itemize}
  \item exposure: CAD; outcome: M.LDL.C.
\end{itemize}

\begin{table}[h]
\caption{MR results of ``CAD - M.LDL.C''}
\footnotesize
\centering
\begin{tabular}{cccc}
\hline
method& $\hat{\beta}$& $\hat{se}$& $p$-value\\
\hline
BWMR& 0.0668& 0.0978& 0.4946\\
Egger& 0.5445& 0.5994& 0.3690\\
GSMR& -0.0393& 0.0647& 0.5433\\
RAPS& 0.3317& 0.1734& 0.0557\\
\hline
\end{tabular}
\end{table}

\begin{figure}[H]
\begin{centering}
\includegraphics[scale=0.3]{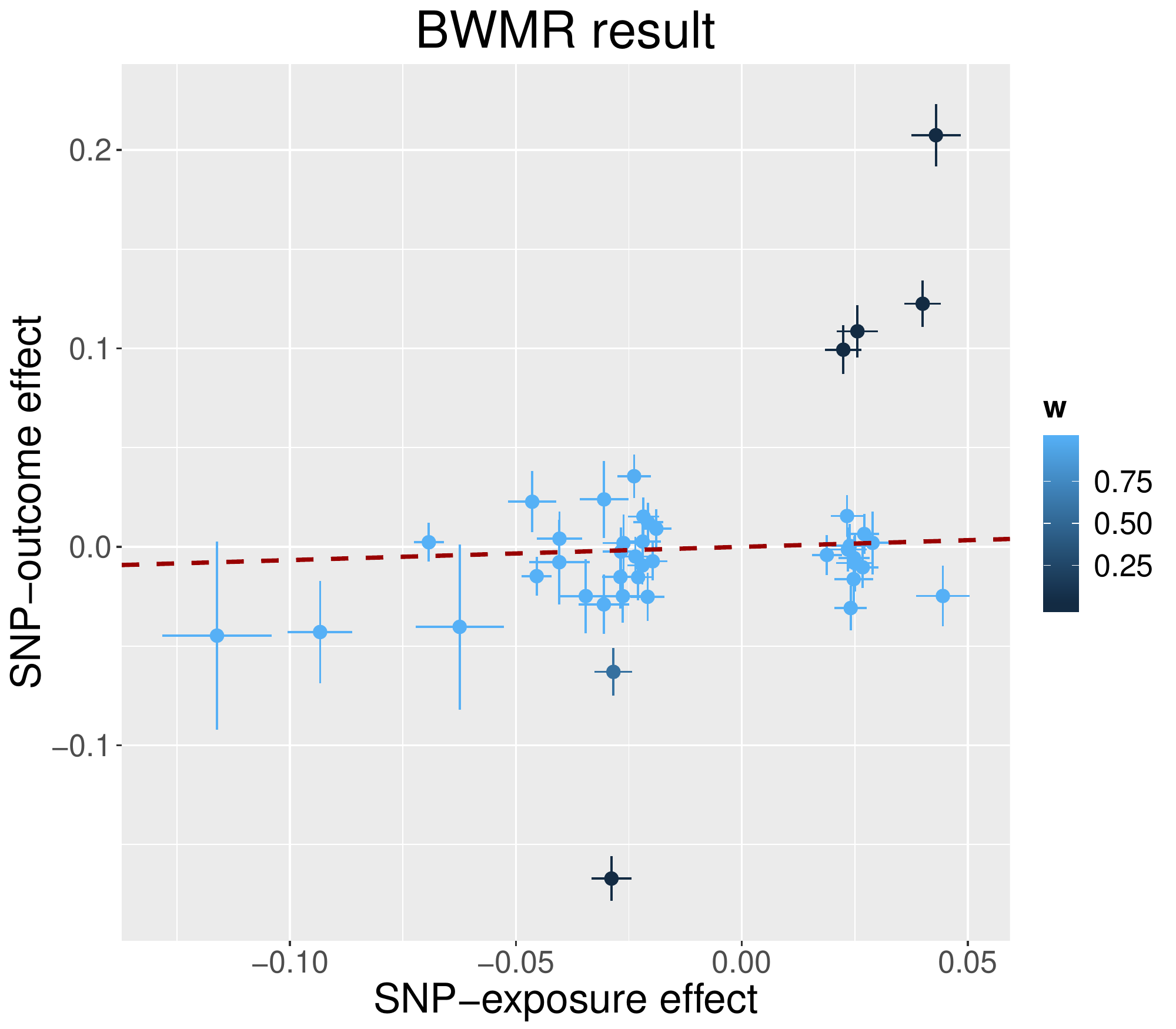}
\includegraphics[scale=0.3]{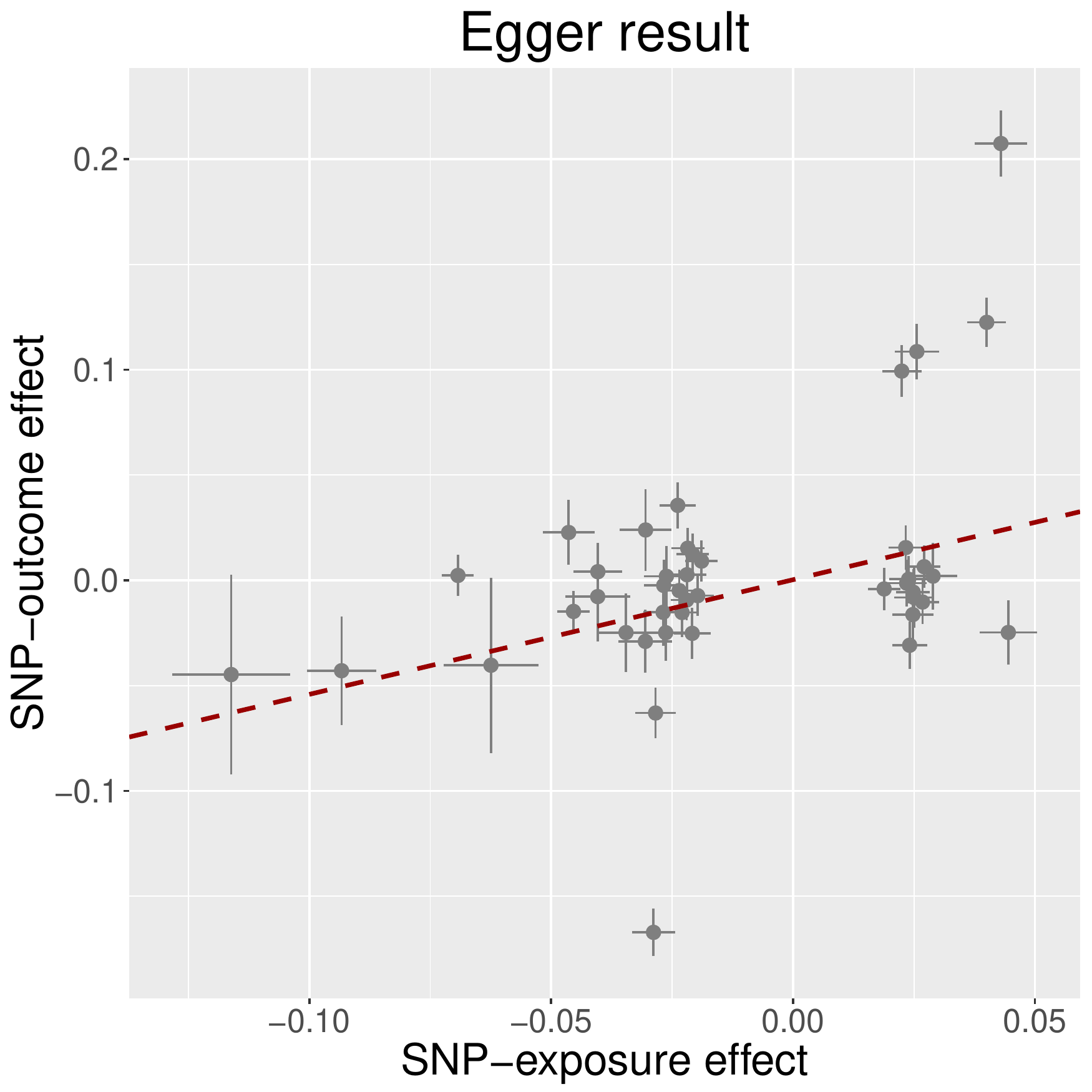}

\includegraphics[scale=0.32]{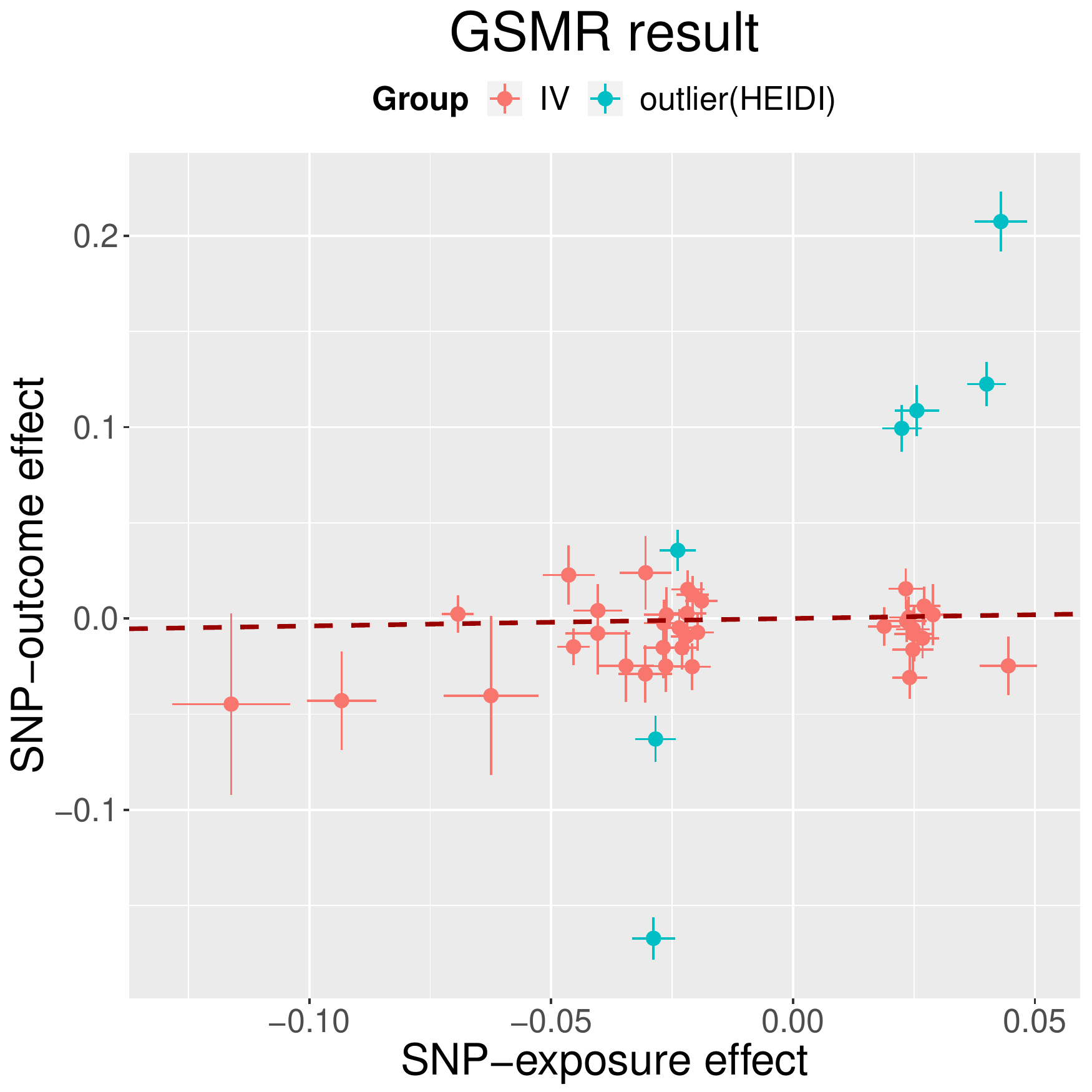}
\includegraphics[scale=0.3]{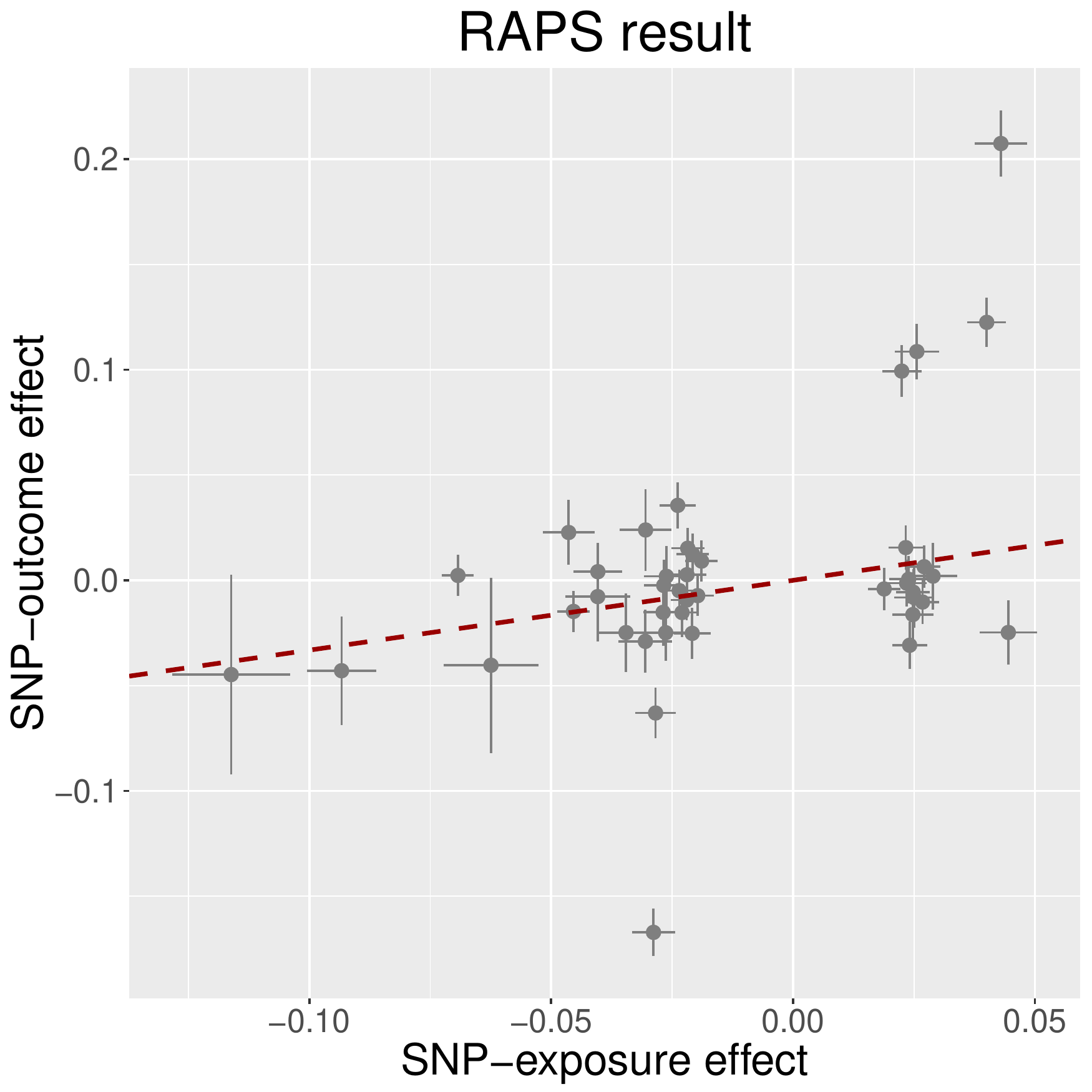}
\par\end{centering}
\caption{Comparisons of causal inference results when estimating the causal effect of CAD on M.LDL.C. We use this example to show that RAPS may not be robust to pleiotropic IVs when the horizontal pleiotropic effects have correlation with SNP-exposure effects.}
\end{figure}

\newpage{}
\subsubsection{MR-based causal inference between metabolites and complex traits}

\begin{itemize}
  \item Detailed results from BWMR
\end{itemize}
\begin{figure}[H]
\begin{centering}
\includegraphics[scale=0.2]{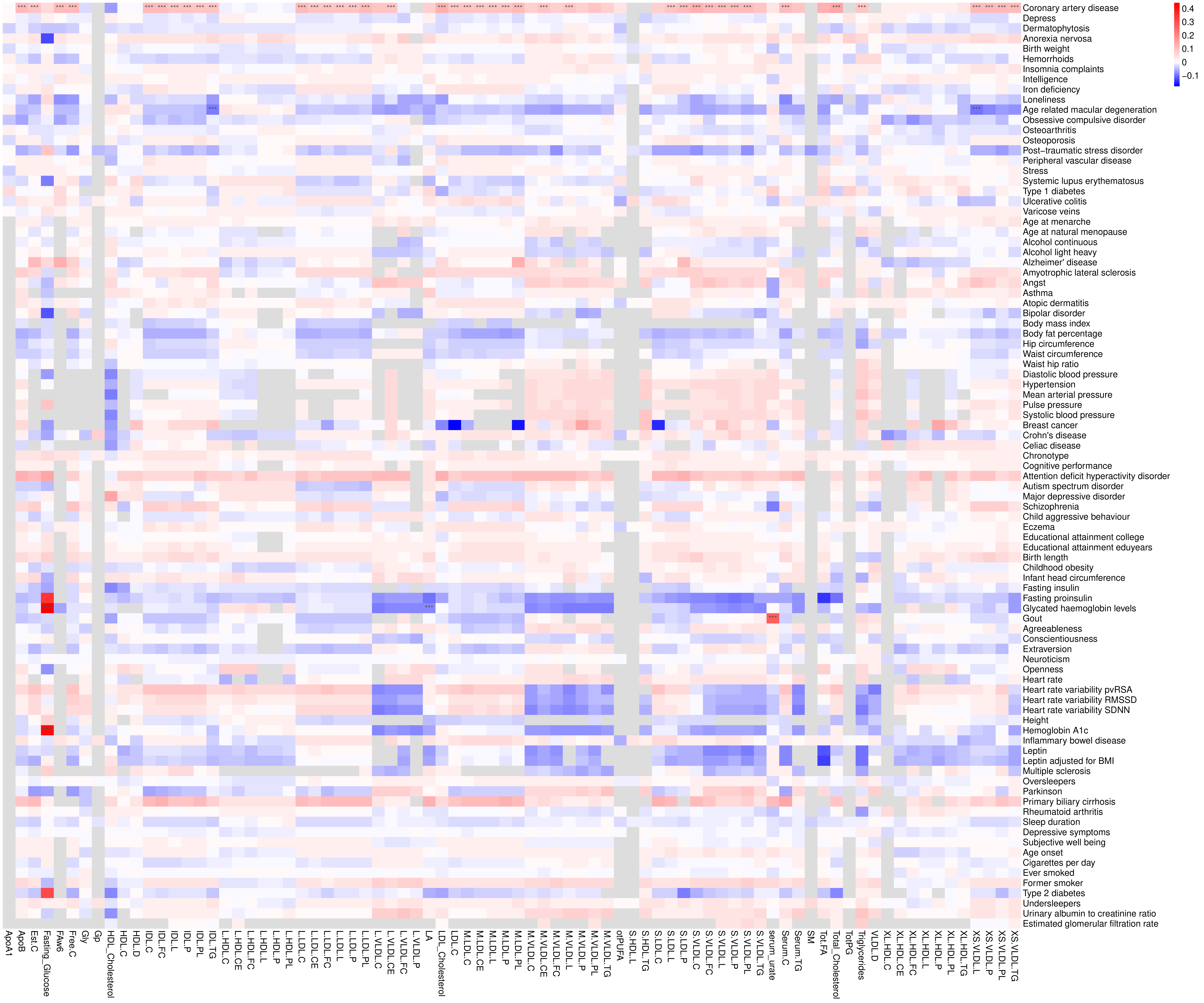}
\par\end{centering}
\caption{Causal effects of $130$ metabolites (exposures) on $92$ complex human traits (outcomes) estimated by method BWMR. We selected SNPs as IVs by the $p$-value threshold $5\times 10^{-8}$. The symbol "***" means the $p$-value is significant after Bonferroni correction at level $0.05$.}
\end{figure}
\newpage{}
\begin{figure}[H]
\begin{centering}
\includegraphics[scale=0.25]{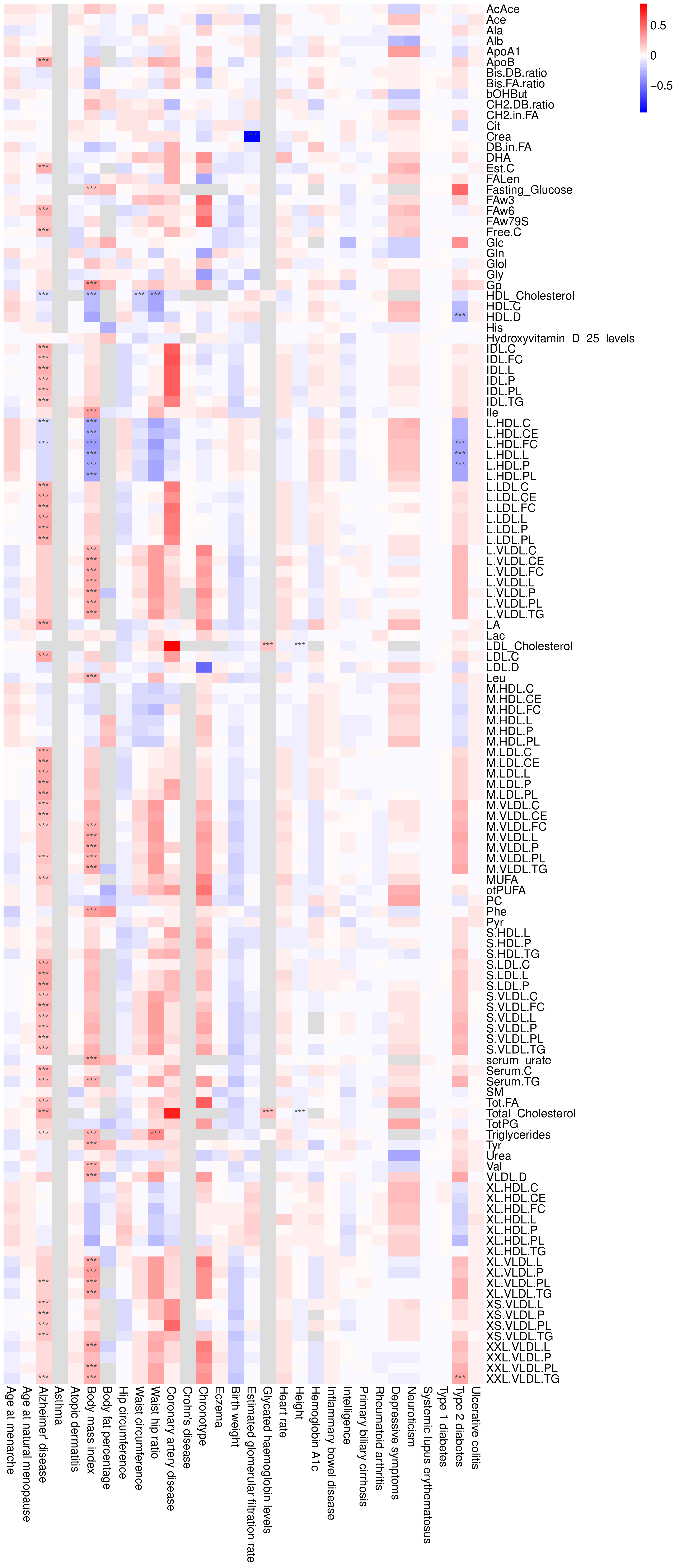}
\par\end{centering}
\caption{Causal effects of $92$ complex human traits (exposures) on $130$ metabolites and the trait dyslipidemia (outcomes) estimated by method BWMR. We selected SNPs as IVs by the $p$-value threshold $5\times 10^{-8}$. The symbol "***" means the $p$-value is significant after Bonferroni correction at level $0.05$.}
\end{figure}

\begin{figure}[H]
\begin{centering}
\includegraphics[scale=0.65]{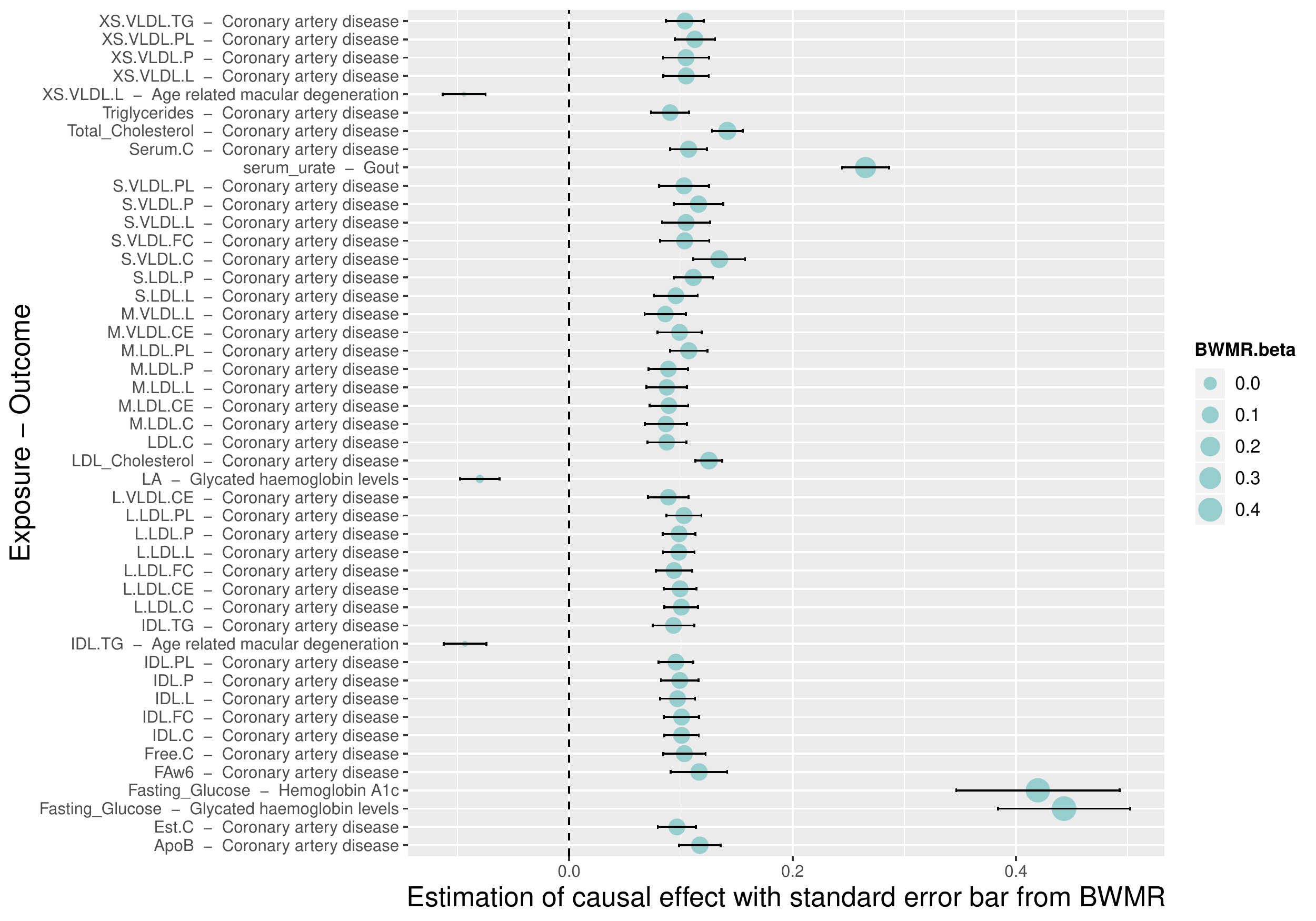}
\par\end{centering}
\caption{Significant causal effects of metabolites on complex human traits estimated by BWMR. We selected SNPs as IVs by the $p$-value threshold $5\times 10^{-8}$ and controlled the type I error rate after Bonferroni correction at level 0.05.}
\end{figure}

\begin{figure}[H]
\begin{centering}
\includegraphics[scale=0.57]{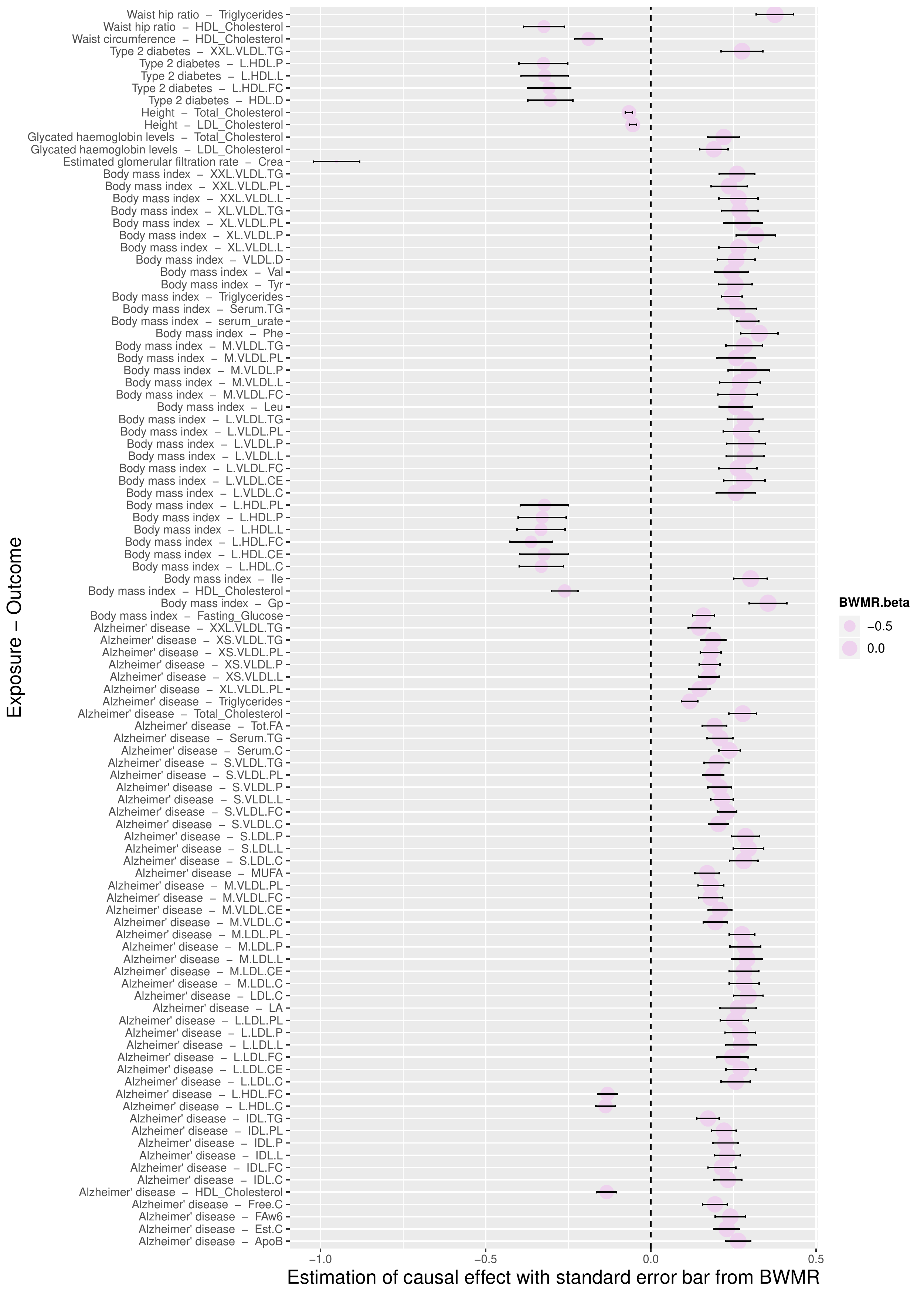}
\par\end{centering}
\caption{Significant causal effects of complex human traits on metabolites estimated by BWMR. We selected SNPs as IVs by the $p$-value threshold $5\times 10^{-8}$ and controlled the type I error rate after Bonferroni correction at level 0.05.}
\end{figure}

\newpage{}
\begin{itemize}
  \item Detailed results from RAPS
\end{itemize}

\begin{figure}[H]
\begin{centering}
\includegraphics[scale=0.65]{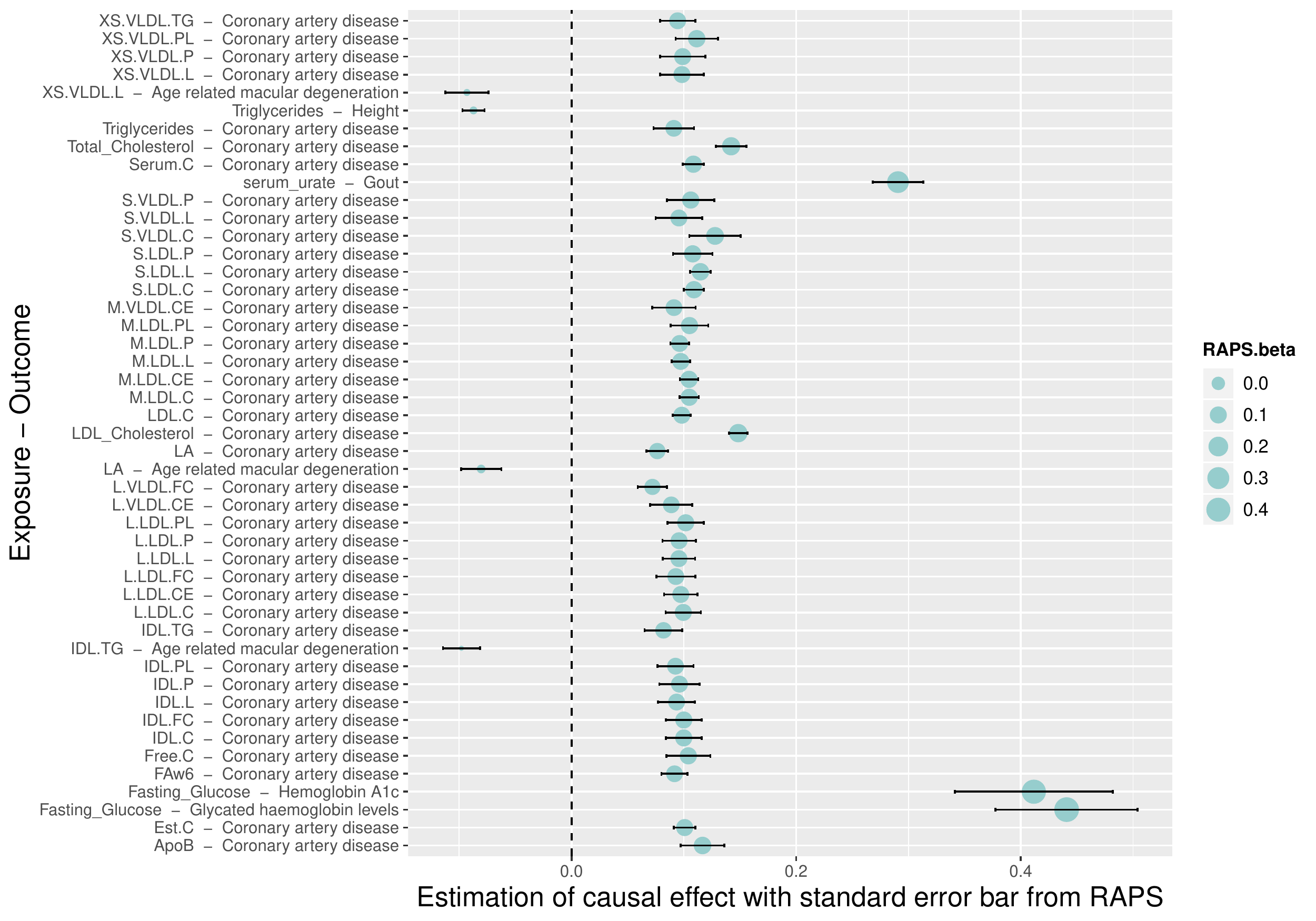}
\par\end{centering}
\caption{Significant causal effects of metabolites on complex human traits estimated by RAPS. We selected SNPs as IVs by the $p$-value threshold $5\times 10^{-8}$ and controlled the type I error rate after Bonferroni correction at level 0.05.}
\end{figure}

\begin{figure}[H]
\begin{centering}
\includegraphics[scale=0.57]{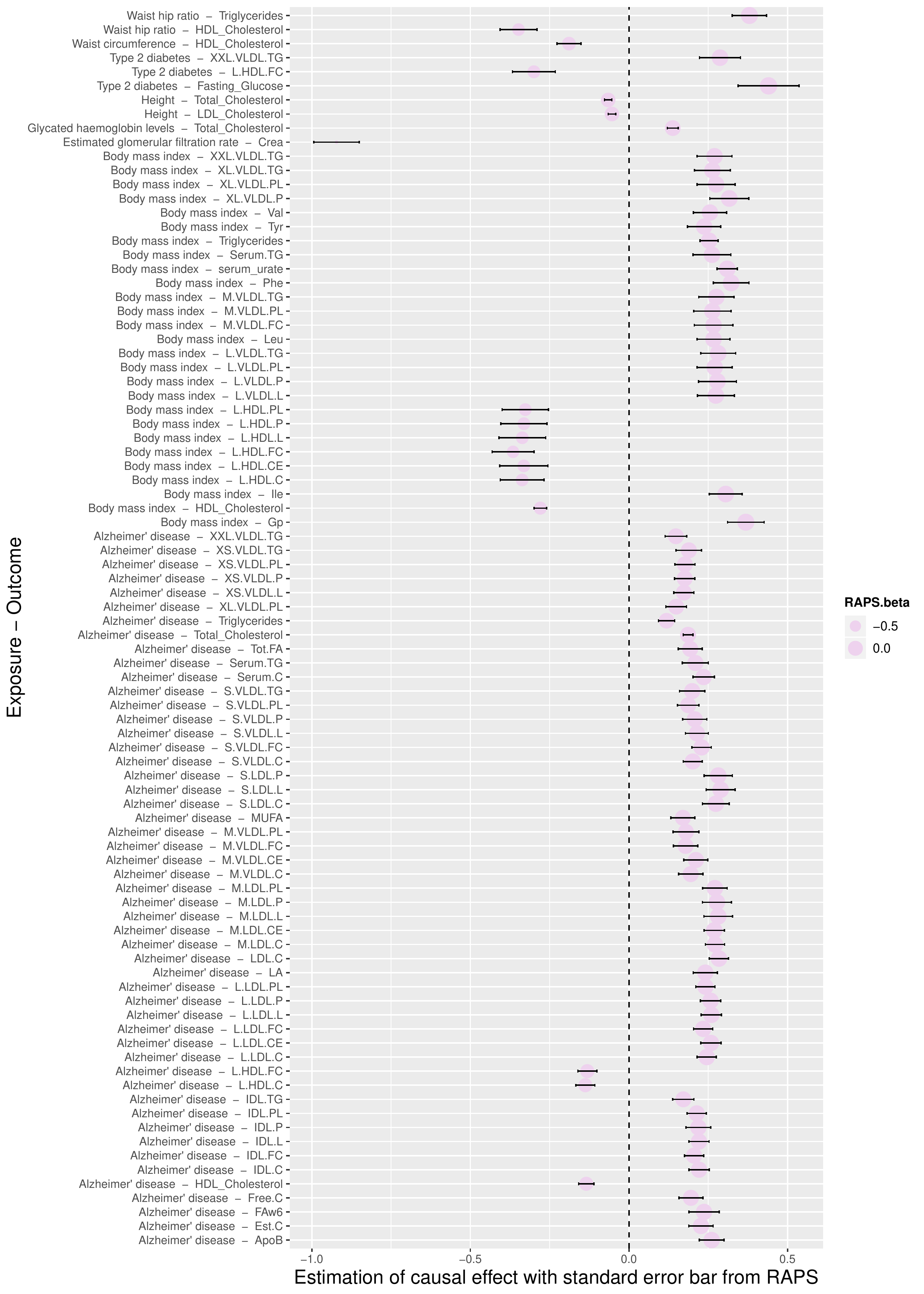}
\par\end{centering}
\caption{Significant causal effects of complex human traits on metabolites estimated by RAPS. We selected SNPs as IVs by the $p$-value threshold $5\times 10^{-8}$ and controlled the type I error rate after Bonferroni correction at level 0.05.}
\end{figure}

\begin{figure}[H]
\begin{centering}
\includegraphics[scale=0.6]{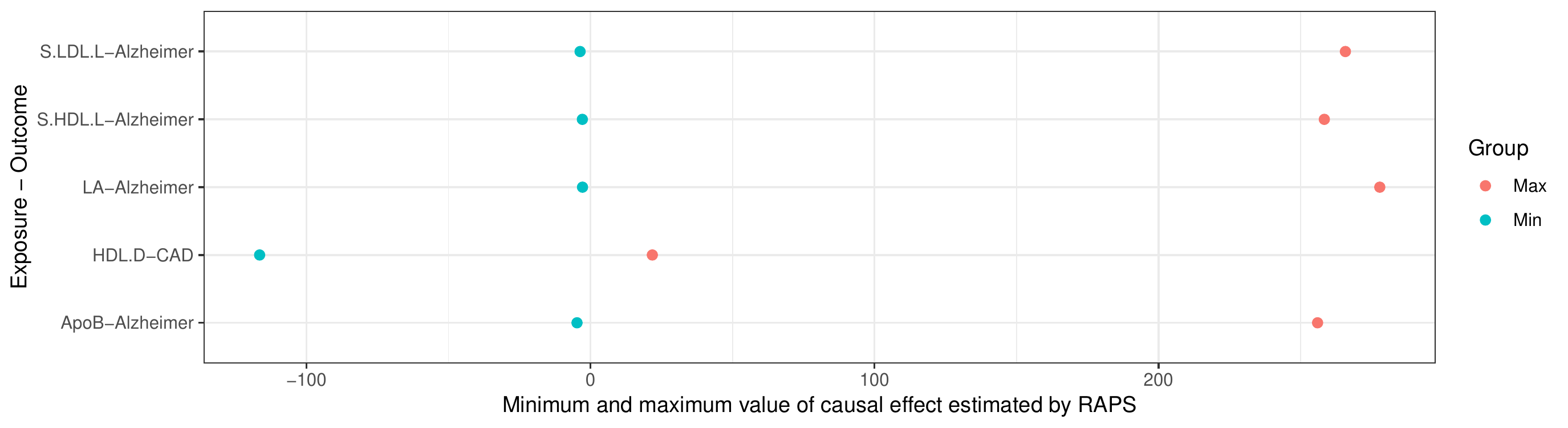}
\par\end{centering}
\caption{Numerical unstability of RAPS. We implemented the method RAPS to estimate the causal effects of metabolites on human traits. To test the stability of the method, we repeated 300 times for each pair of exposure and outcome. The plot shows the cases that esimations from RAPS are not the same (maximum value $-$ minimum value $>$ $50$).}
\end{figure}

\begin{figure}[H]
\begin{centering}
\includegraphics[scale=0.6]{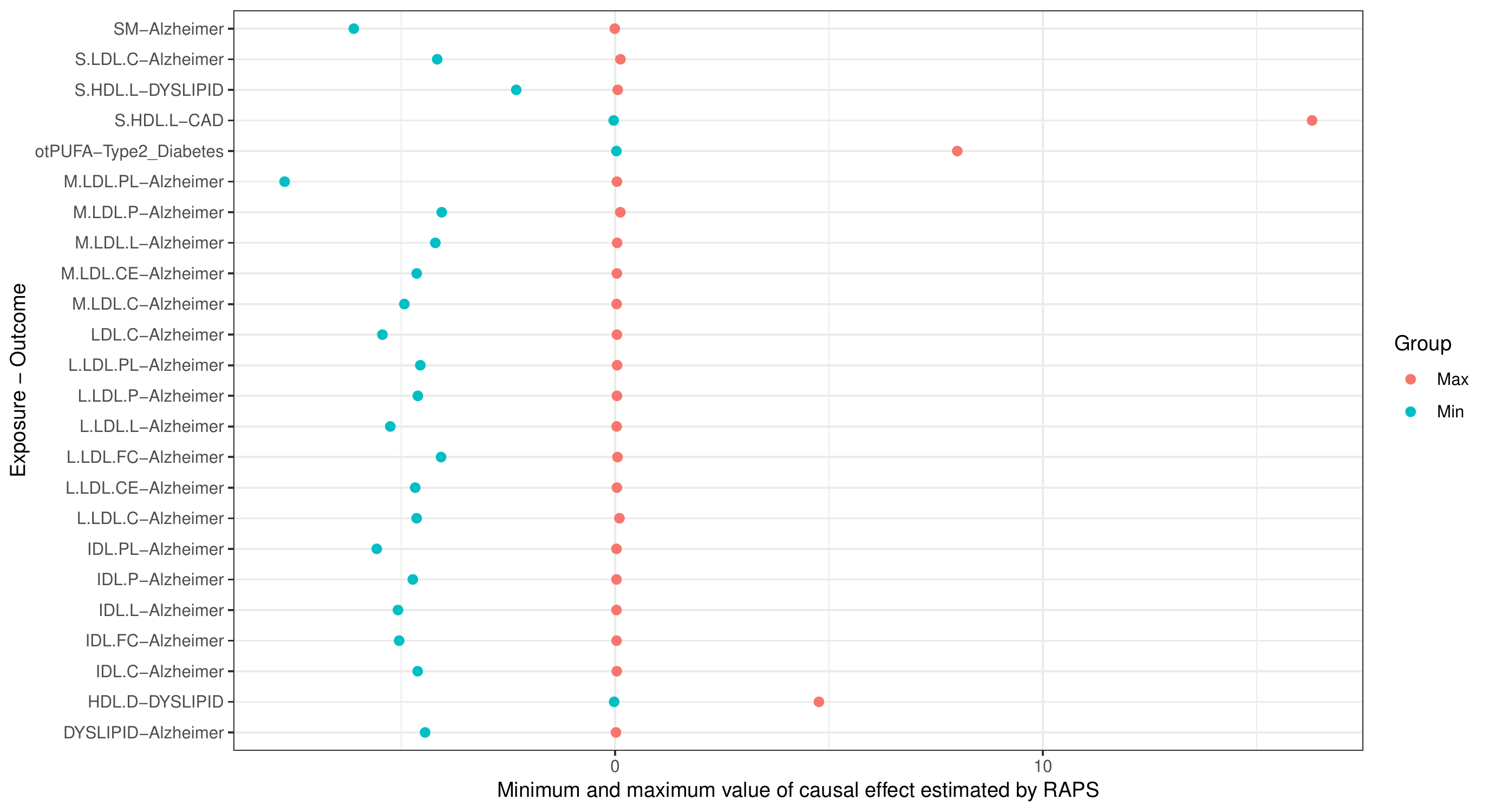}
\par\end{centering}
\caption{Numerical unstability of RAPS. We implemented the method RAPS to estimate the causal effects of metabolites on human traits. To test the stability of the method, we repeated 300 times for each pair of exposure and outcome. The plot shows the cases that esimations from RAPS are not the same ($2$ $<$ maximum value $-$ minimum value $<$ $50$).}
\end{figure}

\newpage{}
\begin{figure}[H]
\begin{centering}
\includegraphics[scale=0.6]{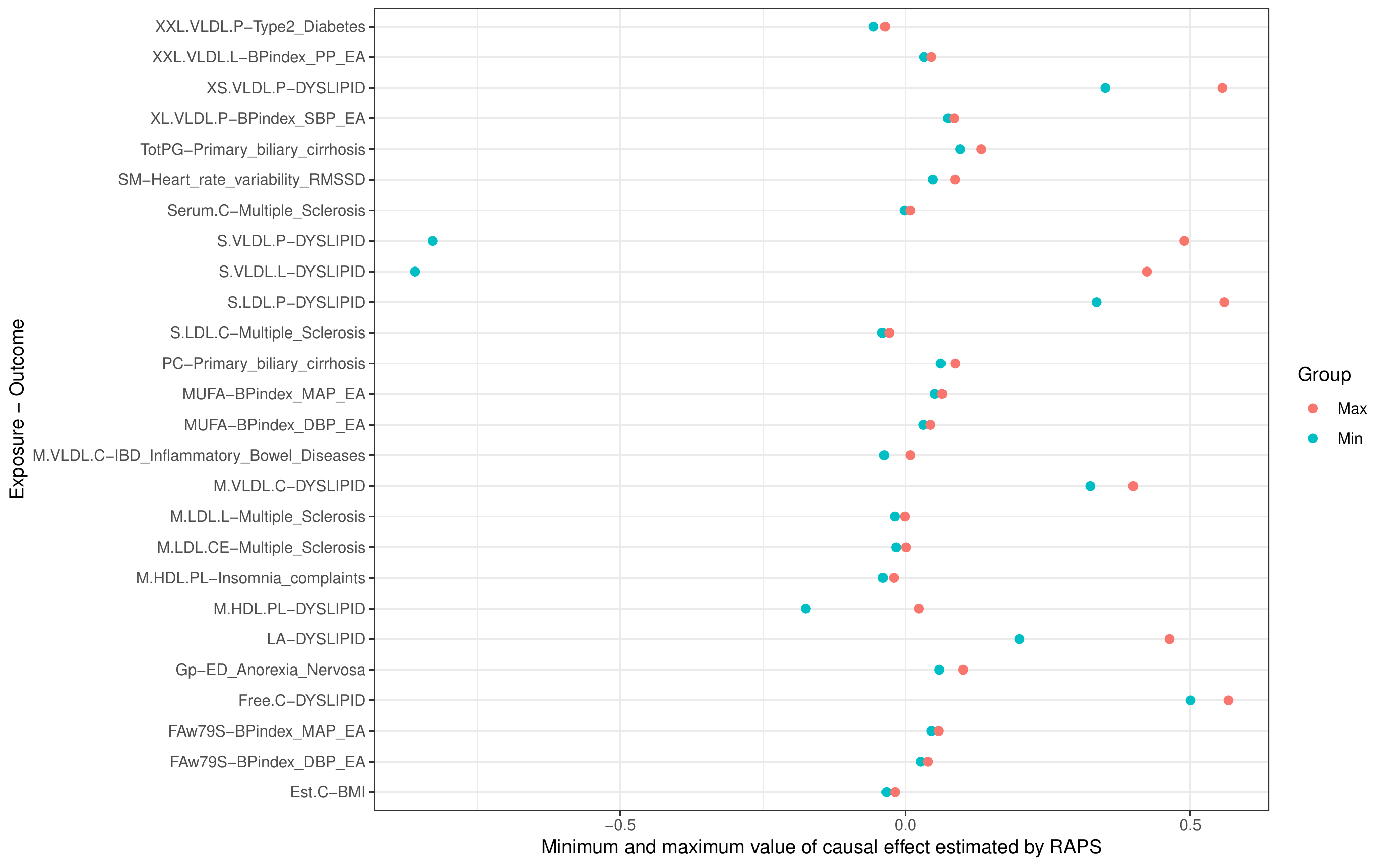}
\par\end{centering}
\caption{Numerical unstability of RAPS. We implemented the method RAPS to estimate the causal effects of metabolites on human traits. To test the stability of the method, we repeated 300 times for each pair of exposure and outcome. The plot shows the cases that esimations from RAPS are not the same ($0.01$ $<$ maximum value $-$ minimum value $<$ $2$).}
\end{figure}

\newpage{}
\begin{itemize}
  \item Detailed results from GSMR
\end{itemize}
\begin{figure}[H]
\begin{centering}
\includegraphics[scale=0.49]{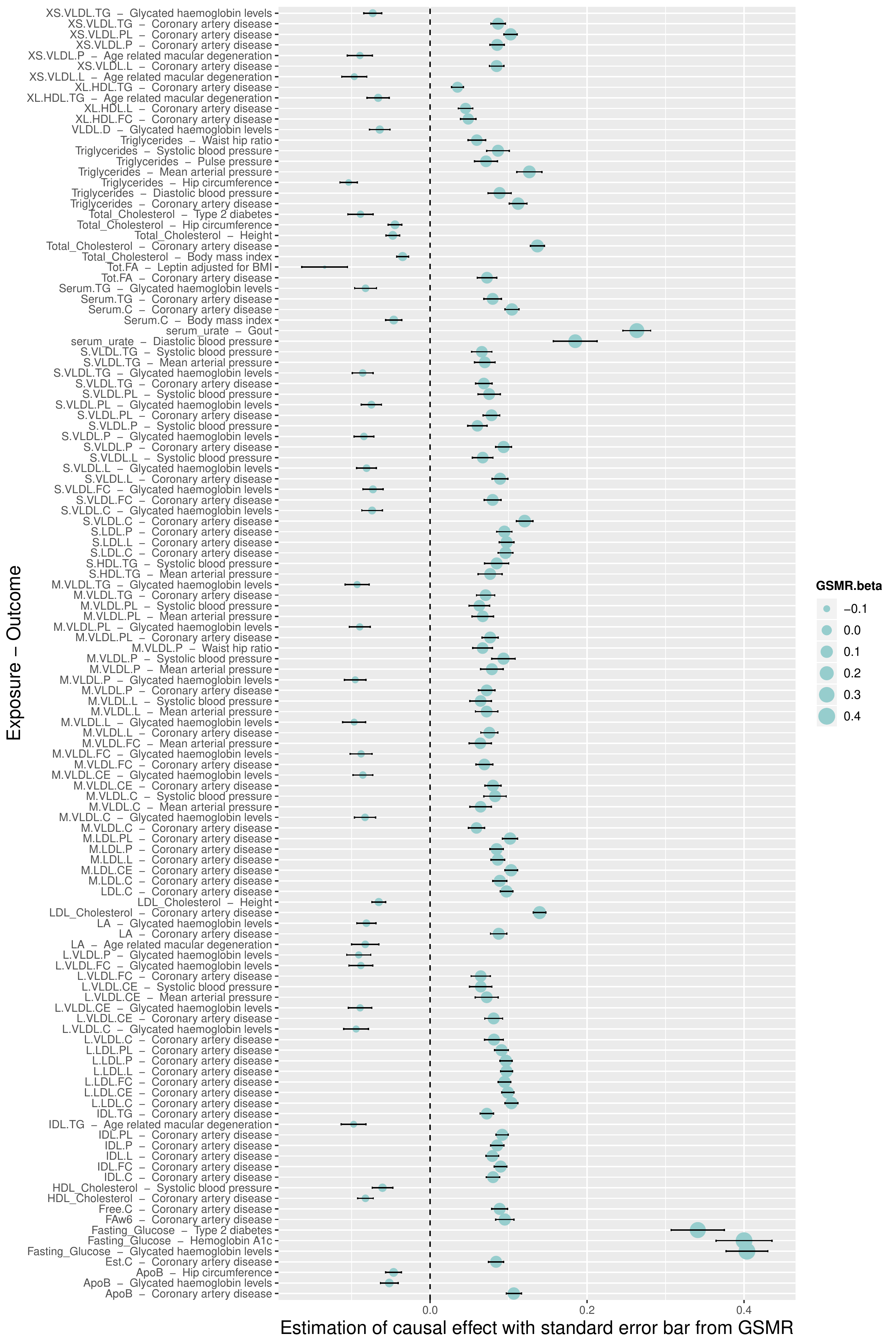}
\par\end{centering}
\caption{Significant causal effects of metabolites on complex human traits estimated by GSMR. We selected SNPs as IVs by the $p$-value threshold $5\times 10^{-8}$ and controlled the type I error rate after Bonferroni correction at level 0.05.}
\end{figure}

\begin{figure}[H]
\begin{centering}
\includegraphics[scale=0.4]{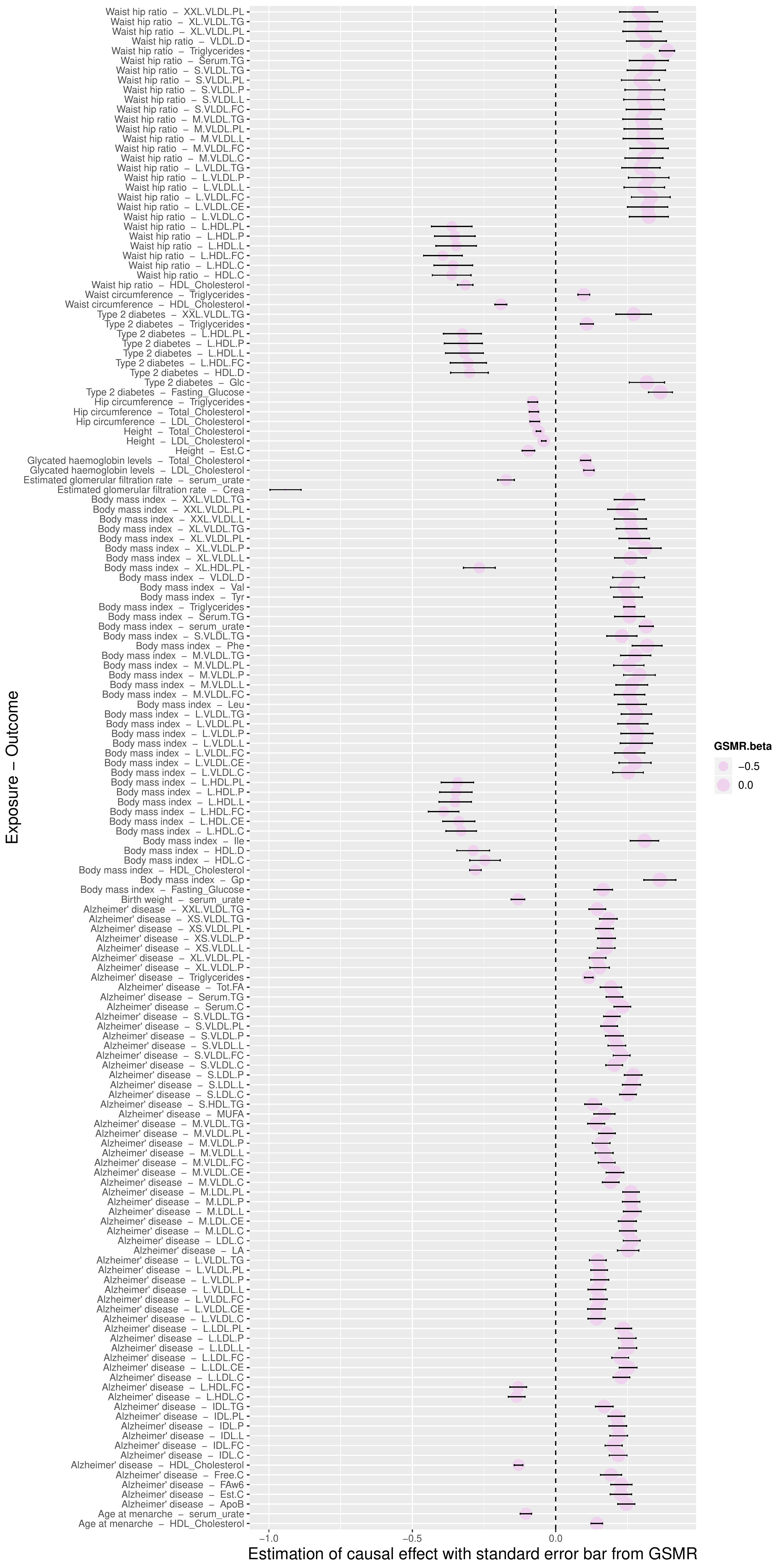}
\par\end{centering}
\caption{Significant causal effects of complex human traits on metabolites estimated by GSMR. We selected SNPs as IVs by the $p$-value threshold $5\times 10^{-8}$ and controlled the type I error rate after Bonferroni correction at level 0.05.}
\end{figure}

\newpage{}
\begin{itemize}
  \item Detailed results from Egger
\end{itemize}
\begin{figure}[H]
\begin{centering}
\includegraphics[scale=0.53]{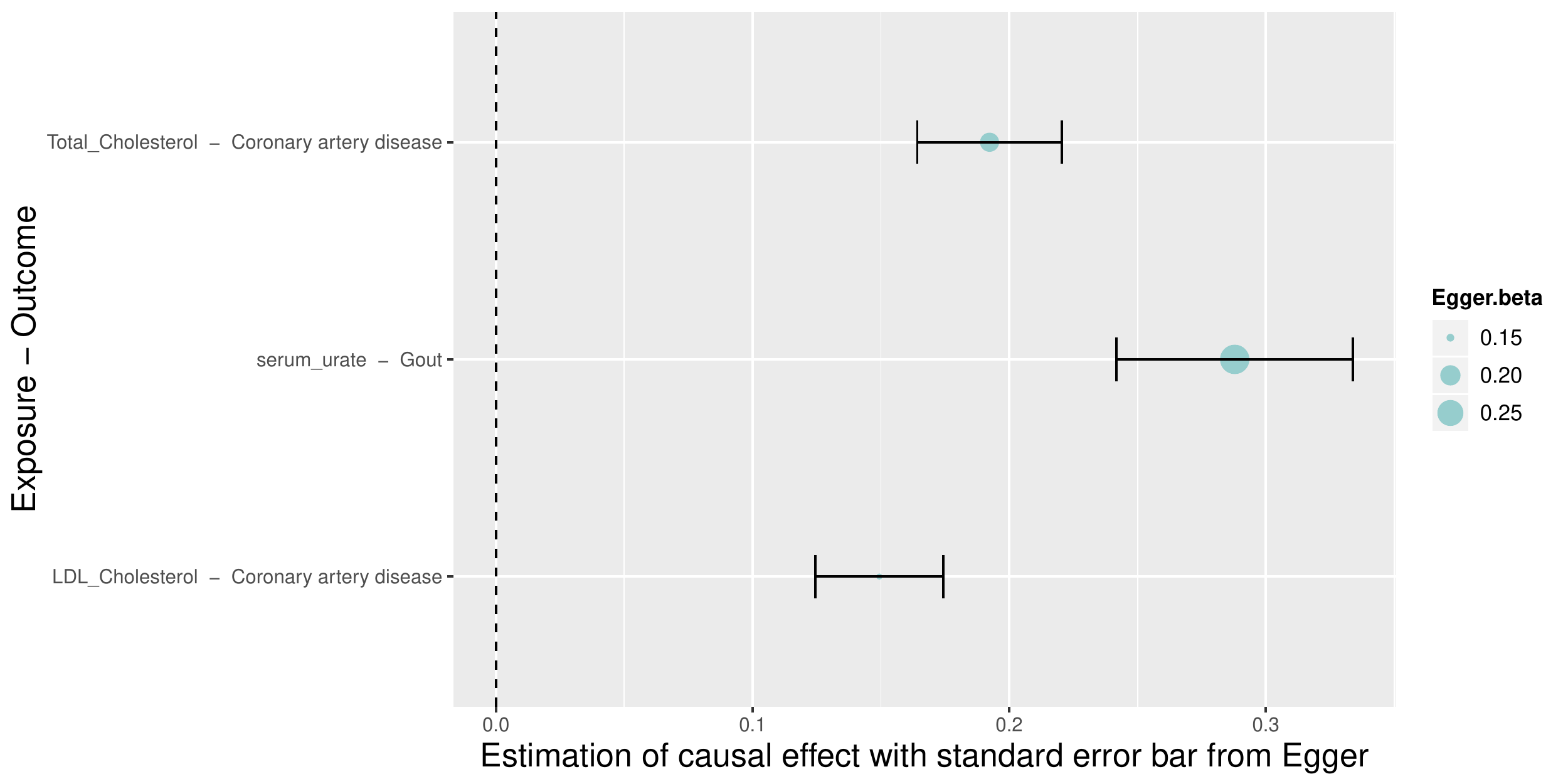}
\par\end{centering}
\caption{Significant causal effects of metabolites on complex human traits estimated by Egger. We selected SNPs as IVs by the $p$-value threshold $5\times 10^{-8}$ and controlled the type I error rate after Bonferroni correction at level 0.05.}
\end{figure}

\begin{figure}[H]
\begin{centering}
\includegraphics[scale=0.52]{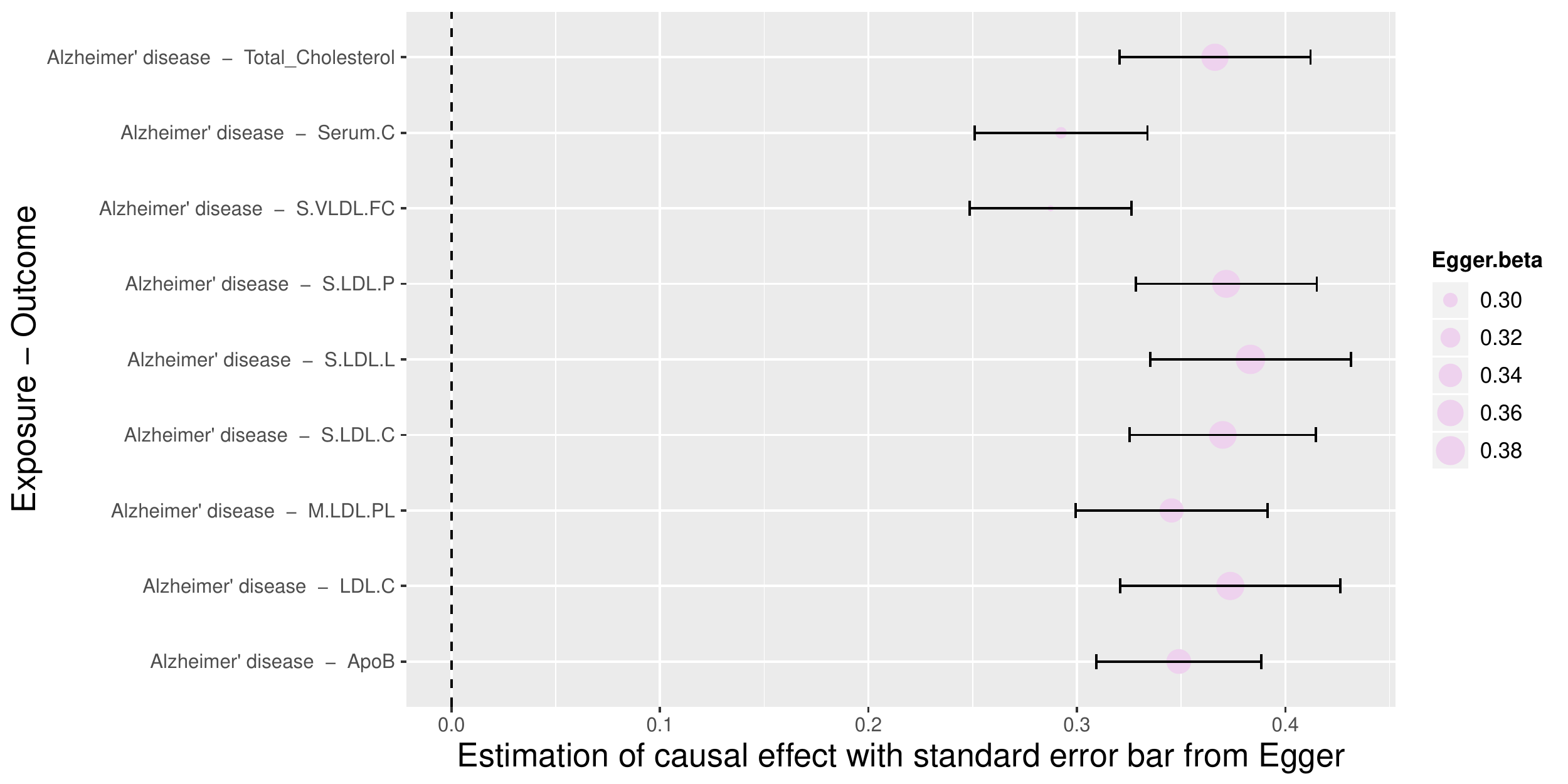}
\par\end{centering}
\caption{Significant causal effects of complex human traits on metabolites estimated by Egger. We selected SNPs as IVs by the $p$-value threshold $5\times 10^{-8}$ and controlled the type I error rate after Bonferroni correction at level 0.05.}
\end{figure}

\bibliographystyle{chicago}
\bibliography{BWMR}

\end{document}